\documentclass{aa}
\usepackage{graphicx}
\usepackage[flushleft]{threeparttable}
\usepackage{amsmath}

\def\la{\lower.5ex\hbox{$\; \buildrel < \over \sim \;$}}
\def\ga{\lower.5ex\hbox{$\; \buildrel > \over \sim \;$}}

\begin{document}      

   \title{Predicting HCN, HCO$^{+}$, multi-transition CO, and dust emission of star-forming galaxies}
   \subtitle{From local spiral and ultraluminous infrared galaxies to high-z star-forming and submillimeter galaxies}

   \author{B.~Vollmer\inst{1}\and P.~Gratier\inst{2}\and J.~Braine\inst{2}\and C.~Bot\inst{1}}

   \offprints{B.~Vollmer, e-mail: Bernd.Vollmer@astro.unistra.fr}

   \institute{CDS, Observatoire astronomique de Strasbourg, Universit\'e de Strasbourg, CNRS, UMR 7550, 
              11 rue de l'Universit\'e, F-67000 Strasbourg, France \and
              Laboratoire d'astrophysique de Bordeaux, Univ. Bordeaux, CNRS, B18N, all\'e Geoffroy Saint-Hilaire, 33615 Pessac, France
              }

   \date{Received / Accepted}

   \authorrunning{Vollmer et al.}

\abstract{
High-z star-forming galaxies have significantly higher gas fractions and star-formation efficiencies per molecular gas mass than local star-forming galaxies. 
In this work, we take a closer look at the gas content or fraction and the associated star-formation rate 
in main sequence and starburst galaxies at $z=0$ and $z \sim 1$--$2$ by applying an analytical model of galactic clumpy gas disks
to samples of local spiral galaxies, ULIRGs, submillimeter (smm), and high-z star-forming galaxies. 
The model simultaneously calculates the total gas mass, H{\sc i}/H$_2$ mass, the gas velocity dispersion, IR luminosity, IR spectral energy distribution, 
CO spectral line energy distribution (SLED), HCN(1--0) 
and HCO$^+$(1--0) emission of a galaxy given its size, integrated star formation rate, stellar mass radial profile, rotation curve, and Toomre $Q$ parameter. 
The model reproduces the observed CO luminosities and SLEDs of all sample galaxies within the model 
uncertainties ($\sim 0.3$~dex). Whereas the CO emission is robust against the variation of model parameters, the
HCN and HCO$^+$ emissions are sensitive to the chemistry of the interstellar medium.
The CO and HCN mass-to-light conversion factors, including CO-dark H$_2,$ are given and compared to the values found in the literature. 
All model conversion factors have uncertainties of a factor of two. 
Both the HCN and HCO$^+$ emissions trace the dense molecular gas to a factor of approximately two for the local spiral galaxies, ULIRGs and smm-galaxies.
Approximately $80$\,\% of the molecular line emission of compact starburst galaxies
originates in non-self-gravitating gas clouds. The effect of HCN infrared pumping is small but measurable ($10$--$20$\,\%). 
The gas velocity dispersion varies significantly with the Toomre $Q$ parameter. The $Q=1.5$ model yields high-velocity dispersions
($v_{\rm disp} \gg 10$~km\,s$^{-1}$) consistent with available observations of high-z star-forming galaxies and ULIRGs. 
However, we note that these high-velocity dispersions are not mandatory for starburst galaxies.
The integrated  Kennicutt-Schmidt law has a slope of approximately 1 for the local spirals, ULIRGs, and smm-galaxies, whereas the
slope is $1.7$ for high-z star-forming galaxies.
The model shows Kennicutt-Schmidt laws with respect to the molecular gas surface density with slopes of approximately 1.5 for local spiral galaxies,
high-z star-forming galaxies.  The relation steepens for compact starburst galaxies.  
The model star-formation rate per unit area is, as observed, proportional to the molecular gas surface density divided by the dynamical timescale.
Our relatively simple analytic model together with the recipes for the molecular line emission appears to capture the essential physics of galactic clumpy gas disks.}

\maketitle

\section{Introduction \label{sec:intro}}

Star formation within galactic disks was proceeding much faster in the first half of the history of the Universe: the cosmic star-formation
rate density declined by a factor of approximately ten since $z=1$ (Madau et al. 1998, Hopkins \& Beacom 2006). The mean star-formation rate with respect to
the total stellar mass also decreases with decreasing redshift: star-forming galaxies form a `main sequence' in the star-formation/stellar-mass 
space (e.g., Brinchmann et al. 2004; Daddi et al. 2007; Elbaz et al. 2007; Noeske et al. 2007a,b; Salim et al. 2007; Whitaker et al. 2012). 
In addition, galactic starbursts (e.g., ultraluminous infrared galaxies (ULIRGs) or smm-galaxies) represent outliers from the main sequence.
The slope and offset of the $\dot{M}_*$--$M_*$ powerlaw main sequence relation change with redshift (Speagle et al. 2014). The most prominent change is an increasing offset
with increasing redshift, that is, the specific star-formation rate ($\dot{M}_*/M_*$) increases significantly with increasing $z$ (by a factor of $\sim 6$
for galaxies with masses of $M_* \sim 3 \times 10^{10}$~M$_{\odot}$; Pannella et al. 2015).
The slope and scatter of this correlation, the evolution of its normalization with cosmic time, contain crucial and still poorly known information on galaxy evolution
(e.g., Karim et al. 2011; Rodighiero et al. 2011; Wuyts et al. 2011; Sargent et al. 2012).

Two factors can be invoked to explain the higher star-formation efficiency: (i) higher gas fractions (see, e.g., Combes et al. 2013) and (ii)
dynamical trigger of interactions, whose frequency increases with redshift (e.g., Conselice et al. 2009, Kartaltepe et al. 2012).

In this work, we take a closer look at the gas content or fraction and the associated star-formation rate in main sequence and starburst galaxies
at $z=0$ and $z \sim 1$--$2$. We look preferentially at local starburst galaxies, ULIRGs, and high-redshift starbursts smm-galaxies.

Genzel et al. (2010) and Tacconi et al. (2012) showed that star-forming galaxies at $z=1$--$2$ have higher gas fractions ($\sim 33$\,\%) and
higher star-formation efficiencies with respect to the molecular gas ($SFE=SFR/M_{\rm H_2} \sim 1/0.7$~Gyr$^{-1}$) than local spiral galaxies
($\sim 10$\,\% and $SFE \sim 1/2.0$~Gyr$^{-1}$; e.g., Bigiel et al. 2008, Leroy et al. 2008).
Local ULIRGs and high-redshift smm-galaxies have the highest star-formation efficiencies (e.g., Pope et al. 2013).

The uncertainties of the determined star-formation rates are typically $\sim 50$\,\% (e.g., Leroy et al. 2008). 
Since molecular hydrogen at temperatures below $100$~K is not directly detectable, one has to rely on CO, HCN, or HCO$^+$ observations
to derive H$_2$ gas masses. Unfortunately, the associated conversion factors have high uncertainties (approx. a factor of two, e.g., Bolatto et al. 2013).
Our understanding of the gas content and thus the star-formation efficiency is limited by these uncertainties.

A complementary way to determine galactic gas masses is the direct modeling of molecular emission.
Narayanan \& Krumholz (2014) combined numerical simulations of disc galaxies and galaxy mergers with molecular line radiative transfer 
calculations to develop a model for the physical parameters that drive variations in CO spectral line energy distributions (SLEDs) in galaxies in terms of the star-formation-rate density.  Their model was able to reproduce the SLEDs of galaxies over a dynamic range of approximately 200 in star-formation-rate surface density.
However, the CO high-$J$ transitions ($J > 8$) of ULIRGs are difficult to reproduce within the model (Fig.~2 of Kamenetzky et al. 2016).

Bournaud et al. (2015) modeled the intensity of CO emission lines, based on hydrodynamic simulations of spirals, mergers, and high-redshift galaxies with very high resolutions 
($3$~pc and $10^3$~M$_{\odot}$) and detailed models for the phase-space structure of the interstellar gas, including shock heating, stellar feedback processes, and galactic winds. 
The simulations were analyzed with a large velocity gradient (LVG) model to compute the local emission in various molecular lines in each resolution element, 
radiation transfer, opacity effect, and the intensity emerging from galaxies to generate synthetic spectra for various CO transitions. 
This model reproduced the known properties of CO spectra and CO-to-H$_2$ conversion factors in nearby spirals and starbursting major mergers. 

Alternatively, galactic gas disks can be modeled analytically, assuming axis-symmetry. Krumholz and Thompson (2007) provided a simple model for understanding how Kennicutt-
Schmidt laws, which relate the star-formation rate to the mass or surface density of gas as inferred from some particular line, depend on the line chosen to define 
the correlation. They assume a probability distribution for the mass fraction of gas at a given density and calculate the molecular emission
with an escape probability formalism. The model gas clouds have constant temperature, Mach number, and optical depth.
Their results showed that for a turbulent medium, the luminosity per unit volume in a given line, provided that this line can be excited at temperatures lower than 
the mean temperature in a galaxy's molecular clouds, increased faster than linearly with the density for molecules with critical
densities larger than the median gas density. The star-formation rate also rose superlinearly with the gas density, and the combination 
of these two effects produced a close to linear correlation between star-formation rate and line luminosity.
        
Kazandjian et al. (2015) investigated the effect of mechanical heating on atomic fine-structure and molecular lines and on their ratios. They tried to use 
those ratios as a diagnostic to constrain the amount of mechanical heating in an object and also study its significance on estimating the H$_2$ mass. 
Equilibrium photodissociation models (PDRs) were used to compute the thermal and chemical balance for the clouds. The equilibria were solved for numerically 
using the optimized version of the Leiden PDR-XDR code. Large velocity-gradient calculations were done as post-processing on the output of the PDR models using 
RADEX (van der Tak et al. 2007). 
They showed that high-$J$ CO line ratios and ratios involving HCN are very sensitive to mechanical heating.

In this work, we investigate a different analytical approach, where we extend the model of galactic clumpy gas disks presented in Vollmer \& Leroy (2011).
The model has a large-scale and a small-scale part. The large-scale part gives the surface density, turbulent velocity, disk height, and gas viscosity.
The small-scale part begins at densities where gas clouds become self-gravitating. The non-self-gravitating and self-gravitating clouds obey different
scaling relations, which are set by observations. For clouds of a certain density, the area filling factor is calculated. 
The gas and dust temperatures are calculated
by the heating and cooling equilibrium. Dense clouds are heated by turbulent, mechanical, and cosmic-ray heating. For all model clouds, the size, 
density, temperature, and velocity dispersion are known. The molecular abundances of individual gas clouds are determined by a detailed chemical network
involving the cloud lifetime, density, and temperature.
Molecular line emission is calculated with an escape probability formalism. 
The model is applied to samples of local spiral galaxies, ULIRGs, high-z star-forming galaxies, and smm-galaxies.
The model simultaneously calculates the total gas mass, H$_2$ mass, the gas velocity dispersion, H{\sc i} mass, IR luminosity, IR SED, CO SLED, HCN(1--0), 
and HCO$^+$(1--0) emission of a galaxy given its size, integrated star-formation rate, stellar mass radial profile, rotation curve, and Toomre $Q$ parameter. 
In addition, the temperature, density, velocity dispersion, and molecular abundance of a gas cloud at a given density can be retrieved.

This article presents a sophisticated model and justifies the physics behind it, shows the results we obtain for different choices of input parameters, 
and compares with observations. The physical processes included in the model are described in detail in Sect.~\ref{sec:model}.  
The steps involved in the calculations and in determining the uncertainties are described respectively in Sects.~\ref{sec:method} and \ref{sec:uncertain}.  
Not all spirals are the same so rather than define a `typical' member of each of the four classes (local spirals, local ULIRGs, submillimeter (smm) galaxies,
and high-z star-forming galaxies), we use true samples of real objects for which we think we can estimate the appropriate values of the input parameters.  
The samples and their origin are described in Sect.~\ref{sec:samples} and the results of the calculations for the samples are shown in detail in Sect.~\ref{sec:results}.  
Hence, the reader interested in how well the model reproduces the observations can go directly to Sect.~\ref{sec:samples}  (short) or even Sect.~\ref{sec:results}.  
Sect.~\ref{sec:variation} evaluates the influence of the choice of the chemical network, the Toomre $Q$ parameter, and the length scale parameter $\delta$ in terms 
of the effect on line and continuum emission.  The importance of each of the heating and cooling processes is described in Sect.~\ref{sec:discussion} and that of 
the assumed gas properties (mass, velocity dispersion, and free-fall time) in Sect.~\ref{sec:physpar}. Finally, we give our conclusion in Sect.~\ref{sec:conclusions}.

\section{The analytical model \label{sec:model}}

\begin{table*}
\begin{center}
\caption{Model Parameters.\label{tab:parameters}}
\begin{tabular}{lll}
\hline\hline
Parameter & Unit & Explanation \\
\hline
$G=5 \times 10^{-15}$ & pc$^{3}$yr$^{-1}$M$_{\odot} ^{-1}$ & gravitation constant \\
$\kappa$ & yr$^{-1}$ & \it epicyclic frequency \\
$\bf Q$ & & \bf Toomre parameter \\
$R$ & pc &galactocentric radius \\
$H$ & pc & thickness of the gas disk\\
$H_{*}$ & pc & thickness of the stellar disk \\
$l_{\rm cl}$ & pc & cloud size \\
$v_{\rm rot}$ & pc\,yr$^{-1}$ & \it rotation velocity \\
$\Omega=v_{\rm rot}/R$ & yr$^{-1}$ & \it angular velocity \\
$\Phi_{\rm V}$ & & volume-filling factor \\
$\Phi_{\rm A}=\Phi_{\rm V}\,H/l_{\rm cl}$ & & area-filling factor \\
$\rho$ & M$_{\odot}$pc$^{-3}$ & disk midplane gas density\\
$\rho_{\rm CNM}$ & M$_{\odot}$pc$^{-3}$ & cool neutral medium density \\
$\rho_{\rm cl}=\rho/\Phi_{\rm V}$ & M$_{\odot}$pc$^{-3}$ & cloud density \\
$\dot{\rho}_{*}$ & M$_{\odot}$pc$^{-3}$yr$^{-1}$ & star-formation rate \\
$\Sigma$ & M$_{\odot}$pc$^{-2}$ & gas surface density \\
$\Sigma_{*}$ & M$_{\odot}$pc$^{-2}$ & \it stellar surface density \\
$\dot{\Sigma}_{*}$ & M$_{\odot}$pc$^{-2}$yr$^{-1}$ & \it star-formation rate \\
$\xi=4.6 \times 10^{-8}$ & pc$^2$yr$^{-2}$ & constant relating SN energy input to SF \\
$\bf \dot{M}$ & M$_{\odot}$yr$^{-1}$ & \bf disk mass accretion rate (radial, within the disk) \\
$v_{\rm turb}$ & pc\,yr$^{-1}$ & gas turbulent velocity dispersion \\
$v_{\rm rad}$ & pc\,yr$^{-1}$ & gas radial velocity \\
$v_{\rm disp}^{*}$ & pc\,yr$^{-1}$ & \it stellar vertical velocity dispersion \\
$c_{\rm s}$ & pc\,yr$^{-1}$ & sound speed \\
$\cal{M}$ & & Mach number \\
$\nu$ & pc$^{2}$yr$^{-1}$ & viscosity \\
$f_{\rm mol}=\Sigma_{\rm H_{2}}/(\Sigma_{\rm HI}+\Sigma_{\rm H_{2}})$ &  & molecular fraction \\
$\alpha$ & yr\,M$_{\odot}$pc$^{-3}$ & constant of molecule formation timescale \\
$l_{\rm driv}$ & pc & turbulent driving length scale \\
$\bf \delta=5$ & & {\bf scaling between the driving length scale and the size of the} \\
& & {\bf  largest self-gravitating structures} \\
$SFE=\dot{\Sigma}_{*}/\Sigma$ & yr$^{-1}$ & star-formation efficiency \\
$t_{\rm ff}^{l}$ & yr & cloud free fall timescale at size $l$ \\
$t_{\rm turb}^{l}$ & yr & cloud turbulent timescale at size $l$ (turbulent crossing time) \\
$t_{\rm mol}^{l}$ & yr & cloud molecule formation timescale at size $l$ \\
$T_{\rm g}$ & K & gas temperature \\
$T_{\rm d}$ & K & dust temperature \\
\hline
\end{tabular}
\begin{tablenotes}
      \item {\bf boldface}: free parameters; {\it italic}: parameters determined from observations. 
    \end{tablenotes}
\end{center}
\end{table*}

Compared to the model described in Sect.~2 of Vollmer \& Leroy (2011) that is based on Vollmer \& Beckert (2003; VB03), the present model
does not include a break radius, where the star-formation timescale changes from
the free fall timescale to the molecular formation timescale. In addition, we included in this more advanced model
(i) the determination of the dense gas fraction, (ii) ISM scaling relations, (iii) the determination of dust and gas temperatures,
(iv) a chemical network for the determination of molecular abundances, (v) a formalism for the
photodissociation of molecules by the interstellar radiation field, and (vi) the determination of dust and molecular line emission.

The model considers the warm, cool neutral, and molecular phases of the ISM as a
single, turbulent gas.  We assume this gas to be in vertical hydrostatic
equilibrium with the midplane pressure balancing the weight of the gas and stellar disk.
The gas is assumed to be clumpy, so that the local density is enhanced relative
to the average density of the disk. Using this local density, we calculate two
timescales relevant to star formation: the free-fall timescale of an individual
clump and the characteristic timescale for H$_2$ to form on grains. The free-fall timescale 
is taken as the governing timescale for star formation. The star-formation rate is used to calculate the
rate of energy injection by supernovae. This rate is related to the turbulent
velocity dispersion and the driving scale of turbulence. These quantities,
in turn, provide estimates of the clumpiness of gas in the disk (i.e., the
contrast between local and average density) and the rate at which viscosity
moves matter inward.

The model relies on several empirical calibrations: e.g., the relationship
between star-formation rate and energy injected into the ISM by supernovae,
the H$_2$ formation timescale (and its dependence on metallicity), and the
turbulent dimension of the ISM (used to relate the driving length scale to the
characteristic cloud size modulo a free parameter $\delta,$ which is constrained by observations). As far as possible, these
are drawn from observations of the Milky Way.

The model only contains two free parameters. First, there is an unknown scaling
factor relating the driving length of turbulence to the size of
gravitationally bound clumps, which we call $\delta$. Second, the mass accretion
rate, $\dot{M}$, which is related to the driving length and turbulent velocity,
is a free parameter. From a detailed comparison of local spiral galaxies from the
THINGS survey, Vollmer \& Leroy (2011) found $\delta=5 \pm 3$. For simplicity, 
we assume $\delta=5$ in this work.
Moreover, the Toomre $Q$ parameter of the gas is set to the observed values for the 
local spirals ($2$--$8$, e.g., Leroy et al. 2008), and to $Q=1.5$ for the ULIRGs, high-z star-forming and submillimeter galaxies.

In the remainder of this section, we discuss our assumptions in slightly more
detail, justify them via comparison to observation and theory, and note the
physics that we neglect.

\subsection{The interstellar medium \label{sec:ISM}}

Following, for example, Mac Low \& Klessen (2004), the warm, cool neutral, and molecular phases of the
ISM are viewed as a single entity. Locally, the exact phase of the gas depends on the
local pressure, metallicity, stellar radiation field, stellar winds, and
shocks. Here, we view these factors as secondary, making a few simplifying
assumptions.
The equilibrium between the different phases of the ISM 
and the equilibrium between turbulence and star formation depends 
on three local timescales:
the turbulent crossing time $t_{\rm turb}^{l}$, the molecule formation timescale $t_{\rm mol}^{l}$, 
and the local free-fall timescale $t_{\rm ff}^{l}$ of a cloud.
In addition, photodissociation of molecules is taken into account.

\subsubsection{The fraction of dense gas \label{sec:gasfrac}}

To calculate the mass fraction between two gas densities, we
use the density probability distribution function of Padoan et al. (1997) for overdensities $x$:
\begin{equation}
p(x){\rm d}x=\frac{1}{x \sqrt{2\pi \sigma^2}}{\rm exp}\big(-\frac{({\rm ln}\,x+\sigma^2/2)^2}{2 \sigma^2}\big) {\rm d}x
\end{equation}
where the standard deviation, $\sigma$, is given by
\begin{equation}
\sigma^2 \simeq {\rm ln}\big(1+({\cal{M}}/2)^2\big)
\end{equation}
and ${\cal{M}}=v_{\rm turb}/c_{\rm s}$ is the Mach number with the sound speed $c_{\rm s}$.
The mass fraction of gas with overdensities exceeding $x$ is then
\begin{equation}
\frac{\Delta M}{M} = \frac{1}{2} \big(1+{\rm erf}(\frac{\sigma^2-2{\rm ln}(x)}{2^{\frac{3}{2}}\sigma})\big) \ .
\end{equation}
The overdensity for a given density $\rho_1$ is calculated with respect to the midplane gas density 
$x=\rho_1/\rho$.

For the calculation of the molecular line emission we divide the ISM into two density bins:
(i) densities $\rho_1$ equal to or higher than that of the self-gravitating clouds (see Sect.~\ref{sec:clumpiness}): 
$\rho_1 \geq \rho/\Phi_{\rm V}$ and
(ii) densities $\rho_2$ equal to or higher than that of the cool neutral medium: $\rho_{\rm CNM} \leq \rho_2 \leq \rho/\Phi_{\rm V}$,
where $\Phi_{\rm V}$ is the volume-filling factor of the largest self-gravitating structures in the disk, computed via
a procedure described in Sect.~\ref{sec:clumpiness}.
Following Wolfire et al. (2003), we set the minimum density of the cool neutral medium to
\begin{equation}
n_{\rm CNM}=\frac{31 \dot{\Sigma_*}/(10^{-8}~{\rm M}_{\odot}{\rm pc}^{-2}{\rm yr}^{-1})}{\big(1+3.1(2.2 \times 10^7~{\rm yr\,M_{\odot}pc^{-3}}/\alpha)^{0.365}\big)}\ {\rm cm^{-3}}.
 \label{eq:cnmdens}
\end{equation}
With respect to Eq.~35 of Wolfire et al. (2003), we set the normalized FUV radiation field 
$G_0'=\dot{\Sigma_*}/(10^{-8}~{\rm M}_{\odot}{\rm pc}^{-2}{\rm yr}^{-1})$ and the normalized dust abundances and gas metallicities
$Z'_{\rm d}=Z'_{\rm g}= Z/Z_{\odot}=2.2 \times 10^7~{\rm yr\,M_{\odot}pc^{-3}}/\alpha$ (Eq.~\ref{eq:zzodot}), 
where $\alpha$ is the constant of the molecule-formation timescale (Eq.~\ref{eq:molform}).
Moreover, we set the normalized total ionization rate by cosmic rays and EUV/X-rays $\zeta_{\rm t}'=1$.
If the CNM density exceeds the midplane density, we set $\rho_{\rm CNM}=\rho$.

The mass fraction of the self-gravitating clouds with respect to the diffuse clouds is a major unknown.
Using the lognormal pdf of Padoan et al. (1997) neglects self-gravitation, which can change the shape of the pdf
significantly (e.g., Schneider et al. 2015). Based on the findings of the latter authors, we adopt the following recipe:
\begin{itemize}
\item
density bin (ii): 
\begin{equation}
\frac{\Delta M}{M}(R) = y\,\frac{\Delta M}{M}(x_{\rm sg}(R))\ ,
\end{equation}
\item
density bin (i): 
\begin{equation}
\frac{\Delta M}{M} = \frac{\Delta M}{M}(x_{\rm CNM}(R)) - y\,\frac{\Delta M}{M}(x_{\rm sg}(R))\ ,
\end{equation} 
where $x_{\rm CNM}$ and $x_{\rm sg}$ are the overdensities of the cool neutral medium and self-gravitating clouds.
\end{itemize}
The normalization factor is
\begin{equation}
y=0.3\, R_0/\big(\int_0^{R_0} \frac{\Delta M}{M}(x_{\rm CNM}(R)) {\rm d}R\big)\ .
\end{equation} 
Within the density bin, the mass fraction of clouds of overdensity between $x_1$ and $x_2$ is calculated as the difference between the
mass fractions. 
For the determination of the Mach number, we adopt the temperature of the cool neutral medium ($\sim 100$~K) to calculate the sound speed.
For the self-gravitating, clouds we adopt the temperature of the molecular cloud ($10$--$30$~K).
This prescription conserves mass, that is, $\sum_{i=1}^N  \big( \frac{\Delta M}{M} \big)_i = 1$.

\subsubsection{ISM scaling relations \label{sec:scaling}}

We assume different scaling relations for the two density regimes: (i) for non-selfgravitating clouds, we adopt the 
scaling relations found for galactic H{\sc i} by Quiroga (1983): $\rho_{\rm cl} \propto l^{-2}$, $v_{\rm turb,cl} \propto l^{1/3}$, and 
thus $v_{\rm turb,cl} = v_{\rm turb}(\rho_{\rm cl}/\rho)^{-1/6}$, where $v_{\rm turb}$ and $\rho$ are the turbulent velocity and the density 
of the disk, respectively. Since the minimum density considered in this work is $100$~cm$^{-2}$, the maximum turbulent velocity of diffuse clouds 
is $\sim v_{\rm turb}/2 \sim 5$~km\,s$^{-1}$.
(ii) For self-gravitating clouds, we adopt the scaling relations of Lombardi et al. (2010): $\rho_{\rm cl} \propto l^{-1.4}$,
$v_{\rm turb,cl} \propto l^{1/2}$.
As described in Sect.~\ref{sec:clumpiness}, the scale of the largest self-gravitating clouds $l_{\rm cl}$ is smaller
than the turbulent driving length scale $l_{\rm driv}$ by a factor $\delta=l_{\rm driv}/l_{\rm cl}$.
We assume that the turbulent velocity dispersion of the largest self-gravitating clouds of density $\rho_{\rm sg}$
is $v_{\rm turb,cl} = v_{\rm turb}/\sqrt{\delta}$, where $v_{\rm turb}$ is the velocity dispersion of the disk.
For the assumed value of $\delta=5$ (see Sect.~\ref{sec:clumpiness}), this yields a velocity dispersion of the largest self-gravitating clouds 
of $v_{\rm turb,cl} \sim 5$~km\,s$^{-1}$, which is consistent with observations (e.g., Solomon et al. 1987).
We thus obtain  
$v_{\rm turb,cl} = v_{\rm turb}/\sqrt{\delta}(\rho_{\rm cl}/\rho_{\rm sg})^{-1/3}$ for clouds with $\rho_{\rm cl} > \rho_{\rm sg}$.
Alternatively, we assume $\rho_{\rm cl} \propto l^{-1}$ and 
$v_{\rm turb,cl} = v_{\rm turb}/\sqrt{\delta}(\rho_{\rm cl}/\rho_{\rm sg})^{-1/2} \propto l^{\frac{1}{2}}$ (Solomon et al. 1987).

It turned out that the different scaling relations for self-gravitating clouds result in consistent molecular line luminosities within
$\sim 10$\,\%, except for the HCN emission of local spirals where the difference is $\sim 20$\,\%. 
The models including the Solomon et al. (1987) scaling always reproduce observations slightly better. 
Therefore, the following results are based on the Solomon et al. (1987) scaling.

\subsubsection{Gas and dust temperatures \label{sec:gdtemp}}

Neufeld \& Kaufman (1993) and Neufeld et al. (1995) considered the radiative cooling of fully shielded 
molecular astrophysical gas over a wide range of temperatures ($10~{\rm K} \leq T_{\rm g} \leq 2500$~K) and 
H$_2$ densities ($10^3$~cm$^{-3} \leq n({\rm H}_2) \leq 10^{10}$ cm$^{-3}$). 
Their model for the radiative cooling of molecular gas includes a detailed treatment of the interstellar 
chemistry that determines the abundances of important coolant molecules, and a detailed treatment of the 
excitation of the species H$_2$, CO, H$_2$O, HCl, O$_2$, C, O, and their isotopic variants where important.

For simplicity, we only take the main cooling agents, CO, H$_2$, and H$_2$O, into account.
We assume CO and H$_2$O abundances of $x_{\rm CO}=10^{-4} (Z/Z_{\odot})$ and 
$x_{\rm H_2O}=10^{-6} (Z/Z_{\odot})$.
According to Fig.~2 of Neufeld et al. (1995), we may underestimate, in this way, the cooling rates
by approximately a factor of 2. However, for densities $n({\rm H}_2) > 10^{5}$~cm$^{-3}$ and low temperatures 
($T \sim 20$~K), the discrepancy increases up to a factor $3$--$4$.

Neufeld \& Kaufman (1993) and Neufeld et al. (1995) defined the molecular cooling rate as
$\Lambda_{\rm g}=L n(H_2)n(M)$, where $n({\rm H}_2)$ and $n({\rm M})$ are the H$_2$ and coolant particle densities.
The rate coefficient $L$ depends on $n({\rm H}_2$), the gas temperature $T$, and
\begin{equation}
\tilde{N}({\rm M})=\frac{g n({\rm M})}{|{\rm d}v_{\rm turb}/{\rm d}l|}\ ,
\end{equation} 
where $g=1$ is a dimensionless geometrical factor and ${\rm d}v_{\rm turb}/{\rm d}l$ is the turbulent velocity gradient.
Neufeld \& Kaufman (1993) and Neufeld et al. (1995) provided an analytical expression for $L$ (Eq.~5 of Neufeld \& Kaufman 1993)
as a function of a set of parameters which depend on $n({\rm H}_2$), $T$, and $\tilde{N}({\rm M})$. 
Since for each model cloud, $n({\rm H}_2$), $T$, and $\tilde{N}({\rm M})$ are known,
we calculated the molecular line cooling $\Lambda_{\rm g}$ by interpolating the tabulated values of this parameter set.

\begin{figure}
  \centering
  \resizebox{\hsize}{!}{\includegraphics{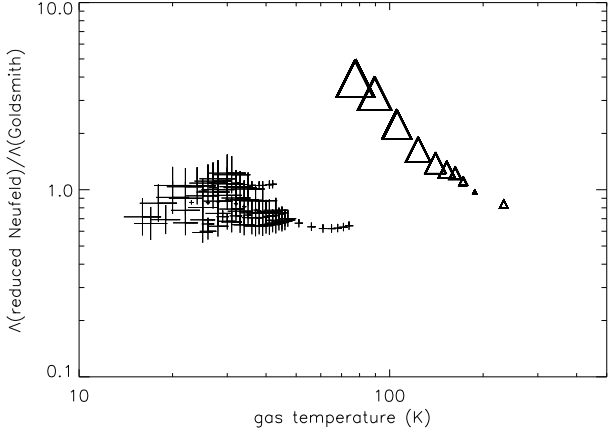}}
  \caption{Ratio between the cooling by CO, H$_2$, and H$_2$O (Neufeld \& Kaufman 1993; Neufeld et al. 1995)
    and the cooling function proposed by Goldsmith (2001) for light (plus signs) and massive (triangles)
    self-gravitating molecular clouds. The symbol sizes are proportional to the scaling factor $\zeta$ (see text).  
  \label{fig:goldsmith}}
\end{figure}

To investigate the differences between our gas cooling and that proposed by Goldsmith (2001),
we calculated these quantities for light and massive self-gravitating molecular clouds of 
different sizes, densities, column densities, and velocity dispersions: (i) light clouds
 with $M_{\rm cl}=10^4$~M$_{\odot}$: $l_{\rm cl}=\zeta^{-1} 10$~pc, $n_{\rm cl}=\zeta^3 \, 380$~cm$^{-3}$, 
$N_{\rm cl}=\zeta^2 \, 10^{22}$~cm$^{-2}$, and $v_{\rm turb}^{\rm cl}= \zeta^{\frac{1}{2}} 5.4$~km\,s$^{-1}$ 
(plus signs in Fig.~\ref{fig:goldsmith}), and 
(ii) massive clouds with $M_{\rm cl}=6 \times 10^4$~M$_{\odot}$: 
$l_{\rm cl}=\zeta^{-1} 10$~pc, $n_{\rm cl}=\zeta^3 \, 2350$~cm$^{-3}$, 
$N_{\rm cl}=\zeta^2 \, 7 \times 10^{22}$~cm$^{-2}$, and $v_{\rm turb}^{\rm cl}= \zeta^{\frac{1}{2}} 13.6$~km\,s$^{-1}$ 
(triangles in Fig.~\ref{fig:goldsmith}) with $1 \leq \zeta \leq 10$. 

Our simplified cooling prescription is in good agreement with that of Goldsmith (2001) ($\sim 0.2$~dex)
for the less massive clouds (Fig.~\ref{fig:goldsmith}). For the more massive clouds, our cooling prescription gives values up to
a factor $4$ ($0.6$~dex) higher than those of Goldsmith (2001) for the highest cloud densities. Overall, the
ratio between our cooling and that of Goldsmith (2001) is approximately $0.3$~dex.
Since the dependence of cooling on temperature is approximately $\Gamma \propto T^{2.5-3.0}$ (Goldsmith 2001),
the corresponding uncertainty on the gas temperature is $0.24$~dex (a factor of $1.7$) at most and 
$0.12$~dex (a factor of $1.3$) overall.

We thus conclude that our cooling prescription and that of Goldsmith (2001) are the same within a factor of $2$.
We prefer to use the reduced Neufeld line cooling instead of the cooling function proposed
by Goldsmith (2001), because it takes into account the cloud column density and velocity dispersion.

For the calculation of the thermal balance within molecular clouds, one needs 
to consider processes affecting the gas and the dust in addition to the radiative gas cooling discussed above.

We assume gas heating via turbulence and cosmic rays:
\begin{equation}
\Gamma_{\rm g} = \Gamma_{\rm turb} + \Gamma_{\rm CR} = \frac{1}{3} \rho \frac{(v_{\rm turb}^{\rm cl})^3}{r_{\rm cl}} + \eta \rho \dot{\Sigma_*}\ .
\end{equation}
Photoelectric heating by UV photons within photodissociation regions is neglected because the
local FUV field plays a minor role for the CO luminosity of a giant molecular cloud (Wolfire et al. 1993; see Sect.~\ref{sec:pdr}).
The factor of $\frac{1}{3}$ in the expression for the turbulent heating is somewhat lower
than the factor of $0.42$ advocated by Mac Low (1999).
Following Nelson \& Langer (1997), the constant $\eta$ is chosen such that for 
$\dot{\Sigma_*}=10^{-8}~{\rm M}_{\odot}{\rm pc}^{-2}{\rm yr}^{-1}$
$, \Gamma_{\rm CR}=6.4 \times 10^{-28} (n({\rm H_2})/{\rm cm^{-3}})~{\rm ergs\,cm^{-3}s^{-1}}$.
Furthermore, $\eta$ includes the attenuation factor $(\Sigma/(0.9~{\rm M_{\odot}pc^{-2}}))^{0.021}\exp(-\Sigma/(9 \times 10^4~{\rm M_{\odot}pc^{-2}}))$ 
described by Padovani \& Galli (2013) and the CR advection by a galactic wind ($1$ for local spiral and high-z star-forming galaxies;
$1$ for non-self-gravitating clouds and $140$ for the self-gravitating clouds in ULIRGs; $140$ for smm-galaxies; see Sect.~\ref{sec:winds}).

The dust is heated by the interstellar UV and optical radiation field:
\begin{equation}
\Gamma_{\rm d} = n_{\rm d} \sigma_{\rm d} F\ ,
\end{equation}
where $n_{\rm d}$ is the density of dust grains and $\sigma_{\rm d}$ the absorption cross section of a grain.
Following Goldsmith (2001), we set $n_{\rm d} \sigma_{\rm d}=7.4 \times 10^{-22} (n({\rm H_2})/{\rm cm^{-3}})$~cm$^{-1}$.
The ratio between the interstellar UV/optical and total radiation field is assumed to be
\begin{equation}
\frac{F}{F_0}=k \times \big( \frac{\dot{\Sigma_*}}{10^{-8}~{\rm M}_{\odot}{\rm pc}^{-2}{\rm yr}^{-1}} + \frac{\Sigma_*}{40~{\rm M_{\odot}pc^{-2}}} \big)\ , 
\end{equation}
where $F_0=5.3 \times 10^{-3}$~ergs\,cm$^{-2}$s$^{-1}$ (Goldsmith 2001).
We assume that the UV radiation is emitted by young massive stars whose surface density is proportional 
to the star-formation rate per unit area $\dot{\Sigma_*}$. The optical light stems from the majority of disks stars 
(Mathis et al. 1983, Draine 2011) whose surface density is $\Sigma_*$. The normalizations of $\dot{\Sigma_*}$ and $\Sigma_*$ are set by
observations of the ISRF at the solar radius: $6.7$\,\% of the total stellar light is emitted in the UV (Mathis et al. 1983, Draine 2011).
This implies that the local Galactic star-formation rate is $\dot{\Sigma_*}=6.7 \times 10^{-10}~{\rm M}_{\odot}{\rm pc}^{-2}{\rm yr}^{-1}$,
which is (i) approximately a factor of two lower than the value given by Kennicutt \& Evans (2012), (ii) consistent with the local star-formation 
rate at $\sim 0.75 \times R_{25}$ in the sample of nearby spiral galaxies of Leroy et al. (2008),
and (iii) a reasonable value for a gas disk at $R=8$~kpc with $v_{\rm rot}=200$~km\,s$^{-1}$, $Q \sim 3$, and $\dot{M}=0.2$~M$_{\odot}$yr$^{-1}$.
Furthermore, we allow for an additional factor $k$, which plays the role of $U_{\rm min}$ in the Draine \& Li (2007) models.
We set $k=1$ for all galaxies except the local spirals where $k=2$. This additional factor is (i) needed to reproduce the observed infrared
spectra energy distributions and (ii) consistent with the distribution found for nearby spiral galaxies by Dale et al. (2012).

In the presence of dust and gas, the interstellar radiation field is attenuated.
For this attenuation we adopted the mean extinction of a sphere of constant density
\begin{equation}
\label{eq:attenuationfactor}
I/I(0)= 3\,(\tau^{-1}-2\,\tau^{-2}+2\,\tau^{-3})-6\,\exp(-\tau)\,\tau^{-3}\ ,
\end{equation}
with $\tau=(Z/Z_{\odot})\,\Sigma/(15~{\rm M_{\odot}pc^{-2}})$.
For high optical depths, Eq.~\ref{eq:attenuationfactor} becomes 
$I/I(0) \sim 3 \tau_{\rm V}^{-1}$, that is, the attenuation decreases to very low values. However, when the molecular clouds become optically 
thick in the near-infrared (at $\tau_{\rm V} \sim 10$), radiative transfer effects become important; a significant infrared radiation 
field builds up which heats the dust until deep into the molecular clouds. To take this additional heating term into account,
we set $I/I(0)=0.246$ if $3\,(\tau^{-1}-2\,\tau^{-2}+2\,\tau^{-3})-6\,\exp(-\tau)\,\tau^{-3} < 0.246$ (see Appendix~\ref{sec:atten}).

The model dust temperature in the absence of collisional dust heating for the local Galactic interstellar radiation field with $I/I(0)$ is 
$T_{\rm dust}=18.4$~K. This temperature lies between the equilibrium temperature of silicate ($16.4$~K) and graphite ($22.3$~K)
for the local Galactic interstellar radiation field (Draine 2011).

The expression for the radiative heating of dust grains yields
\begin{equation}
\label{eq:dustheating}
\Gamma_{\rm d} = 3.9 \times 10^{-24} (\frac{F}{F_0}) (\frac{I}{I_0}) (n({\rm H_2})/{\rm cm^{-3}})~{\rm ergs\,cm^{-3}s^{-1}}\ .
\end{equation}

We assume the dust mass absorption coefficient of the following form:
\begin{equation}
\label{eq:kappa}
\kappa(\lambda)=\kappa_0\,(\lambda_0/\lambda)^{\beta}\ ,
\end{equation}
with $\lambda_0=250~\mu$m, $\kappa_0=0.48~{\rm m}^2{\rm kg}^{-1}$ (Dale et al. 2012), and a gas-to-dust ratio of $GDR=M_{\rm gas}/M_{\rm dust}=\frac{Z}{Z_{\odot}} \times 100$ 
(including helium; R{\'e}my-Ruyer et al. 2014).  Our gas-to-dust ratio is a factor of $1.5$ lower than the solar gas-to-dust ratio (Sofia \& Meyer 2001a, b).
We set the slope $\beta=1.5$ for the local spiral (Dale et al. 2012) and high-z star-forming galaxies, and $\beta=2.0$ for
the ULIRGs (Klaas et al. 2001) and smm-galaxies.
Adapting the dust cooling rate of  Goldsmith (2001) yields
\begin{equation}
\label{eq:dustcooling}
\Lambda_{\rm d}=7.5 \times 10^{-31} (Z/Z_{\odot})\,(T_{\rm d}/{\rm K})^{5.5} \big(n({\rm H_2})/{\rm cm^{-3}} \big)~{\rm ergs\,cm^{-3}s^{-1}}\ .
\end{equation}
Following Goldsmith (2001), the dust cooling energy transfer between dust and gas due to collisions is
\begin{equation}
\Gamma_{\rm gd}= 2 \times 10^{-33} \big(\frac{n({\rm H_2})}{{\rm cm^{-3}}} \big)^2 (\frac{\Delta T}{\rm K}) \sqrt{\frac{T_{\rm g}}{10~{\rm K}}}~{\rm ergs\,cm^{-3}s^{-1}}\ ,
\end{equation}
where $\Delta T=T_{\rm g}-T_{\rm d}$~K.

To determine the thermal balance of gas and dust, coupled together by the gas-dust collisions, we solve  the following equations
simultaneously:\begin{equation}
\Gamma_{\rm g} - \Lambda_{\rm g} - \Lambda_{gd} = 0
\end{equation}
and 
\begin{equation}
\label{eq:dusttemp}
\Gamma_{\rm d} - \Lambda_{\rm d} + \Lambda_{gd} = 0\ .
\end{equation}

\subsubsection{CO, HCN, and HCO$^+$ abundances from chemical network \label{sec:network}}

Chemical modeling is carried out using the {\tt Nautilus} gas-grain code
presented in detail in Hersant et al. (2009), Semenov et al. (2010), and Ruaud et al. (2015).
This code computes the abundances of chemical species (atoms and
molecules) as a function of time by solving the rate equations for a
network of reactions.
For gas-phase reactions, we use the kida.uva.2014 network
(Wakelam et al. 2015\footnote{the network is available online on the
KIDA website {\rm http://kida.obs.u-bordeaux1.fr}}) comprising $489$
species and $6992$ reactions. For grain surface reactions, we use the
desorption, diffusion, activation barrier energies along with a set of
grain surface reactions, all from the KIDA database. Both, thermal and
non-thermal desorption processes are taken into account, the latter
consisting mainly of CR-induced desorption following the formalism
presented by Hasegawa et al. (1993).

The model parameters are time, density, gas temperature, grain
temperature, UV flux, cosmic ray ionization rate, and the elemental
abundances of the elements C, O, and N (C/H$=1.7 \times 10^{-4}$, O/H$=2.4 \times 10^{-4}$, N/H$=6.2 \times 10^{-5}$). 

Grids of models were obtained by varying the cloud lifetime (20 log spaced steps
between $10^3$ and $10^8$~yr), the cloud density (20 log spaced steps
between $10^3$ and $10^8$~cm$^{-3}$), and cloud gas temperatures (20 log
spaced steps between $10$ and $300$~K for local spirals and $10$
and $800$~K for ULIRGs). For each type of cloud, the CO, HCN, and HCO+ abundances
were interpolated on the grid given the lifetime, density, and temperature of the cloud. 

The ISM chemistry also depends on the dust temperature and the cosmic ray ionization rate, which were assumed to be constant for 
each galaxy sample:
\begin{itemize}
\item
local spiral galaxies: $T_{\rm dust}=15$~K, $\zeta_{\rm CR}=3 \times 10^{-17}$~s$^{-1}$,
\item
ULIRGs, smm, high-z star-forming galaxies: $T_{\rm dust}=30$~K, $\zeta_{\rm CR}=1.3 \times 10^{-15}$~s$^{-1}$,
\end{itemize}
Testing of the influence of these parameters on the molecular line emission showed that
the dust temperature plays a minor role. For the choice of the CR ionization rates, we refer to Sect.~\ref{sec:winds}.
To take the gas metallicity into account, all abundances are multiplied by $(Z/Z_{\odot})$.

\subsection{Supernova-driven turbulence \label{sec:SNturbulence}} 

First, we assume that the gas is turbulent,
so that the turbulent velocity is the relevant one throughout the disk 
(making the exact temperature of the gas largely irrelevant; for simplicity, we assume a constant sound
speed of $c_{\rm s}=6$~km\,s$^{-1}$ for the warm neutral medium). We assume that
this turbulence is driven by SNe and that they input their energy in turbulent eddies
that have a characteristic length scale, $l_{\rm driv}$, and a characteristic
velocity, $v_{\rm turb}$.  This driving length scale may be the characteristic
length scale of a SN bubble, but it does not have to be so. It may be set by
the interaction of multiple SN bubbles or of a SN with the surrounding ISM. We
note that based on simulations, the assumption of a single driving scale may
be a simplification (Joung \& Mac Low 2006). The VB03 model does not address the
spatial inhomogeneity of the turbulent driving nor the mechanics of turbulent
driving and dissipation. It is assumed that the energy input rate into the ISM due
to SNe, $\dot{E}_{\rm SN}$, is cascaded to smaller scales without losses by turbulence.
At scales smaller than the size of the largest self-gravitating clouds, the energy is dissipated 
via radiation from cloud contraction and star formation. We refer to Mac Low \& Klessen (2004) for a review of these topics.
We limit our analytical model to the first
energy sink, which is the scale where the clouds become self-gravitating. 

We can connect the energy input into the ISM by SNe directly to the star-formation rate. With the assumption of a constant initial mass function (IMF)
independent of environment one can write

\begin{equation}
  \label{eq:energyflux}
  \frac{\dot{E}_{\rm SN}}{\Delta A}=\xi\,\dot{\Sigma}_{*} 
  = \xi\,\dot{\rho}_{*} l_{\rm driv}=\Sigma \nu \frac{v_{\rm turb}^{2}}{l_{\rm driv}^{2}}\ ,
\end{equation}

where $\Delta A$ is the unit surface element of the disk and the CO disk thickness is assumed to be $l_{\rm driv}$. 
The gas disk viscosity is defined as $\nu=l_{\rm driv} v_{\rm turb}$ (VB03 and Sect.~\ref{sec:accretiondisk}).
Following Vollmer \& Beckert (2001), the turbulent energy dissipation rate is 
$\Delta E/(\Delta A \Delta t)=\rho \nu v_{\rm turb}^2/l_{\rm driv}=\rho v_{\rm turb}^3$.
The turbulent dissipation timescale is 
\begin{equation}
\Delta t=\frac{\Sigma v_{\rm turb}^2}{\Delta E/(\Delta A \Delta t)}=\frac{\rho H v_{\rm turb}^2}{\Delta E/(\Delta A \Delta t)}=
\frac{H}{v_{\rm turb}} \sim \Omega^{-1}\ .
\end{equation}
This result is in agreement with numerical simulations of turbulence that show a decay of turbulence on an approximately
crossing timescale (e.g., Stone et al. 1998; Mac Low et al. 1998).

The factor of proportionality $\xi$ relates the local SN energy input to the
local star-formation rate and is assumed to be independent of local
conditions.  $\xi$ is normalized using Galactic observations by
integrating over the Galactic disk and results in $\xi=4.6 \times
10^{-8}$~(pc/yr)$^{2}$ (see VB03).  The adopted energy that is
  injected into the ISM is $E^{\rm kin}_{\rm SN}=10^{50}$~ergs based
  on numerical studies by Thornton et al. (1998).  The final two parts of
Eq.~\ref{eq:energyflux} assume that stars form over a
characteristic scale equal to the driving length and equate energy
output from SNe with the energy transported by turbulence (see VB03).

In the outer galactic disk, where the star-formation activity is very low,
turbulence can be maintained by the energy gained via accretion within the
gravitational potential of the galaxy (Vollmer \& Beckert 2002).
In this case, the energy injection rate is 
\begin{equation}
\frac{\dot{E}_{\rm acc}}{\Delta A}=\frac{\dot{M}}{2\pi} \Omega^2 \ .
\end{equation}
This energy source represents an addition to the model described in Vollmer \& Leroy (2011).
The total energy injection rate is 
\begin{equation}
\frac{\dot{E}_{\rm tot}}{\Delta A}=\frac{\dot{E}_{\rm SN}}{\Delta A}+\frac{\dot{E}_{\rm acc}}{\Delta A}\ .
\end{equation}

\subsection{Star formation in molecular clouds \label{sec:SFRGMC}} 

Second, we assume that stars form
out of gravitationally bound clouds. We take the local gravitational free-fall time, given by

\begin{equation}
  t^{\rm l}_{\rm ff}=\sqrt{\frac{3\pi}{32G\rho_{\rm cl}}}\ ,
  \label{eq:localff}
\end{equation}

\noindent to be the relevant timescale for star formation. Here, $G$ is the
gravitational constant and $\rho_{\rm cl}$ the density of a single cloud.

Cloud collapse, and thus star formation can only proceed if enough molecules
form during the cloud collapse to allow the gas to continue cooling\footnote{However, based on
theoretical arguments and numerical simulations Krumholz et al. (2011, 2012) and Glover \& Clark (2012a,b)
argue that C$^+$ cooling is sufficiently strong for gas to from stars as long as it is
sufficiently shielded from the interstellar radiation field.}. Vollmer \& Leroy (2011)
assumed that, where the timescale for H$_2$ formation is long (compared to the
free-fall time), the relevant timescale for star formation is the H$_2$ formation timescale.
Since we aim at reproducing the ISM properties within the optical disks, we decided not to
include this complication in our present model.

\subsection{Molecular fraction \label{sec:molfrac1}}

We follow two approaches to calculate the fraction of molecular hydrogen:
(i) based on the lifetimes of the molecular clouds and (ii) based on photodissociation of molecules by the
interstellar radiation field. In Sect.~\ref{sec:molfrac}, we show that both formalisms
lead to consistent results for the molecular fraction.

\subsubsection{Molecular fraction based on the lifetimes of the molecular clouds}

This approach assumes that molecular clouds are relatively
short-lived, appearing and disappearing over approximately a free-fall time
(equivalently, by our construction, a turbulent crossing time); otherwise they
might reach chemical equilibrium even when the H$_2$ formation time is long
compared to the free-fall time. Accordingly, we estimate the molecular
ratio in the disk from the ratio of a cloud lifetime (i.e., the crossing or
free-fall time) to the H$_2$ formation time scale:
\begin{equation}
R_{\rm mol}=\frac{\Sigma_{\rm H_{2}}}{\Sigma_{\rm HI}}=t_{\rm turb}^{l}/t_{\rm mol}^{l}\ .
\end{equation}
The molecular fraction is
\begin{equation}
f_{\rm mol}=\frac{\Sigma_{\rm H_{2}}}{\Sigma_{\rm HI}+\Sigma_{\rm H_{2}}}=\frac{R_{\rm mol}}{1+R_{\rm mol}}\ .
\end{equation}

We take the characteristic time to form H$_2$ out of H to be approximately
\begin{equation} \label{eq:molform}
t_{\rm mol}^{l}=\alpha / \rho_{\rm cl}\ ,
\end{equation} 
  where $\alpha$ is a coefficient that depends on the gas phase metallicity and
  temperature (Draine \& Bertoldi 1996) and $\rho_{\rm cl}$ is the
  density of a single cloud. 

The coefficient of the molecular formation timescale $\alpha_0$ is
assumed to be metallicity dependent (Tielens \& Hollenbach 1985). Because we
  admit external gas accretion, the metallicity of the star-forming
  ISM mainly depends on the ratio of accretion to star formation rate
  $a$.  Small $a<1$ lead to a metallicity derived from a closed box
  model, whereas in the case of $a>1,$ the metallicity is equal to the true
  stellar yield $y_{\rm true}$ (K\"{o}ppen \& Edmunds 1999). For gas fractions
  higher than 0.04, the difference between the two solutions is less
  than a factor of two.  Moreover, Dalcanton (2007) showed that the
  effective yield $y_{\rm eff}=Z_{\rm gas}/\ln(1/f_{\rm gas})$, where
  $Z_{\rm gas}$ is the gas metallicity and $f_{\rm gas}$ the gas
  fraction, for disk galaxies with a rotation velocity higher than
  $100$~km\,s$^{-1}$ is approximately constant, that is, for these
  galaxies, a closed box model can be applied. We thus feel confident
  to estimate the gas phase metallicity based on a closed box model using the
gas fraction:
\begin{equation} \label{eq:alphacb}
\alpha=\alpha_0 \times \big( \ln(\frac{\Sigma_{*}+\Sigma}{\Sigma})\big)^{-1}\ ,
\end{equation}
where $\Sigma_{*}$ is the stellar surface density and $\alpha_0=3.6 \times 10^{7}~{\rm yr\,M_{\odot}pc^{-3}}$.  
Adopting a stellar and gas surface density of
$\Sigma_{*}=40$~M$_{\odot}$pc$^{-2}$ and $\Sigma_{\rm
  gas}=10\,$~M$_{\odot}$pc$^{-2}$ at the solar radius of the Galaxy yields $\alpha_{\odot}=2.2 \times
10^{7}$~yr\,M$_{\odot}$pc$^{-3}$, which corresponds to the value used by
Hollenbach \& Tielens (1997).
Within this framework, the metallicity is 
\begin{equation} \label{eq:zzodot}
Z/Z_{\odot}=(2.2 \times 10^7~{\rm yr\,M_{\odot}pc^{-3}})/\alpha\ .
\end{equation}

It turned out that the gas phase metallicities of the ULIRG and high-z star-forming galaxy samples are
underestimated by up to a factor $10$ with our simple closed-box model (see Sect.~\ref{sec:metals}). 
To remedy the situation, we adopted the following heuristic recipe for all galaxies:
\begin{equation}
\label{eq:metulirgs}
\alpha=\frac{3.6 \times 10^{7} \times \big( \ln(\frac{\Sigma_{*}+\Sigma}{\Sigma})\big)^{-1}}{{\rm max}\big( (2 \times 10^9~{\rm yr}\,\frac{\dot{M_*}}{M_{\rm gas}})^{\frac{1}{3}},1.0\big)}\ {\rm yr\,M_{\odot}pc^{-3}}\ .
\end{equation}
A possible explanation for this recipe is accretion of pre-enriched gas onto or into the galactic disks in which case the
closed-box model underestimates the metallicity. Less gas depletion in starburst galaxies might also play a role.
Since we expect starburst galaxies to host galactic winds (see, e.g., Veilleux et al. 2005), the ejection of metals due to these
outflows (leaky box model) is assumed to be much less significant than the addition of metals from accretion.

\subsubsection{Molecular fraction based on photodissociation \label{sec:dissociation}}

For the determination of the H$_2$ column density of a gas cloud, we take into account 
(i) photo-dissociation of H$_2$ molecules and (ii) the influence of the finite cloud lifetime on the H$_2$ formation.

For the photo-dissociation of H$_2$ molecules, we follow the approach of Krumholz et al. (2008, 2009). 
These authors solved the idealized problem of determining the location of the atomic-to-molecular transition in a uniform spherical 
gas cloud bathed in a uniform isotropic dissociating radiation field. It is assumed that the transition from atomic to molecular gas 
occurs in an infinitely thin shell. The cloud has a constant inner molecular and outer atomic gas density.
The inner molecular core and the outer atomic shell are assumed to be in thermal pressure equilibrium.
The atomic gas density is taken to be the density of the cool neutral medium (Eq.~\ref{eq:cnmdens}).
The H$_2$ to H{\sc i} ratio is 
\begin{equation}
R_{\rm H_2} \simeq \big(1+(s/11)^3(\frac{125+s}{96+s})^3)^{\frac{1}{3}}\big)-1
,\end{equation}
with $s=(\Sigma_{\rm cl}/1~{\rm M}_{\odot})(Z/Z_{\odot})/(4\,\tau_{\rm H{\sc i}})$. The H{\sc i} optical depth is
\begin{equation}
\tau_{\rm H{\sc I}}=\frac{\chi}{4} \frac{2.5+\chi}{2.5+\chi {\rm e}}\ ,
\end{equation}
with the dimensionless radiation field strength $\chi$ , which we set to
$\chi=3.1\,(\dot{\Sigma_*}/(10^{-8}~{\rm M}_{\odot}{\rm pc}^{-2}{\rm yr}^{-1}))/(n_{\rm cl}/(100~{\rm cm}^{-3}))$.
Here, we assume a constant ratio between the inner molecular and outer atomic gas density, which is of the order of $10$.
For $\tau_{\rm H{\sc I}}=\frac{1}{4}$ and solar metallicity, the transition between a molecular- and atomic-dominated cloud occurs
at $\Sigma_{\rm cl} \simeq 20$~M$_{\odot}$pc$^{-2}$.
The H$_2$ fraction of the cloud is $f_{\rm H_2}=R_{\rm H_2}/(1+R_{\rm H_2})$.
This treatment insures a proper separation of H{\sc i} and H$_2$ in spiral galaxies, that is, 
clouds of low density ($\sim 100$~cm$^{-3}$) and low column density 
($\sim 10^{21}$~cm$^-2$) are fully atomic, whereas clouds of high density, that is, GMCs, ($\geq 1000$~cm$^{-3}$) and high column 
density ($\geq 10^{22}$~cm$^-2$) are fully molecular. In starburst regions (e.g., in ULIRGs), where gas densities and surface 
densities are much higher, this treatment has no effect, since the gas will be fully molecular.

In a second step, we take into account the molecular fraction due to the finite lifetime of the gas cloud 
$f_{\rm mol}^{\rm life}=t_{\rm ff}^{\rm cl}/t{\rm mol}^{\rm cl}/(1+t_{\rm ff}^{\rm cl}/t_{\rm mol}^{\rm cl})$.
The total molecular fraction of a cloud is $f_{\rm mol}=f_{\rm mol}^{\rm life} \times f_{\rm mol}^{\rm diss}$. The molecular fraction
due to the finite lifetime $f_{\rm mol}^{\rm life}$ has the highest influence on $f_{\rm mol}$ at large galactic radii.

We now go from the H$_2$ mass fraction to the CO mass fraction.
In an externally irradiated gas cloud, a significant H$_2$ mass may lie outside
the CO region, that is, it is dark in the outer regions of the cloud where the gas phase carbon resides in C or C$^+$. 
In this region, H$_2$ self-shields or is shielded by dust from UV photodissociation, whereas CO is photodissociated.
Following Wolfire et al. (2010), the dark gas mass fraction for a cloud of constant density is 
\begin{equation}
\label{eq:fdg}
f_{\rm DG}=\frac{M_{\rm H_2}-M_{\rm CO}}{M_{\rm H_2}}=1-\big(1 - \frac{2 \Delta A_{\rm V, DG}}{A_{\rm V}}\big)^3
,\end{equation}
with
\begin{equation}
\begin{split}
\Delta A_{\rm V,DG}=&0.53-0.045\,{\rm ln}\big(\frac{\dot{\Sigma_*}/(10^{-8}~{\rm M}_{\odot}{\rm pc}^{-2}{\rm yr}^{-1})}{n_{\rm cl}}\big)-\\
&0.097\,{\rm ln}\big(\frac{Z}{Z_{\odot}}\big)
\end{split}
,\end{equation}
and $A_{\rm V}=2\,(Z/Z_{\odot})N_{\rm cl}/(1.9 \times 10^{21}~{\rm cm}^{-2})$ where $N_{\rm cl}$ is the H$_2$ column density.
The CO mass fraction is then $f_{\rm CO}=f_{\rm H_2} \, \big(1 - \frac{2 \Delta A_{\rm V, DG}}{A_{\rm V}}\big)^3$.

Since the attenuation of the UV radiation field leading to Eq.~\ref{eq:fdg} is mainly caused by dust,
we expect HCN to survive everywhere where the ISRF is attenuated enough to permit a high CO abundance.
Thus, the HCN abundance should approximately follow the CO abundance, unless there is a strong X-ray/cosmic ray flux
that is not attenuated by dust.
In the absence of a proper theoretical model for the HCN dissociation, we thus assume the same dissociation rate for HCN as for CO.

\subsection{Vertical disk structure \label{sec:vertical}}

In the model, the disk scale height is
determined unambiguously by the assumption of hydrostatic equilibrium and the turbulent pressure
(Elmegreen 1989):

\begin{equation}
p_{\rm turb}=\rho v_{\rm  turb}^{2} = \frac{\pi}{2} G \Sigma ( \Sigma + \Sigma_{*} \frac{v_{\rm turb}}{v_{\rm disp}^{*}})~,
\label{eq:pressure}
\end{equation}

\noindent where $\rho$ is the average density, $v_{\rm turb}$ the gas turbulent
velocity in the disk, $v_{\rm disp}^{*}$ the stellar vertical velocity dispersion, 
and $\Sigma$ the surface 
density of gas and stars. The stellar velocity dispersion is calculated
by $v_{\rm disp}^{*}=\sqrt{2 \pi G \Sigma_{*} H_{*}}$, where
the stellar vertical height is taken to be $H_{*}=l_{*}/7.3$ with  $l_{*}$ being the
stellar radial scale length (Kregel et al. 2002). 
We neglect thermal, cosmic ray, and magnetic pressure.

\subsection{Treatment as an accretion disk \label{sec:accretiondisk}}

The turbulent motion of clouds is
expected to redistribute angular momentum in the gas disk, like an effective
viscosity would do. This allows accretion of gas towards the center and makes
it possible to treat the disk as an accretion disk (e.g., Pringle 1981).
This gaseous turbulent accretion disk rotates in a given gravitational
potential $\Phi$ with an angular velocity $\Omega=\sqrt{R^{-1}\frac{{\rm
      d}\Phi}{{\rm d}R}}$, where $R$ is the disk radius. The disk has an
effective turbulent viscosity that is responsible for mass accretion and
outward angular momentum transport. In this case, the turbulent velocity is
driven by SN explosions, which stir the disk and lead to viscous transport of
angular momentum. In addition, star formation removes gas from the viscous evolution.
Following Lin \& Pringle 1987, the evolution of the gas surface density 
is given by
\begin{equation}
\frac{\partial \Sigma}{\partial t}=-\frac{1}{R}\frac{\partial}{\partial R}\left(
\frac{(\partial/\partial R)[\nu \Sigma R^3 ({\rm d}\Omega/{\rm }dR)]}{({\rm d}/{\rm d}R)(R^2 \Omega)}\right)
-\dot{\Sigma}_{*}+\dot{\Sigma}_{\rm ext}\ ,
\label{eq:linpringle}
\end{equation}
where $\nu$ is the gas disk viscosity, $\Omega$ the angular velocity, 
and $\dot{\Sigma}_{\rm ext}$ is the external mass
accretion rate. In contrast to Lin \& Pringle (1987), we assume a continuous and non-zero external gas mass 
accretion rate.

Forbes et al. (2014) presented an analytical approach based on Eq.~\ref{eq:linpringle}.
They showed that galaxies tend to be in a slowly evolving equilibrium state
wherein new accretion is balanced by star formation, galactic winds, and radial transport of
gas through the disc by gravitational instability-driven torques.
For a stationary gas disk in such an equilibrium, where star formation is balanced by
external accretion, the local mass and momentum conservation together with $\dot{\Sigma}_{*}=\dot{\Sigma}_{\rm ext}$ yield:
\begin{equation}
\nu \Sigma=\frac{\dot{M}}{2\pi}\ ,
\label{eq:transport}
\end{equation}
where $\dot{M}$ is the mass-accretion rate within the disk.
In the absence of external mass accretion,
the gas disk can be assumed to be stationary as long as the star-formation timescale $t_*$ exceeds the 
viscous timescale $t_{\nu}=R^2/\nu$.
For $\dot{\Sigma}_{\rm ext} < \dot{\Sigma}_{*}$ and $t_* < t_{\nu}$ , Eq.~\ref{eq:transport}
is not valid. In this case the gas disk
is rapidly turned into stars within the gas consumption time ($2$~Gyr, Evans 2008).
Since most spiral galaxies still have a significant amount of gas, we think that 
spiral galaxies are generally not in this state.
Solving the time dependent Eq.~\ref{eq:linpringle} is beyond the scope of this 
work and we apply Eq.~\ref{eq:transport}.

The viscosity is related to the driving length scale and
characteristic velocity of the SN-driven turbulence by $\nu=v_{\rm turb}l_{\rm
  driv}$ (VB03). Because the lifetime of a collapsing and star-forming cloud
($t_{\rm ff}^{l} < t_{\rm turb}^{l}$) is smaller than the turnover time of the
large-scale eddy ($l_{\rm driv}/v_{\rm turb}$), the turbulent and clumpy ISM
can be treated as one entity for the viscosity description.

\subsection{Clumpiness \label{sec:clumpiness}}

A critical factor in the model is the relationship between
the density of individual clouds, $\rho_{\rm cl}$, and the average density of
the disk, $\rho$. It is the density of individual clouds that is relevant to
the timescale for star formation. In this model, the two are related by the
volume filling factor, $\Phi_{\rm V,}$ so that $\rho_{\rm cl}=\Phi_{\rm
  V}^{-1}\rho$.

Here, $\rho_{\rm cl}$ refers to the density of the largest
self-gravitating structures in the disk, so that for these structures,
the turbulent crossing time and gravitational free-fall time are
equal. The scale of such a cloud, $l_{\rm cl}$ , is smaller than the
driving length scale, $l_{\rm driv}$ , by a factor $\delta$, which we do
not know {\em a priori}. Following Vollmer \& Leroy (2011), we set $\delta=5$.

Shear, due to differential galactic rotation, could stabilize
  clouds, modifying the timescale for collapse. However, this effect
  is mainly important when the ratio of the cloud to disk surface
  density is lower than the ratio of cloud to disk velocity
  dispersion, which is not the case over most of the disk in a typical
  spiral. Typical GMC surface densities are $\sim
  200$~M$_{\odot}$pc$^{-2}$ (Solomon et al. 1987), whereas disk surface
  densities only exceed $100$~M$_{\odot}$pc$^{-2}$ in the very center
  of spiral galaxies (Leroy et al. 2008). 

We can calculate the turbulent timescale for the cloud, $t_{\rm turb}^l$, for
a fractal ISM:

\begin{equation}
t_{\rm turb}^{l}=\delta^{-\frac{2}{3}-\frac{3-D}{3}}
\,l_{\rm driv}/v_{\rm turb}\ ,
\end{equation}

where $D$ is the fractal dimension (see, e.g., Frisch 1995) of the ISM.
We assume $D=2$ for a compressible, self-gravitating fluid, which is close to
the findings of Elmegreen \& Falgarone (1996). Once $\delta$ and thus $t_{\rm turb}^l$ are
specified, we can solve for the density of the corresponding scale by setting
$t_{\rm ff}^l = t_{\rm turb}^l$. The volume filling factor is then defined by
comparing $\rho_{\rm cl}$ and $\rho$.
Once the volume filling factor is known (from $\delta$ or $l_{\rm cl}$), we
can calculate the local star-formation rate, $\dot{\rho_*}$, via

\begin{equation}\label{eq:starform}
\dot{\rho}_{*} = \Phi_{\rm V} \frac{\rho}{t_{\rm ff}^{\rm l}}\ ,
\end{equation}

\noindent where $t_{\rm ff}^l$ is the local free-fall timescale
determined by $t_{\rm ff}^l = t_{\rm turb}^l$ corresponds to the contraction timescale $t_{\rm c}=\sqrt{\pi/(G \rho_{\rm cl})}$ (Ostriker et al. 1999)
of clouds of constant density in Virial equilibrium.
Since, in our model, the lifetime of a cloud is the free-fall as suggested by
  Ballesteros-Paredes \& Hartmann (2007), this implies that during the cloud lifetime, approximately 
$\dot{\rho}_{*,{\rm cl}}/(\rho_{\rm cl} t_{\rm ff}^l)=\Phi_{\rm V} \sim 1$\,\% of the cloud mass turns into stars.

\noindent The vertically integrated star-formation rate in the inner disk 
where $t_{\rm sf}^{\rm l}=t_{\rm ff}^{\rm l}=t_{\rm turb}^{\rm l}=\delta^{-1} t_{\rm turb}$ is 

\begin{equation}\label{eq:starformm1}
\dot{\Sigma}_{*} = \Phi_{\rm V} \frac{\rho}{t_{\rm ff}^{\rm l}} l_{\rm driv} = 
\delta \Phi_{\rm V} \rho v_{\rm turb}\ ,
\end{equation}

\noindent that is, it is the mass flux density of the turbulent ISM into the regions
of star formation.

\subsection{Thermal dust emission \label{sec:dustemission}}

The dust temperature $T_{\rm d}$ of a gas cloud of given density and size illuminated by a local mean radiation field is calculated by solving Eq.~\ref{eq:dusttemp}.  
With the dust mass absorption coefficient of Eq.~\ref{eq:kappa}, the dust optical depth is 
\begin{equation}
\tau(\lambda)= \kappa(\lambda)\,\Sigma_{\rm cl} (GDR)^{-1}\ ,
\end{equation}
where $\Sigma_{\rm cl}$ is the cloud surface density in g/cm$^2$. 
The infrared emission at a given wavelength at a given galactic radius $R$ is calculated in the following way:
\begin{equation}
\label{eq:idust}
I_{\rm dust}(\lambda)=\sum_{\rm i=1}^{\rm N}  \big(\Phi_{\rm A} \big)_i\, (f_{\rm mass})_i\, \big(1-\exp(-\tau(\lambda))\big)_i B(\lambda,T_{\rm d})_i\ ,
\end{equation}
where $B(\lambda,T_{\rm d})$ is the Planck function and $\Phi_{\rm A}=1.5\,(\Delta M/M)\,(\Sigma/\Sigma_{\rm cl})$ the area filling factor.
The factor $1.5$ takes into account that the mean cloud surface density is $1.5$ times lower than the surface density in the cloud center 
$\Sigma_{\rm cl}=\rho_{\rm cl}l_{\rm cl}$.
The integration of the Eq.~\ref{eq:idust} yields the total infrared emission at a given galactic radius $R$:
\begin{equation}
I_{\rm TIR}(R)=\int_{10\,\mu{\rm m}}^{1\,{\rm mm}} I_{\rm dust}(\lambda) {\rm d}\lambda\ .
\label{eq:idust1}
\end{equation} 
At a given wavelength, $\lambda, $ the effective background temperature of the thermal dust emission $T_{\rm eff\ dust}$ is determined by 
$I_{\rm dust}(\lambda)=B(\lambda,T_{\rm eff\ dust})$.
We do not subtract the cosmic infrared background, because it is always much smaller than the dust emission within the region of interest.
The total infrared luminosity is given by
\begin{equation}
L_{\rm TIR}=2\pi \int_0^{\rm R_0} I_{\rm TIR}(R)\,R\,{\rm d}R\ .
\label{eq:idust2}
\end{equation}
To reproduce the observed total infrared luminosity, it is essential to take into account the diffuse warm neutral medium, which is
not taken into account for the molecular line emission, because the gas is in atomic form.
We do so by explicitly calculating the dust infrared emission based on the proper dust temperature
(Eq.~\ref{eq:dusttemp}), density of $\rho/2$, area filling factor
$\Phi^{\rm WNM}_{\rm A}=(1-\Phi_{\rm A}^{\rm CNM})$, and gas mass fraction of $(\Delta M/M)_{\rm WNM}=\big(1-(\Delta M/M)_{\rm CNM+mol}\big)$.

\subsection{Molecular line emission \label{sec:lineemission}}

A molecular line source is usually observed by chopping the telescope's beam between
on- and off-source positions and measuring the difference in antenna temperatures.  
In general, the difference in brightness 
temperatures is
\begin{equation}
\Delta T^*_{\rm A}=\big(1-{\rm e}^{-\tau}\big)\frac{h\nu}{k}\big(\frac{1}{{\rm e}^{h\nu/kT_{\rm ex}}-1}-
\frac{1}{{\rm e}^{h\nu/kT_{\rm bg}}-1}\big)\ ,
\end{equation}
where $\tau$ is the optical depth of the line, $\nu$ the frequency of the observations, $h$ and $k$ the
Planck and Boltzmann constants, and $T_{\rm ex}$ and $T_{\rm bg}$ the excitation and background brightness
temperatures, respectively. 

Considering only a single collider (H$_2$) for simplicity, the excitation temperature is
\begin{equation}
\frac{1}{T_{\rm ex}}=\big(\frac{1}{T_{\rm g}}+(\frac{A_{ul}}{n q_{ul}}\frac{T_{\rm bg}}{T_*})\frac{1}{T_{\rm bg}}\big)/(1+\frac{A_{ul}}{nq_{ul}}\frac{T_{\rm bg}}{T_*})\ ,
\end{equation}
where $T_*=h \nu_{ul}/k$, $n$ is the gas density, $nq_{ul}$ the collisional de-excitation rate, and $A_{ul}$ the Einstein coefficients of 
the transition $ul$. The background brightness temperature $T_{\rm bg}$ is the sum
of the effective emission temperatures of the galaxy's dust $T_{\rm eff\ dust}$ and the cosmic background at the 
galaxy redshift $T_{\rm CMB}$ (see Eq.~17 of da Cunha et al. 2013).
For optically thin transitions, the ratio of the radiative and collisional rates is just the ratio of 
the density to the critical density for the transition
\begin{equation}
n_{\rm crit}=\frac{A_{ul}}{q_{ul}}\ .
\end{equation}
We use $T_{\rm CMB}=2.73$~K for the local spiral galaxies and ULIRGs,
$T_{\rm CMB}=6.0$~K for the high-z star-forming galaxies, and $T_{\rm CMB}=8.19$~K for the submillimeter galaxies.
This corresponds to a cosmic microwave background of $T_{\rm CMB}=2.73\,(1+z)$~K (see, e.g., Carilli \& Walter 2013)
and mean redshifts of $\langle z \rangle =0,\ 1.2,\ 2$, respectively.

For optically thick transitions, the upper-level population can be enhanced due to absorption of line photons, 
leading to excitation temperatures higher than those expected simply due to H$_2$ collisions, 
since the line photons 
emitted upon spontaneous decay cannot easily escape the cloud. This so-called radiative trapping of the line 
photons builds up the radiation field at the frequency of the line, leading to enhanced excitation of the upper 
state via photon absorption. The escape probability formalism can be used to treat this optically thick situation 
(see, e.g., Scoville 2013).

This formalism is applicable to situations in which systematic velocity gradients are large compared to the 
small-scale thermal motions. The line photons from one region of the cloud are then incoherent with other 
regions due to the Doppler shift; they can then only interact with molecules in the local region near where 
they were emitted. 
In the photon trapping regime, the spontaneous decay rates ($A_{ul}$) and thus the critical density
($n_{\rm crit}$) used in analyzing the equilibrium molecular 
excitation are reduced by a factor $\beta$ equal to the effective probability for escape of line photons from the 
emission region (Scoville \& Solomon 1974; Goldreich \& Kwan 1974). 
For a spherical cloud of uniform density, Draine (2011) gives 
\begin{equation}
\beta=\frac{1}{1+0.5 \tau}\ .
\end{equation}
The critical density in the optically thick case is then 
\begin{equation}
n_{\rm crit}=\beta\,\frac{A_{ul}}{q_{ul}}\ .
\end{equation}
For our analytic analysis, we follow Scoville et al. (2015) and use the sum of the collision rate coefficients 
out of the upper level $J$ to any other rotational level (both below and above $J$)
since all of these transitions couple the level to the gas kinetic temperature. 

For the determination of the optical depth of a molecular emission line, we follow Draine (2011).
The line-center optical depth, from cloud center to edge, for a transition from level $J+1$ to level $J$ is
\begin{equation}
\tau_{(J+1),J}=n_J r_{\rm cl} \big(1-\frac{n_{(J+1)}}{n_J}\frac{g_J}{g_{(J+1)}}\big) \frac{\lambda^3}{8 \pi^{\frac{3}{2}} v_{\rm turb}^{\rm cl}}\frac{g_{(J+1)}}{g_J}A_{(J+1),J}\ ,
\end{equation}
where $r_{\rm cl}=l_{\rm cl}/2$ is the cloud radius, $\lambda$ the wavelength of the observations, and $g_J=2J+1$
the transition weights.
Following Draine (2011), we adopt the following expression for the CO line optical depth:
\begin{equation}
\begin{split}
\tau_{(J+1),J}=&281\,n_3R_{19}\frac{Z}{Z_{\odot}}\big(\frac{n({\rm CO})/n_{\rm H}}{7\times 10^{-5}}\big)\big(\frac{n({\rm CO},J)}{n({\rm CO})}\big)\\
&\big(\frac{2~{\rm km\,s^{-1}}}{v_{\rm turb}^{\rm cl}}\big)\big(1-\frac{n_{(J+1)}}{n_J}\frac{g_J}{g_{(J+1)}}\big)\ ,
\end{split}
\end{equation}
where $n_3=n/(10^3~{\rm cm^{-3}})$ and $R_{19}=r_{\rm cl}/(10^{19}~{\rm cm})$.
The fraction of molecules of species $X$ in a given rotational level is 
\begin{equation}
\begin{split}
\frac{n({\rm X},J)}{n({\rm X})}=&\frac{(2J+1){\rm e}^{-B_0J(J+1)/kT_{\rm ex}}}{\sum_J (2J+1){\rm e}^{B_0j(J+1)/kT_{\rm ex}}}\simeq \\
&\simeq \frac{(2J+1){\rm e}^{-B_0J(J+1)/kT_{\rm ex}}}{\big(1+(kT_{\rm ex}/B_0)^2\big)^{\frac{1}{2}}}\ ,
\end{split}
\end{equation}
where $B_0$ is the rotation constant of a molecule of species $X$.

In summary, we consider two-level molecular systems in which the level populations are determined by
a balance of collisions with H$_2$, spontaneous decay and line photon absorption, and stimulated emission with
$\tau > 1$. Our final expression for the CO line optical depth reads
\begin{equation}
\begin{split}
&\tau_{(J+1),J}^{\rm CO}=393\,n_3R_{19}\big(\frac{2~{\rm km\,s^{-1}}}{v_{\rm turb}^{\rm cl}}\big)\frac{Z}{Z_{\odot}}\\
&\big(1-{\rm e}^{-B_0(J+1)(J+2)/kT_{\rm ex}`}\big)\big(\frac{(2J+1){\rm e}^{-B_0J(J+1)/kT_{\rm ex}}}{(1+(\frac{kT_{\rm ex}}{B_0})^2)^{\frac{1}{2}}}\big)\\
&\big(\frac{2(J+1)+1}{3(2J+1)}\big) \frac{A_{(J+1),J}^{\rm CO}}{A_{1,0}^{\rm CO}} \big(\frac{5.53~{\rm K}}{B_0(J+1)(J+2)/k}\big)^3\ ,
\end{split}
\end{equation}
where the rotation temperature is $B_0/k=2.77$~K.
We use a normalization, which is different from that of Draine (2011), because we assume the canonical $x({\rm CO})=10^{-4}$.
In a second step, the HCN abundances are calculated using a chemical network (see Sect.~\ref{sec:network}).

For simplicity, we neglected the hyperfine structure of HCN. In this simplified treatment, we can write
\begin{equation}
\begin{split}
&\tau_{(J+1),J}^{\rm HCN}=87\,n_3R_{19}\big(\frac{2~{\rm km\,s^{-1}}}{v_{\rm turb}^{\rm cl}}\big)\frac{Z}{Z_{\odot}}\\
&\big(1-{\rm e}^{-B_0(J+1)(J+2)/kT_{\rm ex}}\big)\big(\frac{(2J+1){\rm e}^{-B_0J(J+1)/kT_{\rm ex}}}{(1+(\frac{kT_{\rm ex}}{B_0})^2)^{\frac{1}{2}}}\big)\\
&\big(\frac{2(J+1)+1}{3(2J+1)}\big)\frac{A_{(J+1),J}^{\rm HCN}}{A_{1,0}^{\rm HCN}} \big(\frac{4.25~{\rm K}}{B_0(J+1)(J+2)/k}\big)^3\ ,
\end{split}
\end{equation}
where we assumed a HCN abundance of $x({\rm HCN})=2 \times 10^{-8}$. In a second step, the HCN abundances are calculated 
using a chemical network (see Sect.~\ref{sec:network}).
The rotation constant ($B_0/k=2.13$~K) and
Einstein coefficients are those of HCN. In the present work, we only investigate the HCN(1--0) transition.

The rotation constants, Einstein coefficients, and collision rates were taken from the Leiden Atomic and 
Molecular Database (LAMDA; Sch\"{o}ier et al. 2005). The CO collision rates were provided by Yang et al. (2010).
The HCN collision rates were taken from the He--HCN rate coefficients calculated by Dumouchel et al. (2010), 
scaled by a factor of $1.36$ to go to HCN--H$_2$ (see Green \& Thaddeus 1976). 

The HCO$^+$ emission was calculated in the same way.

We verified that the model brightness temperature
is consistent (within $10$-$20$\,\%) with the brightness temperature calculated by RADEX (van der Tak et al. 2007) 
for densities, gas kinetic temperatures, column densities, and linewidths typical for giant molecular clouds.

\subsection{HCN infrared pumping \label{sec:irpumping}}

HCN has a large dipole moment and therefore does not trace dense gas if there is another excitation mechanism
that is faster than the H$_2$ collisions and independent of gas density. One such excitation path is through a vibrationally excited
state,  to  which  molecules  can  be  pumped  by  infrared  radiation (Carroll \& Goldsmith 1981).
The first vibrationally excited state of HCN is its bending state ($v_2=1$) $1024$~K above the ground  with an emitting wavelength of $\lambda=14~\mu$m
(Sakamoto et al. 2010). Following Sakamoto et al. (2012), we define an equivalent gas density 
\begin{equation}
n_{\rm equiv}=\exp(-T_0/T_{\rm vib})\,A_{\rm vib}/\gamma_{J,J-1}\ ,
\end{equation}
where $T_0=1024$~K corresponds to the energy gap between the two vibrational levels $v=0$ and $1$,
$A_{\rm vib}=3.7$~s$^{-1}$ is the Einstein coefficient for the vibrational transition, and $\gamma_{J,J-1}$ is the collisional rate coefficient. 
As already stated in Sect.~\ref{sec:lineemission}, we follow Scoville et al. (2015) and use the sum of the collision rate coefficients 
out of the upper level $J$ to any other rotational level (both below and above $J$)
since all of these transitions couple the level to the gas kinetic temperature. 
$T_{\rm vib}$ is the equivalent blackbody temperature at $\lambda=14~\mu$m of the local and global background radiation. 
HCN IR-pumping is implemented in the model by replacing the cloud density $n_{\rm cl}$ by $n_{\rm equiv}$ if $n_{\rm equiv} > n_{\rm cl}$ in
the HCN emission calculations (Sect.~\ref{sec:lineemission}).

HCO$^+$ has a similar vibrational bending state at $\lambda=12~\mu$m. The associated excitation through radiative pumping is less than
half as significant as that provided by the HCN molecule (Imanishi et al. 2016). For this work, we did not take IR-pumping of HCO$^+$ into account.

\subsection{Photon-dominated regions \label{sec:pdr}}

Photodissociation regions (PDRs) are regions of a gas cloud where the physics and chemistry is dominated by penetrating FUV photons.
The structure of PDRs can be described by a plane-parallel, semi-infinite slab illuminated by an intense FUV field $G_0$, measured in 
units of the Habing (1968) interstellar radiation field ($=1.6 \times 10^{-3}$~ergs\,cm$^{-2}$s$^{-1}$). 
At the surface of the cloud, an atomic surface layer is created by the incoming FUV photons. The transition from atomic to molecular hydrogen
occurs at the depth approximately $A_{\rm V} \sim 2$. As the FUV photons are attenuated by the dust, the phase of carbon shifts from C$^+$
to C and CO at $A_{\rm V} \sim 4$. Deep inside the cloud ($A_{\rm V} > 10$), HCN and HCO$^+$ are formed. 

Wolfire et al. (1993) found that the integrated CO(1--0) luminosity of
giant molecular clouds increases by only $\sim 10$\,\% between $G_0=1$ and $G_0=100$ models. The luminosities are similar because the higher incident FUV
field forces clumps to become optically thick in the CO(1--0) transition deeper into the cloud where dust extinction lowers the radiation field
to a value near $G_0 \sim 1$. This results in similar gas temperatures in the optically thick clumps for both models.
Thus, the local FUV radiation field plays a minor role as long as $G_0 > 1$. 

Loenen et al. (2008) subdivided PDRs into two types according to the cycle of star formation; 
UV-dominated high-density ($n \ge 10^{5}$~cm$^{-3}$) PDRs from deeply embedded young stars, and lower-density ($n=10^{4.5}$~cm$^{-3}$) PDRs
that are dominated by mechanical feedback from supernova shocks. Due to the short duty cycle of the evolutionary stage involving young stars
compared to the second stage involving supernovae, most of the luminous infrared galaxies of their sample are observed to be in the later stage
of their evolution. Since we can reproduce the multi-transition CO, HCN(1--0), and HCO$^+$ luminosities without the inclusion of PDRs (Sect.~\ref{sec:results}),
we suggest that the conclusion of Loenen et al. (2008) is also valid for the ULIRGs, smm, and high-z star-forming galaxies. 

Within our model framework, PDRs are taken into account through the dissociation of molecules (Sect.~\ref{sec:dissociation}), but not through their
molecular line emission.

\subsection{Galactic winds \label{sec:winds}}

Galactic winds are driven by multiple supernova explosions during a starburst or by AGN activity.
The following statements are based on the review by Veilleux et al. (2005).
The minimum star-formation rate that creates a galactic wind is $\dot{M}_* \sim 5$~M$_{\odot}$yr$^{-1}$ or 
$\dot{\Sigma}_* \sim 10^{-3}$~M$_{\odot}$kpc$^{-2}$yr$^{-1}$. 
Galactic winds remove mass and, if they are magnetized, angular momentum from the galactic gas disk.
Mass outflow rates range between $0.1$~M$_{\odot}$yr$^{-1}$ and $10$~M$_{\odot}$yr$^{-1}$, with a trend for the rate to increase with increasing
star-formation rate. The mass-to-light conversion factors are highly uncertain for galactic winds. 
The ratio between the mass outflow and star-formation rate varies between $0.01$ and $10$.

By writing $\nu \Sigma=\dot{M}/(2\,\pi)$ (Eq.~\ref{eq:diskeq}), we ignored galactic external accretion and outflows (wind).
Including galactic winds, Eq.~\ref{eq:linpringle} becomes
\begin{equation}
\frac{\partial \Sigma}{\partial t}=-\frac{1}{R}\frac{\partial}{\partial R}\left(
\frac{(\partial/\partial R)[\nu \Sigma R^3 ({\rm d}\Omega/{\rm }dR)]}{({\rm d}/{\rm d}R)(R^2 \Omega)}\right)
-\dot{\Sigma}_{*}-\dot{\Sigma}_{\rm wind}+\dot{\Sigma}_{\rm ext}\ .
\label{eq:linpringle1}
\end{equation}
Within the framework of our equilibrium model, we thus assumed $\dot{\Sigma}_{\rm ext}=\dot{\Sigma}_{*}+\dot{\Sigma}_{\rm wind}$.
This might not be true for the compact starbursts as the ULIRGs. In this case, the constant $\dot{M}/(2\,\pi)$ links the
turbulent dissipation timescale $t_{\rm turb}$ to the star-formation rate per unit surface $\dot{\Sigma}_*$
(combining the mass, momentum, and energy conservation equations of Eq.~\ref{eq:diskeq}):
\begin{equation}
\dot{M}=2 \pi \xi \dot{\Sigma}_* t_{\rm turb}^2\ .
\end{equation}

Galactic winds also remove cosmic ray particles from the galactic disk. 
Following Suchkov et al. (1993; see also Papadopoulos 2010), the CR energy densities in starburst galaxies driving a galactic wind 
scale with respect to that of the Galaxy as
\begin{equation}
\frac{U_{\rm CR}}{U_{\rm CR, Gal}} \sim \frac{\dot{\Sigma}_*}{\dot{\Sigma}_{\rm *, Gal}} \times \frac{v_{\rm diff}}{v_{\rm wind}}\ ,
\end{equation}
where $v_{\rm diff}$ is the diffusion velocity at which CRs escape from quiescent disks such as the Milky Way while
$v_{\rm wind}$ is the velocity of a star-formation-induced wind at which CRs are advected out of the star-forming regions.
The diffusion velocity is set by the Alfven velocity $v_{\rm A}=B/\sqrt{4\pi \rho}$, where $B$ is the magnetic field.
Here, we assumed that the SN explosion rate is proportional to the star-formation rate.
For the chemical network (Sect.~\ref{sec:network}), we assumed $\frac{U_{\rm CR}}{U_{\rm CR, Gal}}=40$ for the ULIRGs, smm, and high-z star-forming galaxies. 
Since our model yields $\langle \frac{\dot{\Sigma}_{\rm *}}{\dot{\Sigma}_{\rm *, local\ spirals}} \rangle \sim 6000$ for the ULIRGs and smm-galaxies,
the mean ratio is $\langle \frac{v_{\rm diff}}{v_{\rm wind}} \rangle \sim 7 \times 10^{-3}$. Assuming $v_{\rm diff}=10$~km\,s$^{-1}$ gives a mean wind velocity
of $v_{\rm wind} \sim 1500$~km\,s$^{-1}$. This represents approximately four times the escape velocity, two times the observed mean velocity of a molecular wind
(Feruglio et al. 2015, Sakamoto et al. 2015), and is close to terminal wind velocities (e.g., Veilleux et al. 2005).

For the ULIRG model, we decided to apply the full CR heating $\frac{U_{\rm CR}}{U_{\rm CR, Gal}} = \frac{\dot{\Sigma}_*}{\dot{\Sigma}_{\rm *, Gal}}$
to the non-self-gravitating clouds. This modification did not
change the CO flux luminosities, but increased the HCN(1--0) luminosities by a factor of $\sim 1.5$ putting them closer
to the observed luminosities. This treatment implies that (i) the initially spherical wind escapes mostly vertically from the star-forming regions without
touching many of the non-self-gravitating clouds (``champagne effect''), (ii) the SN wind that blows in the direction of the galactic disk
is absorbed by the non-self-gravitating clouds, and (iii) the solid angle around the star-forming region occupied by non-self-gravitating clouds is substantial
(i.e., a wind opening angle $\ga 90^{\odot}$). Further testing of this hypothesis on local starburst galaxies
is needed to clarify the situation.

In the present model, the assumed CR ionization rate of the high-z star-forming galaxies is consistent with an absence of
a galactic wind. In a future project, we will investigate the influence of a lower CR energy density due to a wind on the chemical network and thus the
molecular line emission of these galaxies.

\section{Method \label{sec:method}}

\begin{figure*}
  \centering
  \resizebox{\hsize}{!}{\includegraphics{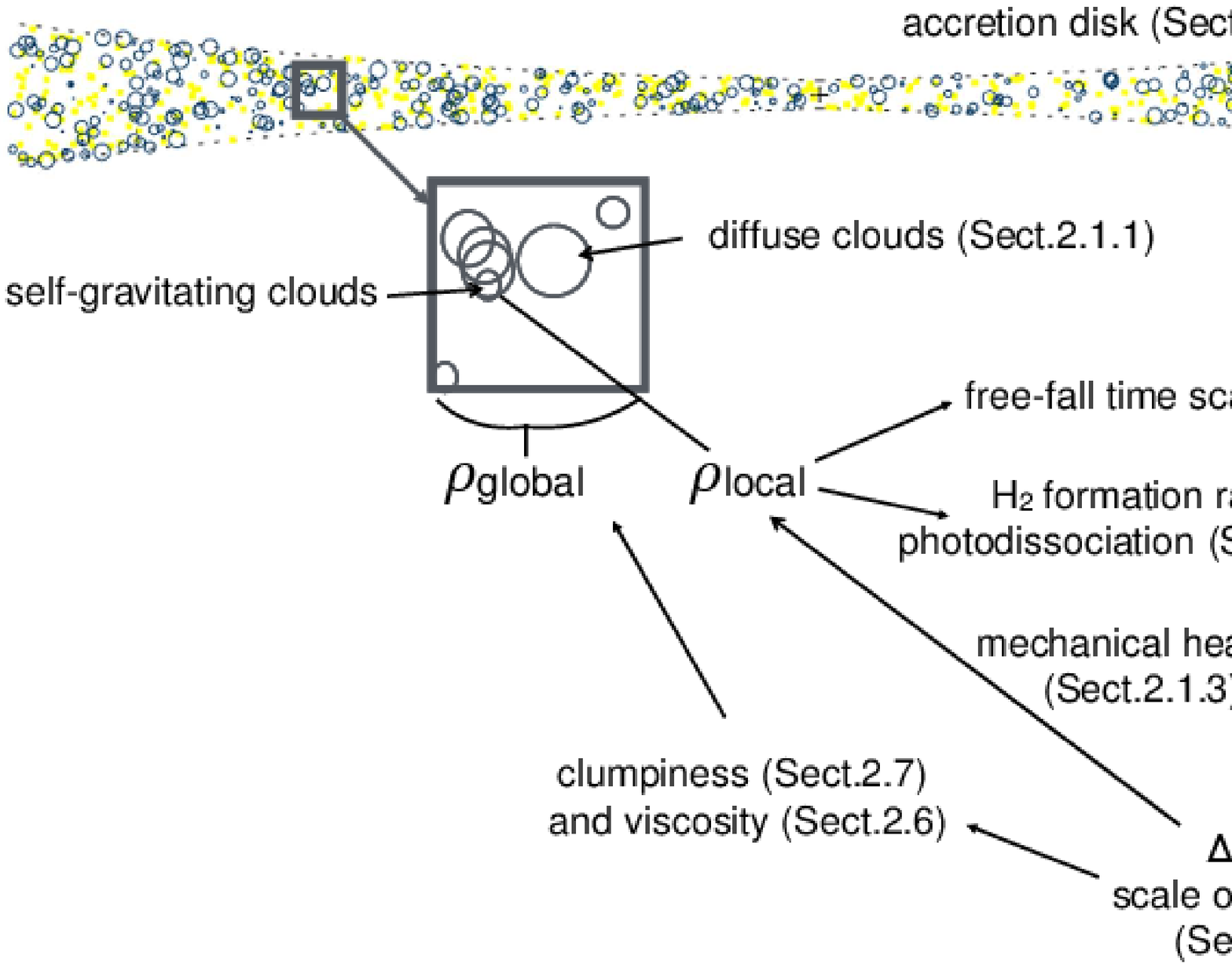}}
  \resizebox{\hsize}{!}{\includegraphics{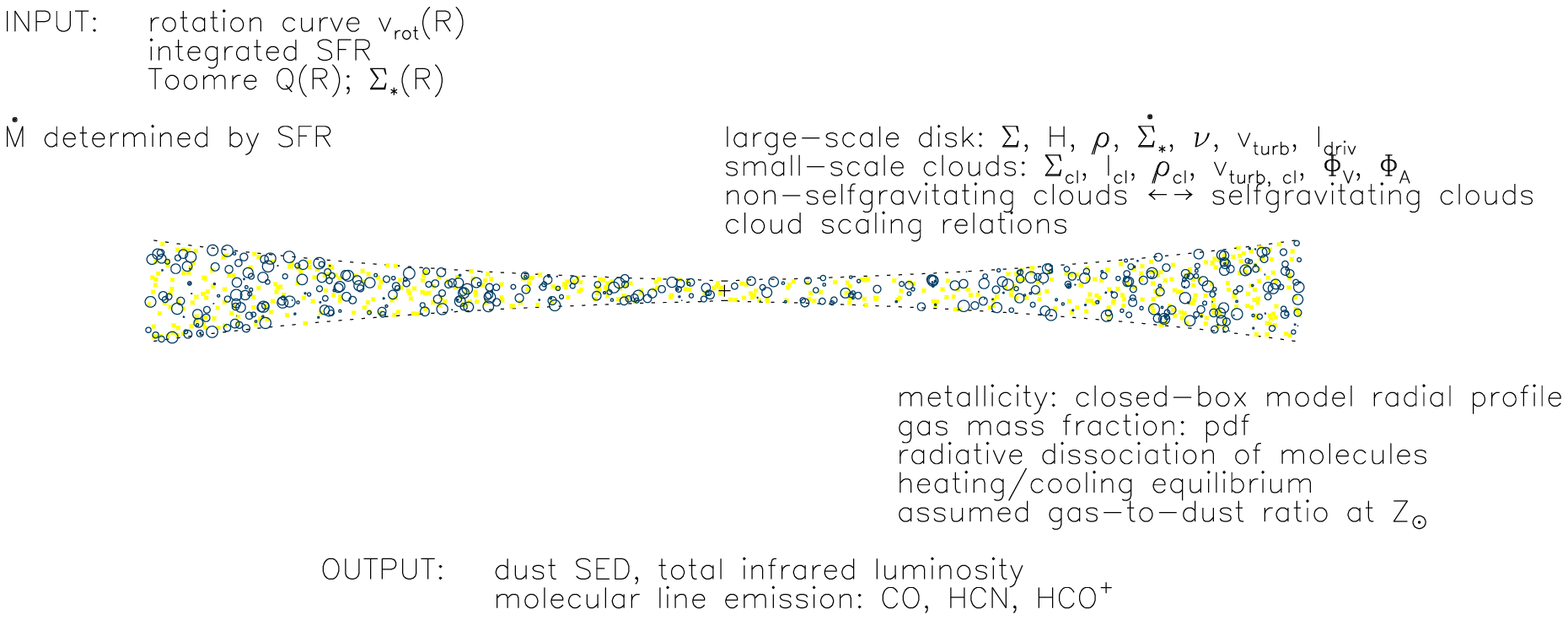}}
  \caption{Schematic outline of our clumpy star-forming galactic disk model.
  \label{fig:bild1}}
\end{figure*}

The outline of our model/method is presented in Fig.~\ref{fig:bild1}.

The VB03 model yields the following system of equations to describe a turbulent clumpy galactic accretion disk:
\begin{equation}
\label{eq:diskeq}
\begin{gathered}
\nu  =  v_{\rm turb} l_{\rm driv}\ ,\\
\nu \Sigma  =  \frac{\dot{M}}{2\pi}\ ,\\
\Sigma  =  \rho\,H\ ,\\
p_{\rm turb}=\rho v_{\rm  turb}^{2} = \frac{\pi}{2} G \Sigma ( \Sigma + \Sigma_{*} \frac{v_{\rm turb}}{v_{\rm disp}^{*}})~,\\
Q  =  \frac{v_{\rm turb} \Omega}{\pi G \Sigma}\ ,\\
\Sigma \nu \frac{v_{\rm turb}^{2}}{l_{\rm driv}^{2}} =  \xi\,\dot{\Sigma}_{*} + \frac{\dot{M}}{2\pi} \Omega^2\ ,\\
\dot{\Sigma}_{*} =  \Phi_{\rm V} \frac{\rho}{t_{\rm SF}^{\rm l}} l_{\rm driv}\ ,\\
t_{\rm SF}^{\rm l}=\sqrt{\frac{3\pi}{32G\rho_{\rm cl}}}=t_{\rm turb}^{\rm l}\ .
\end{gathered}
\end{equation}
The meaning of the variables is given in Table~\ref{tab:parameters}.
For the global comparison between the observed and the model radial profiles, we solve the set of equations
given above numerically.

The free parameters of the analytical model are the Toomre parameter $Q$ and the disk mass accretion rate $\dot{M}$.
For the local spirals, we set $Q$ to values derived in Vollmer \& Leroy (2011) (see Table~\ref{tab:gleroy}). For the ULIRGs, high-z star-forming galaxies,
and submillimeter galaxies, we assume a constant $Q=1.5$ (see Tables~\ref{tab:gulirg}, \ref{tab:gphibbs}, \ref{tab:gbzk}). The mass accretion rate $\dot{M}$
is determined by the total star-formation rate of the galactic disks (see Eqs.~\ref{eq:diskeq}).

For the calculation of the infrared dust emission and molecular line emission from the galactic disk, we divide the ISM into two density bins
(see Sect.~\ref{sec:scaling}):
(i) non-self-gravitating clouds with densities $\rho_2$ equal or higher than that of the cool neutral medium: 
${\rm max}(n_{\rm CNM},100~{\rm cm^{-3}}) \leq \rho_2 \leq \rho/\Phi_{\rm V}$ and
(ii) self-gravitating clouds with densities $\rho_1 \geq \rho/\Phi_{\rm V}$.
The calculation of the molecular emission is done in the following steps:
\begin{enumerate}
\item
Calculation of the temperatures of the dust and cool neutral medium $T_{\rm CNM}$ and the self-gravitating clouds $T_{\rm cl}$ according 
to Sect.~\ref{sec:gdtemp}.
Even if the gas is not molecular at densities of $100$~cm$^{-3}$, we use the CO, H$_2$, H$_2$O cooling function for model consistency,
because we want to apply the same code for local spirals and ULIRGs. For clouds of lowest densities in local galaxies,
this yields temperatures between $80$ and $150$~K, well in the range of observed CNM temperatures (Kulkarni \& Heiles 1987; Dickey \& Lockman 1990).
In addition, in these clouds, almost all H$_2$ and CO is dissociated (Sect.~\ref{sec:dissociation}) and the molecular line emission 
is very weak (Sect.~\ref{sec:lineemission}) due to the low gas density in these low-density clouds. Therefore, the high uncertainty
of the calculated temperature of these clouds does not affect the total molecular line emission of the galactic disk.
\item
We assume different scaling relations for the two density regimes: 
(i) non-self-gravitating clouds and (ii) self-gravitating clouds (Sect.~\ref{sec:scaling}).
\item
Within the two density regimes, the properties (size, column density, turbulent velocity, temperature) of clouds with over-densities 
$2^N$ with $N=1,2,...$ are calculated according to the scaling relations. The mass fraction $\frac{\Delta M}{M}$ 
of each density bin is calculated via the pdfs described in Sect.~\ref{sec:gasfrac}. 
This procedure insures mass conservation, that is, $\sum_{i=1}^N  \big( \frac{\Delta M}{M} \big)_i =1$.
\item
The fraction of molecular mass $f_{\rm H_2}$  and mass emitting in the molecular line $f_{\rm CO}$ is determined according 
to Sect.~\ref{sec:dissociation}. In the absence of a detailed theory of HCN dissociation, we use the same dark gas fraction 
for HCN as for CO.
\item
The molecular line emission ($T^*_{\rm A}$) is calculated according to Sect.~\ref{sec:lineemission} using the density, size, 
and temperature of the clouds in each density bin.
\item
The total dust infrared emission at a given galactic radius $R$ is determined by Eq.~\ref{eq:idust},
the TIR luminosity is calculated via Eqs.~\ref{eq:idust1} and \ref{eq:idust2}, and the total CO or HCN molecular line emissions are calculated in the following way:
\begin{equation}
I_{\rm mol}(R)=\sum_{\rm i=1}^{\rm N}  \big(\Phi_{\rm A} \big)_i\, (f_{\rm CO})_i\, (\Delta T^*_{\rm A})_i\, 2.35\,(v_{\rm turb,cl})_i\,,
\end{equation}
where the factor $2.35$ links the turbulent velocity to the linewidth.
The cosmic and local dust infrared backgrounds are taken into account. 
The total flux density is calculated by integrating the radial profile
\begin{equation}
F=2\,\pi \int_{0}^{R_0} I(R)\,R\,{\rm d}R\ .
\end{equation}
The molecular fraction is
\begin{equation}
\frac{\Sigma_{\rm H_2}}{\Sigma}=\sum_{\rm i=1}^{\rm N} \big( \frac{\Delta M}{M} \big)_i\,(f_{\rm H_2})_i
\end{equation} 
The total molecular gas mass is 
\begin{equation}
M_{\rm H_2}=2\,\pi \int_{0}^{R_0} \Sigma_{\rm H_2}\,R\,{\rm d}R\ .
\end{equation}
\end{enumerate}
The H{\sc i} surface density is $\Sigma_{\rm HI}=\Sigma-\Sigma_{\rm H_2}$ and the total H{\sc i} mass is
\begin{equation}
M_{\rm HI}=2\,\pi \int_{0}^{R_0} \Sigma_{\rm HI}\,R\,{\rm d}R\ .
\end{equation}
The upper limit of the galactic radius is $R_0=4.5 \times R_*$ for the local spiral galaxies and high-z star-forming galaxies 
(van der Kruit \& Searle 1982) and $R_0=3.4 \times l_*$ for the ULIRGs and submillimeter galaxies, where $l_*$ is the exponential scalelength of the stellar disk.
The latter normalization was chosen to reproduce the spatial extents of ULIRGs (Downes \& Solomon 1998).

\section{Uncertainties \label{sec:uncertain}}

All the different steps of our model calculations have associated uncertainties. In the following, we discuss the
main uncertainties for each step.
\begin{enumerate}
\item
Analytical model (Sect.~\ref{sec:model}, Eq.~\ref{eq:diskeq}): the constant $\xi$ that links the star-formation rate to the energy injection rate
is calibrated to the Galaxy. We do not know if this calibration also holds for the densest starburst regions; for example, in ULIRGs.
In the presence of outbreaking supernovae giving rise to a galactic wind, one might expect a lower value of $\xi$.
Since the total and molecular gas masses are only weakly dependent on $\xi$, we expect an uncertainty of the order of a few $10$\,\% on
the determination of the gas mass associated to the uncertainty of $\xi$. For our calculations, we regard $\xi$ as constant. 

The mass accretion rate $\dot{M}$ is determined by the total star-formation rate
which is typically uncertain to a factor of $1.5$ (Leroy et al. 2008, Genzel et al. 2015). Due to an uncertain AGN contribution to
the IR luminosity, the uncertainty is expected to be higher for ULIRGs. Since $\dot{M}_* \propto \dot{M}^{0.4\ {\rm to}\ 0.6}$ (Vollmer \& Leroy 2011),
we estimate the uncertainty associated to the determination of $\dot{M}$ from the integrated star-formation rate to be of the
order of a few tens of percent. The uncertainty on the derived total gas mass is of the same order. 

Another uncertainty comes from the unknown velocity dispersion of the disk stars or the thickness of the galactic stellar disk.
The applied relation $H_*/l_*=7.3$ for local spiral galaxies (Kregel et al. 2002) has a dispersion of $\sim 30$\,\%. We estimate the uncertainty 
on the gas pressure and thus the gas surface density and total gas mass to be small for local spirals. However, it is expected to be
higher for ULRIGs, high-z star-forming galaxies, and submillimeter galaxies that might deviate from this relation.

The mass and momentum conservation equation $\dot{M}/(2\,\pi)=\nu \Sigma$ represents another source of uncertainty. 
This equation implies that the mass loss through star formation is balanced by radial or external gas accretion.
In most cases, radial accretion within the disk is negligible.
In the absence of this external accretion, we expect that the gas surface density in the central part of the galaxy decreases
because of star formation. We take this phenomenon into account by increasing $Q$ towards the galaxy center in some of the local 
spiral galaxies. For ULIRGs, high-z star-forming galaxies and submillimeter galaxies we assume a constant $Q$. This might be justified
by a large gas mass and star-formation rate of these galaxies. Vollmer \& Leroy (2011) have shown that the system of equations 
including the mass and momentum conservation equation describes the radial H$_2$, star formation, turbulent velocity, and molecular 
fraction profiles of local spiral galaxies in a satisfying way.

For the metallicity, we use a simple closed-box model.
As long as the star-formation timescale is smaller than the gas accretion timescale, this assumption is justified.
In ULIRGs and submillimeter galaxies, this might not be the case, leading to an overestimation of the metallicity and thus the molecular 
abundances $X$. This will increase the molecular line emission $\propto X^{0.4}$ (Scoville \& Solomon 1974).

We are thus confident that the model equations describe the physics of a star-forming galactic disk within a factor of $1.5$-$2$.

\item
Density probability distribution functions (pdfs) (Sect.~\ref{sec:gasfrac}): the role of gas self-gravitation is not included in the pdf of Padoan et al. (1997).
In the presence of self-gravitation, the shape of the density pdf is altered at high densities, that is, the mass fraction of high-density
gas is higher in a pdf including self-gravitation than in a pdf without self-gravitation (see, e.g., Schneider et al. 2015).
We estimate the uncertainty of the mass calculation due to different pdfs to be a factor of $2$.

\item
Gas and dust temperature calculations (Sect.~\ref{sec:gdtemp}): the main uncertainty comes here from the gas cooling function, which is
assumed to be dominated by CO, H$_2$, and H$_2$O line cooling. Based on Neufeld et al. (1995), this leads to a possible underestimation of
the total cooling by up to a factor of $3$ to $4$. On the other hand, the comparison with the cooling function proposed by Goldreich (2001)
shows an overall discrepancy of a factor of $2$. We believe that, on average, our cooling function is uncertain to a factor of $\sim 2$.
The resulting gas temperature is uncertain by approximately a factor of $\sim 1.3$.

\item
Molecular line emission (Sect.~\ref{sec:lineemission}): sources of uncertainties are manyfold: (i) we only calculate a single transition at a time
and not the full system of transitions of a given molecule, (ii) we assume spherical geometry of all clouds, whereas a sheet- or filament-like geometry 
is not excluded for non-self-gravitating clouds. This directly affects the area-filling factor $\Phi_{\rm A}$, (iii) the area-filling factor
is small enough so that line emission of a smaller cloud is not absorbed by a larger cloud. This effect might only play a role in ULIRGs,
(iv) we assume abundances that are proportional to the closed-box metallicity; we therefore neglect gas depletion and chemistry effects.
We had to modify our simple closed-box model for starburst galaxies to take these effects into account (Eq.~\ref{eq:metulirgs}).
Of all these sources of uncertainty we estimate (iv) to be most important, leading to an uncertainty associated to the molecular 
line emission of a factor of a $2$ for CO and of a few for HCN.

\item
H$_2$ and CO dissociation (Sect.~\ref{sec:dissociation}): we treat the photo-dissociated region in molecular gas clouds in a crude way.
It is assumed that the outer H{\sc i} envelope has the density of the CNM, which has not to be the case. The dark gas fraction (H$_2$ 
without CO) also depends on the density of the cloud envelope or the density profile of the cloud. We estimate the uncertainty of the
H$_2$ and CO-emitting mass calculation to be approximately a factor of $2$.

\item
Scaling relations (Sect.~\ref{sec:scaling}): whereas the scaling relations for self-gravitating GMCs are relatively robust, those for diffuse
(non-self-gravitating) clouds are less well established. This will lead to higher uncertainties for galaxies with very dense gas disks,
especially for ULIRGs, where the molecular line emission from the diffuse clouds dominates the total emission. We estimate the uncertainty 
on the calculation of the mass fraction at a given density due
to the adopted scaling relations to be a factor of $2$ for the ULIRGs and $30$-$50$\,\% for the high-z star-forming and submillimeter galaxies.

\end{enumerate}
In summary, the most important uncertainties are due to the analytical model, the choice of the pdf, and the adopted molecular abundances.
All uncertainties are of the order of a factor of $2$.
In addition, the uncertain scaling relations of ULIRGs might add an additional factor to the molecular line emission calculation of ULIRGs.

\section{Galaxy samples \label{sec:samples}}

The input parameters of our model are: the exponential scale-length of the stellar disk $l_*$, the stellar mass $M_*$,
the rotation velocity $v_{\rm rot}$, and the star-formation rate $\dot{M}_*$. We apply our model to four galaxy samples 
for which these parameters are determined observationally in a uniform way: local spiral galaxies, local ultraluminous 
infrared galaxies (ULIRGs), high-z star-forming galaxies, and submillimeter galaxies. Following Boissier et al. (2003), for all galaxies, we assume 
a rotation curve of the form
\begin{equation}
v_{\rm rot}(R)=v_{\rm flat}\big(1-\exp(-\frac{R}{l_{\rm flat}})\big)\ ,
\end{equation}
where $v_{\rm flat}$ and $l_{\rm flat}$ represent the velocity at
which the rotation curve is flat and the length scale over which it
approaches this velocity, respectively. Moreover, we assumed $\delta=5$ for all galaxies.
The stellar surface density is assumed to be of the form:
\begin{equation}
\Sigma_*=\Sigma_{*,0} \exp(-\frac{R}{l_*})
,\end{equation}
with $M_*=2\,\pi \int_0^{R_0} \Sigma_* R\,{\rm d}R$.

\subsection{Local spiral galaxies}

The sample of local spiral galaxies (Table~\ref{tab:gleroy}) is taken from Leroy et al. (2008).
For this work, we did not consider dwarf galaxies, which will be the subject of a subsequent article. 
The spiral galaxies have total stellar masses in excess of $10^{10}$~M$_{\odot}$. The gas masses are derived from
IRAM 30m CO(2--1) HERACLES (Leroy et al. 2009) and VLA H{\sc i} THINGS (Walter et al. 2008) data. The star-formation rate
was derived from Spitzer MIR and GALEX UV data (Leroy et al. 2008). The total infrared luminosities are taken from Dale et al. (2012).
Following Vollmer \& Leroy (2011), the Toomre parameter of NGC~628, NGC~3198, NGC~5194, and NGC~7331 was set to $Q(R)=Q+3\,\exp(-2\,R/l_*)$, that of
NGC~3351 by $Q(R)=Q-4\,\exp\big(-(2\,R/l_*)^2\big)$. The Toomre parameter $Q$ was assumed to be constant for all other galaxies (see Table~\ref{tab:gleroy}).

\subsection{Local ULIRGs}

The ULIRG sample (Table~\ref{tab:gulirg}) is taken from Downes \& Solomon (1998). These authors derived the spatial extent, rotation velocity, gas mass, and dynamical mass 
$M_{\rm dyn}$ for local ULIRGs from PdB interferometric CO-line observations. The total infrared luminosities are taken from Graci\'{a}-Carpio et al. (2008).
We adopted the star-formation rates based on FIR data from Graci\'{a}-Carpio et al. (2008).
We calculated the stellar mass as $M_*=M_{\rm dyn}-M_{\rm gas}$ and assumed that the stellar scale length is approximately equal to the observed CO scale length.

\subsection{High-z star-forming galaxies}

The high-z star-forming sample (Table~\ref{tab:gphibbs}) is taken from PHIBSS (Tacconi et al. 2013), the IRAM PdB high-z blue sequence CO(3--2) survey of the molecular 
gas properties in massive, main-sequence star-forming galaxies at $z=1$-$1.5$. For our purpose, we only took the disk galaxies from PHIBSS. The stellar masses given by 
Tacconi et al. (2013) were derived from SED fitting, assuming a Chabrier IMF. Following Genzel et al. (2010), we calculated the star-formation rate from the total infrared luminosity using $\dot{M_*}(1~{\rm M}_{\odot}{\rm yr}^{-1})=10^{-10} L_{\rm TIR}({\rm L}_{\odot})$.
Their star-formation rates are based on the sum of the observed UV- and IR-luminosities, or an extinction-corrected H$\alpha$ luminosity. 
Their half-light radii were derived  from Sersic  fits  to  the HST ACS  and/or WFC3  CANDELS data (Grogin et al. 2011).
To estimate the characteristic circular velocities, Tacconi et al. (2013) took the isotropic virial estimate $v_{\rm circ}=\sqrt{3/(8\,\ln 2)}\Delta v_{\rm FWHM}$,
where $\Delta v_{\rm FWHM}$ is the CO(3--2) linewidth for unresolved galaxies without a velocity gradient, and $v_{\rm circ}=1.3\,\big(\Delta v_{\rm blue-red}/(2 \sin(i))\big)$
if a velocity gradient ($\Delta v_{\rm blue-red}$) indicative of rotation is detected in a galaxy with an inclination $i$.
Since the inclination angle is difficult to determine in these high-z star-forming disk galaxies we adopted the following strategy for
the determination of the rotation velocity: if $v_{\rm rot} < \sqrt{(M_{\rm gas}+M_*)\,G/(2\,l_*),}$ the assumed rotation velocity is 
$v_{\rm rot}=\sqrt{(M_{\rm gas}+M_*)\,G/(2\,l_*)}$; otherwise $v_{\rm rot}=v_{\rm circ}$. In this way, the rotation velocity of $15$ out of $45$ galaxies
was increased by more than $50$\,\%. 

\subsection{Submillimeter galaxies}

The smm-galaxy sample (Table~\ref{tab:gbzk}) was drawn from Genzel et al. (2010). 
The total infrared luminosities are based on the $850$~$\mu$m flux densities (Genzel et al. 2010). We calculated the star-formation rate
from the total infrared luminosity using $\dot{M_*}({\rm M}_{\odot}{\rm yr}^{-1})=1.7 \times 10^{-10} L_{\rm TIR}({\rm L}_{\odot})$.
Stellar masses are from the SED fits in Erb et al. (2006) 
and F\"{o}rster Schreiber et al. (2009) and rotation velocities and half-light radii
are from the data in the same references with the methods discussed in F\"{o}rster Schreiber et al. (2009).
When the stellar mass was not available, it was set to $10^{11}$~M$_{\odot}$.
The CO-line observations were made in the CO(2--1), CO(3--2), and CO(4--3) lines.

\section{Results \label{sec:results}}

The aim of the present work is to directly compare infrared and molecular line luminosities. 
Before doing so, the distribution of the galaxy metallicities, dust SEDs, TIR luminosities, and dust temperatures are presented.

\subsection{Metallicity \label{sec:metals}}

As described in Sect.~\ref{sec:model}, we use a simple close-box model for the metallicity (Eq.~\ref{eq:zzodot}).
The oxygen abundance is then calculated by 12+log(O/H)=$\log(Z/Z_{\odot})+8.7$, with 12+log(O/H)=8.7
being the solar oxygen abundance (Asplund et al. 2005).

\begin{figure}
  \centering
  \resizebox{\hsize}{!}{\includegraphics{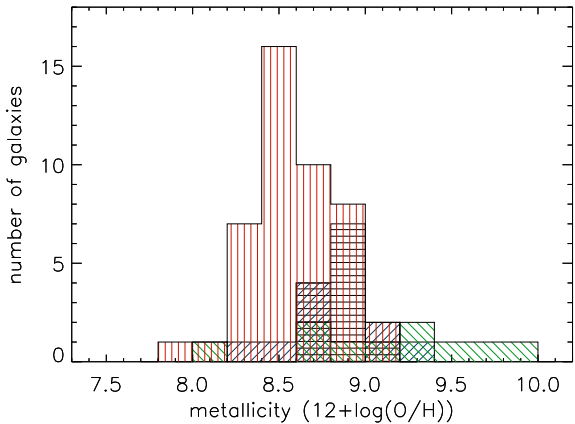}}
  \caption{Distribution of galaxy metallicities. Black solid line: local spiral galaxies. Blue dotted line: ULIRGs. Green dashed line: smm-galaxies. 
    Red dash-dotted line: high-z star-forming galaxies. The vertical black line corresponds to the solar metallicity of 12+log(O/H)=8.7.
  \label{fig:metallicities}}
\end{figure}

For the global mean metallicity, the average is weighted by optical emission from a disk where 
a uniform distribution of stars is mixed homogeneously with dust (``mixed'' model of McLeod et al. 1993):
\begin{equation}
\langle Z \rangle=\frac{\int_0^{R_{0}} Z\,\dot{\Sigma}_*\,\frac{1-\exp(-\tau)}{\tau}\,{\rm d}R}{\int_0^{R_{0}} \dot{\Sigma}_*\,\frac{1-\exp(-\tau)}{\tau}\,{\rm d}R}\ ,
\end{equation}
with $\tau=\Sigma/(7.5~{\rm M}_{\odot}{\rm pc}^{-2})$.
The mean oxygen abundances resulting from our modeling $\langle$12+log(O/H)$\rangle$ are presented in Fig.~\ref{fig:metallicities}.

The oxygen abundances of our local spiral galaxy sample range is $8.7 \leq $12+log(O/H)$\leq 9.0$. 
This is close to the findings of Moustakas \& Kennicutt (2006) who studied
$14$ nearby disk galaxies with integrated spectrophotometry and observations of more than $250$ individual H{\sc ii}
regions; their oxygen abundances based on the McGaugh (1991) calibration also fall in the range $8.6 \leq $12+log(O/H)$\leq 9.0,$ with
most of the galaxies having an oxygen abundance of $8.8$-$9.0$.

The oxygen abundances of our ULIRG sample fall in the range $8.6 \leq $12+log(O/H)$\leq 9.4$.
The mean oxygen abundance of the sample is 12+log(O/H)$=9.0 \pm 0.3$.
Compared to the oxygen abundances determined by Rupke et al. (2008) and Kilerci Eser et al. (2014), with lower and upper limits 
of $8.4$ and $9.0$, respectively, four out of nine ULIRGs show model oxygen abundances in excess of $9.0$.
The model metallicities are thus up to a factor of $2.5$ higher than the observed metallicities derived from optical emission line
diagnostics. 

The oxygen abundances or metallicities of the smm-galaxy sample are in the range $8.7 \leq $12+log(O/H)$\leq 9.4$, that is, the 
metallicities are mostly supersolar. Three out of ten galaxies show metallicities in excess of 12+log(O/H)$=9.4$.
We decided not to arbitrarily modify these obviously too high model metallicities.
The mean oxygen abundance of the smm-galaxy sample is 12+log(O/H)$=9.3 \pm 0.4$.
Swinbank et al. (2004) found slightly subsolar metallicities in their sample of 30 smm-galaxies at a median redshift of 
$z \sim 2.4$. Tecza et al. (2004) found a supersolar oxygen abundance of 12+log(O/H)=9.0 for SMM J14011+0252 at z=2.57.
Nagao et al. (2012) found a solar metallicity of the submillimeter galaxy LESS J033229.4--275619 at z=4.76. The model
metallicities of our smm-galaxy sample are thus a factor $2$-$4$ higher than the observed metallicities.

The oxygen abundances of our high-z star-forming galaxy sample range between $8.0 \leq $12+log(O/H)$\leq 9.0$.
The mean oxygen abundance of the sample is 12+log(O/H)$=8.6 \pm 0.2$, thus very close to solar metallicity.
This is consistent with the results of Shapley et al. (2004) who found solar and possibly supersolar metallicities
in high-z star-forming galaxies. It is also consistent with the mean metallicity of 12+log(O/H)$=8.7 \pm 0.2$ of the
sample of 50 galaxies at $z \sim 1.2$ in the MASSIV survey (Queyrel et al. 2012).

We conclude that the integrated model metallicities of the local spiral and high-z star-forming galaxies are in
good agreement with observations. On the other hand, the model metallicities of half of the galaxies of the
ULIRG and smm-galaxy sample show metallicities that are $2$-$4$ times higher than the observed metallicities.
This leads to a potential overestimation of the molecular line emission ($\propto X^{0.4}$; Scoville \& Solomon 1974) by a factor of $1.3$-$1.7$.

\subsection{Dust SED, TIR luminosity, and dust temperature \label{sec:sedtir}}

For the direct comparison between the model and observed dust IR spectral energy distributions (SED), we extracted
all available photometric data points for our galaxy samples from the CDS VizieR database\footnote{\tt http://vizier.u-strasbg.fr/viz-bin/VizieR}. Since the flux densities
in the different catalogs are determined within different apertures, we only take the highest flux densities for
a given wavelength range around a central wavelength $\lambda_0$ ($0.75 \leq \lambda/\lambda_0 \leq 1.25$).
In this way, only the outer envelope of the flux density distribution is selected. We did not attempt to remove flux densities
below this outer envelope, which most probably corresponds to apertures that do not include the whole object or are erroneous. 
Since our dust model does not include stochastically heated small grains and PAHs, the observed IR flux densities
for $\lambda \la 50~\mu$m cannot be reproduced by the model. Flux densities at wavelengths $\ge 70~\mu$m are available for all local spirals/ULIRGs, 9 out of 10 smm-galaxies, and 30 out of 44 high-z star-forming galaxies.

The model dust IR SEDs of the local spiral galaxies reproduce the observed SEDs  very well (Fig.~\ref{fig:IRspectra_spirals}).
Only the flux densities at $\lambda > 200~\mu$m NGC~4736, NGC~4535, NGC~6946, and NGC~3627 are somewhat overestimated by the model.

The models of the ULIRGs reproduce the existing observations  very well (Fig.~\ref{fig:IRspectra_ulirgs}). The comparison
between the model and observed SED is difficult for Arp~220, because the observed SED contains the whole system,
whereas the model SEDs are made separately for the Disk, Western, and Eastern nuclei. For the comparison, we assumed
that $30$\,\%, $20$\,\%, and $30$\,\% of the observed total flux densities are emitted by the Disk, Western, and Eastern nuclei, respectively.

The models of the smm-galaxies reproduce the existing observations in a satisfactory way (Fig.~\ref{fig:IRspectra_smm}). 
Only for SMM~J123549+6215 are the infrared flux densities underestimated by approximately a factor of $2$.

As for the local spiral galaxy sample, the model dust IR SEDs of the high-z star-forming galaxies reproduce the observed 
SEDs  well (Fig.~\ref{fig:IRspectra_phibss1}) at almost all wavelengths, especially those of the $z \sim 1.5$ (EGS) sample.
The model SEDs of the $z \sim 2.5$ sample underestimate the observed SEDs by up to a factor of $2$ for BzK~4171 and BzK~16000 and
overestimate the observed SEDs by approximately a factor of $2$ for BzK~17999.

\begin{figure}
  \centering
  \resizebox{\hsize}{!}{\includegraphics{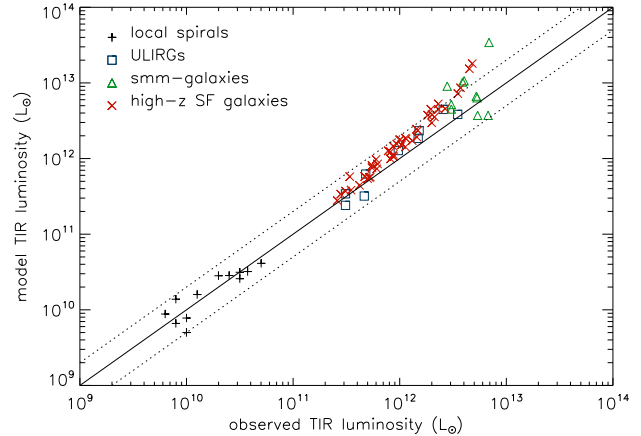}}
  \caption{Model total infrared luminosity as a function of the observed TIR luminosity of the galaxies. Black plus symbols represent local spiral galaxies,
blue boxes represent ULIRGs, green triangles represent smm-galaxies and red crosses represent high-z star-forming galaxies.
  \label{fig:tirlum}}
\end{figure}

The comparison of the model total IR luminosities (from $10~\mu$m to $1000~\mu$m) to the observed total IR luminosities is presented
in Fig.~\ref{fig:tirlum}. The model reproduces the total IR luminosities within a factor of $2$. Only for two
smm-galaxies and two high-z star-forming galaxies does the model underestimate the total IR luminosities by more than a factor of $\sim 2.5$.

\begin{figure}
  \centering
  \resizebox{\hsize}{!}{\includegraphics{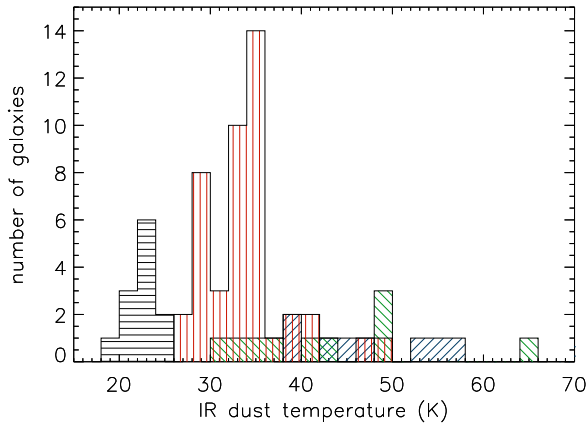}}
  \caption{Model distribution of dust temperatures. The black solid line represents local spiral galaxies, the blue dotted line represents ULIRGs the green dashed line represents smm-galaxies and the 
    red dash-dotted line represents high-z star-forming galaxies. 
  \label{fig:tdust}}
\end{figure}

We fitted modified Planck functions with $\beta=1.5$ (see Sect.~\ref{sec:dustemission}) to the model dust IR SEDs to derive dust temperatures.
The modified Planck functions are shown as red dashed lines in Figs.~\ref{fig:IRspectra_spirals} to \ref{fig:IRspectra_phibss4}.
The resulting distribution of dust temperatures for the different galaxy samples are presented in Fig.~\ref{fig:tdust}.

The dust temperatures of the local spiral galaxies range between $19$ and $24$~K. This is in excellent agreement with Dale et al. (2012; Fig.~10),
which is not surprising given the good fit to the observed IR dust SEDs.

The dust temperatures of the ULIRGs range between $39$ and $72$~K with a mean of $50 \pm 11$, 
in reasonable agreement with the results of Symeonidis et al. (2013) who found that
the majority of (U)LIRGs at all redshifts have mean dust temperatures between $25$ and $45$~K using IRAS- and Herschel-selected samples,
and Hwang et al. (2012) who found a dust temperature range between $35$ and $43$~K based on Herschel IR SEDs.
Our dust temperatures are somewhat lower than the temperature distribution ($61 \pm 9$) found by Klaas et al. (2001) for local ULIRGs.

The dust temperatures of the smm-galaxies range between $31$ and $64$~K. The smm-galaxy dust temperatures cover approximately the same
range as the ULIRG dust temperatures, but their mean dust temperature is somewhat smaller $43 \pm 10$~K compared to $50 \pm 11$~K for
the ULIRG sample. This is in good agreement with the dust temperatures ($30$-$45$~K) of smm-galaxies with IR luminosities $< 10^{13}$~L$_{\odot}$
of Hwang et al. (2010; Fig.~3).

The dust temperatures of the high-z star-forming galaxies range between $27$ and $48$~K with a mean of $33 \pm 4$~K.
This is in good agreement with (i) the mean temperature of the stacked $z \sim 1$ sample ($32 \pm 2$) of Magdis et al. (2012; Table~2)
and (ii) the mean temperature of $30$~K found for $z \sim 1$ sources by Magnelli et al. (2014; Eq.~4) based on Herschel observations.

We conclude that our model reproduces the dust IR SEDs of all galaxy samples.
The IR flux densities of one smm-galaxy and two high-z star-forming galaxies at $z \sim 2.5$ underestimated by a factor of $\sim 2$.
The derived dust temperatures of all galaxy samples are consistent with MIR and FIR observations.

\subsection{Molecular fraction and H{\sc i} mass \label{sec:molfrac}}

\begin{figure}
  \centering
  \resizebox{\hsize}{!}{\includegraphics{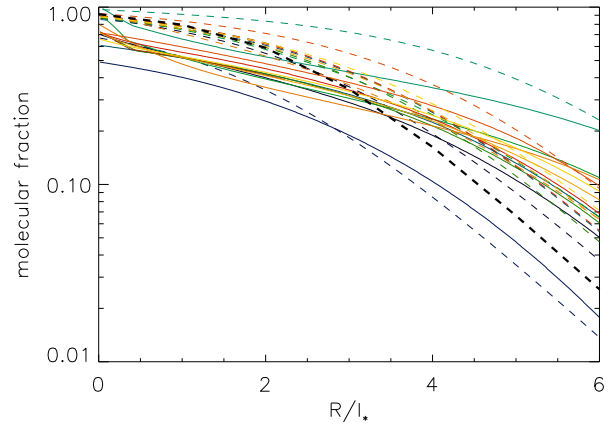}}
  \caption{Radial profiles of the molecular fraction $f_{\rm mol}$ for the local spiral galaxies. The solid line shows this model and the
dashed line shows $f_{\rm mol}=t_{\rm ff}/t{\rm mol}/(1+t_{\rm ff}/t_{\rm mol})$, where the free-fall and molecule-formation times
    are those of self-gravitating clouds (Vollmer \& Leroy 2011). The profile for each galaxy is shown in a different color.
    The observed relation $R_{\rm mol}=10.6 \exp(-R/0.21R_{25})$ (Leroy et al. 2008) is shown as a thick dashed line.
  \label{fig:ffmmooll}}
\end{figure}

In the present model, the molecular fraction is calculated for each cloud of scale $l_{\rm cl}$. It is determined by
the photodissocation of H$_2$ molecules or the finite lifetime of a cloud (see Sect.~\ref{sec:method}).
As a first step, we compare this molecular fraction to the molecular fraction defined by Vollmer \& Leroy (2011)
as $f_{\rm mol}=t_{\rm ff}/t_{\rm mol}/(1+t_{\rm ff}/t_{\rm mol})$, where the free-fall and molecule formation times
are those of self-gravitating clouds at a galactic radius $R$ (Fig.~\ref{fig:ffmmooll}). 
The mean deviation between the two molecular fractions is approximately $30$\,\%, with a maximum deviation of approximately $50$\,\%
at large galactic radii. At small galactic radii, the molecular fraction of the present model is $\sim 30$\,\% smaller, 
whereas at large galactic radii, it is up to $\sim 50$\,\% higher than that of Vollmer \& Leroy (2011).
We thus conclude that both prescriptions are comparable.
This is quite surprising, because the Vollmer \& Leroy (2011) prescription is only based on the properties of the
self-gravitating clouds. We interpret this result as evidence for the dominant role of self-gravitating clouds for
the formation of molecular hydrogen in local spiral galaxies.

\begin{figure}
  \centering
  \resizebox{\hsize}{!}{\includegraphics{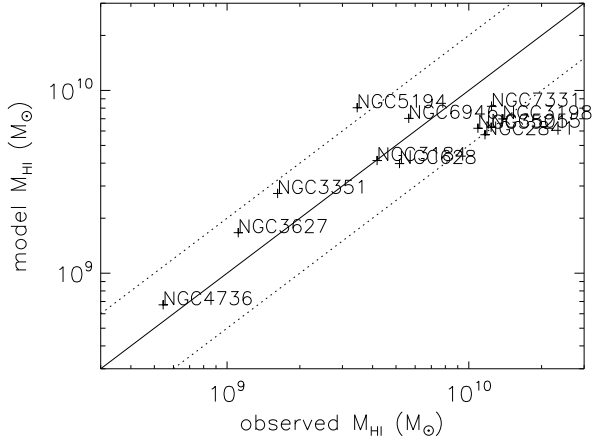}}
  \caption{Model H{\sc i} mass as a function of the observed H{\sc i} mass for the sample of local spiral galaxies from Walter et al. (2008).
    The solid line corresponds to equality, the dotted lines to factors of $1/2$ and $2$.
  \label{fig:HImasses}}
\end{figure}

Since the model yields the molecular fraction of the ISM, we can calculate the H{\sc i} surface density radial profile
and the H{\sc i} mass. The comparison between the observed and model H{\sc i} masses is
presented in Fig.~\ref{fig:HImasses}. The model reproduces the observed H{\sc i} masses within a factor of $2$.
Especially the H{\sc i} masses of the galaxies with $M_{\rm HI} > 10^{10}$~M$_{\odot}$ are underestimated by a factor of $\sim 2$.
This is intrinsic to the model and cannot be compensated by a modification of the free model parameters ($Q$ and $\delta$).

\begin{figure}
  \centering
  \resizebox{\hsize}{!}{\includegraphics{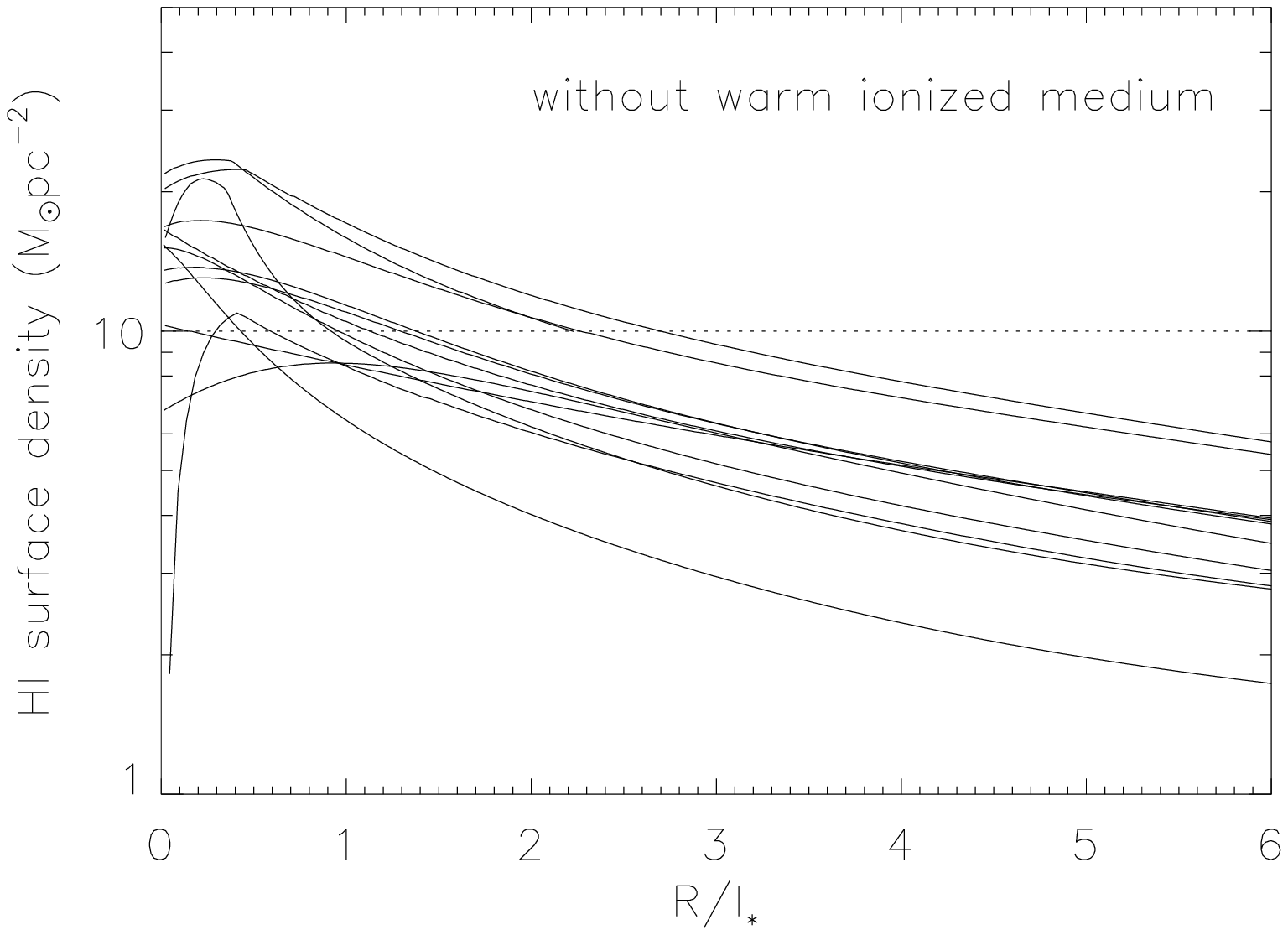}}
   \resizebox{\hsize}{!}{\includegraphics{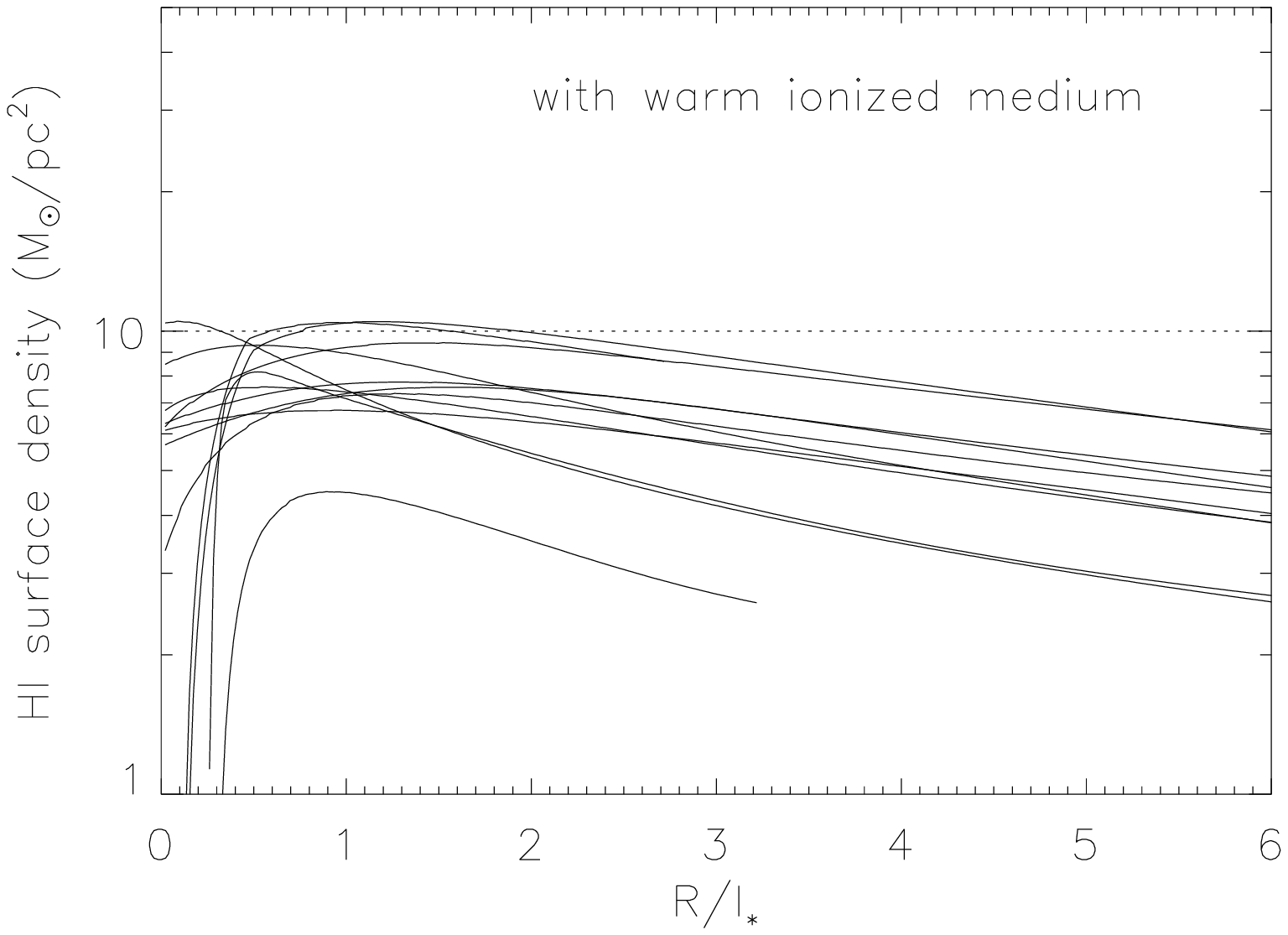}}
  \resizebox{\hsize}{!}{\includegraphics{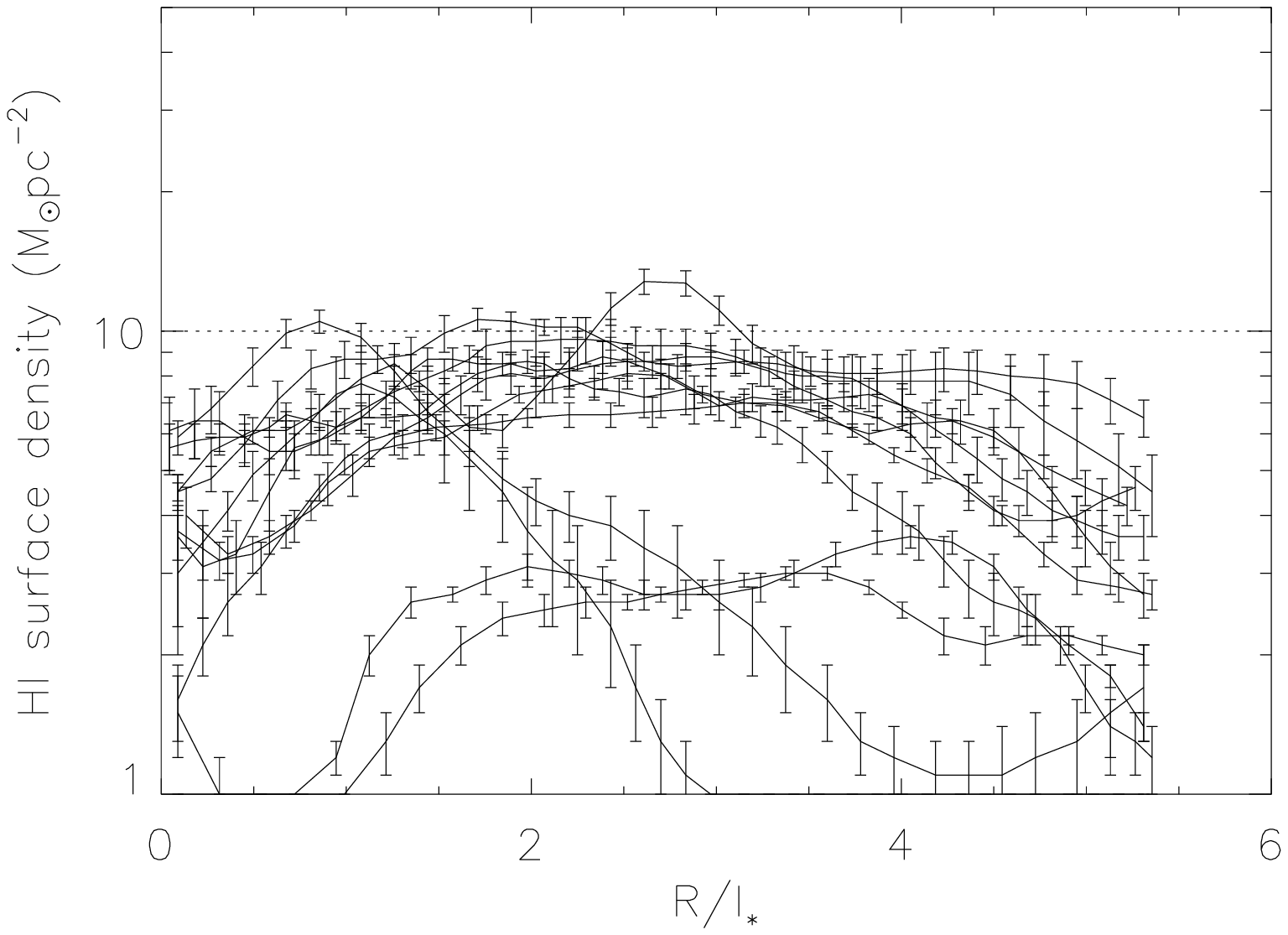}}
  \caption{Upper panel: model H{\sc i} radial profiles of the local spiral galaxies. The radial distance is normalized by the
    stellar scale length $l_*$. The middle panel shows model H{\sc i} including UV ionization with a constant H{\sc i} volume density, while the lower panel shows THINGS radial H{\sc i} profiles from Leroy et al. (2008) where we have assumed $R_{25}=4.5 \times l_*$.
    The atomic gas surface density saturates at $\sim 10$~M$_{\odot}$pc$^{-2}$ (dotted line).
  \label{fig:HIprofiles}}
\end{figure}

The associated radial H{\sc i} surface density profiles together with the observed H{\sc i} profiles from Leroy et al. (2008)
are presented in the upper and lower panels of Fig.~\ref{fig:HIprofiles}. Whereas the model profiles are all monotonically declining, most of the observed profiles
are constant in the range $1 \leq R/l_* \leq 4$. In addition, most of the observed profiles show a depression in the central part
of the galactic disks, whereas the models often show a maximum toward the galaxy center. The latter difference
can be explained by (i) an underestimation of the model molecular fraction or (ii) by the ionization of atomic hydrogen
in the inner part of the galactic disks, that is, the inclusion of a warm ionized medium into the model. 

In order to investigate the effect of ionization by
the UV radiation of massive stars in the galactic disk, we use the ionization-recombination equilibrium to calculate the
number surface density of ionized gas following Maloney (1993):
\begin{equation}
N_{\rm ion}=7.7 \times 10^{18} \frac{(\varphi/10^4)}{(n/10^{-2})}~{\rm cm}^{-2}\ ,
\end{equation}
where $\varphi$ is the flux of ionizing photons and $n$ the gas number density in cm$^{-3}$.
The surface density of ionized gas is then $\Sigma_{\rm ion}=1.36 \times m_{\rm p} \times N_{\rm ion}\ .$
Newborn massive stars are preferentially located in high-density regions, which they illuminate. As long as the
area filling factor of the high-density gas surrounding these stars is not too large, the UV photons can escape the H{\sc ii}
regions and ionize the warm neutral medium. This scenario is supported by H$\alpha$ observations of Thilker et al. (2002) and
Oey et al. (2007) who found a fraction of diffuse H$\alpha$ emission of $0.5$-$0.6$ for local spiral galaxies.
Based on Galactic H{\sc i} observations (Dickey \& Lockman 1990), we used a constant gas number density of $n=5$~cm$^{-3}$ for the warm neutral medium, which
corresponds to the observed midplane density of $0.6$~cm$^{-3}$ (Dickey \& Lockman 1990) and a volume-filling factor of $0.12$.
For the relation between the ionizing photon flux and the star-formation rate we use
\begin{equation}
\varphi = 2.3 \times 10^{-7} \big( \dot{\Sigma}_*/(1~{\rm M}_{\odot}{\rm pc}^{-2}{\rm yr}^{-1}) \big)~{\rm photons}\,{\rm cm}^{-2}{\rm sec}^{-1}\ ,
\end{equation}
based on the star-formation rate calibration of Kennicutt (1998).
We then calculated the H{\sc i} surface density as
\begin{equation}
\Sigma_{\rm HI}=(1-f_{\rm mol}) \times \Sigma - \Sigma_{\rm ion} \ . 
\end{equation}
The resulting H{\sc i} surface density is presented in the middle panel of Fig.~\ref{fig:HIprofiles}.
As the observed H{\sc i} surface density, the model H{\sc i} surface density is approximately constant $\Sigma_{\rm HI} \sim 10$~M$_{\odot}$pc$^{-2}$ between
$R=l_*$ and $R=4 \times l_*$.
We conclude that the inclusion of the warm ionized medium in the model leads to radial H{\sc i} surface density profiles that are well
comparable to observations. However, for the galaxies with the highest H{\sc i} masses (NGC~2841, NGC~3198, NGC~3521, and NGC~5055), 
the model H{\sc i} surface density profiles and total H{\sc i} masses are underestimated by approximately a factor of two.

\subsection{CO(1--0) and HCN(1--0) radial profiles for local spiral galaxies \label{sec:profiles}}

\begin{figure}
  \centering
  \resizebox{\hsize}{!}{\includegraphics{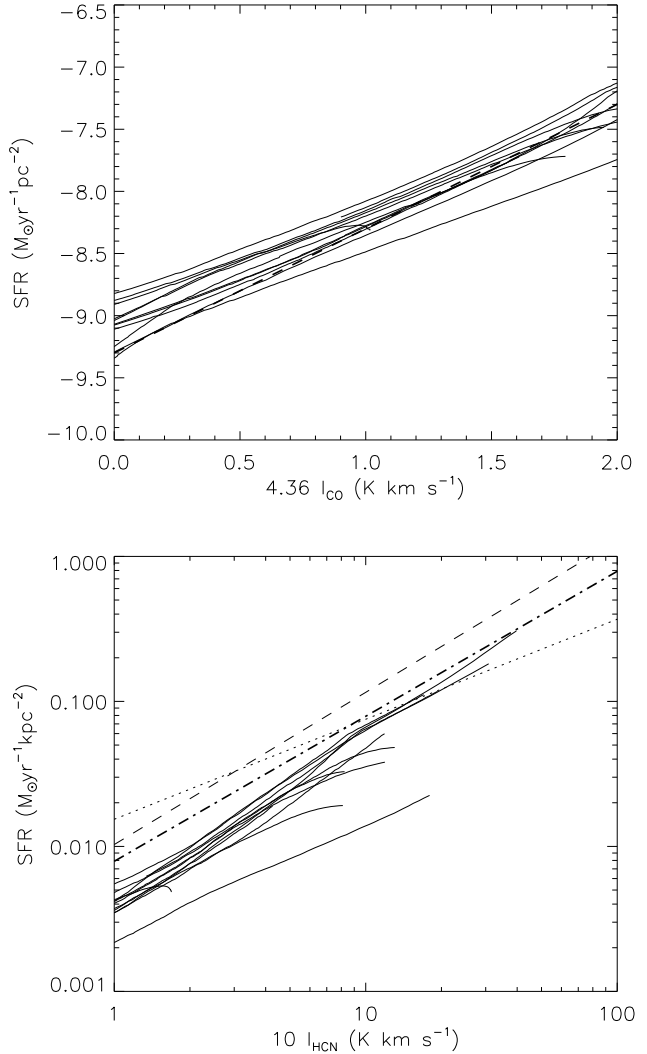}}
  \caption{Upper panel: model radial profiles of the star-formation rate surface density (thin solid lines) as a function of molecular gas surface density 
    $\Sigma_{\rm H_2}=4.36 \times I_{\rm CO}$ for the THINGS local spiral galaxies. The thick dashed line corresponds to a star-formation rate timescale of 
    $2 \times 10^{9}$~yr (Leroy et al. 2008). The lower panel shows model radial profiles of the star-formation rate surface density (thin solid lines) as a function of dense molecular 
    gas surface density $\Sigma_{\rm dense}=10 \times I_{\rm HCN}$ for the THINGS local spiral galaxies. The dashed line corresponds to
    the relation $\log(\dot{\Sigma}_*)=1.12 \times \log(\Sigma_{\rm dense})-2.10$ found by Graci\'a-Carpio et al. (2008).
    The dotted line corresponds to the relation $\log(\dot{\Sigma}_*)=0.69 \times \log(\Sigma_{\rm dense})-1.58$ found by Usero et al. (2015).
    The dash-dotted line corresponds to the Gao \& Solomon (2004) relation $\log(\dot{\Sigma}_*)=\log(\Sigma_{\rm dense})-2.05$.
    All observed relations are corrected for the model conversion factor between the total infrared luminosity and the star-formation rate,
    which is different from the value assumed in the literature (see Sect.~\ref{sec:profiles}).
  \label{fig:THINGS_profiles}}
\end{figure}

The CO and HCN emission can be spatially resolved in local spiral galaxies only (at $D=10$~Mpc, the beam sizes are $\sim 20''$ or $\sim 1$~kpc).
Whereas CO maps are frequently found in the literature (e.g., Wong \& Blitz 2002, Leroy et al. 2008), HCN maps are
rare (e.g. Chen et al. 2015, Bigiel et al. 2016). The H$_2$ and dense gas surface densities are usually calculated with a CO-H$_2$ conversion factor
of $\alpha_{\rm CO}=4.36$~M$_{\odot}$pc$^{-2}$/(K\,km\,s$^{-1}$) (e.g., Bolatto et al. 2013) and  $\alpha_{\rm HCN}=10$~M$_{\odot}$pc$^{-2}$/(K\,km\,s$^{-1}$)
(e.g., Gao \& Solomon 2004). The model SFR-$\Sigma_{\rm H_2}$ relation of the local spiral galaxies is presented in the top panel of Fig.~\ref{fig:THINGS_profiles} 
together with the observed relation $\dot{\Sigma}_*=\Sigma_{\rm H_2}/(2 \times 10^9~{\rm yr})$ (e.g., Bigiel et al. 2008, Leroy et al. 2008),
which corresponds to a constant star-formation rate timescale of $2 \times 10^{9}$~yr.
The model SFR-$\Sigma_{\rm H_2}$(CO) relations are well consistent with the observed relation within a scatter of approximately $0.2$~dex.
Overall, they show a somewhat flatter slope than the observed relation. 
The situation is more complex for the dense gas (HCN) than for H$_2$. The observed $\dot{\Sigma}_*$--$\Sigma_{\rm dense}$ relations show different slopes;
a linear slope was found by Gao \& Solomon (2004) and Garcia-Burillo et al. (2012) for integrated SFR and gas masses, whereas a sub-linear slope of $0.7$ was found
by Usero et al. (2015) using single pointings in nearby spiral galaxies. For a direct comparison between the model and observed
$\dot{\Sigma}_*$--$\Sigma_{\rm dense}$ relations, the conversion factor between the total infrared luminosity and the star-formation rate $c_{\rm TIR}$ must be 
taken into account, $SFR (M_{\odot} {\rm yr}^{-1})=c_{\rm TIR} L_{\rm TIR} ({\rm L}_{\odot})$. The different conversion factors are 
$c_{\rm TIR}=2.0,\ 1.7,\ 1.5 \times 10^{-10}$ used by Gao \& Solomon (2004), Garcia-Burillo et al. (2012), and Usero et al. (2015), respectively.
The model conversion factor for the local spiral galaxy sample is $c_{\rm TIR}=0.9 \times 10^{-10}$, that is, a factor of $\sim 2$ lower than the
values used in the literature. For a consistent comparison between the model and observations, we multiplied the star-formation rates of
the observed relation by $c_{\rm TIR}^{\rm model}/c_{\rm TIR}^{\rm obs}$. The resulting model $\dot{\Sigma}_*$--$\Sigma_{\rm dense}$ relation 
(lower panel of Fig.~\ref{fig:THINGS_profiles})
has a slope consistent with observations, but shows a negative offset of $0.2$--$0.3$~dex with respect to the observed relations.

We conclude that the resolved $\dot{\Sigma}_*$--$\Sigma_{\rm H_2}$ and $\dot{\Sigma}_*$--$\Sigma_{\rm dense}$ relations are consistent with available IR, CO, and HCN observations
within the uncertainties (see Sect.~\ref{sec:uncertain}).

\subsection{Integrated CO, HCN(1--0), and HCO$^+$(1--0) flux densities \label{sec:intflux}}

\begin{figure}
  \centering
  \resizebox{\hsize}{!}{\includegraphics{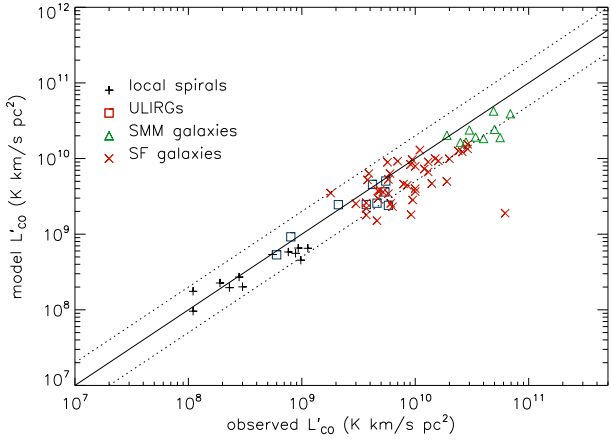}}
  \caption{Model CO luminosity as a function of the observed  luminosity for local spirals (CO(2--1)), local ULIRGs (CO(1--0)), submillimeter (CO(3--2)),
    and high-z star-forming galaxies (CO(3--2)).
  \label{fig:plots_HCNCO_CO}}
\end{figure}

\begin{figure}
  \centering
  \resizebox{\hsize}{!}{\includegraphics{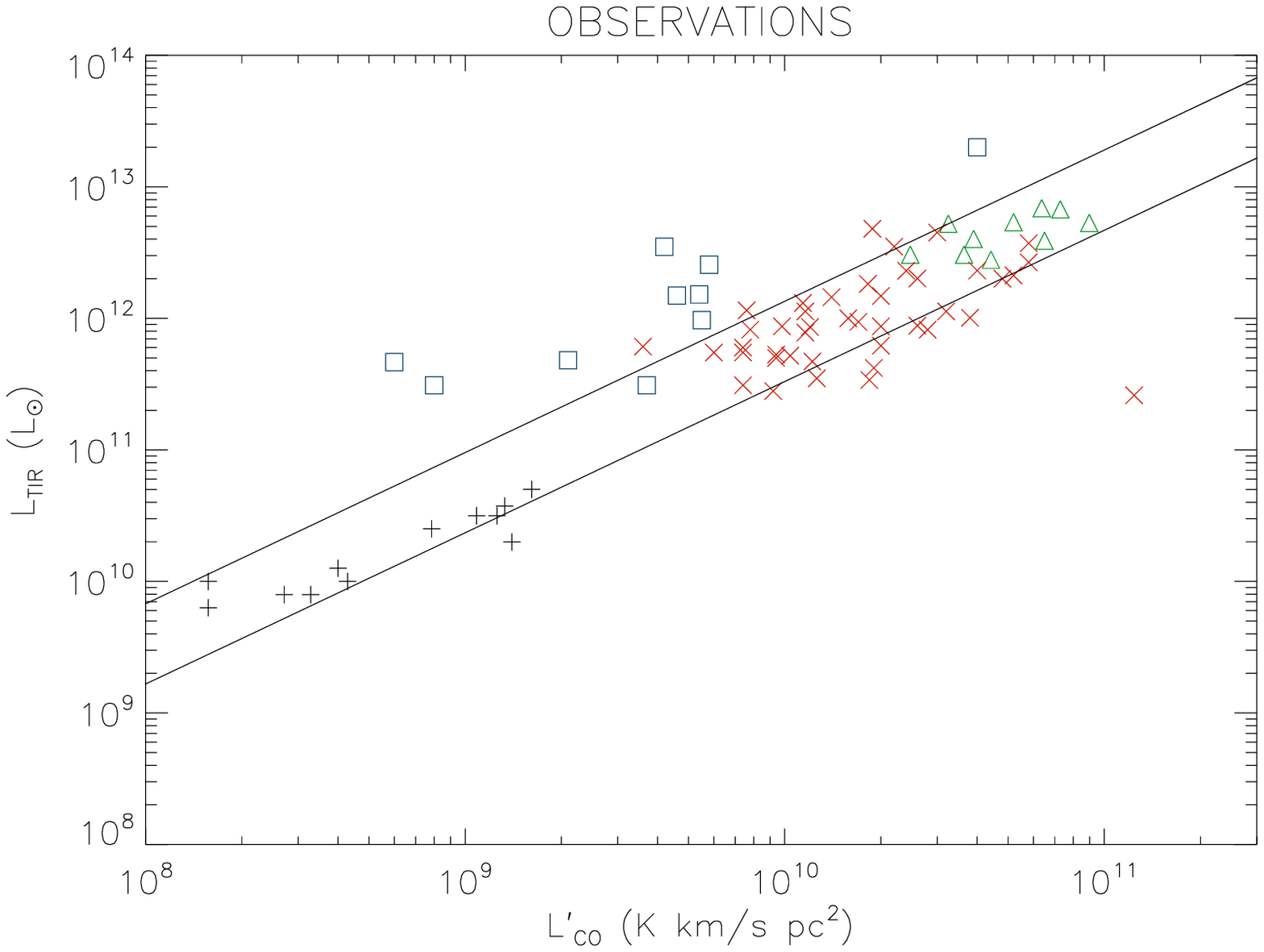}}
  \resizebox{\hsize}{!}{\includegraphics{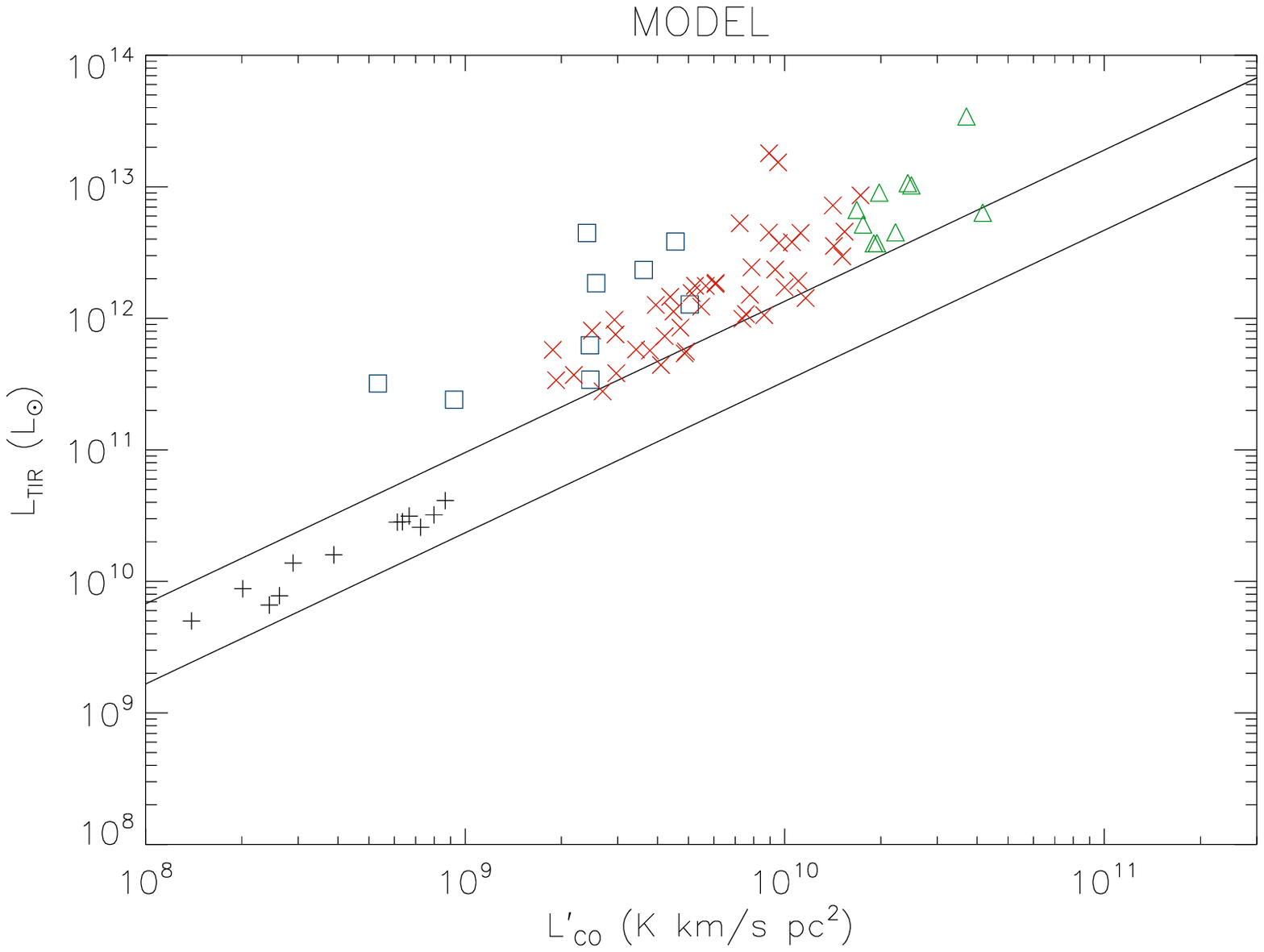}}
  \caption{Total infrared luminosity as a function of CO(1--0) luminosity. Upper panel shows observations while the lower panel shows the model.
    The symbols are the same as in Fig.~\ref{fig:plots_HCNCO_CO}.
  \label{fig:plots_HCNCO_SFRCO}}
\end{figure}

For the ULIRG, smm, and high-z star-forming galaxies, we can only compare the integrated model HCN, CO, and HCO$^+$ luminosities to observations.
The comparison between the model and the observed CO luminosities is shown in Fig.~\ref{fig:plots_HCNCO_CO}, where the observed transitions were used:
CO(2--1) for the local spiral galaxies, CO(1--0) for the ULIRGs, CO(3--2) for the smm-galaxies, and CO(3--2) for the high-z star-forming galaxies.
The corresponding mean ratio $L_{\rm model}/L_{\rm obs}$ and its uncertainty are presented in Table~\ref{tab:correl} (preferred model).
We observe approximately linear correlations between the model and observed CO luminosities. The model CO luminosities of the smm and high-z star-forming
galaxies are $\langle \log(L_{\rm CO,obs}/L_{\rm CO,model}) \rangle \sim 0.2$~dex smaller than observed; the ratio is only $\sim 0.1$~dex for the local spirals and ULIRGs.
The corresponding model and observed $L_{\rm  TIR}$--$L'_{\rm CO}$ relations for the CO(1--0) line are shown in Fig.~\ref{fig:plots_HCNCO_SFRCO}.
Line ratios from the literature are applied to determine the CO(1--0) fluxes 
(CO(2--1)/CO(1--0)$=0.7$, CO(3--2)/CO(1--0)$=0.77$ Genzel et al. 2010, and CO(3--2)/CO(1--0)$=0.5$ Tacconi et al. 2013).
The differences between the model and observed relation are again due to the underestimation ($0.2$~dex) of the model CO luminosities 
of the smm and high-z star-forming galaxies with respect to observations.
Overall, the model $L_{\rm  TIR}$--$L'_{\rm CO}$ relations are consistent with the observed relations.
\begin{table*}
\begin{center}
\caption{Comparison to observed CO, HCN(1--0), and HCO$^+$(1--0) data. Ratio between the model and observed line luminosities. \label{tab:correl}}
\begin{tabular}{lccccc}
\hline
Galaxy sample & CO & HCN(1--0) & HCN(1--0) GS04$^{\rm a}$  & HCN(1--0) GC08$^{\rm b}$  & HCO$^+$(1--0)  \\
\hline
Preferred model & & & & & \\
\hline
local spirals & $0.86 \pm  0.31$ & -- &  $1.72  \pm   0.49$ & $1.13 \pm 0.26$  & -- \\ 
ULIRGs & $0.75  \pm   0.37$ & $0.67 \pm  0.22$ & $0.36   \pm  0.15$ & $0.51 \pm  0.13$ & $0.89 \pm  0.30$ \\
smm-galaxies & $0.67  \pm   0.20$ & -- & $0.37 \pm 0.06$ & $0.69 \pm  0.14$ & -- \\
high-z star-forming galaxies & $0.72 \pm  0.39$ & -- & $0.18 \pm 0.03$ & $0.27 \pm  0.06$ & -- \\
\hline
Constant abundances  & & & & & \\ 
\hline
local spirals & $1.23 \pm 0.38$ & -- & $0.86 \pm 0.24$ &  $0.56 \pm  0.12$ & -- \\
ULIRGs & $0.75 \pm  0.36$ & $0.32 \pm  0.10$ & $0.18 \pm  0.09$ & $0.25 \pm  0.08$ & $0.37 \pm  0.12$ \\
smm-galaxies & $0.63 \pm  0.19$ & -- & $0.20 \pm  0.05$ &  $0.37 \pm  0.09$ & -- \\
high-z star-forming galaxies & $0.73 \pm  0.36$ & -- &  $0.21 \pm  0.07$ & $0.30 \pm  0.07$ & -- \\
\hline
$Q=1$ & & & & & \\
\hline
ULIRGs & $0.55 \pm  0.33$ & $0.61 \pm  0.17$ & $0.33 \pm  0.14$ & $0.46 \pm 0.11$ & $0.91 \pm  0.29$ \\ 
smm-galaxies & $0.63 \pm  0.25$ & -- & $0.23 \pm 0.03$ & $0.44 \pm  0.08$ & -- \\
high-z star-forming galaxies & $0.74 \pm  0.43$ & -- & $0.19 \pm  0.04$ & $0.27 \pm 0.06$ & -- \\
\hline
$\delta=15$ & & & & & \\
\hline
local spirals & $1.09 \pm  0.47$ & -- & $2.20 \pm 1.04$ & $1.43 \pm  0.60$ & -- \\
ULIRGs & $0.67 \pm 0.40$ & $0.78 \pm  0.20$ & $0.40 \pm  0.13$ & $0.57 \pm  0.09$ & $1.13 \pm  0.36$ \\
smm-galaxies & $0.86 \pm  0.33$ & -- & $0.37 \pm  0.04$ & $0.69 \pm  0.08$ & -- \\ 
high-z star-forming galaxies & $0.74 \pm  0.43$ & -- & $0.19 \pm  0.04$ & $0.27 \pm  0.06$ & -- \\
\hline
No cloud substructure & & & & & \\
\hline
local spirals & $1.16 \pm  0.38$ & -- & $1.69 \pm  0.49$ & $1.10 \pm  0.27$ & -- \\
ULIRGs & $0.94 \pm  0.46$ & $0.73 \pm  0.25$ & $0.38 \pm  0.14$ & $0.55 \pm  0.13$ &  $1.01 \pm  0.36$ \\
smm-galaxies & $0.91 \pm 0.30$ & -- & $0.39 \pm  0.06$ & $0.73 \pm  0.15$ & -- \\
high-z star-forming galaxies & $0.93 \pm  0.55$ & -- & $0.13 \pm 0.04$ & $0.19 \pm  0.08$ & -- \\ 
\hline
No CR heating & & & & & \\
\hline
local spirals &  $0.85 \pm  0.31$ & -- & $1.70 \pm  0.48$ & $1.12 \pm  0.26$ & -- \\
ULIRGs & $0.75 \pm  0.37$ & $0.44 \pm  0.20$ & $0.22 \pm  0.07$ & $0.32 \pm  0.08$ & $0.80 \pm 0.32$ \\
smm-galaxies & $0.67 \pm  0.20$ & -- & $0.36 \pm  0.06$ & $0.68 \pm  0.14$ & -- \\
high-z star-forming galaxies & $0.47 \pm  0.27$ & -- & $0.12 \pm 0.02$ & $0.17 \pm 0.03$ & -- \\
\hline
No HCN IR-pumping & & & & & \\
\hline
local spirals &  $0.85 \pm  0.31$ & -- & $1.70 \pm  0.48$ & $1.12 \pm  0.26$ & -- \\
ULIRGs & $0.75 \pm  0.37$ & $0.61 \pm  0.20$ & $0.34 \pm  0.16$ & $0.47 \pm  0.15$ & $0.89 \pm  0.30$ \\
smm-galaxies & $0.67 \pm 0.20$ & -- & $0.33 \pm  0.03$ & $0.63 \pm  0.09$ & -- \\
high-z star-forming galaxies & $0.72 \pm  0.39$ & -- & $0.16 \pm  0.04$ & $0.22 \pm  0.05$ & -- \\
\hline
\end{tabular}
\begin{tablenotes}
\item Ratio between the model and observed line luminosities.
\item local spirals: CO(2--1); ULIRGs: CO(1--0); smm-galaxies: CO(3--2); high-z star-forming galaxies: CO(3--2).
\item $^{\rm a}$ with respect to the Gao \& Solomon (2004) relation.
\item $^{\rm b}$ with respect to the Graci\'a-Carpio et al. (2008) relation.
    \end{tablenotes}
\end{center}
\end{table*}

\begin{figure}
  \centering
  \resizebox{\hsize}{!}{\includegraphics{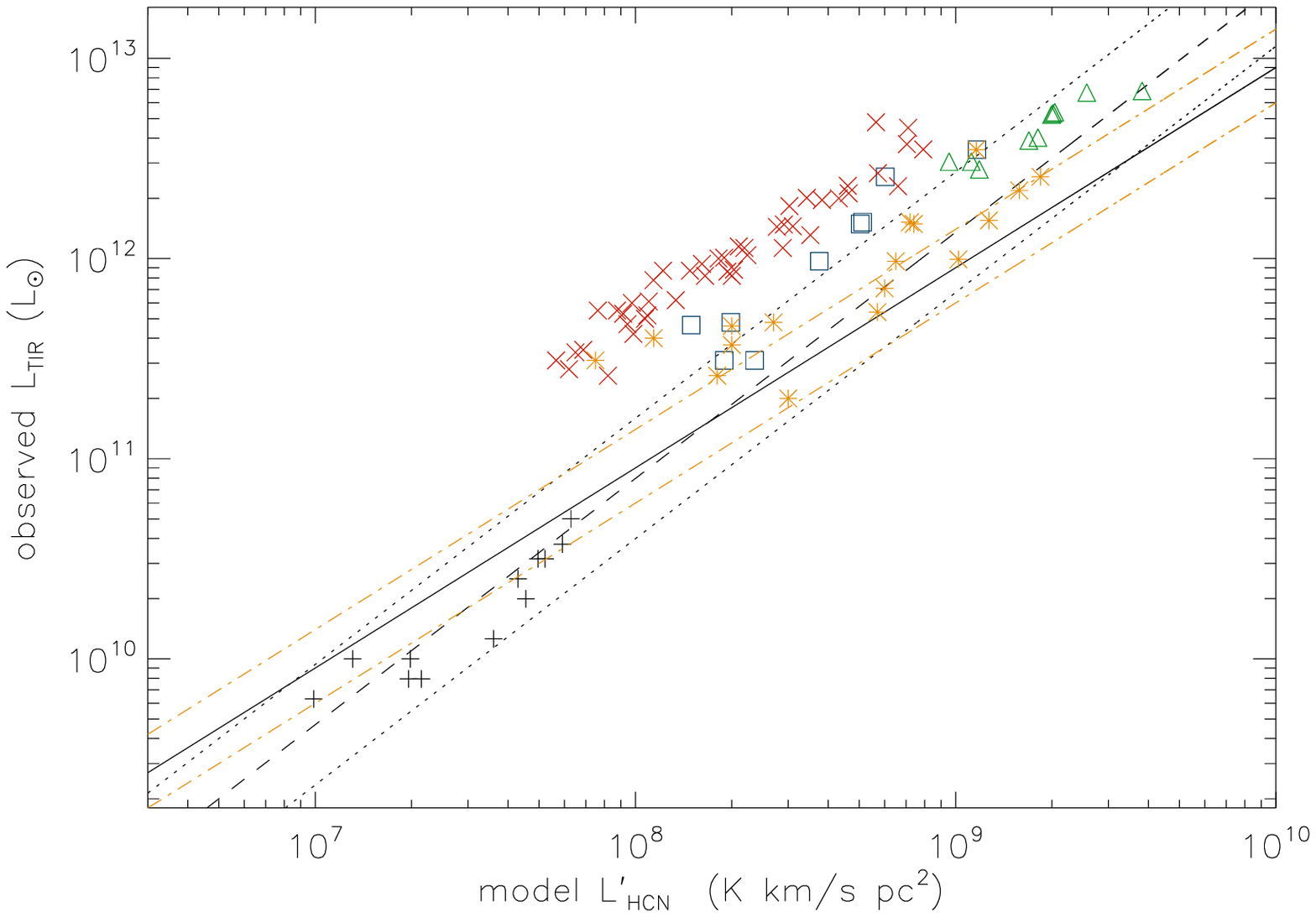}}
  \resizebox{\hsize}{!}{\includegraphics{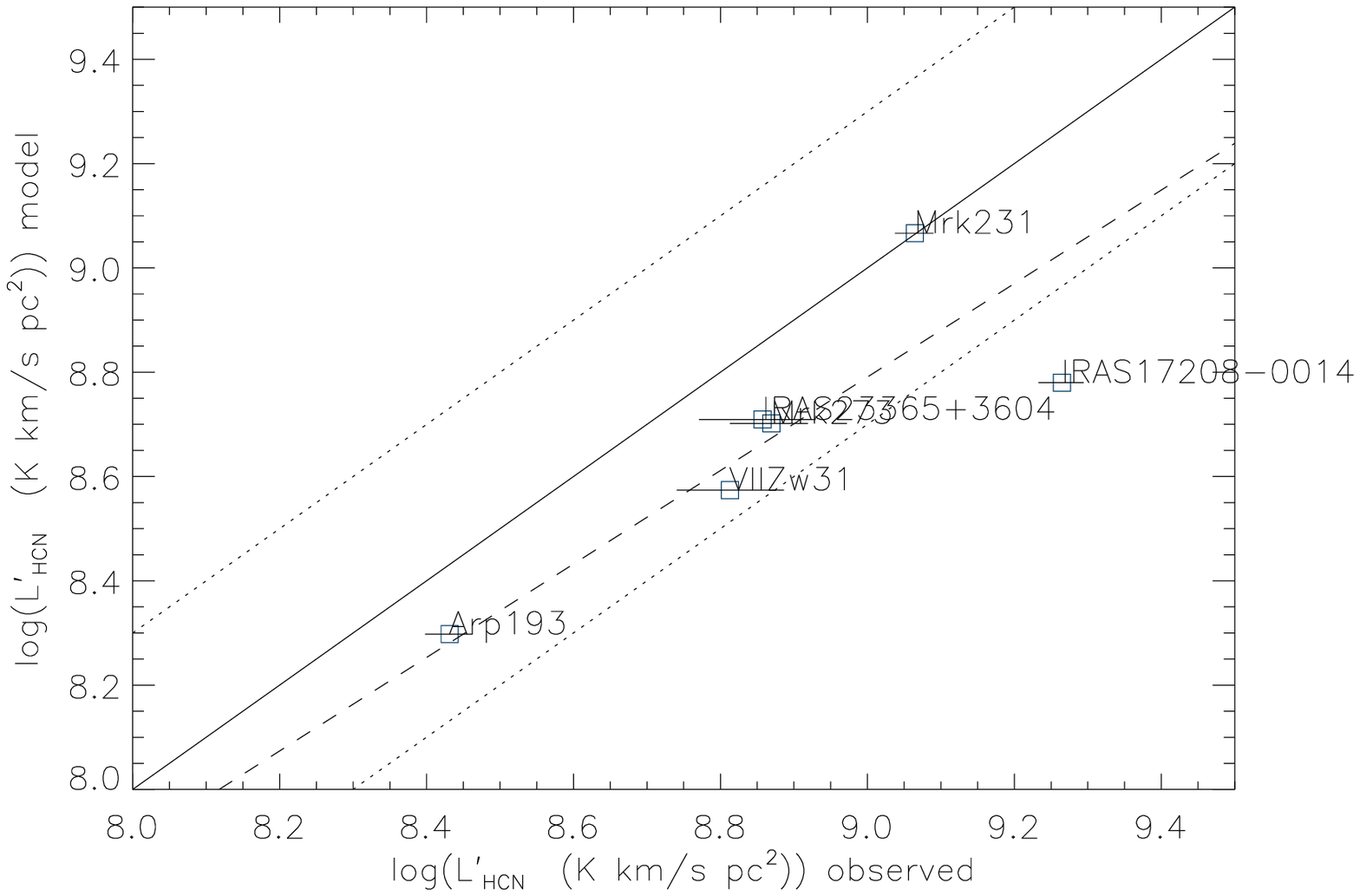}}
  \caption{Upper panel: observed total infrared luminosity as a function of the model HCN(1--0) luminosity. The symbols are the same as in Fig.~\ref{fig:plots_HCNCO_CO}.
    In addition, orange stars represent the observed HCN(1--0) of an ULIRG subsample (from Graci\'a-Carpio et al. 2008). The solid line
    represents the correlation $L_{\rm TIR}=900 \times L'_{\rm HCN}$ found by Gao \& Solomon (2004). The dashed line represents the correlation
    found by Graci\'a-Carpio et al. (2008) with $L_{\rm TIR} = 1.28 \times L_{\rm FIR}$: $\log(L_{\rm TIR}=1.23 \times \log(L'_{\rm HCN})+1.06$.
    Lower panel shows the model HCN(1--0) luminosity as a function of the observed HCN(1--0) luminosity  (Graci\'a-Carpio et al. 2008) for individual galaxies.
    The dashed line corresponds to a robust bisector fit.
  \label{fig:plots_HCNCO_HCN}}
\end{figure}

For the HCN(1--0) luminosities, the situation is more complex. We only compare $L_{\rm  TIR}$--$L'_{\rm HCN}$ relations (upper panel of Fig.~\ref{fig:plots_HCNCO_HCN})
for the local spirals, smm, and high-z star-forming galaxies, because the overlap between the model and observed samples is very small.
For the ULIRG sample, there are six out of nine galaxies in common between the Downes \& Solomon (1998) and Graci\'a-Carpio et al. (2008) samples,
which permits a direct comparison of the HCN(1--0) luminosities (lower panel of Fig.~\ref{fig:plots_HCNCO_HCN}).
Since the observed relations have different slopes, we compare the model relation to the relations observed by Gao \& Solomon (2004) and Graci\'a-Carpio et al. (2008).
Based on the $L_{\rm  TIR}$--$L'_{\rm HCN}$ relation, the model overestimates the HCN(1--0) luminosities of the local spiral galaxies by $\sim 0.2$~dex and underestimates
those of ULIRGs by $\sim 0.3$~dex. However, the direct comparison of model and observed HCN(1--0) luminosities for ULIRGs yields an underestimation of
only $0.13$~dex. This means that the Downes \& Solomon (1998) ULIRG sample contains mainly HCN-bright galaxies.
Concerning the HCN emission of smm-galaxies, Gao \& Solomon (2007) claimed that the FIR/HCN ratios in these high-redshift sources lie systematically above the FIR/HCN 
correlation established for nearby galaxies by approximately a factor of $2$. This behavior is well reproduced by the model.
Since there are no HCN detections of high-z star-forming galaxies in the literature, we can only suggest that their HCN emission might be a factor of 
$3$ lower than expected from observed $L_{\rm  TIR}$--$L'_{\rm HCN}$ relations.

\begin{figure}
  \centering
  \resizebox{\hsize}{!}{\includegraphics{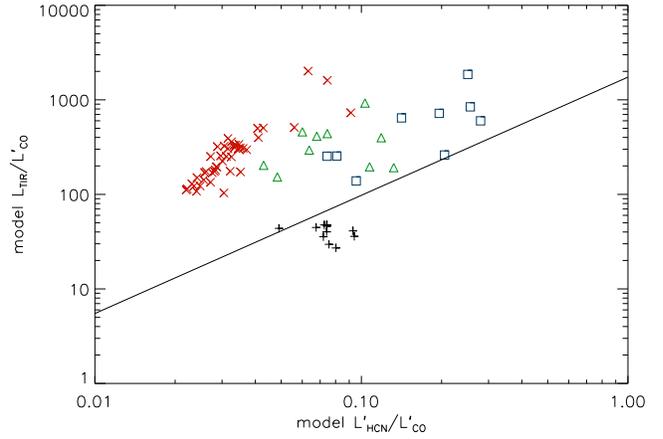}}
  \caption{Ratio between total infrared luminosity and CO(1--0) luminosity as a function of the ratio between HCN(1--0) and CO(1--0) luminosity, that is,
    the dense gas fraction. The symbols are the same as in Fig.~\ref{fig:plots_HCNCO_CO}. The solid line is the relation observed by Gao \& Solomon (2004).
  \label{fig:plots_HCNCO_HCNCO}}
\end{figure}

The model HCN/CO ratio is compared to observations in Fig.~\ref{fig:plots_HCNCO_HCNCO} for the four galaxy samples.
As expected, the model points for the local spiral galaxies lie below whereas those of the ULIRGs lie above the observed correlation.
Within the model, the smm-galaxies follow the same correlation as the ULIRGs, whereas the HCN emission of the high-z star-forming galaxies 
is significantly lower than expected by the observed correlation.

\begin{figure}
  \centering
  \resizebox{\hsize}{!}{\includegraphics{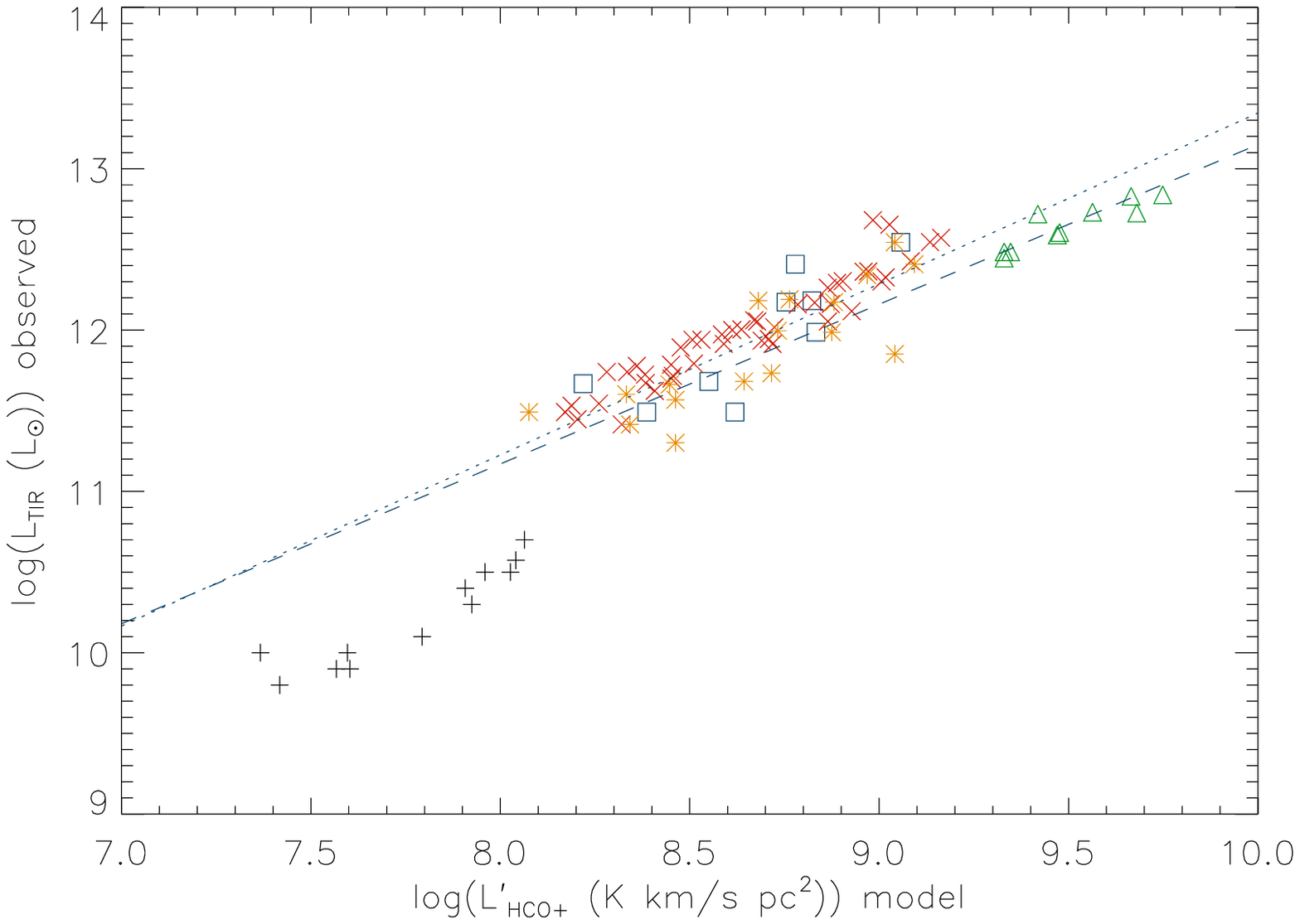}}
  \resizebox{\hsize}{!}{\includegraphics{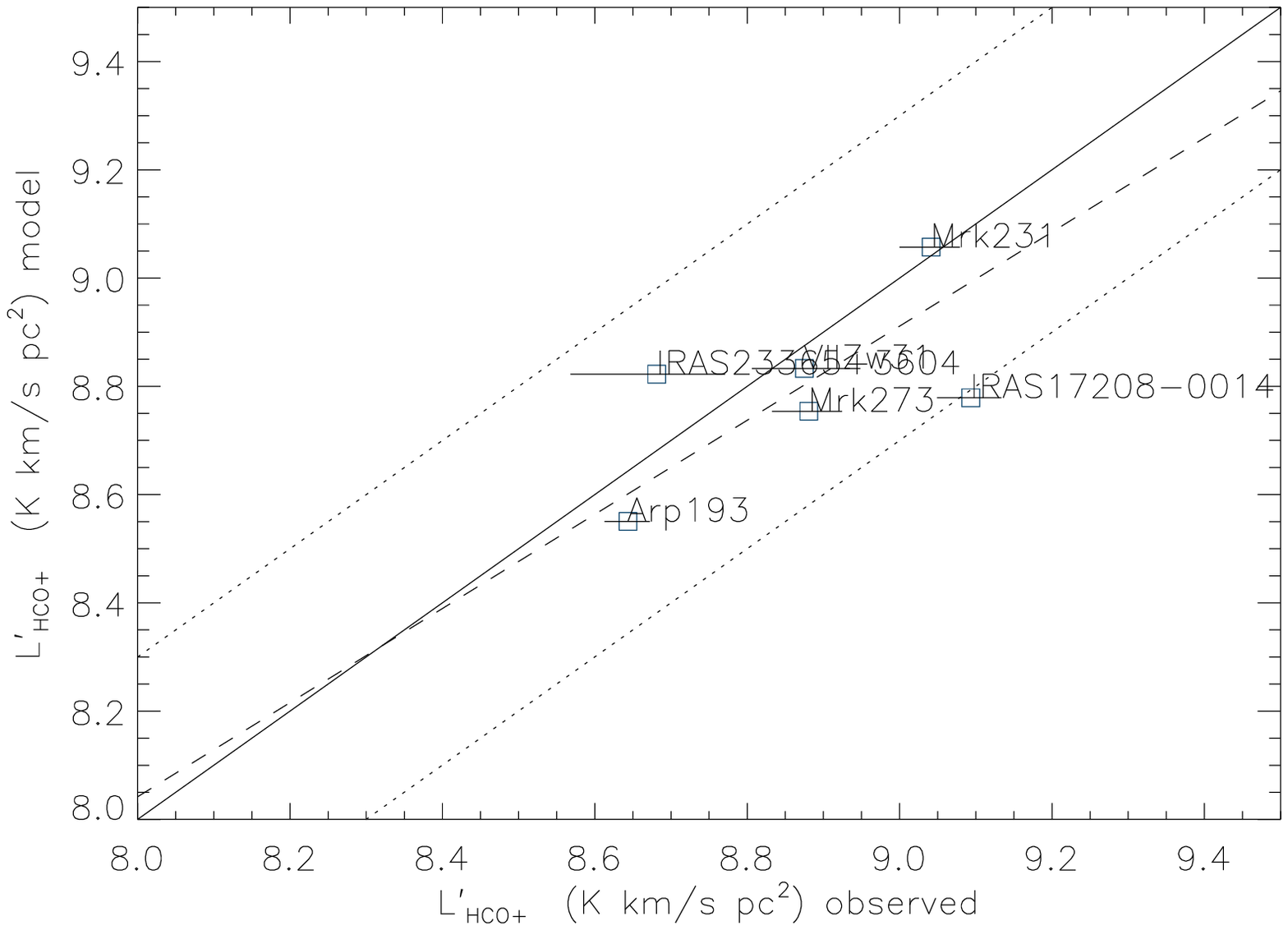}}
  \caption{Upper panel: Observed total infrared luminosity as a function of the model HCO$^+$(1--0) luminosity. 
    The symbols are the same as in Fig.~\ref{fig:plots_HCNCO_CO}.
    The dashed line represents the relation $\log(L_{\rm TIR})=0.99 \times \log(L'_{\rm HCO+})+3.25$ found by Juneau et al. (2009).
    The dotted line represents the relation $\log(L_{\rm TIR}=1.06 \times \log(L'_{\rm HCO+})+2.75$ found by Garcia-Burillo et al. (2012) with 
    $L_{\rm TIR} = 1.28 \times L_{\rm FIR}$. Lower panel: model HCO$^+$ (1--0) luminosity as a function of the observed HCO$^+$(1--0)
    luminosity. The solid line corresponds to equality, the dotted lines to factors of $1/2$ and $2$. The dashed line represents a
    robust bisector fit to the data.
  \label{fig:plots_HCNCO_HCO}}
\end{figure}

The model integrated HCO$^+$ luminosity can be compared to observations via the $L_{\rm TIR}$--$L'_{\rm HCO+}$ relation (upper panel of Fig.~\ref{fig:plots_HCNCO_HCO}).
The observed relations (Juneau et al. 2009, Garcia-Burillo et al. 2012) do not differ significantly. Within the model, the ULIRGs and high-z star-forming  
galaxies follow the observed relations. The smm-galaxies lie somewhat below the observed relation. The model HCO$^+$ luminosities
are a factor of approximately three higher than the expected HCO$^+$ luminosities assuming  $L'_{\rm HCN}=L'_{\rm HCO+}$ (e.g., Nguyen et al. 1992, Brouillet et al. 2005, Knudsen et al. 2007).
The direct comparison of the model and observed HCO$^+$ luminosities for the ULIRG sample shows good agreement
(lower panel of Fig.~\ref{fig:plots_HCNCO_HCO}). The model thus seems to overestimate the HCO$^+$ luminosities of the local spiral galaxies. 
We note that the HCO$^+$ emission strongly depends on the cosmic ray ionization rate used for the chemical network. 

\begin{figure}
  \centering
  \resizebox{\hsize}{!}{\includegraphics{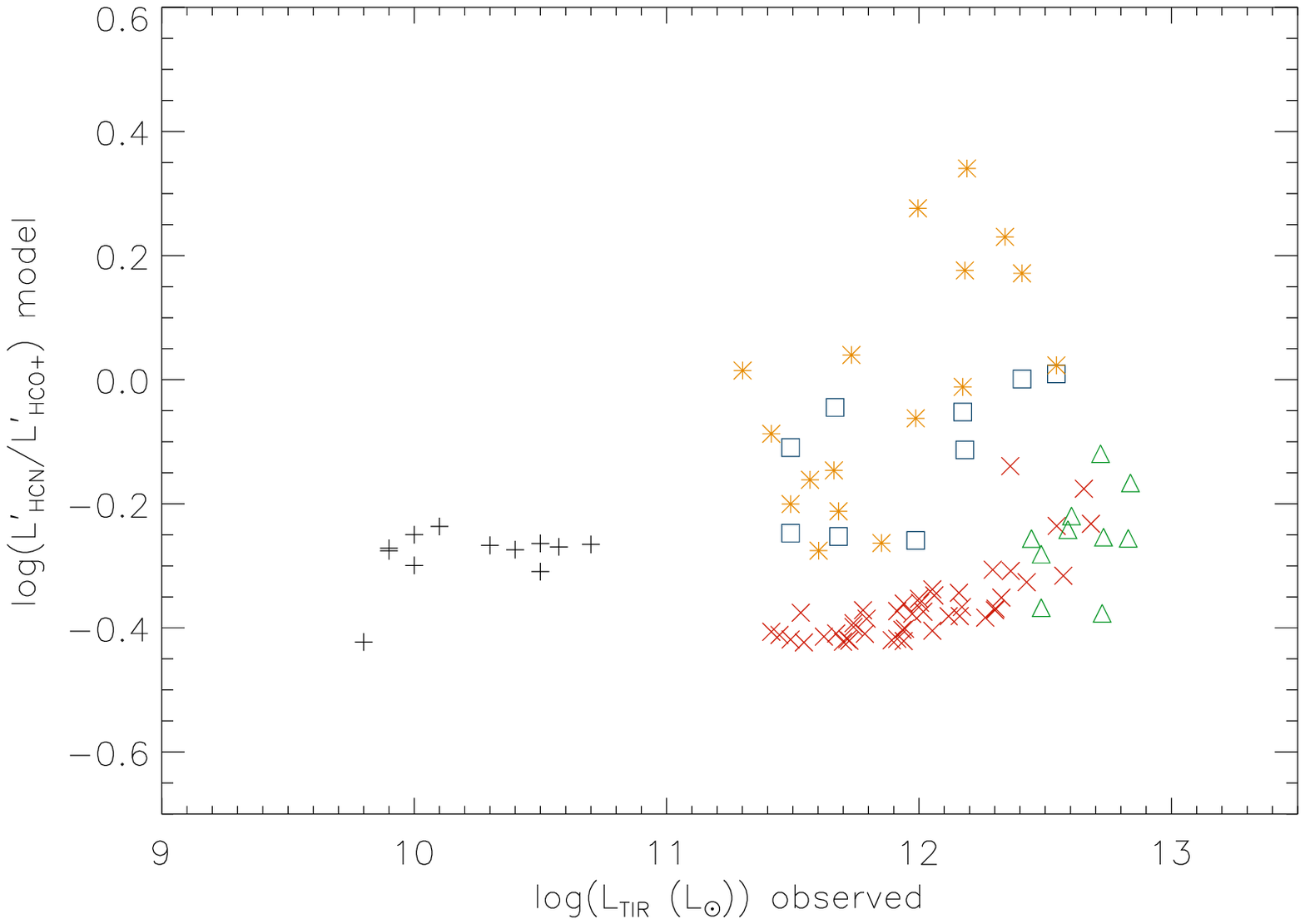}}
  \caption{The ratio between HCN(1--0) and HCO$^+$(1--0) luminosities as a function of the total infrared luminosity.
    The symbols are the same as in Fig.~\ref{fig:plots_HCNCO_CO}. In addition, orange stars represent the observed HCN(1--0) of a ULIRG subsample 
    (from Graci\'a-Carpio et al. 2008).
  \label{fig:plots_HCNCO_HCNHCO}}
\end{figure}

As an additional consistency check, the HCN/HCO$^+$ ratio as a function of the total infrared luminosity is shown in Fig.~\ref{fig:plots_HCNCO_HCNHCO}.
Observations of local spiral galaxies (e.g., Nguyen et al. 1992, Brouillet et al. 2005, Knudsen et al. 2007) show 
$\langle \log(L'_{\rm HCN}/L'_{\rm HCO^+}) \rangle \sim 0.0$.
As expected, the HCN/HCO$^+$ ratio of the local spiral galaxies is smaller by $\sim 0.2$~dex than the observed ratio.
The model HCN/HCO$^+$ ratio for the ULIRGs are well within the observed range (Graci\'a-Carpio et al. 2008).

We conclude that, overall, the CO luminosities are underestimated by up to a factor of $1.5$. The model HCN luminosities for local spirals are 
overestimated by a factor of $1.5$ and those for ULIRGs are underestimated by a factor of $1.5$ with respect to observations. 
The model HCN luminosities are consistent with observations for the smm-galaxies.
The model HCO$^+$ luminosities are consistent observations for all galaxy samples except the local spiral galaxies, where they are significantly overestimated.

\subsection{CO SLEDs}

The lowest three rotational transitions of CO, which trace the cooler gas component, are
relatively easily accessible with ground-based radio and submillimeter telescopes, and have
been observed in many local galaxies. 
It was not until the launch of the Herschel Space Observatory (Pilbratt et al. 2010) that the CO ladder up to $J=13$
was generally available for the ISM within our Galaxy and in nearby galaxies.
Early SPIRE observations showed much brighter high-$J$ CO emission than would be predicted by cool ($T_{\rm kin} < 50$~K) molecular gas in giant molecular clouds, 
the type of gas responsible for the CO(1--0) emission. 
A  warmer, denser (higher pressure) component of molecular gas is responsible  for  the  emission  of  mid- ($J=4$--$3$ to $J=6$--$5$) and 
high-$J$ ($J=7$--$6$ and above) CO lines.
Far-IR CO rotational lines, with $J_{\rm upper} \ge 13$, arise from states $500$--$7000$~K above the 
ground state and have critical densities of $10^6$ to $10^8$~cm$^{-3}$ (Hailey-Dunsheath et al. 2012).

\begin{figure}
  \centering
  \resizebox{\hsize}{!}{\includegraphics{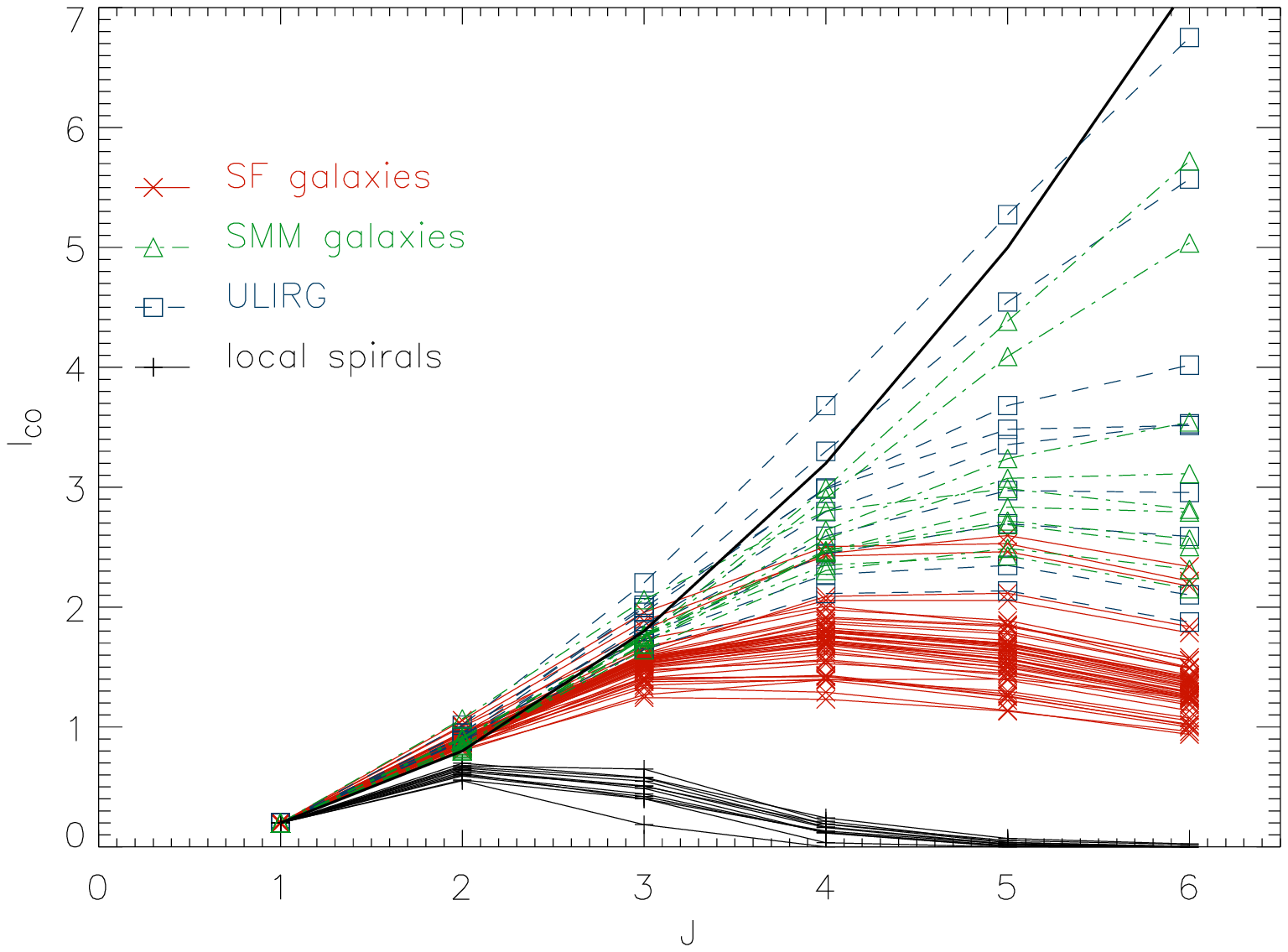}}
  \resizebox{\hsize}{!}{\includegraphics{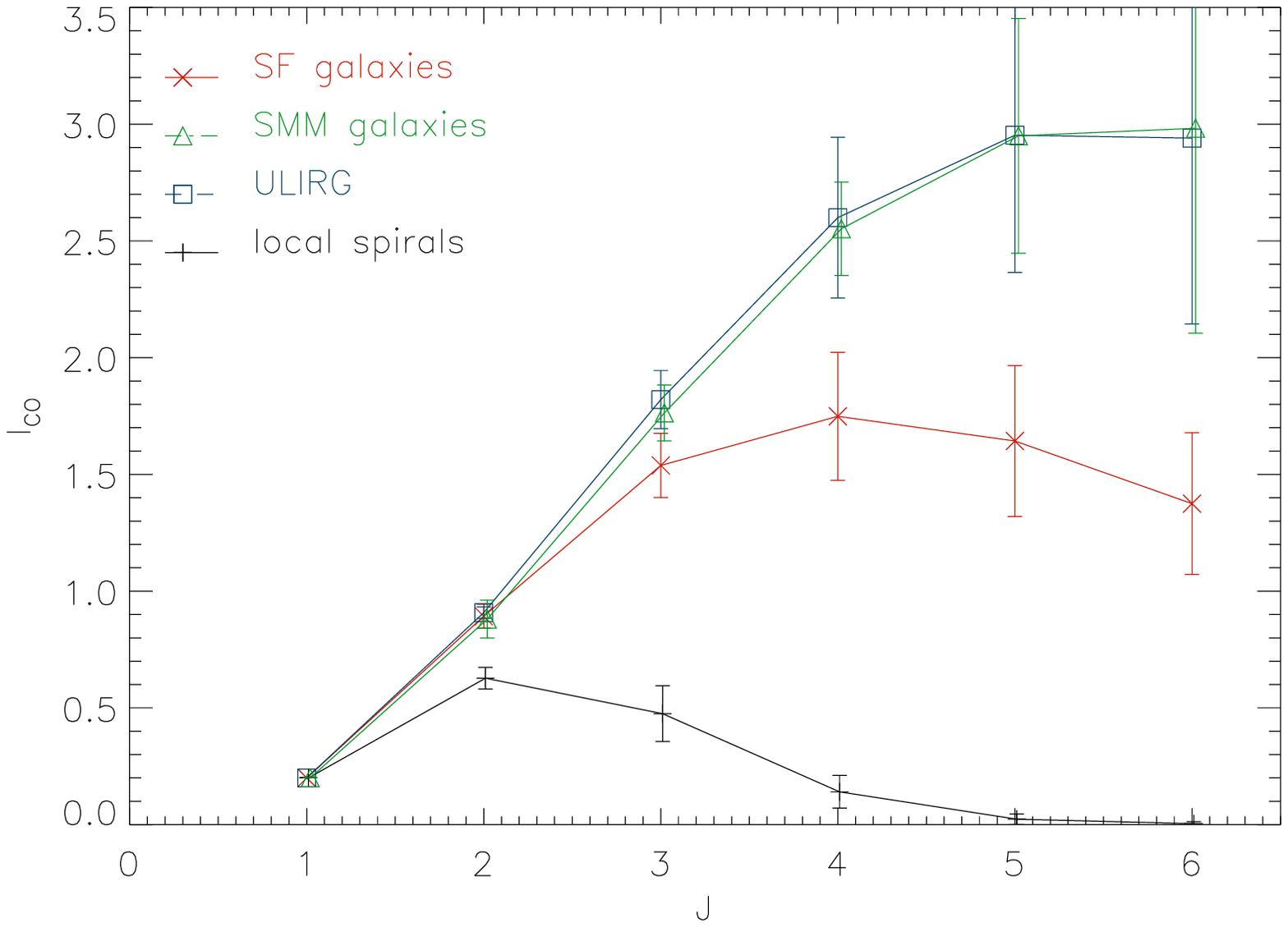}}
  \caption{Upper panel: model CO spectral line energy distributions (SLEDs) of the local spiral, ULIRG, submillimeter, and high-z star-forming galaxies.
    The thick solid line corresponds to a constant brightness temperature.
   Lower panel shows the mean model CO SLEDs of our galaxy samples. For direct comparison with Fig.~8 of Daddi et al. (2015), the CO(1--0) emission of all
   galaxies was set to $I_{\rm CO}=0.2$~Jy\,km\,s$^{-1}$. 
  \label{fig:plots_HCNCO_SLED}}
\end{figure}

It is well known locally that empirically and observationally, different CO excitation properties characterize spiral galaxies as opposed to merging-driven
ULIRGs (e.g., Daddi et al. 2015). The latter are much more highly excited in their high-$J$ CO transitions (Weiss
et al. 2007, Papadopoulos et al. 2012). The model CO  SLEDs of all galaxies of the four samples
and the mean CO SLEDs are presented in Fig.~\ref{fig:plots_HCNCO_SLED}. The latter can be directly compared to the observed mean SLEDs in
Fig.~8 of Daddi et al. (2015). We can only compare the model CO SLEDs of the local spiral galaxies to that of the inner Milky Way, which shows, as expected, higher
intensities for transition with $J_{\rm upper} \ge 3$. The model CO SLEDs of the ULIRGs are consistent in shape and absolute values with observations (Daddi et al. 2015).
The shape of the model CO SLEDs of the smm and high-z star-forming galaxies is different from that of the observed SLEDs;
whereas the shape of the model CO SLEDs is concave, that of the observed CO SLED is convex. This difference is mainly due to the CO(3--2) flux, which
in the model is a factor of two higher than observed.

\begin{figure}
  \centering
  \resizebox{\hsize}{!}{\includegraphics{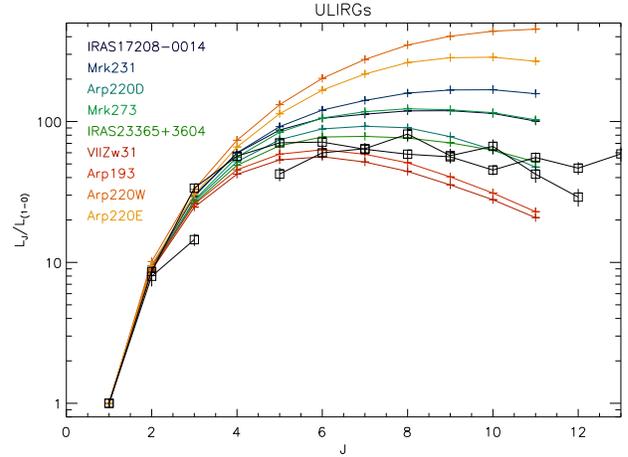}}
  \caption{CO SLEDs of the local ULIRGs. The model CO SLEDs are shown as colored lines.
    The black boxes and triangles linked by black lines represent the observed mean CO SLEDs of ULIRGs with total infrared luminosities
    between $3 \times 10^{11}$ and $10^{12}$~L$_{\odot}$ and higher than $10^{12}$~L$_{\odot}$, respectively (from Kamenetzky et al. 2016). 
  \label{fig:ulirg_ladder}}
\end{figure}

In a sample of ULIRGs, higher CO transitions ($J_{\rm upper} > 6$) were observed by Kamenetzky et al. (2016). Their mean CO SLEDs
for galaxies with  total infrared luminosities between $3 \times 10^{11}$ and $10^{12}$~L$_{\odot}$ and higher than $10^{12}$~L$_{\odot}$
are shown together with the model ULIRG CO SLEDs in Fig.~\ref{fig:ulirg_ladder}.
The two compact residual disks of Arp~220 (E and W) cannot be directly compared to this sample because Arp~220 would be seen as one
entity (Disk + East + West) by the Herschel satellite. The shape and the absolute values of the observed CO SLEDs are well reproduced 
by the model.

\subsection{Integrated CO and HCN conversion factors \label{sec:conversion}}

\begin{figure}
  \centering
  \resizebox{\hsize}{!}{\includegraphics{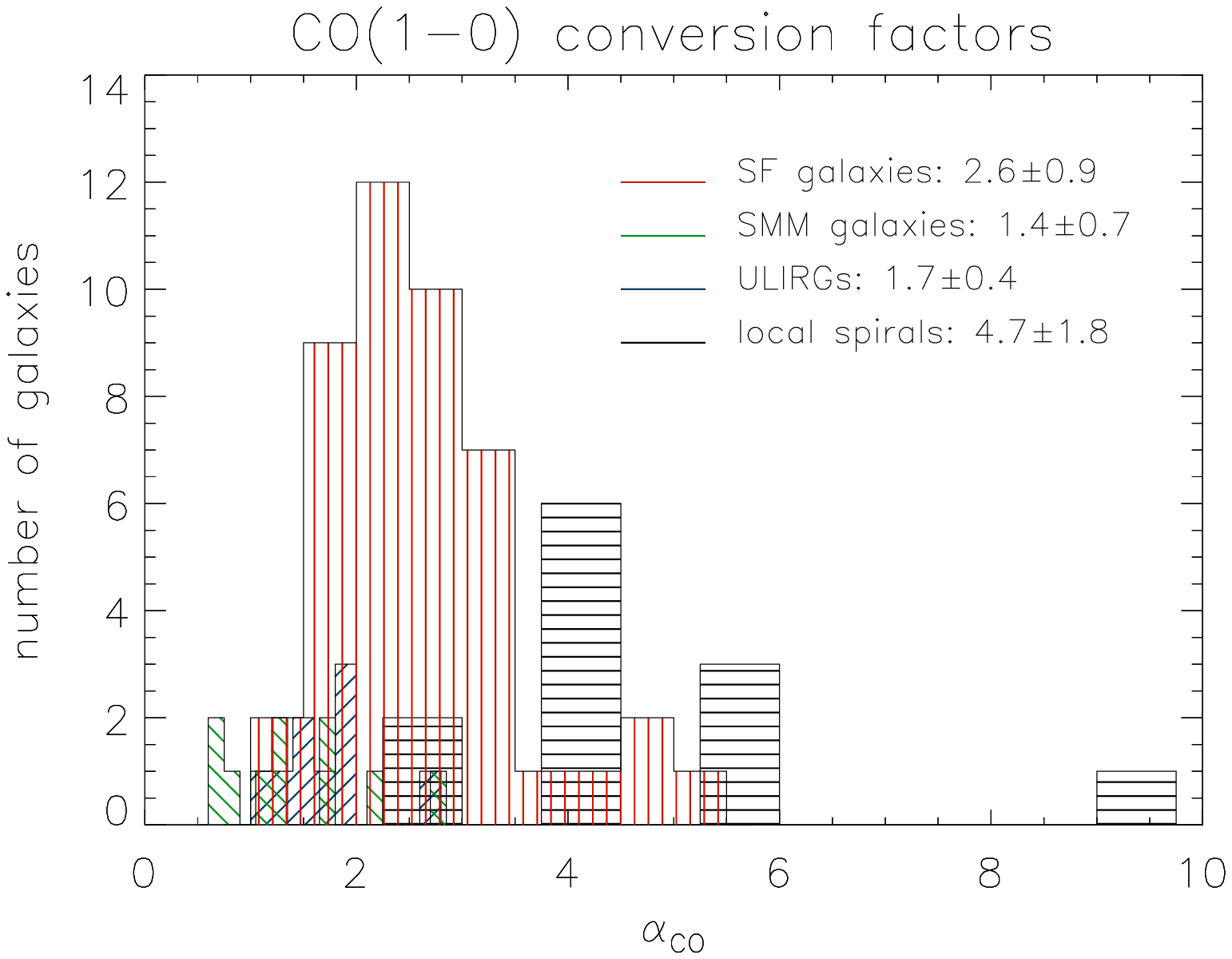}}
  \resizebox{\hsize}{!}{\includegraphics{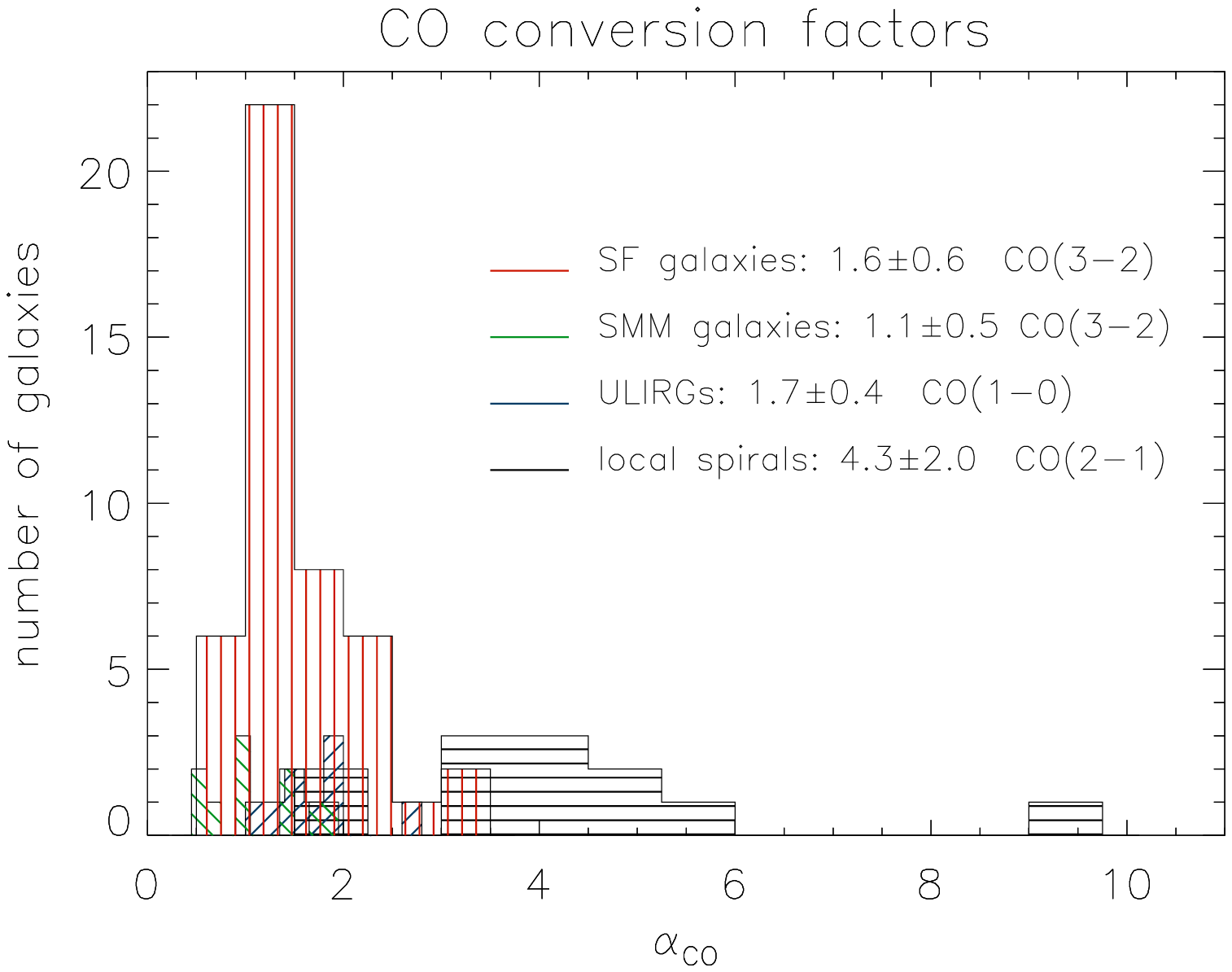}}
  \caption{Upper panel: CO(1--0) conversion factors for our galaxy samples.
    Lower panel shows CO conversion factors based on CO(2--1) (local spirals), CO(1--0) (ULIRGs), CO(4--3) (smm-galaxies), and CO(3--2) (high-z star-forming galaxies).
    These conversion factors imply the following line ratios: CO(2--1)/CO(1--0)$=0.7$, CO(4--3)/CO(1--0)$=0.63$ (Genzel et al. 2010), and CO(3--2)/CO(1--0)$=0.5$ 
    (Tacconi et al. 2013).
  \label{fig:plots_HCNCO_alphaCO}}
\end{figure}

Since the integrated molecular fraction and the line emission are calculated within the model (see Sect.~\ref{sec:molfrac}),
the integrated mass-to-light conversion factors can be determined. For the CO line emission we use two approaches:
(i) the model CO(1--0) is adopted for all galaxies (upper panel of Fig.~\ref{fig:plots_HCNCO_alphaCO}) and
(ii) the model flux of the observed CO line (CO(2--1) for the local spiral galaxies, CO(1--0) for the ULIRGs, CO(3--2) for the smm-galaxies,
and CO(3--2) for the high-z star-forming galaxies) is calculated and a line ratio given in the literature is applied to determine the
CO(1--0) flux (CO(2--1)/CO(1--0)$=0.7$, CO(3--2)/CO(1--0)$=0.77$ Genzel et al. 2010, and CO(3--2)/CO(1--0)$=0.5$ 
Tacconi et al. 2013) (lower panel of Fig.~\ref{fig:plots_HCNCO_alphaCO}). The mass-to-light conversion factor is then calculated by $\alpha_{\rm CO}=M_{\rm H_2}/L'_{\rm CO}$.
The mean conversion factors are given in Table~\ref{tab:convtable}.

The model CO(1--0) conversion factor varies between $2$~M$_{\odot}$(K\,km\,s$^{-1}$pc$^2$)$^{-1}$ and $6$~M$_{\odot}$(K\,km\,s$^{-1}$pc$^2$)$^{-1}$ with
one galaxy showing $\alpha_{\rm CO}=10$~M$_{\odot}$(K\,km\,s$^{-1}$pc$^2$)$^{-1}$. The mean conversion factor of local spirals 
($\langle \alpha_{\rm CO} \rangle =4.7 \pm 1.8$~M$_{\odot}$(K\,km\,s$^{-1}$pc$^2$)$^{-1}$) 
is close to the observed value of $\alpha_{\rm CO}=4.3$~M$_{\odot}$(K\,km\,s$^{-1}$pc$^2$)$^{-1}$ (Bolatto et al. 2013).
The mean conversion factor of the ULIRGs 
($\langle \alpha_{\rm CO} \rangle =1.7 \pm 0.4$~M$_{\odot}$(K\,km\,s$^{-1}$pc$^2$)$^{-1}$) is twice the usually assumed conversion
factor for dense starburst galaxies ($\alpha_{\rm CO}=0.8$~M$_{\odot}$(K\,km\,s$^{-1}$pc$^2$)$^{-1}$; Downes \& Solomon 1998) while that of smm-galaxies is similar to it.
The model CO(1--0) conversion factor of high-z star-forming galaxies is intermediate between those of local spiral galaxies and ULIRGs/smm-galaxies
($\langle \alpha_{\rm CO} \rangle =2.6 \pm 0.9$~M$_{\odot}$(K\,km\,s$^{-1}$pc$^2$)$^{-1}$).

The situation changes only slightly if higher $J$ transitions are used instead of CO(1--0). The only remarkable change is the decrease of the
CO conversion factor of high-z star-forming galaxies by $30$\,\% to ($\langle \alpha_{\rm CO} \rangle =1.6 \pm 0.6$~M$_{\odot}$(K\,km\,s$^{-1}$pc$^2$)$^{-1}$).
However, the distribution of the CO conversion factors for high-z star-forming galaxies shows a tail with conversion factors
comparable to that of the Galaxy.

\begin{figure}
  \centering
  \resizebox{\hsize}{!}{\includegraphics{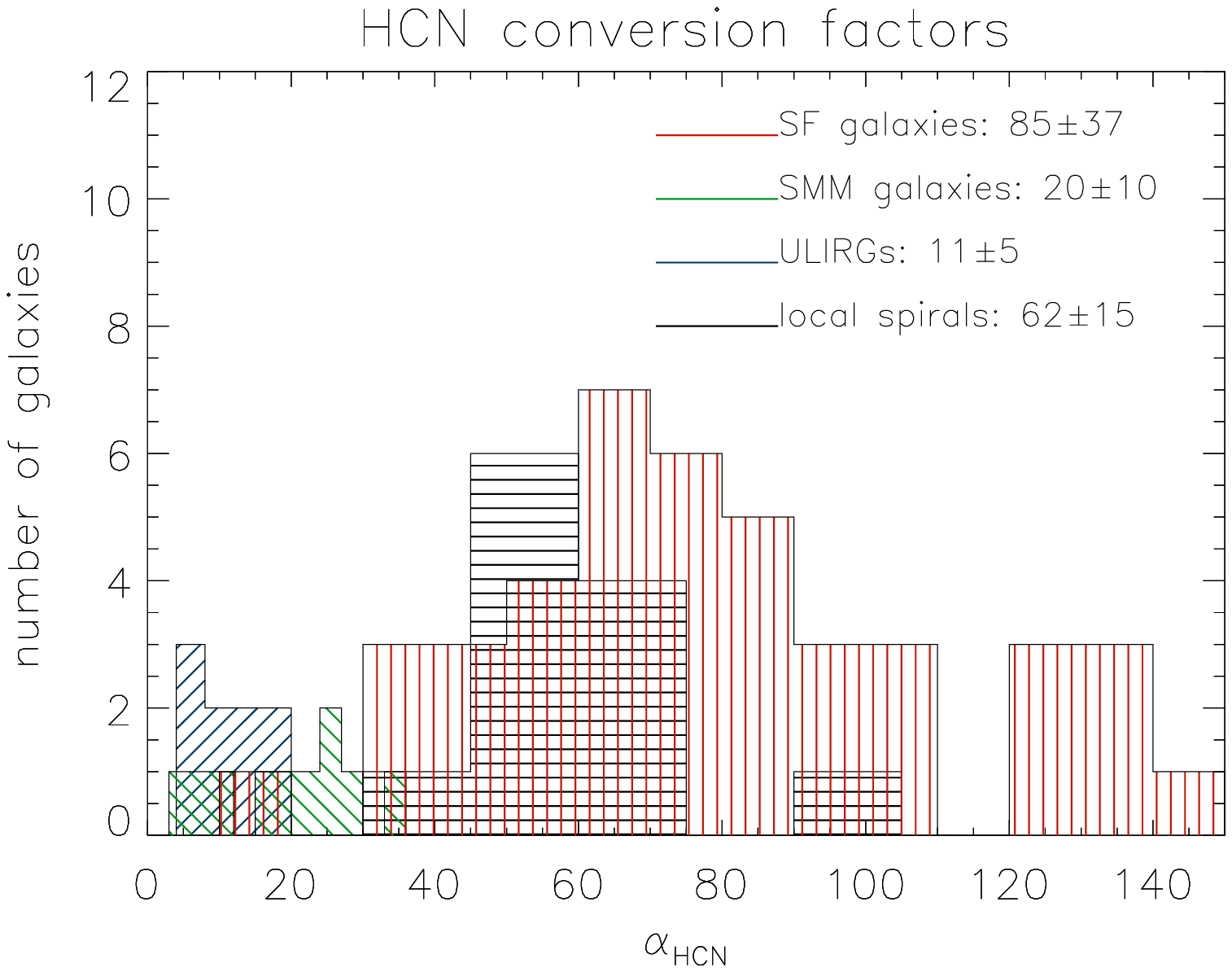}}
  \resizebox{\hsize}{!}{\includegraphics{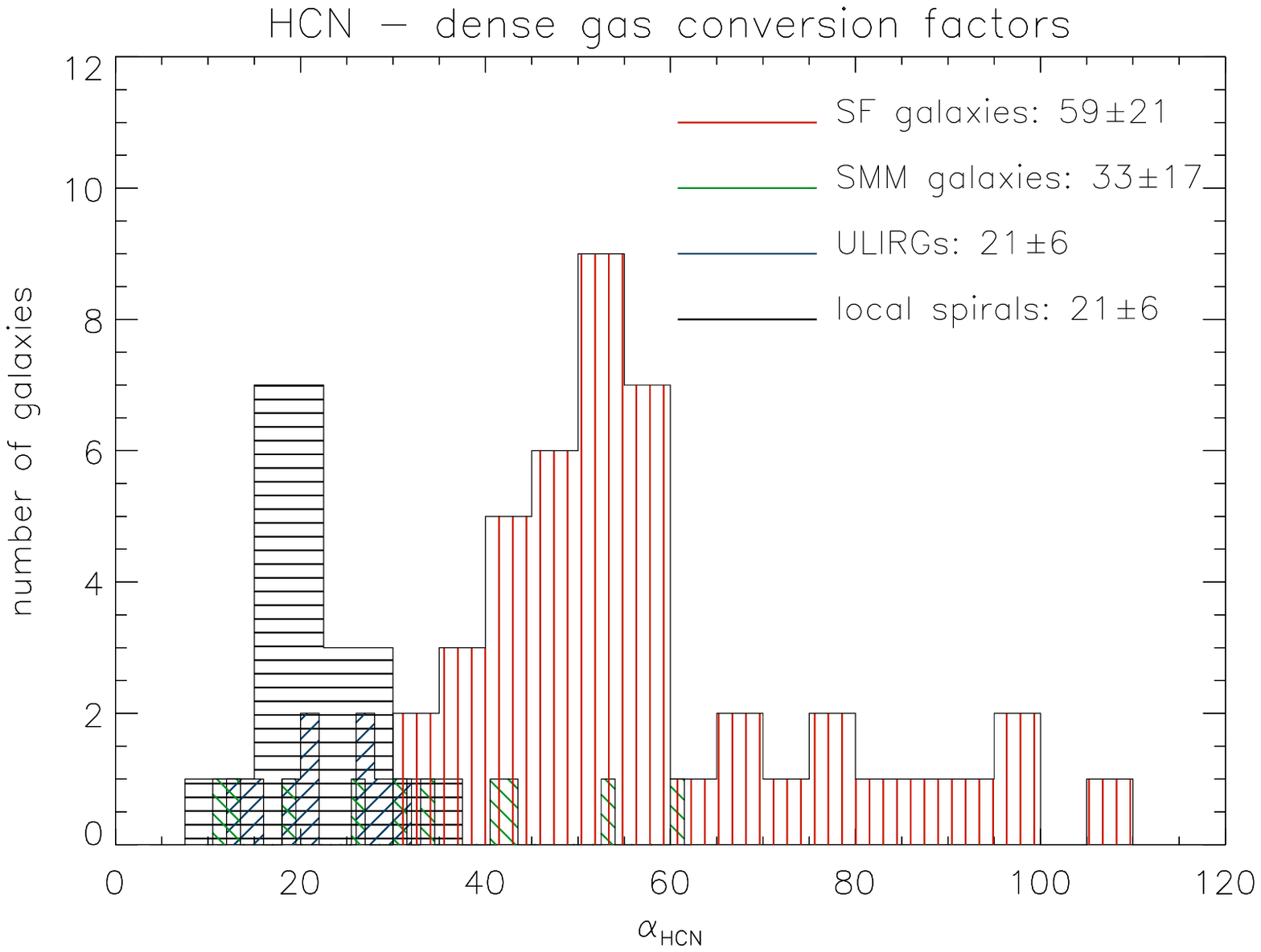}}
  \resizebox{\hsize}{!}{\includegraphics{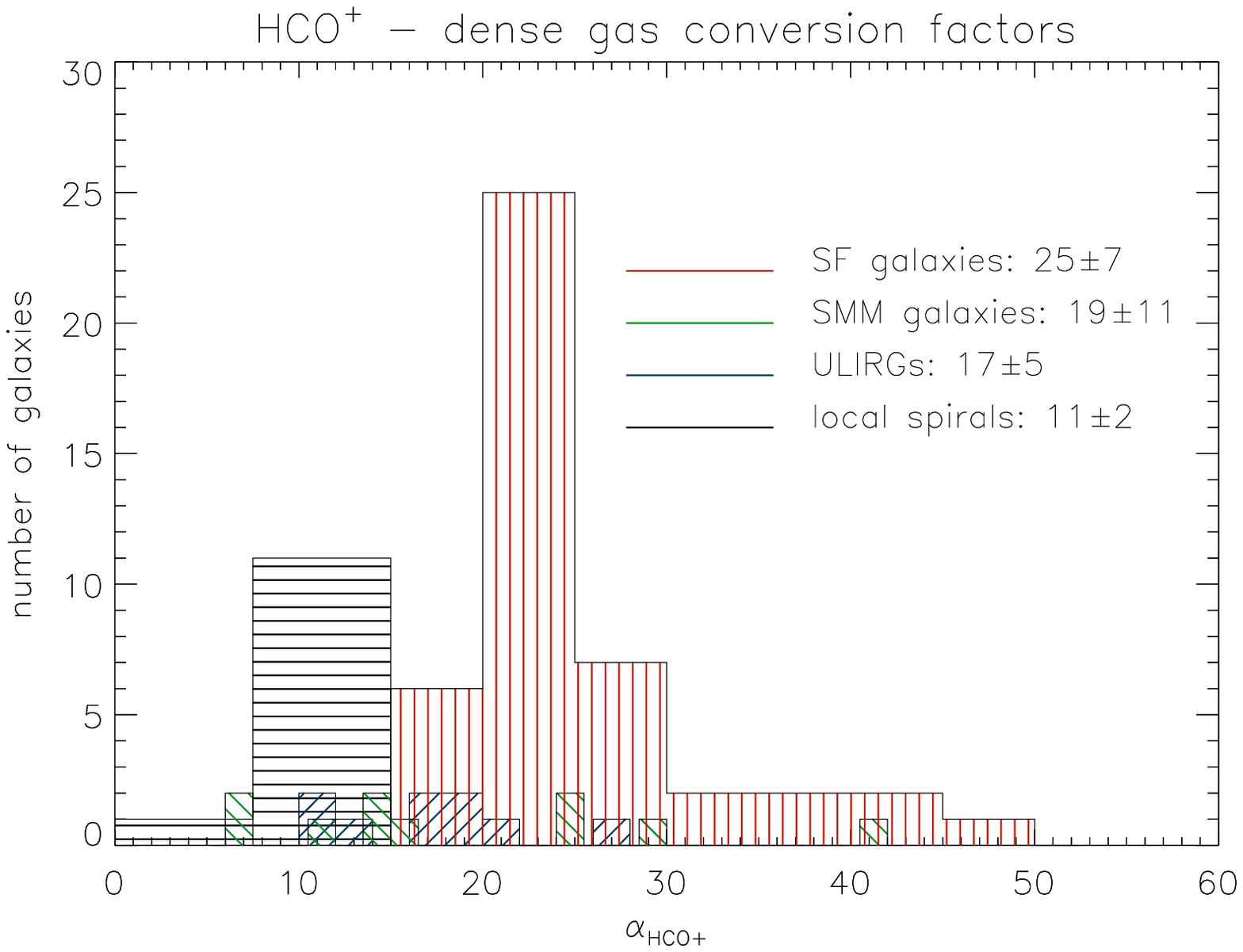}}
  \caption{Upper panel: HCN(1--0)-to-H$_2$ conversion factor for our galaxy samples.
    The middle panel shows HCN(1--0)-to-dense gas ($n > 3 \times 10^{4}$~cm$^{-3}$) conversion factor.
    The lower panel shows the HCO$^+$(1--0)-to-dense gas ($n > 3 \times 10^{4}$~cm$^{-3}$) conversion factor.
  \label{fig:plots_HCNCO_alphaHCN}}
\end{figure}

The HCN conversion factor is usually given with respect to the dense gas, that is, gas with densities exceeding $n=3 \times 10^4$~cm$^{-3}$ (e.g., Gao \& Solomon 2004).
For completeness, we give the HCN(1--0)--$M_{\rm H_2}$ and the HCN(1--0)--$M_{\rm dense}$ conversion factors in Fig.~\ref{fig:plots_HCNCO_alphaHCN}.
Only the latter conversion factor can be compared to the literature (lower panel of Fig.~\ref{fig:plots_HCNCO_alphaHCN}), 
where $\alpha_{\rm HCN}=10$~M$_{\odot}$(K\,km\,s$^{-1}$pc$^2$)$^{-1}$ is
assumed (e.g., Gao \& Solomon 2004). 
The mean HCN conversion factors are $\alpha_{\rm HCN}=21 \pm 6$, $33 \pm 17$, and $59 \pm 21$~M$_{\odot}$(K\,km\,s$^{-1}$pc$^2$)$^{-1}$ for the local spiral galaxies/ULIRGs,
smm-galaxies, and high-z star-forming galaxies, respectively (Table~\ref{tab:convtable}).

The mean HCO$^+$ conversion factors are $\alpha_{\rm HCO+}=11 \pm 2$, $17 \pm 5$, $19 \pm 11$, and $25 \pm 7$~M$_{\odot}$(K\,km\,s$^{-1}$pc$^2$)$^{-1}$ for the 
local spiral galaxies, ULIRGs, smm-galaxies, and high-z star-forming galaxies, respectively (Table~\ref{tab:convtable}).
Since the HCO$^+$(1--0) luminosity of the local spiral galaxies is significantly overestimated by the model, the associated conversion factor is
a lower limit. We thus find a relatively uniform HCO$^+$ conversion factor $\alpha_{\rm HCO+} \sim 20$~M$_{\odot}$(K\,km\,s$^{-1}$pc$^2$)$^{-1}$ for all galaxy samples.

We conclude that all model mass-to-light conversion factors are consistent with the values used in the literature within a factor of two.
Both, the HCN and HCO$^+$ emission trace the dense molecular gas to a factor of approximately two for the local spiral galaxies, ULIRGs and smm-galaxies.
For the high-z star-forming galaxies, HCO$^+$ might be the better tracer, but this needs to be confirmed. 

\begin{table*}
\begin{center}
\caption{CO(1--0) and HCN(1--0) conversion factors in units of M$_{\odot}$(K\,km\,s$^{-1}$pc$^2$)$^{-1}$.\label{tab:convtable}}
\begin{tabular}{lcccc}
\hline
Galaxy sample & CO(1--0) & HCN(1--0) & HCN(1--0) & HCO$^+$(1--0) \\
 & --H$_2$ & --H$_2$ & --dense gas & --dense gas \\
\hline
Preferred model & & & \\
\hline
local spirals &  $4.7 \pm 1.8$ & $62 \pm 15$ & $21 \pm 6$ & $11 \pm 2$ \\
ULIRGs & $1.7 \pm 0.4$ & $11 \pm 5$ & $21 \pm 6$ & $17 \pm 5$ \\
smm-galaxies & $1.4 \pm 0.7$ & $20 \pm 10$ & $33 \pm 17$ & $19 \pm 11$ \\
high-z star-forming & $2.6 \pm 0.9$ & $85 \pm 37$ & $59 \pm 21$ & $25 \pm 7$ \\ 
\hline
No substructure & & & \\
\hline
local spirals & $2.9 \pm 0.9$ & $52 \pm 10$ & -- & -- \\
ULIRGs & $1.3 \pm 0.3$ & $10 \pm 5$ & $8 \pm 3$ & $6 \pm 2$ \\ 
smm-galaxies & $0.9 \pm 0.3$ & $16 \pm 8$ & $6 \pm 2$ & $3 \pm 1$ \\
high-z star-forming & $1.6 \pm 0.6$ & $111 \pm 75$ & $11 \pm 9$ & $4 \pm 3$ \\
\hline
\end{tabular}
\begin{tablenotes}
      \item
        for dense gas $\alpha_{\rm HCN}=10$~M$_{\odot}$(K\,km\,s$^{-1}$pc$^2$)$^{-1}$ (e.g., Gao \& Solomon 2004)
    \end{tablenotes}
\end{center}
\end{table*}

\section{Variation of model parameters \label{sec:variation}}

Only the preferred model is presented in Sect.~\ref{sec:results}. A natural question to ask is 
how our model results depend on the different model assumption and free parameters.
To answer this question, we replace the chemical network by constant molecular abundances and vary the free parameters $Q$ and $\delta$.
In addition, in Sect.~\ref{sec:discussion}, we remove the cloud substructure, cosmic ray heating, and HCN IR-pumping.

\subsection{Importance of the chemical network \label{sec:constabund}}

Initially, we adopted constant molecular abundances for the CO, HCN, and HCO$^+$ molecules, which are then scaled with metallicity: 
$x_{\rm CO}=10^{-4}$, $x_{\rm HCN}=x_{\rm HCO^+}=2 \times 10^{-8}$.
Whereas the CO abundance corresponds to the canonical value (e.g., Draine 2011), the assumption of a constant HCN and HCO$^+$ is not well justified.
The resulting ratios between the model and observed line luminosities are shown in Table~\ref{tab:correl}.

The mean of the ratios between the model and observed CO luminosities are 
$\langle \log(L'_{\rm CO,\ model}/L'_{\rm CO,\ obs}) \rangle =-0.09 \pm 0.15$ for the local spiral,
$-0.10 \pm 0.16$ for the ULIRG, $-0.19 \pm 0.12$ for the smm-galaxy, and $-0.20 \pm 0.22$ for the high-z star-forming galaxy samples, respectively.
Thus, the observed CO luminosities are reproduced for all galaxy samples by the model within $\sim 0.2$~dex or a factor of $\sim 1.6$.

\begin{figure}
  \centering
  \resizebox{\hsize}{!}{\includegraphics{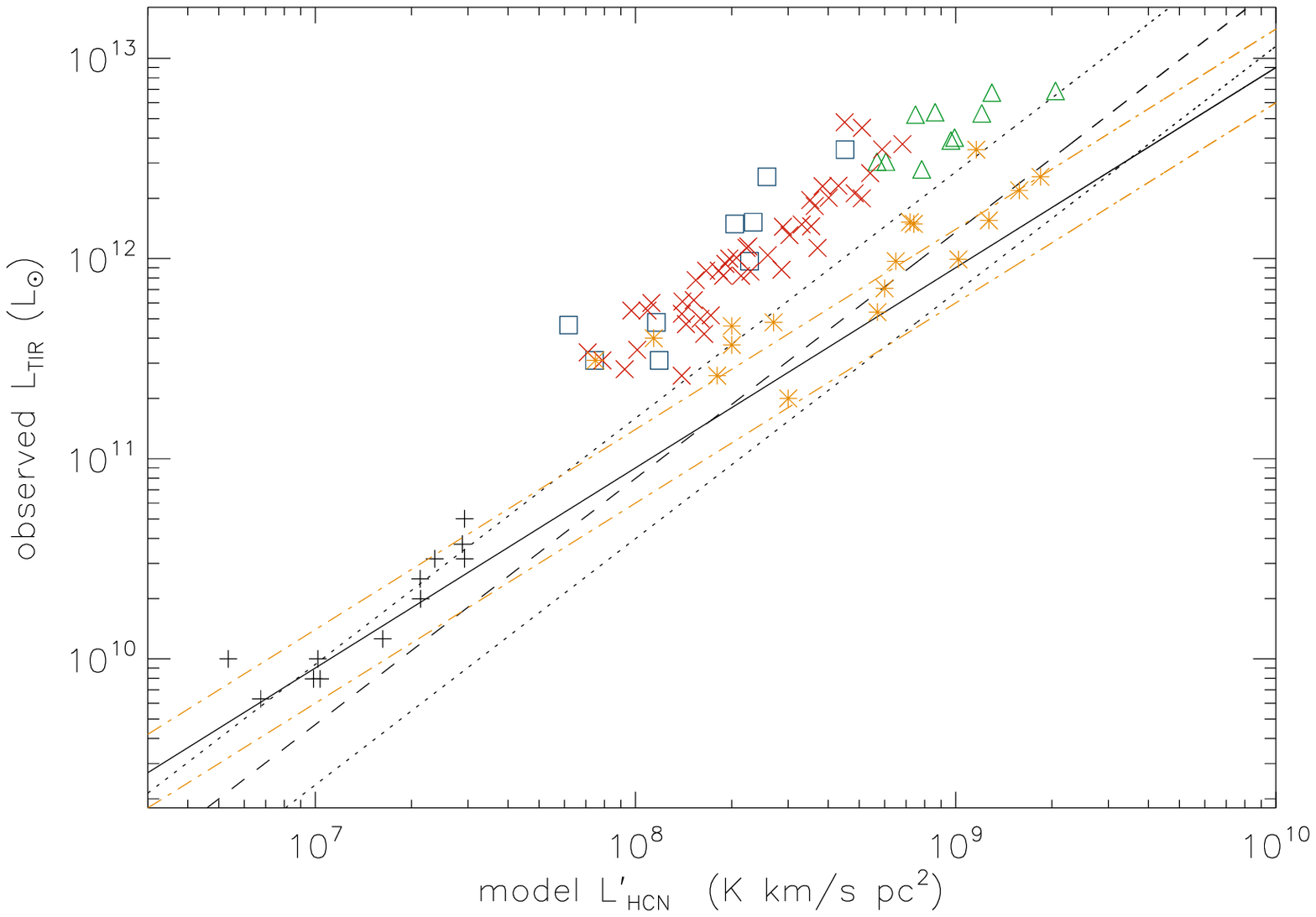}}
  \resizebox{\hsize}{!}{\includegraphics{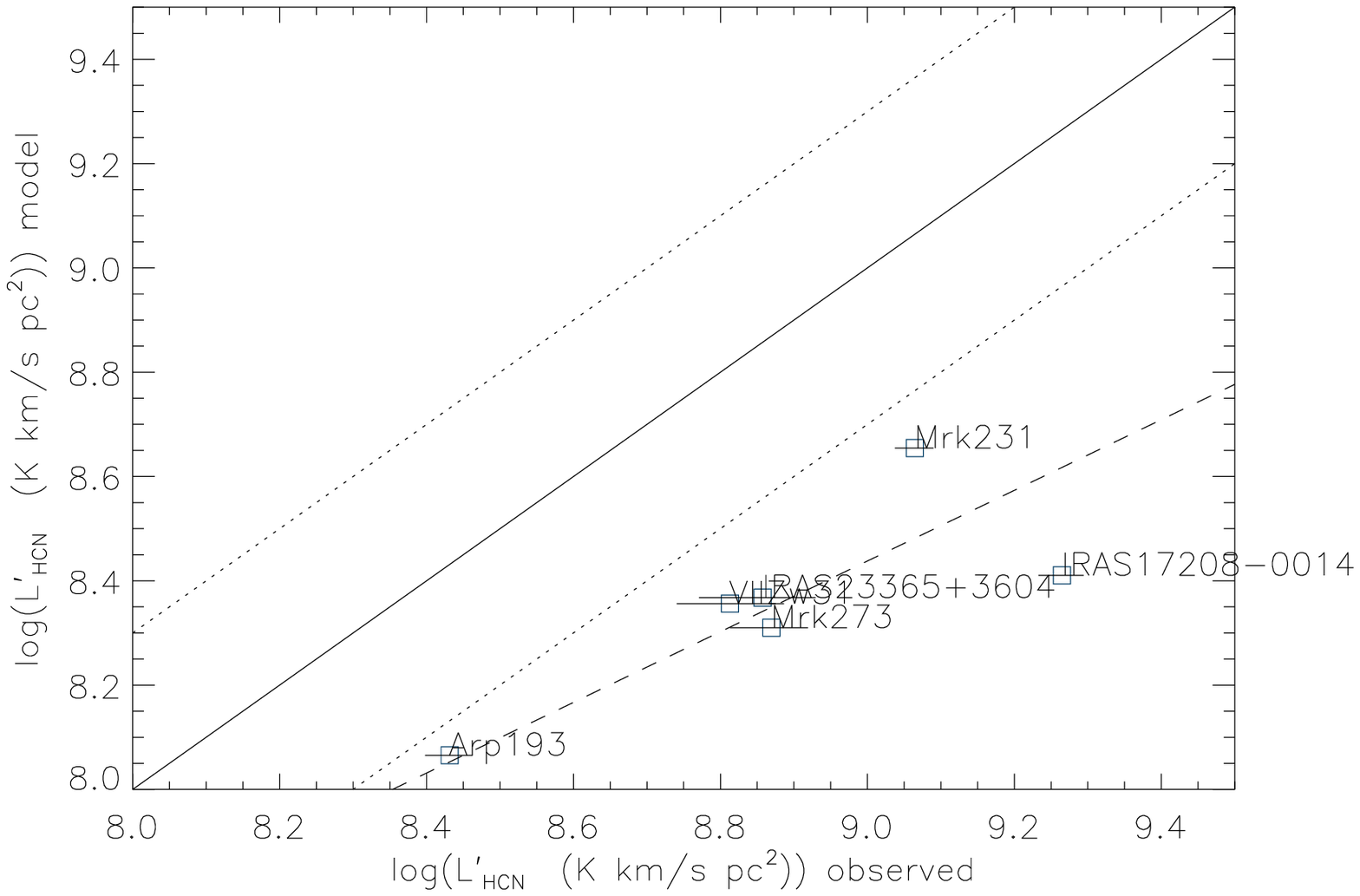}}
  \caption{{\it Constant abundances model.} Upper panel shows observed total infrared luminosity as a function of the model HCN luminosity.
    The solid line corresponds to $L'_{\rm HCN,\ obs}=900 \times L_{\rm TIR,\ obs}$ (Gao \& Solomon 2004). The dashed line corresponds to $L_{\rm TIR,\ obs}=12 \times L_{\rm HCN,\ obs}^{'\ 1.23}$
    (Graci\'a-Carpio et al. 2008) with factors of $0.5$ and $2$ (dotted lines). 
    Lower panel shows the model HCN(1--0) luminosity as a function of the observed HCN(1--0) luminosity. The dashed line represents a robust bisector fit.
    Compare to Fig.~\ref{fig:plots_HCNCO_HCN}.
  \label{fig:plots_HCNCO_constantabundances_HCN}}
\end{figure}

The resulting HCN(1--0) luminosities are compared to the observed CO luminosities in Fig.~\ref{fig:plots_HCNCO_constantabundances_HCN}.
In order to compare all model HCN luminosities to observations even in the absence of HCN measurements, we assumed $L'_{\rm HCN,\ obs}=900 \times L_{\rm TIR,\ obs}$ (Gao \& Solomon 2004).
The mean of the ratios between the model and observed HCN luminosities are $\langle \log(L'_{\rm HCN,\ model}/L'_{\rm HCN,\ obs}) \rangle =0.22 \pm 0.12$ for the local spiral,
$-0.47 \pm 0.15$ for the ULIRG , $-0.44 \pm 0.07$ for the smm-galaxy , and $-0.75 \pm 0.08$ for the high-z star-forming galaxy samples, respectively.
Whereas the model HCN luminosities agree well with the observed HCN luminosities for the local spiral galaxies, the model underestimates the HCN luminosities by up to a factor of four for the ULIRG, smm-galaxy, and high-z star-forming galaxies.

\begin{figure}
  \centering
  \resizebox{\hsize}{!}{\includegraphics{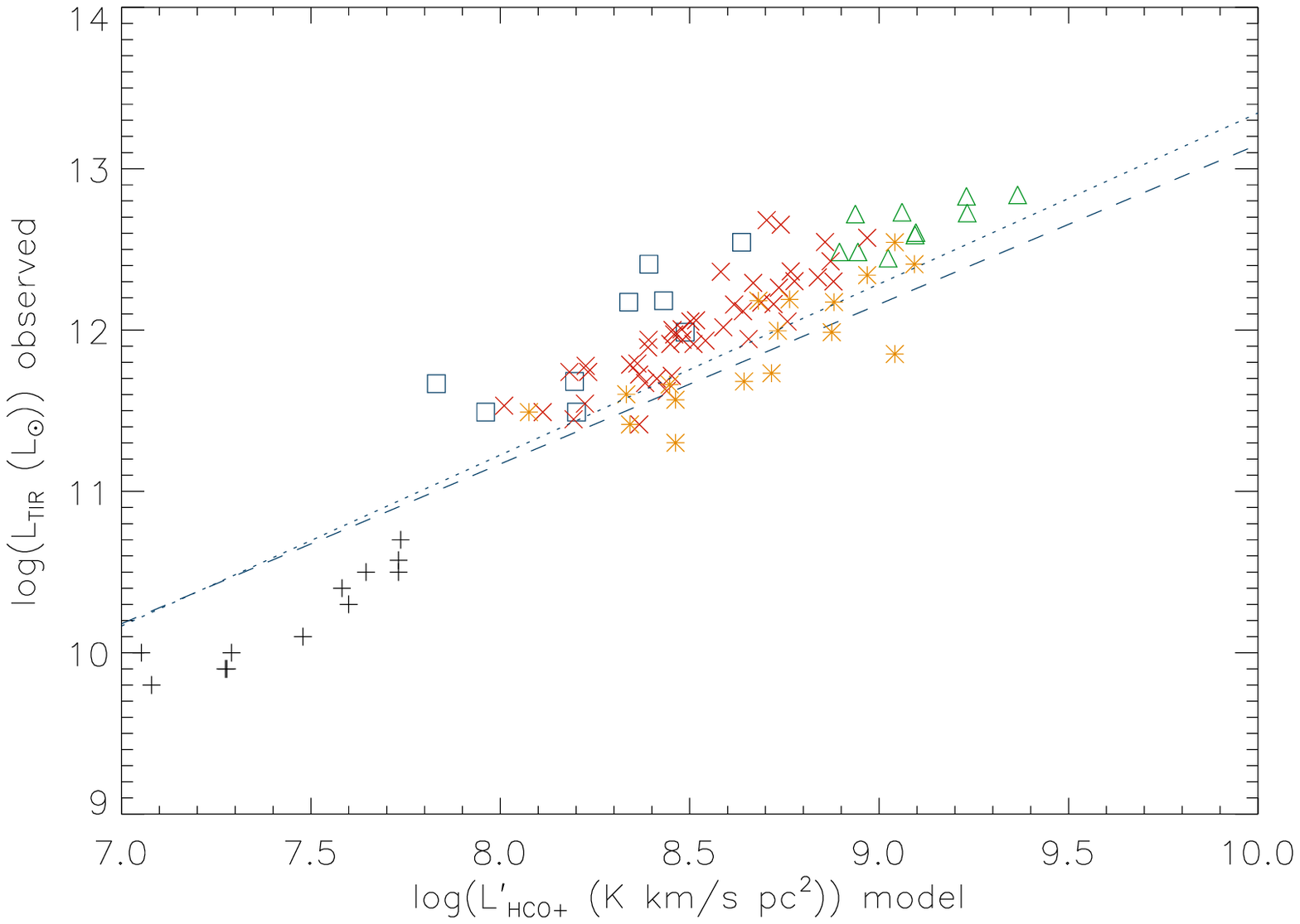}}
  \resizebox{\hsize}{!}{\includegraphics{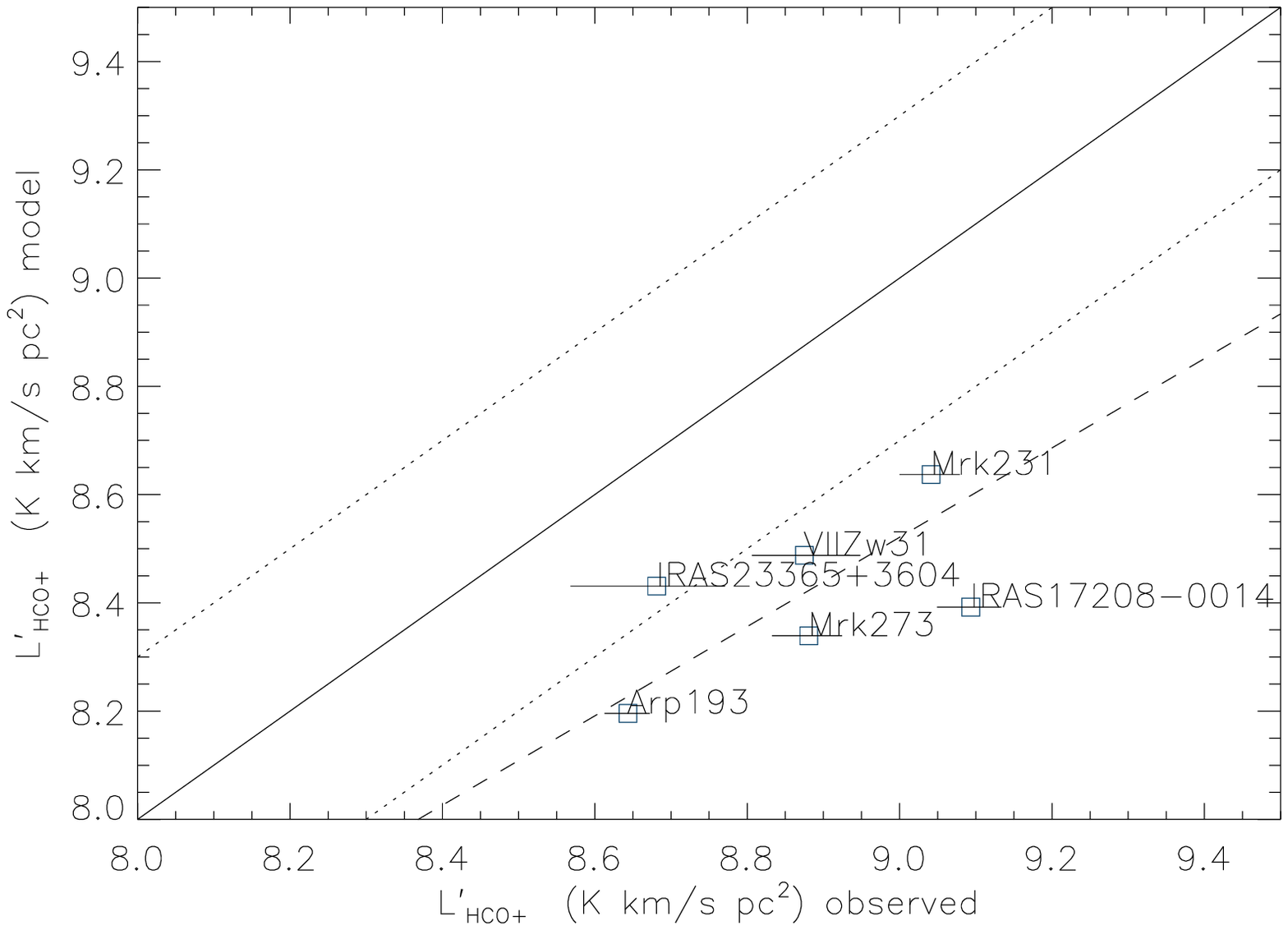}}
  \caption{{\it Constant abundances model.} Upper panel shows observed total infrared luminosity as a function of the HCO$^+$(1--0) luminosity.
    Lower panel shows the model HCN(1--0) luminosity as a function of the observed HCO$^+$(1--0) luminosity. The dashed line represents a robust bisector fit.
    Compare to Fig.~\ref{fig:plots_HCNCO_HCO}.
  \label{fig:plots_HCNCO_constantabundances_HCO+}}
\end{figure}

Observations of HCN(1-0)/HCO$^+$(1-0) ratios of large galaxy samples are rare. Graci\'a-Carpio et al. (2006) and Juneau et al. (2009) found
$\langle \log(L'_{\rm HCN,\ model}/L'_{\rm HCO^+,\ obs}) \rangle \sim 0.2 \pm 0.2$ in local ULIRGs. 
The model with constant abundances systematically underestimates the HCN/HCO$^+$ ratio by approximately a factor of three for all galaxy samples 
(Fig.~\ref{fig:plots_HCNCO_constantabundances_HCO+}).
We conclude that, as expected, our model with constant HCN(1-0) and HCO$^+$(1-0) abundances does not lead to HCN and HCO$^+$ luminosities that are comparable to observations. 

We conclude that models with the canonical CO abundances reproduce observations as well as models using a detailed chemical network.
However, constant HCN and HCO$^{+}$ abundances yield HCN and HCO$^{+}$ line luminosities that are at least a factor of two smaller than
the observed line luminosities.

\subsection{Toomre Q \label{sec:Q}} 

\begin{figure}
  \centering
  \resizebox{\hsize}{!}{\includegraphics{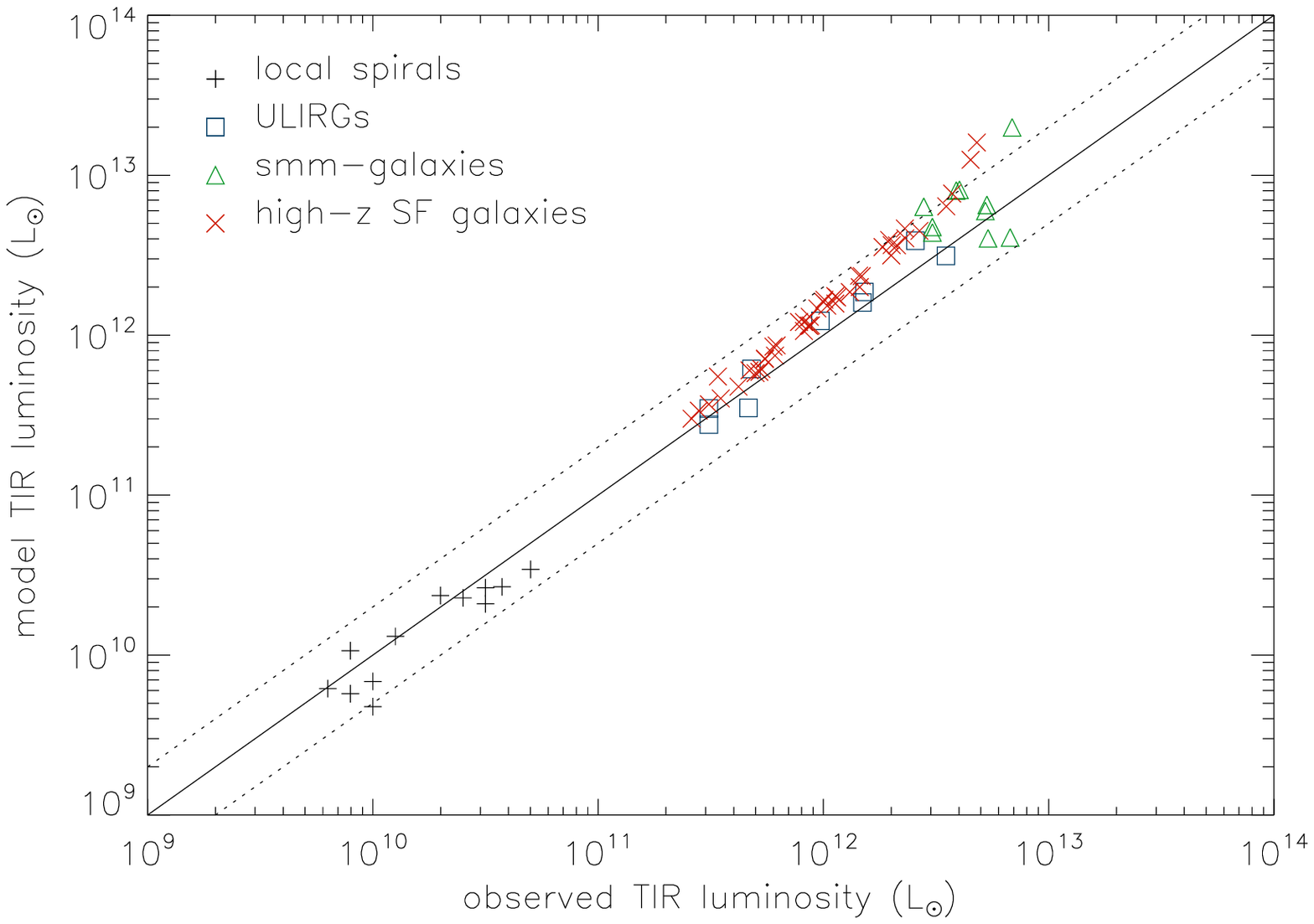}}
  \resizebox{\hsize}{!}{\includegraphics{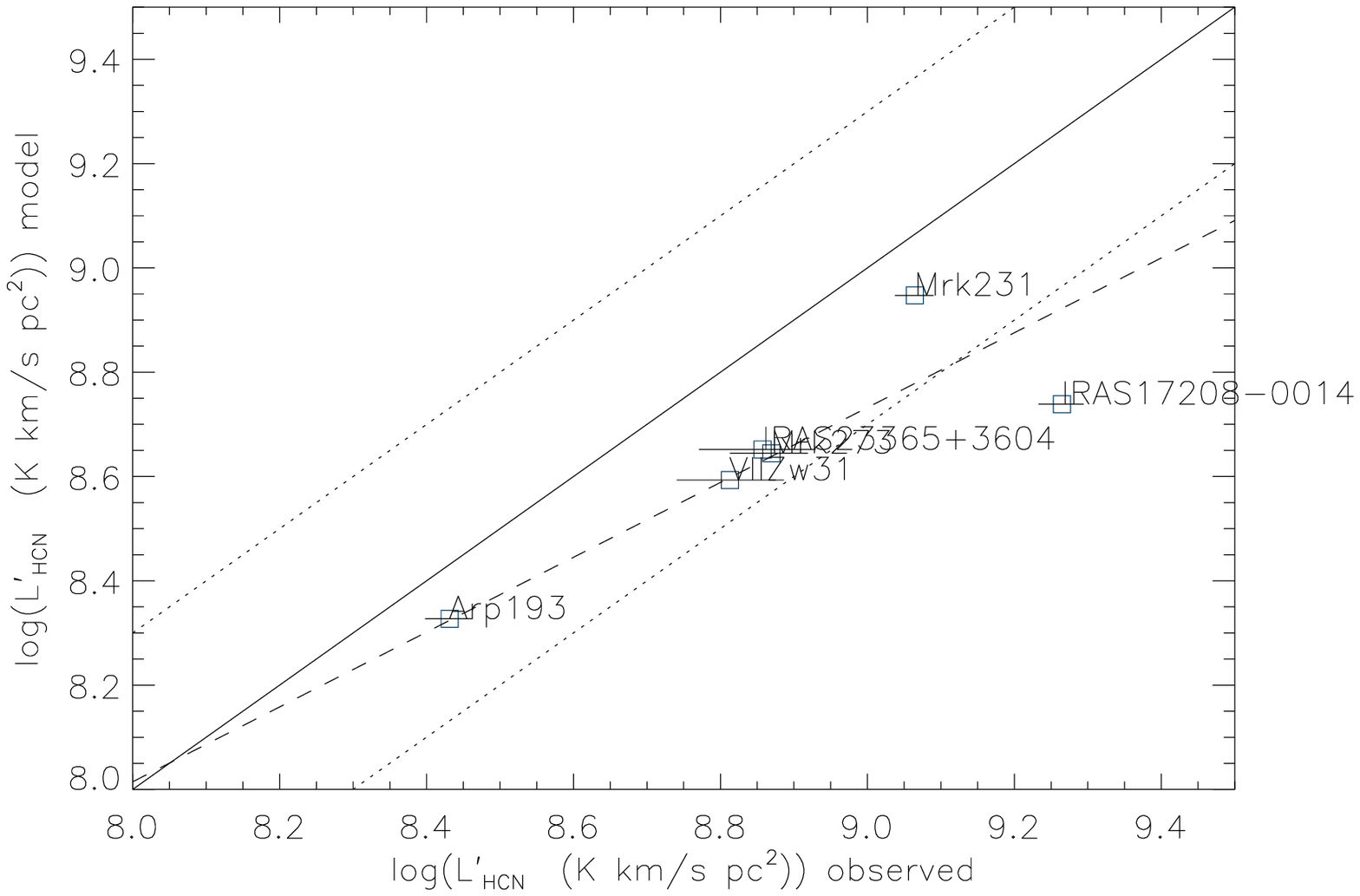}}
  \resizebox{\hsize}{!}{\includegraphics{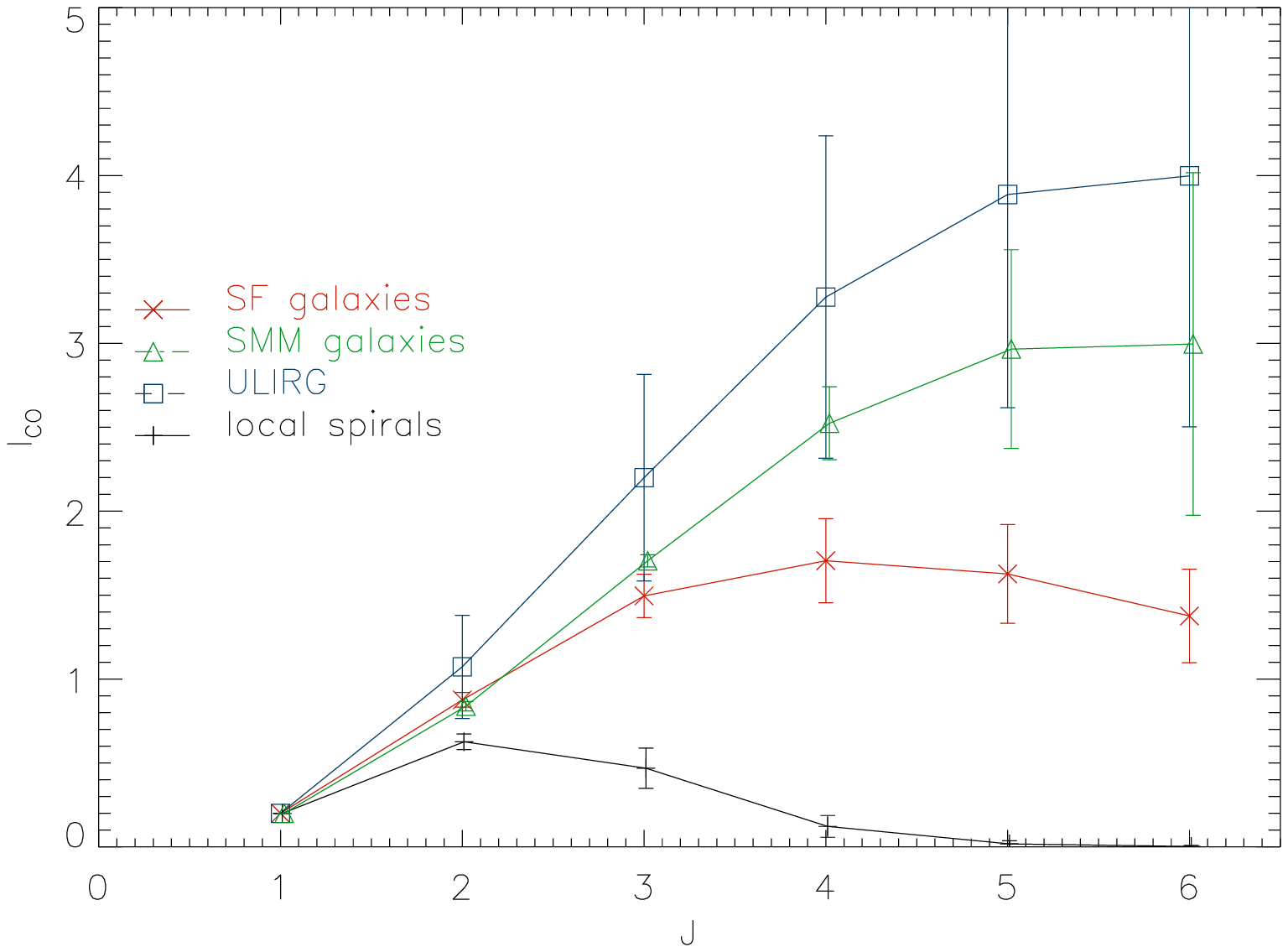}}
  \caption{Same as Fig.~\ref{fig:plots_HCNCO_CO}, upper panel of Fig.~\ref{fig:plots_HCNCO_HCN}, and lower panel of Fig.~\ref{fig:plots_HCNCO_SLED}, 
    but the ULIRG, smm-galaxy, and high-z star-forming galaxy models were calculated with $Q=1$, that is, maximum gas mass.
    \label{fig:plots_HCNCO_Q_CO}}
\end{figure}

To investigate the influence of the Toomre $Q$ parameter on the model CO and HCN line emission, we tested $Q=1$ instead of $Q=1.5$ for
the ULIRGs, smm, and high-z star-forming galaxies (see Table~\ref{tab:correl}). The comparison of the model CO and HCN luminosities and CO SLED with observations
is presented in Fig.~\ref{fig:plots_HCNCO_Q_CO}. The resulting ratios between the model and observed line luminosities are shown in Table~\ref{tab:correl}.
Whereas the model CO and HCO$^+$ line luminosities are barely affected, the model HCN line luminosity decreases by $\sim 15$\,\% when $Q=1$ instead of $Q=1.5$.
The CO SLEDs of the smm and high-z star-forming galaxies do not change significantly with respect to $Q=1.5$,
but the ULIRG CO SLED increases by $\sim 30$\,\% (lower panel of Fig.~\ref{fig:plots_HCNCO_Q_CO}).

We conclude that the low $J$ CO, HCN, and HCO$^+$ line emissions of all galaxies are not significantly affected when $Q$ is decreased 
from $1.5$ to $1$. However, a variation of $Q$ changes the gas velocity dispersion significantly (Sect.~\ref{sec:veldisp}). 

\subsection{The scale parameter $\delta$ \label{sec:delta}}

\begin{figure}
  \centering
  \resizebox{\hsize}{!}{\includegraphics{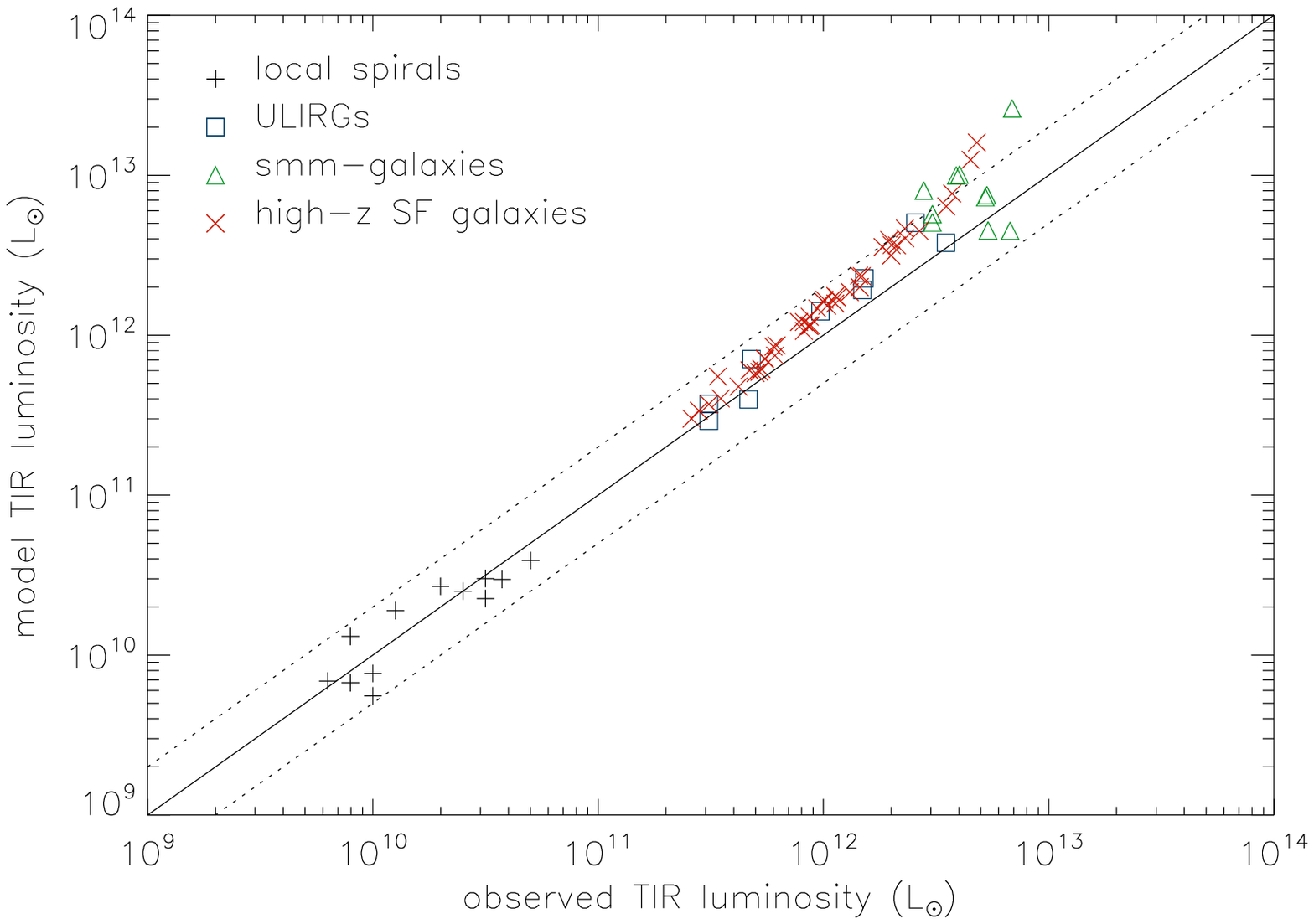}}
  \resizebox{\hsize}{!}{\includegraphics{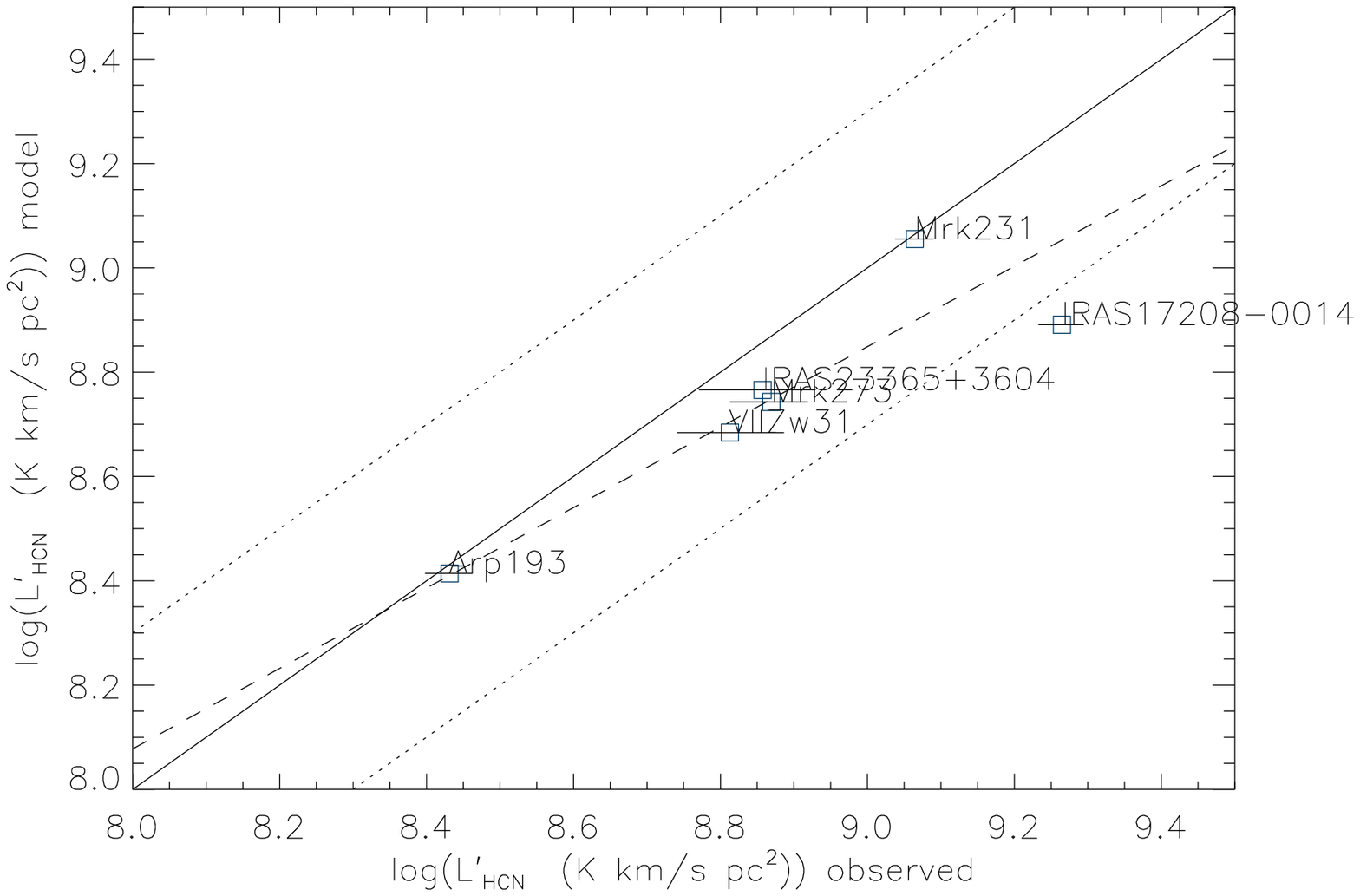}}
  \resizebox{\hsize}{!}{\includegraphics{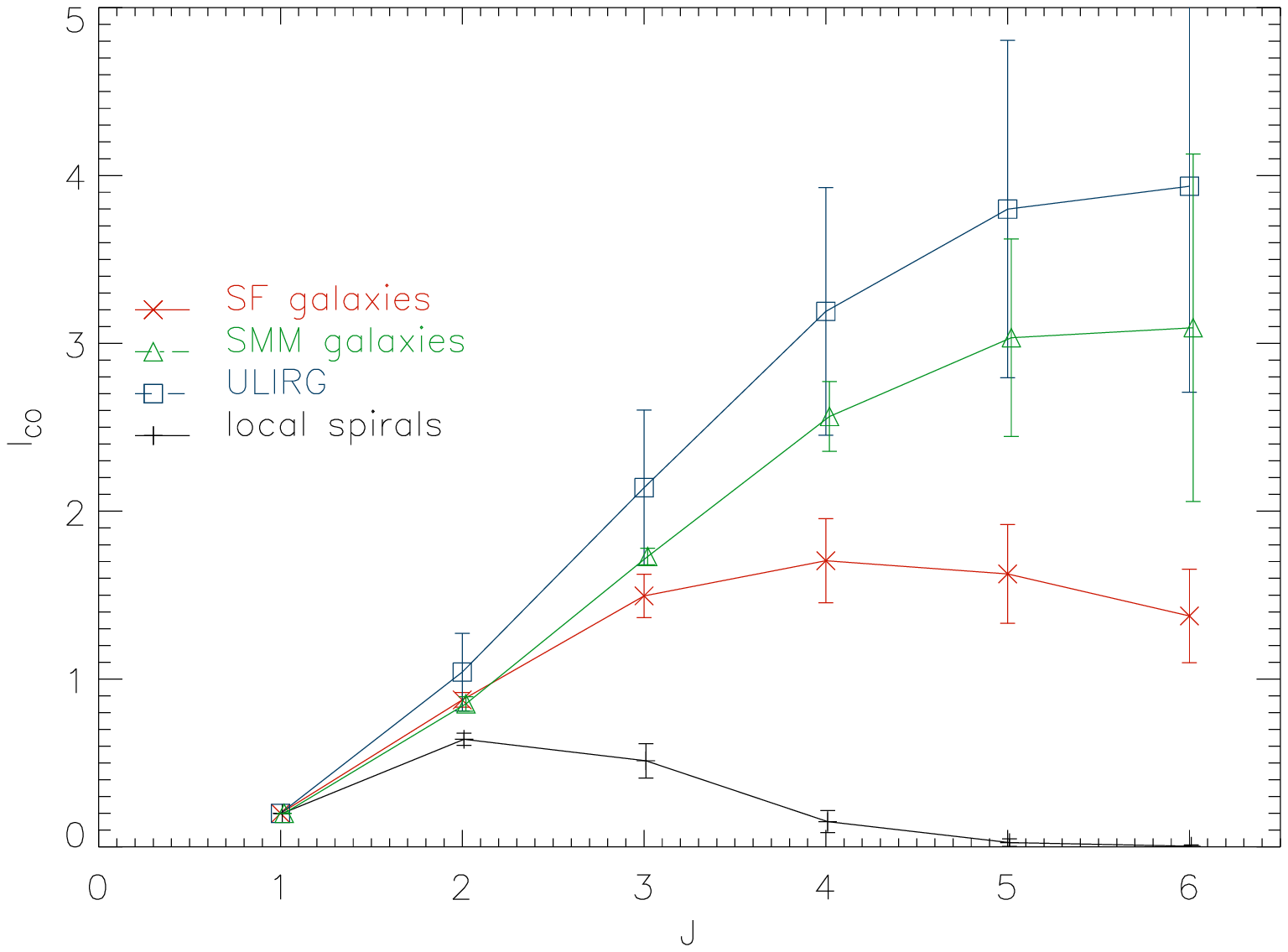}}
  \caption{Same as Fig.~\ref{fig:plots_HCNCO_CO}, upper panel of Fig.~\ref{fig:plots_HCNCO_HCN}, and lower panel of Fig.~\ref{fig:plots_HCNCO_SLED}, 
    but the ULIRG, smm-galaxy, and high-z star-forming galaxy models were calculated with $\delta=15$.
    \label{fig:plots_HCNCO_Q_delta}
}
\end{figure}

As described in Sect.~\ref{sec:clumpiness}, the scale of the largest self-gravitating clouds $l_{\rm cl}$ is smaller
than the turbulent driving length scale $l_{\rm driv}$ by a factor $\delta=l_{\rm driv}/l_{\rm cl}$.
For a typical cloud size of $20$~pc and a driving length scale of $100$~pc, one obtains $\delta=5$, which is the mean value
determined for local spiral galaxies by Vollmer \& Leroy (2011).
To investigate the influence of the scale parameter $\delta$ on the model CO and HCN line emission, we used $\delta=15$ instead of $\delta=5$ for
all galaxies. The comparison of the model CO and HCN luminosities and CO SLED with observations
is presented in Fig.~\ref{fig:plots_HCNCO_Q_delta}. The resulting ratios between the model and observed line luminosities are shown in Table~\ref{tab:correl}.
As for the $Q=1$, the model CO, HCN, and HCO$^+$ line luminosities are barely affected. The effect on the CO SLEDs is also comparable to the $Q=1$ models:
whereas the CO SLEDs of the local spiral galaxies, smm, and high-z star-forming galaxies do not change significantly with respect to $Q=1.5$,
the ULIRG CO SLED increases by $\sim 30$\,\%.

We conclude that the low $J$ CO, HCN, and HCO$^+$ line emissions of all galaxies are not significantly affected, but the CO SLED of the ULIRGs is increased 
by $\sim 30$\,\% when $\delta$ is increased by a factor of three. In contrast to the decrease of $Q$, the increase of $\delta$ does not
result in a significant increase of the gas velocity dispersion (Sect.~\ref{sec:veldisp}).

\section{Galactic physics \label{sec:discussion}}

In this section we examine the role of non-self-gravitating clouds for molecular line emission and investigate
how different recipes for galactic physics influence the model results.

\subsection{The role of non-self-gravitating clouds for molecular line emission \label{sec:nonself}}

\begin{figure}
  \centering
  \resizebox{\hsize}{!}{\includegraphics{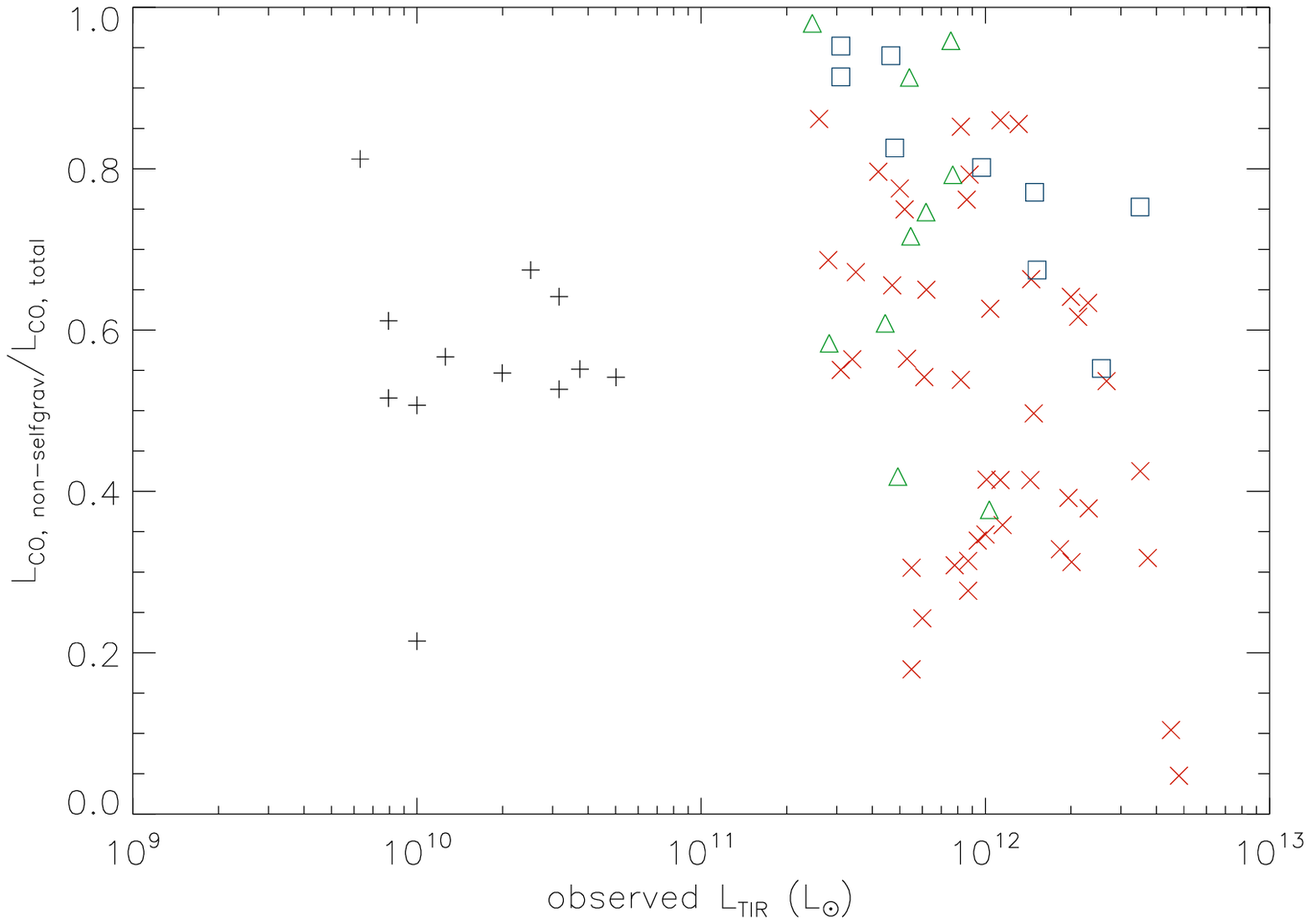}}
  \resizebox{\hsize}{!}{\includegraphics{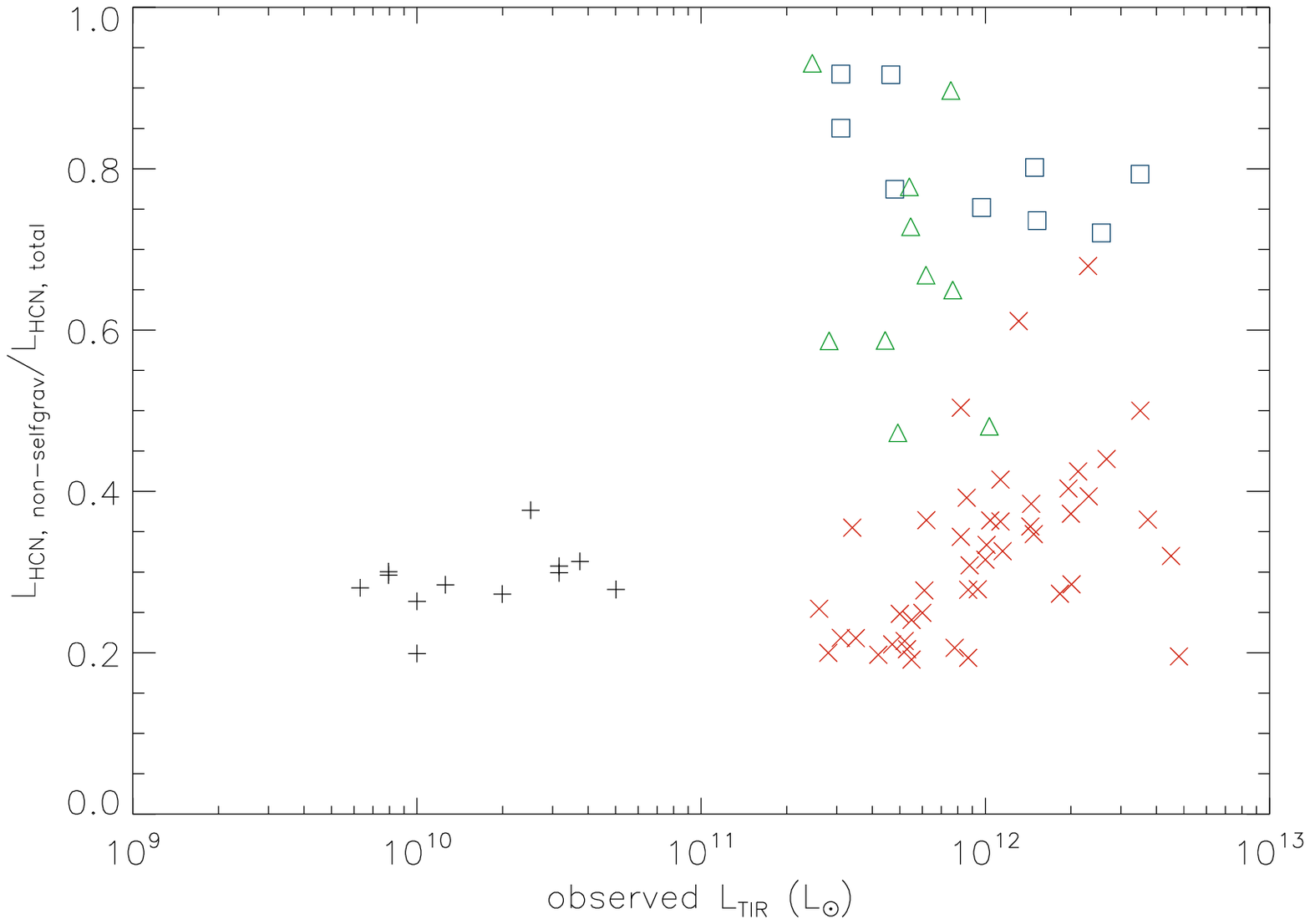}}
   \resizebox{\hsize}{!}{\includegraphics{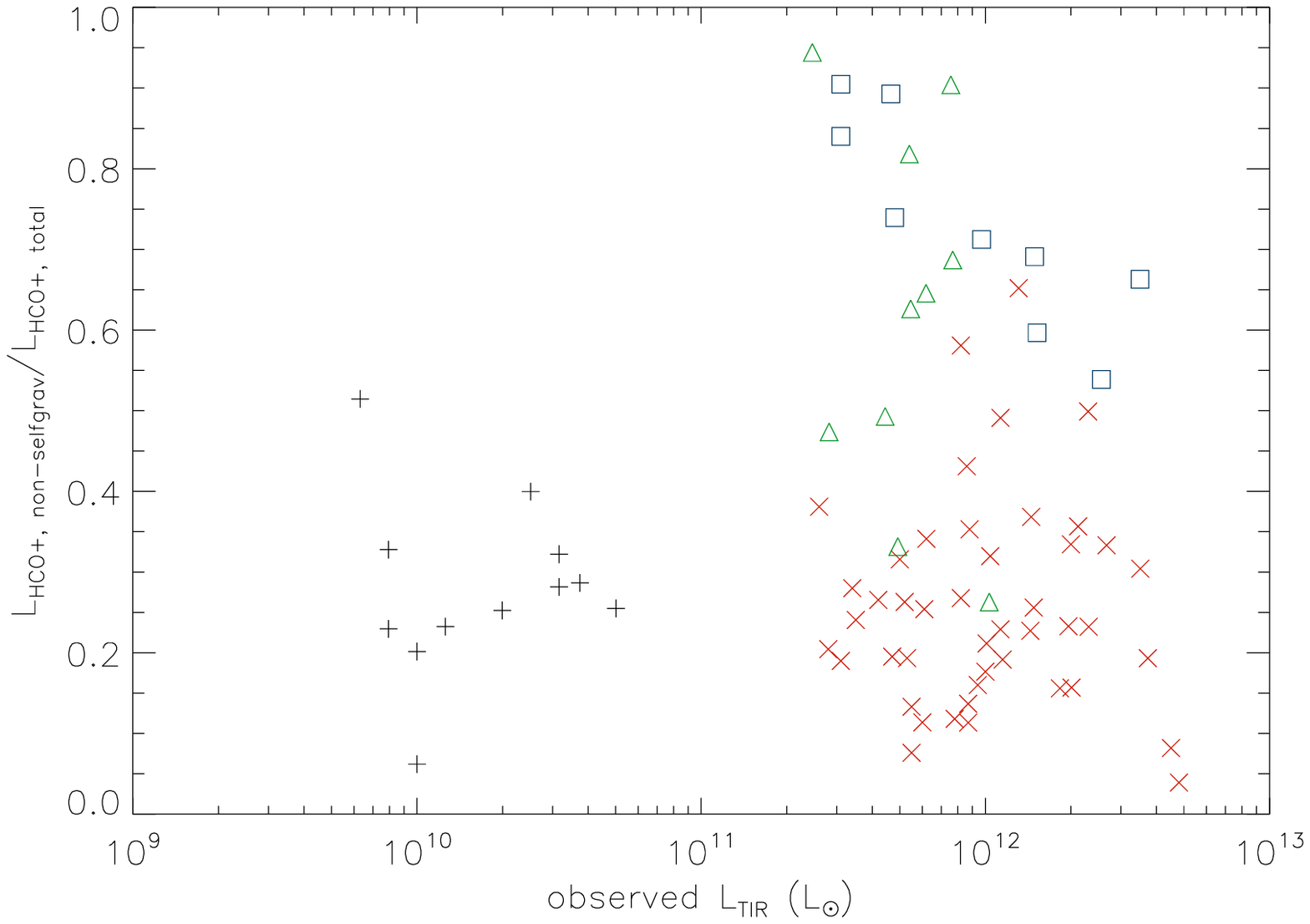}}
  \caption{The role of non-self-gravitating clouds for molecule emission. 
    Ratio between the CO(1--0) (upper panel), HCN(1--0) (middle panel), and HCO$^+$ (lower panel) emission of non-self-gravitating clouds 
    and the total CO(1--0) emission. The symbols are the same as in Fig.~\ref{fig:plots_HCNCO_CO}. 
  \label{fig:newplots_CO}}
\end{figure}

Within the framework of the analytical model (Sect.~\ref{sec:model}), the scale parameter $\delta$ determines the density of
the self-gravitating clouds $\rho_{\rm cl}=\rho/\phi_{\rm V}$ via $t_{\rm ff}^l = t_{\rm turb}^l$. At this characteristic spatial
length scale, the scaling relations for the density and velocity dispersion change (see Sect.~\ref{sec:scaling}).
We would like to know which fractions of molecular line emission originate in self-gravitating and non-self-gravitating gas clouds
in our galaxy samples. To answer this question, the ratio between the molecular line emission of non-self-gravitating clouds 
and the total line emission is shown as a function of the total infrared luminosity in Fig.~\ref{fig:newplots_CO}.
In the local spiral galaxies, approximately half of the total CO emission originates in non-self-gravitating clouds ($f_{\rm nsg, CO} \sim 0.5$). 
This is consistent with the ratio between CO emission from diffuse clouds and the total emission $I_{\rm diff}/I_{\rm tot}=0.6 \pm 0.1$ 
determined by Polk et al. (1988)\footnote{$I_{\rm diff}/I_{\rm tot}=1/(1+F)$ with $F=0.7 \pm 0.3$ from Polk et al. (1988).}

In the ULIRGs, the fraction is $f_{\rm nsg, CO} \sim 0.8$. The smm-galaxies also show high fractions of emission from non-self-gravitating clouds,
however there are also two smm-galaxies with $f_{\rm nsg, CO} \sim 0.4$. The high-z star-forming galaxies show an approximately flat distribution with
$0.3 \le f_{\rm nsg, CO} \le 0.9$. 

For the HCN and HCO$^+$ emission, the situation is different. Since the critical densities for these transitions are much higher than for the
CO emission, it is expected that the denser self-gravitating clouds dominate the HCN and HCO$^+$ emission. 
This is indeed the case for
the local spiral galaxies and the high-z star-forming galaxies. However, in the ULIRGs and smm galaxies, the HCN and HCO$^+$ emission
mostly originates in non-self-gravitating clouds ($0.7 \le f_{\rm nsg, HCN/HCO^+} \le 0.9$).
Even in local spiral galaxies, the model predicts that $\sim 30$\,\% of the HCN(1-0) emission is emitted by
non-self-gravitating clouds. In these clouds, the effective critical density might be as low as $n_{\rm crit}^{\rm HCN} \sim 10^4$~cm$^{-3}$ 
due to radiative trapping, approximately $30$ times lower than the critical density in the optically thin limit (Shirley 2015).
Future combined interferometric and single dish HCN(1-0) observations of local spiral galaxies will be able to test our prediction.

We conclude that compact starburst galaxies (ULIRGs and smm-galaxies) have, on average, significantly higher fractions of
molecular line emission that originates in non-self-gravitating clouds. This effect is more pronounced for HCN(1--0) and HCO$^+$(1--0) emission. 
In compact starburst galaxies, the molecular line emission is most frequently dominated by emission from non-self-gravitating clouds/filaments.
This is due to the highly compact nature of ULIRG and smm-galaxie centers, such that the density of the intercloud medium rivals or even exceeds
the density of Galactic giant molecular clouds.

\subsection{The role of cloud substructure \label{sec:substruct}}

\begin{figure}
  \centering
  \resizebox{\hsize}{!}{\includegraphics{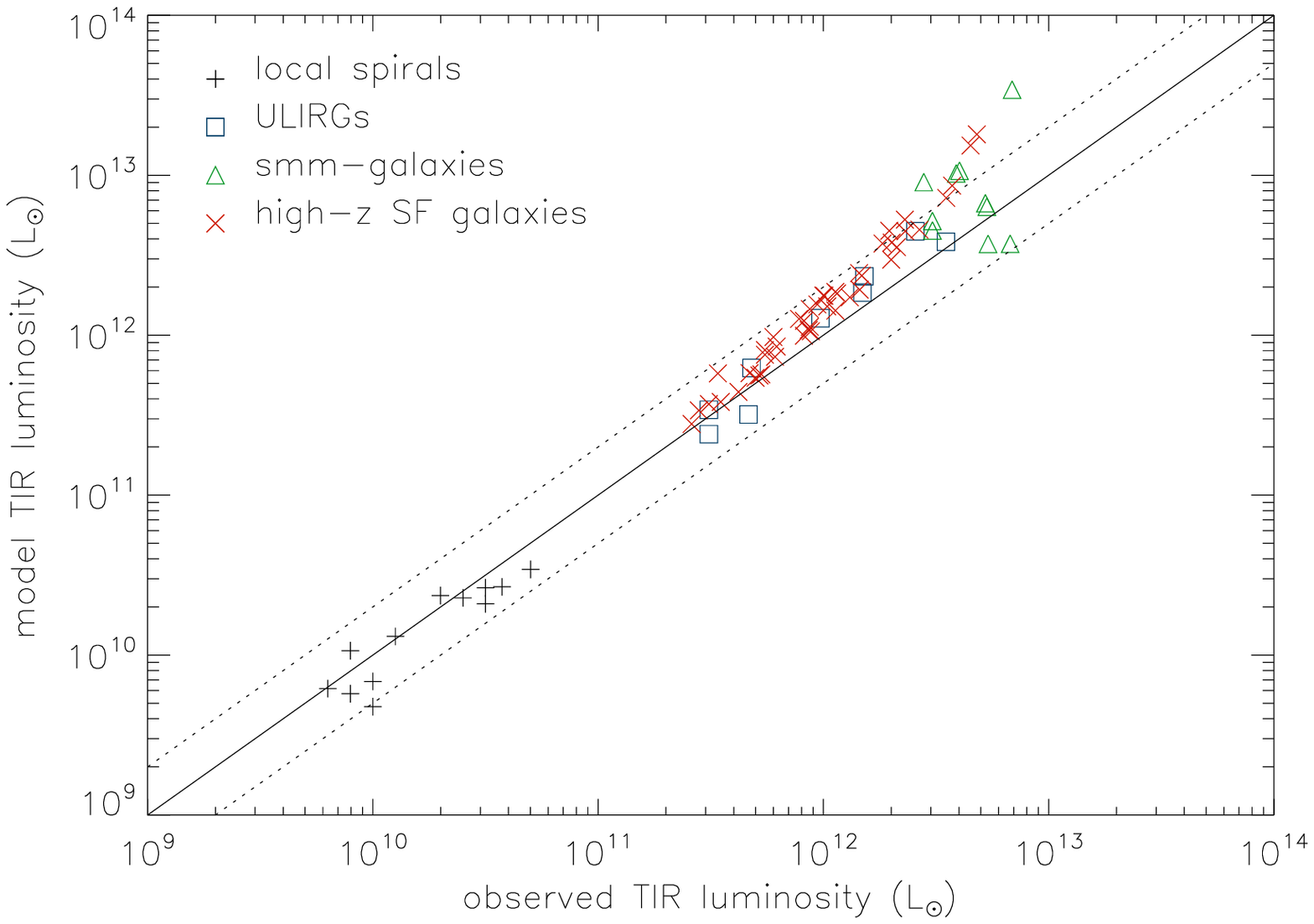}}
  \resizebox{\hsize}{!}{\includegraphics{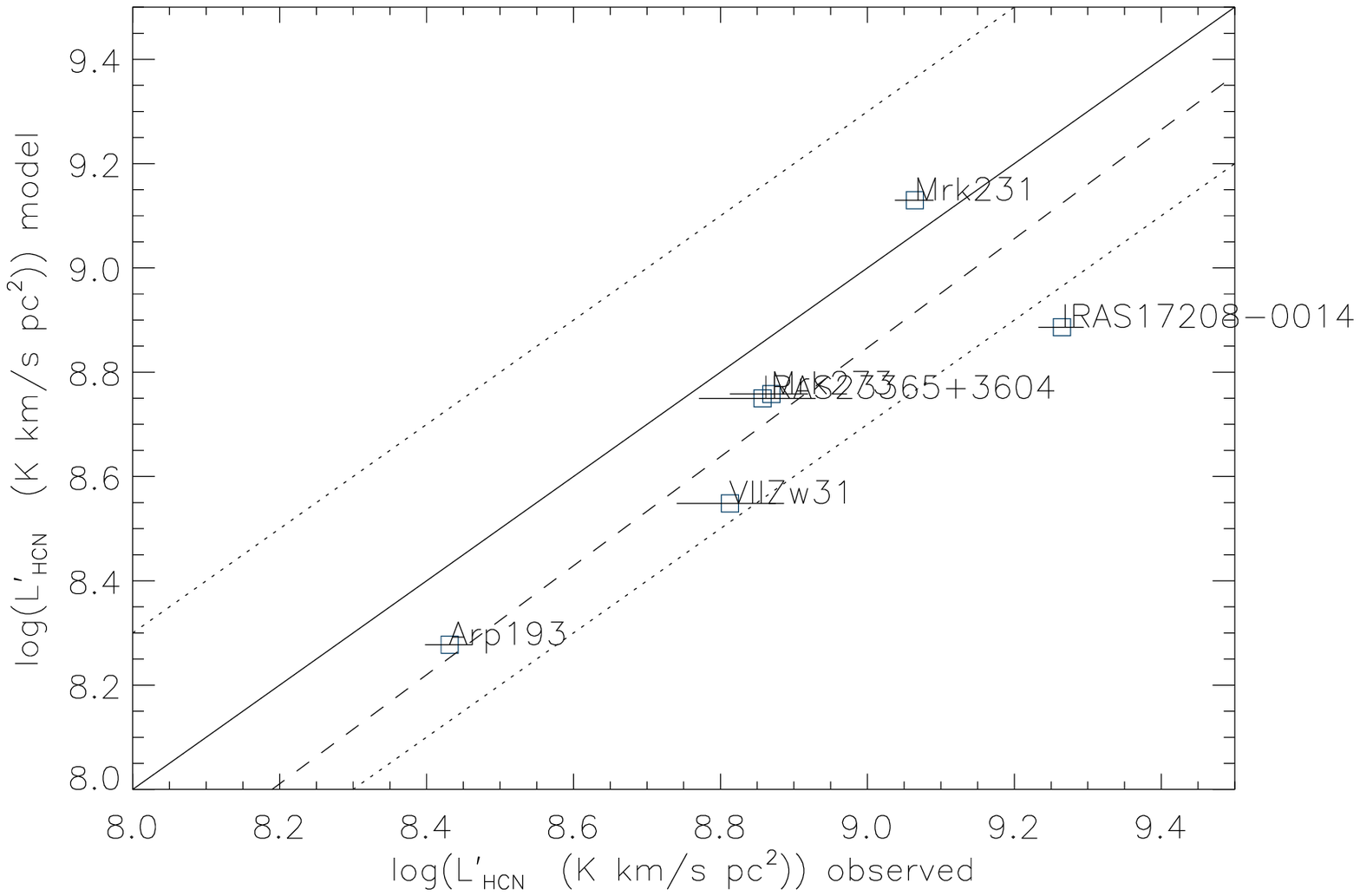}}
  \resizebox{\hsize}{!}{\includegraphics{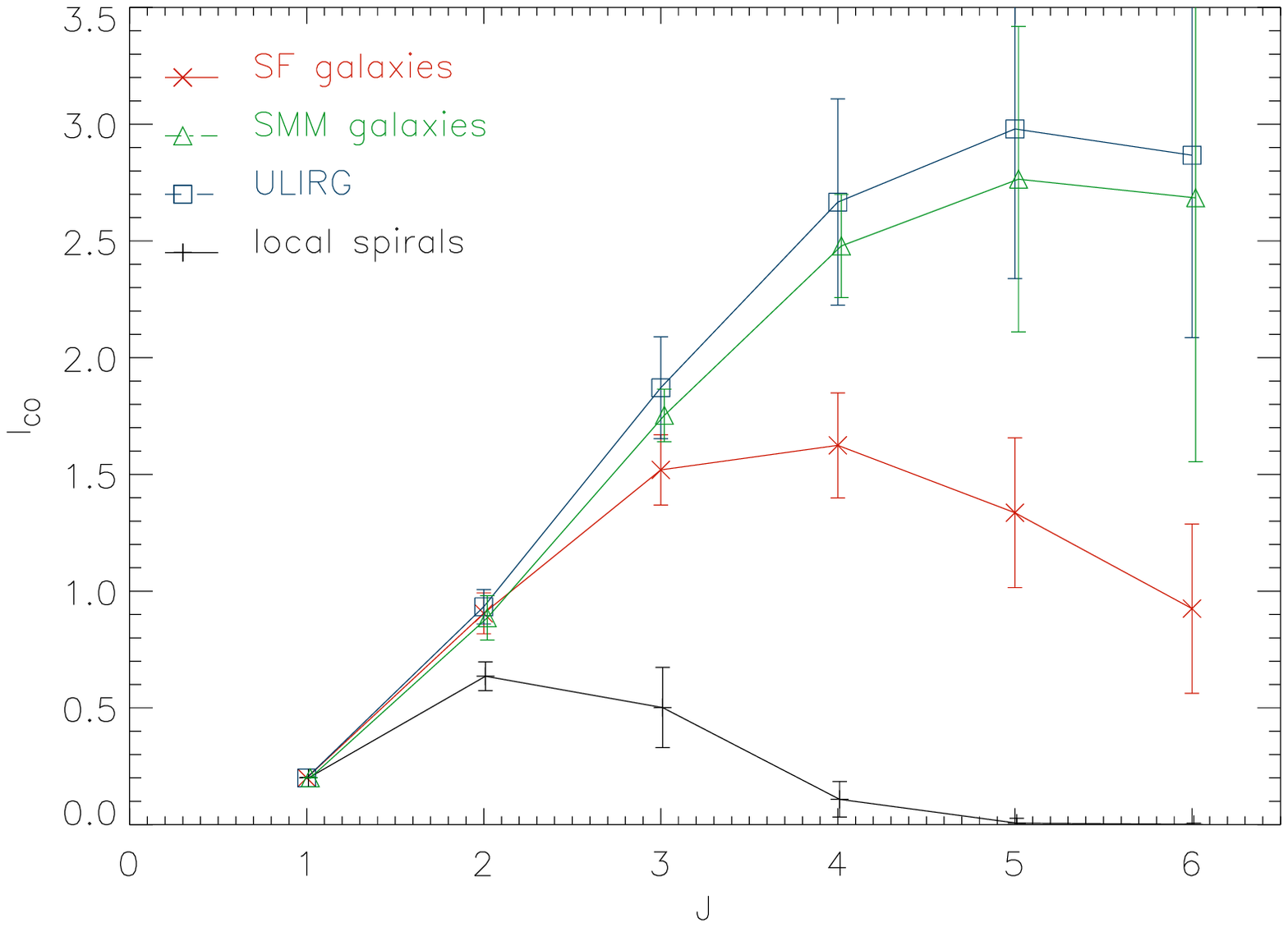}}
  \caption{Same as Fig.~\ref{fig:plots_HCNCO_CO}, upper panel of Fig.~\ref{fig:plots_HCNCO_HCN}, and lower panel of Fig.~\ref{fig:plots_HCNCO_SLED}, 
    but the local spiral galaxies, ULIRG, smm , and high-z star-forming galaxy models were calculated without taking
    into account substructure of the self-gravitating clouds.
    \label{fig:plots_HCNCO_nosub_CO}}
\end{figure}

The molecular line emission of giant molecular clouds shows substructure (e.g., Heyer \& Dame 2015). 
This substructure is taken into account by the model (Sect.~\ref{sec:model}).
To investigate the role of GMC substructure, we tested a model 
where self-gravitating clouds with uniform densities are assumed. The resulting CO and HCN luminosities are presented in 
Fig.~\ref{fig:plots_HCNCO_nosub_CO}.
It might be expected that the higher densities in cloud substructures, that is, cores, increase the molecular line emission. 
To our surprise, the changes, with respect to the model including substructure of self-gravitating clouds, are marginal (Table~\ref{tab:correl}).
As expected, these small changes are most visible at high-$J$ CO transitions.

This effect can be explained by a balance between the absence of high-density gas and the increase of the area-filling factor
of emission from self-gravitating clouds with a uniform gas density.

\subsection{Cosmic ray heating \label{sec:CRheat}}

\begin{figure}
  \centering
  \resizebox{\hsize}{!}{\includegraphics{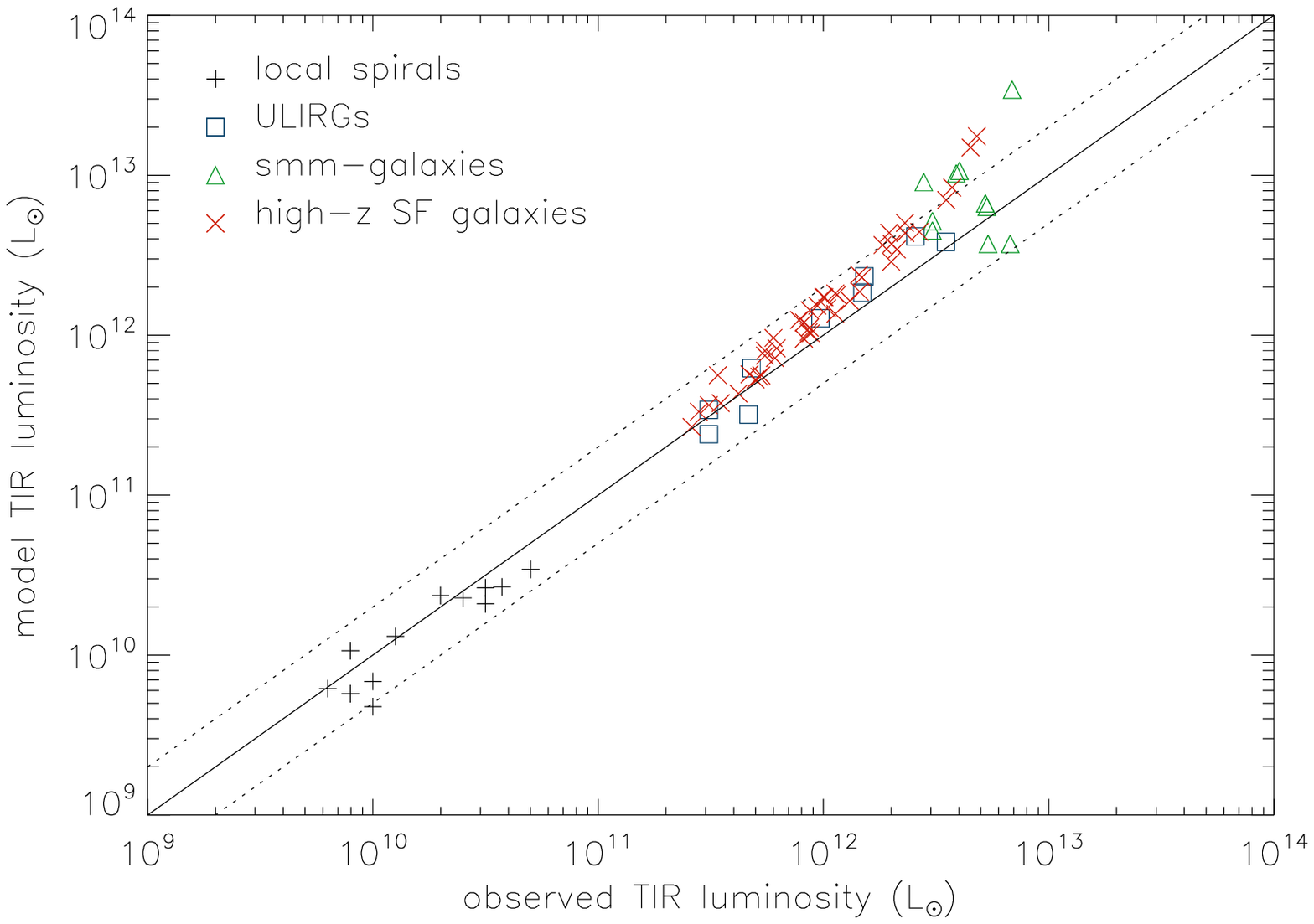}}
  \resizebox{\hsize}{!}{\includegraphics{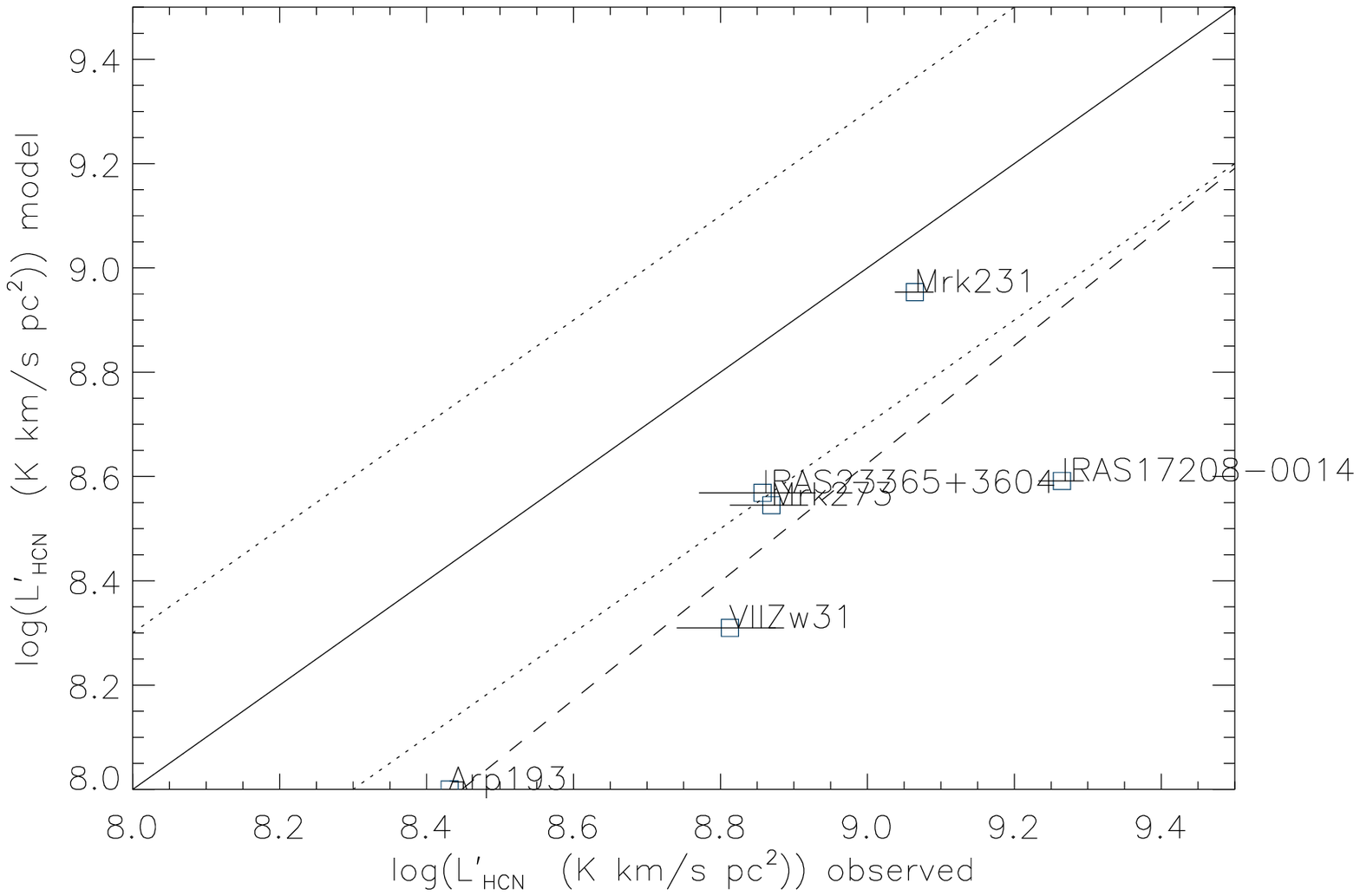}}
  \resizebox{\hsize}{!}{\includegraphics{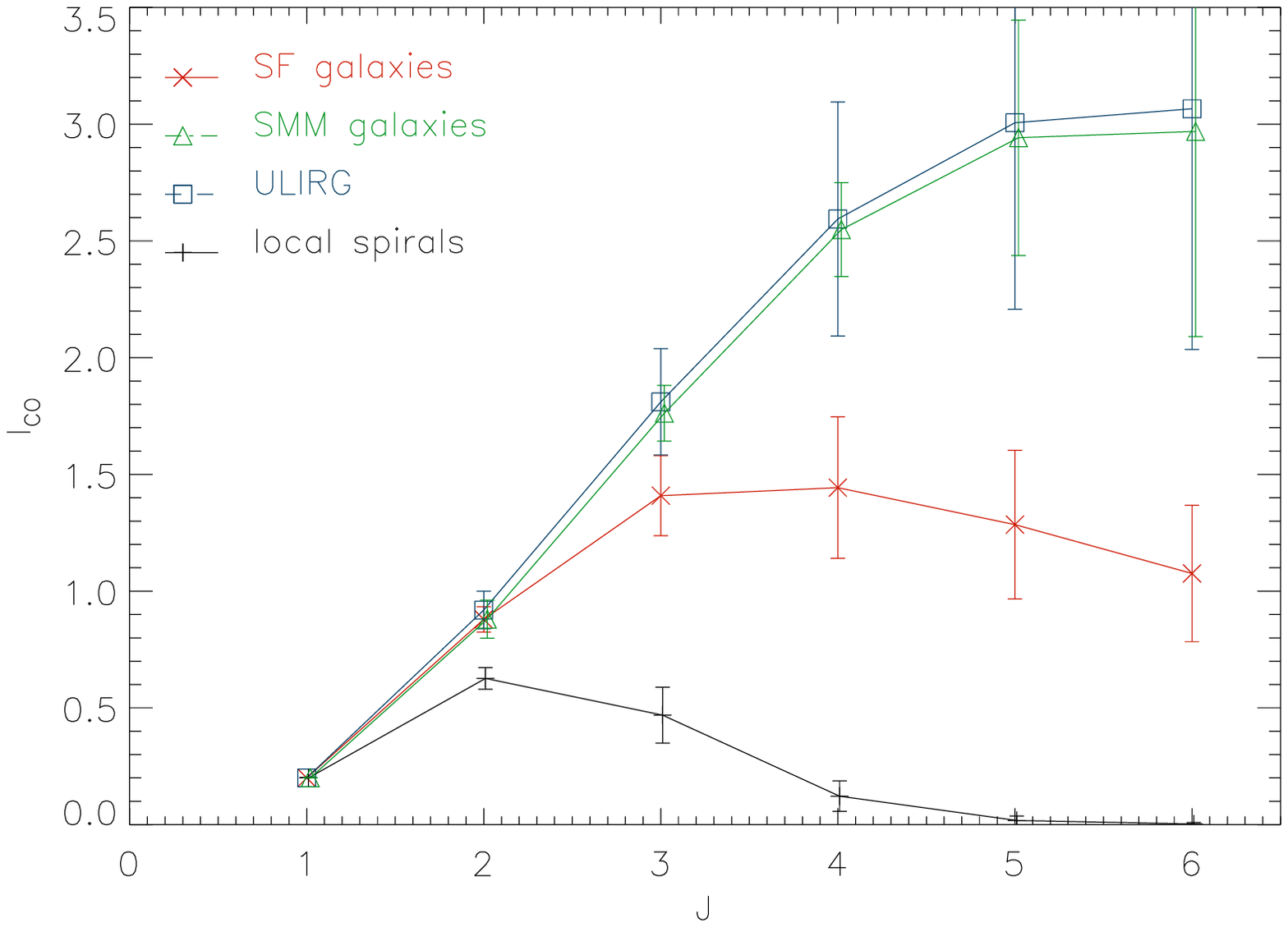}}
  \caption{As for Fig.~\ref{fig:plots_HCNCO_CO}, upper panel of Fig.~\ref{fig:plots_HCNCO_HCN}, and lower panel of Fig.~\ref{fig:plots_HCNCO_SLED}, 
    but the local spiral galaxies, ULIRG, smm, and high-z star-forming galaxy models were calculated without cosmic ray heating.
    \label{fig:plots_HCNCO_noCR_CO}}
\end{figure}

The molecular line emission depends on the gas density, column density, and temperature. Within the model
framework, the gas temperature within the dense gas clouds depends on the turbulent and cosmic ray heating (see Sect.~\ref{sec:gdtemp}).
The cosmic ray ionization rate is a factor of $40$ higher in ULIRGs and smm-galaxies than in local spiral galaxies.
To investigate the influence of the cosmic ray heating, we made a model calculation without this heating term.
The resulting CO and HCN luminosities are presented in Fig.~\ref{fig:plots_HCNCO_noCR_CO}.
The CO(1--0) luminosities of all galaxies are not altered by the absence or presence of cosmic ray heating (Table~\ref{tab:correl}).
This implies that, at the relevant density regime of $300$-$1000$~cm$^{-3}$, turbulent heating dominates over cosmic ray heating.

One would expect that the situation changes for molecular line transitions with higher critical densities because
the cosmic ray heating is proportional to the gas density, whereas the turbulent heating is proportional to
the square root of the gas density (see Sect.~\ref{sec:gdtemp}).
Surprisingly, the model ULIRG CO emission of transitions with upper $J$ of $2 \ge J \ge 11$ only changes by at most $\pm 0.1$~dex.
Four ULIRGs even show a slight increase of CO emission around $J=7$ when the CR ray heating is suppressed.
We suspect that the effect is due to a changing chemistry that requires thorough investigation, something beyond the scope of this work.

There is no effect of CR heating on HCN(1--0) and HCO$^+$(1--0) emission for the local spiral and smm-galaxies.
In the absence of CR heating, the HCN(1--0) line luminosity decreases by a factor of approximately 1.5 for ULIRGs and
high-z star-forming galaxies. The HCO$^+$ (1--0) line emission shows the same behavior as the HCN(1--0) line emission for
local spiral galaxies, smm-galaxies, and high-z star-forming galaxies. For ULIRGs, the situation is more complicated;
whereas the overall comparison with the observed $L_{\rm IR}$--$L_{\rm HCN}$ relation shows a decreasing line emission in the absence of CR heating, 
the direct comparison of the model and observed HCN luminosities is consistent with no change.

We conclude that within the framework of this model, the inclusion of CR heating does not significantly affect the CO emission, but increases the
HCN(1--0), and HCO$^+$(1--0) emission by at most a factor of two. This factor is needed to reproduce the observed HCN(1--0) emission of the ULIRGs.

\subsection{Infrared-pumping \label{sec:IRpumping}}

\begin{figure}
  \centering
  \resizebox{\hsize}{!}{\includegraphics{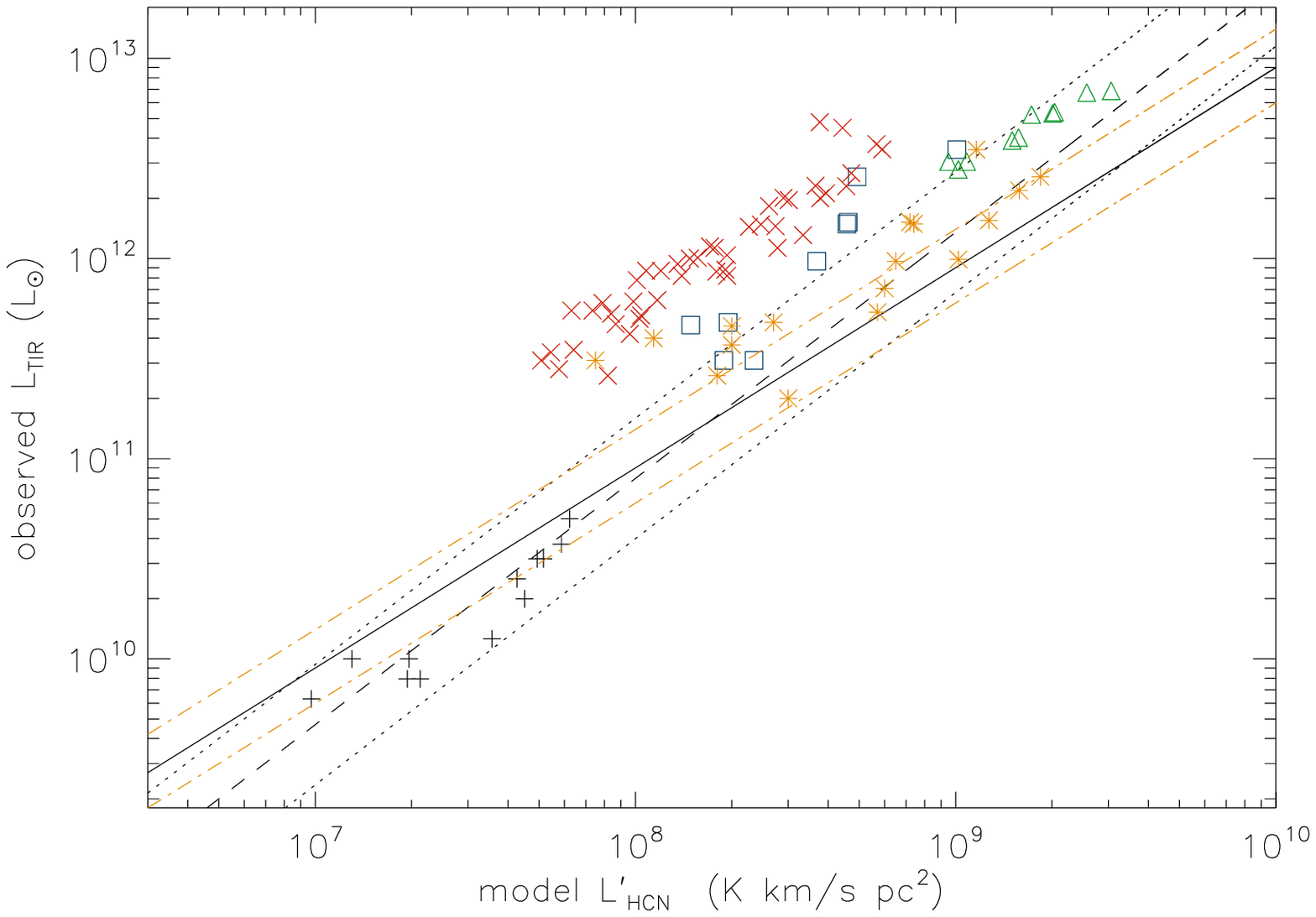}}
  \resizebox{\hsize}{!}{\includegraphics{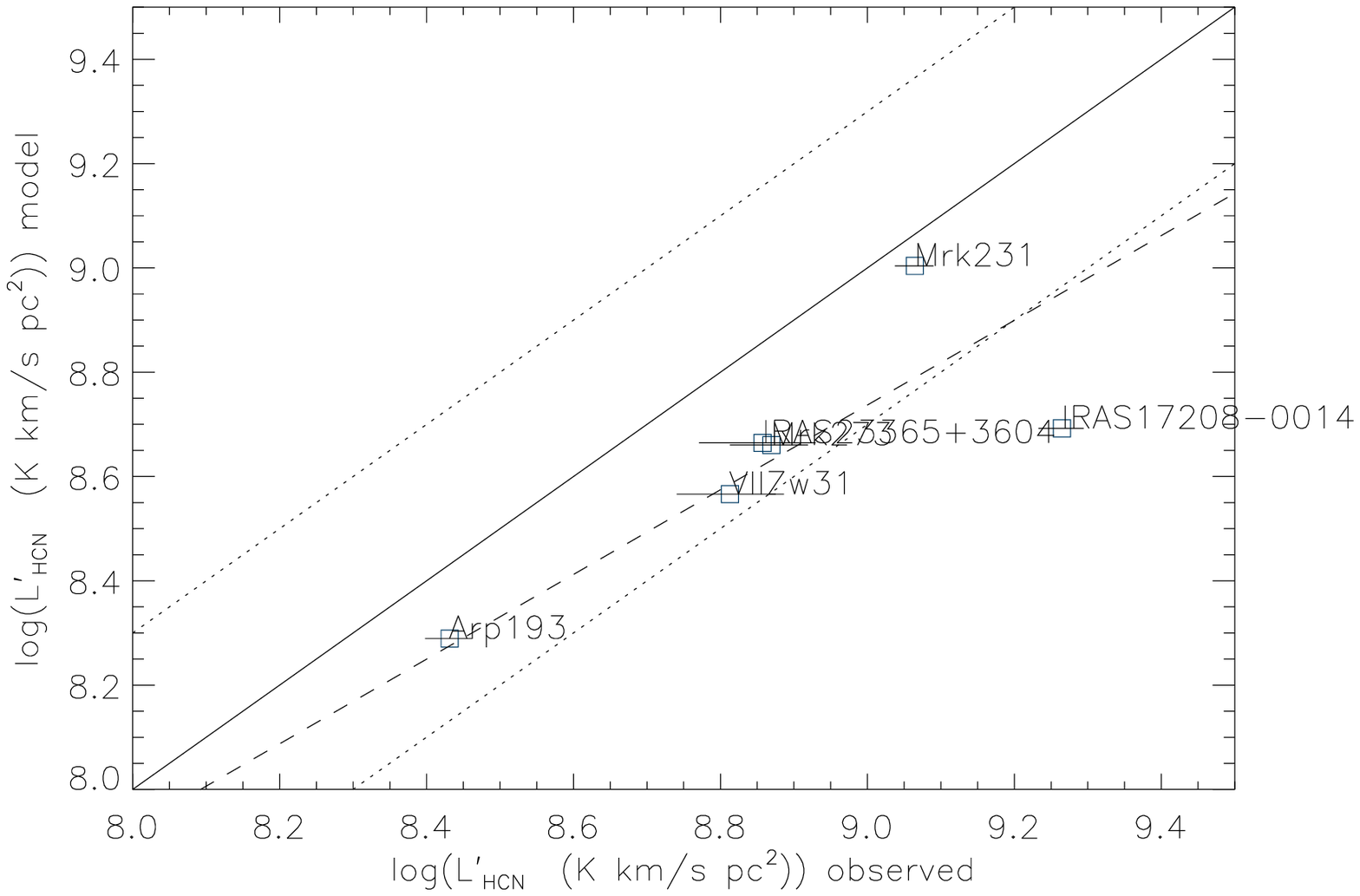}}
  \caption{{\it The local spiral galaxies, ULIRG, smm, and high-z star-forming galaxy models were calculated without IR-pumping for HCN}.
    Upper panel shows observed total infrared luminosity as a function of the model HCN luminosity.
    The solid line corresponds to $L'_{\rm HCN,\ obs}=900 \times L_{\rm TIR,\ obs}$ (Gao \& Solomon 2004). 
    The dashed line corresponds to $L_{\rm TIR,\ obs}=12 \times L_{\rm HCN,\ obs}^{'\ 1.23}$ (Graci\'a-Carpio et al. 2008) with factors of $0.5$ and $2$ (dotted lines).
    Lower panel shows the model HCN(1--0) luminosity as a function of the observed HCN(1--0) luminosity. 
    Compare to Figs.~\ref{fig:plots_HCNCO_CO} and \ref{fig:plots_HCNCO_HCN}.
    \label{fig:plots_HCNCO_noIRpump_HCN}}
\end{figure}

Infrared pumping of HCN (Sect.~\ref{sec:irpumping}) via  the $14~\mu$m bending modes is suggested to play an important role for the HCN(1--0) emission
 of ULIRGs (e.g., Aalto et al. 2015). To investigate the influence of infrared pumping on the HCN(1--0) line emission, we calculated a
model without infrared pumping. The results are presented in Fig.~\ref{fig:plots_HCNCO_noIRpump_HCN} and Table~\ref{tab:correl}.
The HCN(1--0) emission of local spiral galaxies and smm-galaxies is not significantly affected by HCN infrared pumping.
The HCN(1--0) emission of ULIRGs and high-z star-forming galaxies decreases by $20$--$30$\,\% when the infrared pumping is suppressed. 
Thus, within the framework of our model, HCN infrared pumping has a measurable but relatively small effect on the HCN(1--0) of
ULIRGs and high-star-forming galaxies.

\section{Physical properties of the galaxies \label{sec:physpar}}

\subsection{Gas fraction \label{sec:gasfraction}}

The observed high gas fractions ($M_{\rm H_2}/(M_{\rm H_2}+M_*)$) of high-z star-forming galaxies (Tacconi et al. 2013) strongly depend on the assumed CO conversion
factor $\alpha_{\rm CO}$. Tacconi et al. (2013) used a Galactic conversion factor. Our CO conversion factor is a factor of approximately two smaller (Sect.~\ref{sec:conversion}). 
Genzel et al. (2015) claimed that the H$_2$ mass estimates of the high-z star-forming galaxies with a Galactic 
conversion factor agree to better than $50$\,\% with the H$_2$ mass estimates based on the infrared SEDs. 
We suppose that our molecular mass estimate is consistent with that of Tacconi et al. (2013) given the uncertainties of the observed and model (see Sect.~\ref{sec:uncertain})
CO and dust conversion factors of approximately a factor of two. The infrared luminosities of high-z star-forming galaxies ($\sim 10^{11}$~L$_{\odot}$) are between those of 
local spiral galaxies ($\sim 10^{10}$~L$_{\odot}$) and ULIRGs ($\sim 10^{12}$~L$_{\odot}$). It seems thus reasonable to find conversion factors for high-z star-forming galaxies 
that range between the Galactic conversion factor and that for ULIRGs.

\begin{figure}
  \centering
  \resizebox{\hsize}{!}{\includegraphics{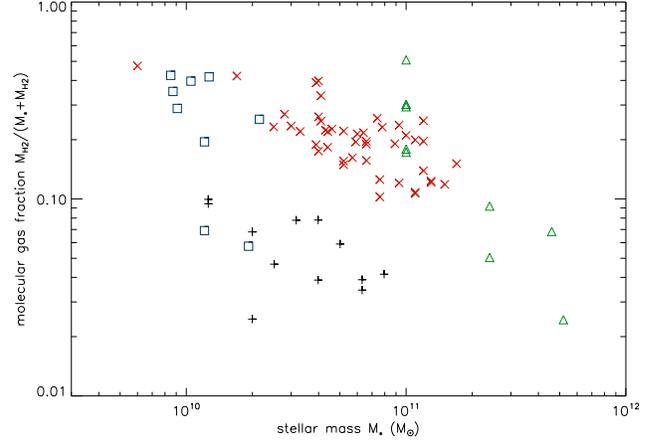}}
  \caption{Model gas fraction as a function of the stellar mass.
  \label{fig:gasfractions}}
\end{figure}

The model H$_2$ gas fractions are presented in Fig.~\ref{fig:gasfractions}.
Whereas the H$_2$ gas fraction of local spirals is approximately $5$\,\%, that of the high-z star-forming galaxies range between $10$ and $30$\,\% with a mean of $\sim 20$\,\%.
Tacconi et al. (2013) found a H$_2$ gas fraction of $0.33$. If we use a Galactic conversion factor, we recover these high gas fractions.
However, we determined a conversion factor for the high-z star-forming galaxies which is half of the Galactic conversion factor.
So, at least in our model counterparts of the high-z star-forming galaxies, we know that the H$_2$ gas fraction is lower.

In our model, the highest H$_2$ gas fractions, exceeding $30$\,\%, are found in ULIRGs. The widely varying H$_2$ gas fractions of the smm-galaxies greatly depend
on the determination of the stellar masses from observations, which have large uncertainties.

\subsection{Gas velocity dispersion \label{sec:veldisp}}

\begin{figure}
  \centering
  \resizebox{\hsize}{!}{\includegraphics{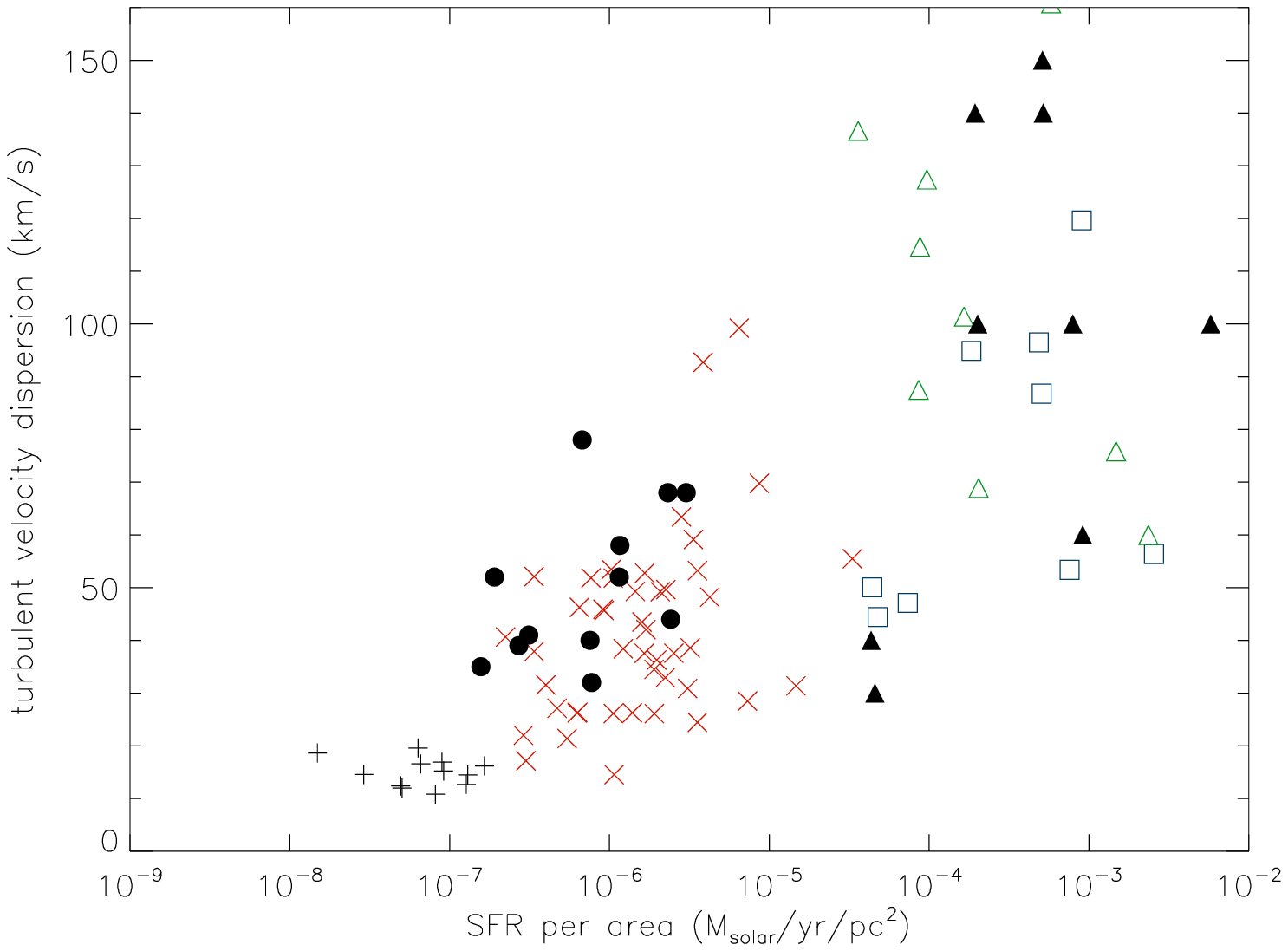}}
  \resizebox{\hsize}{!}{\includegraphics{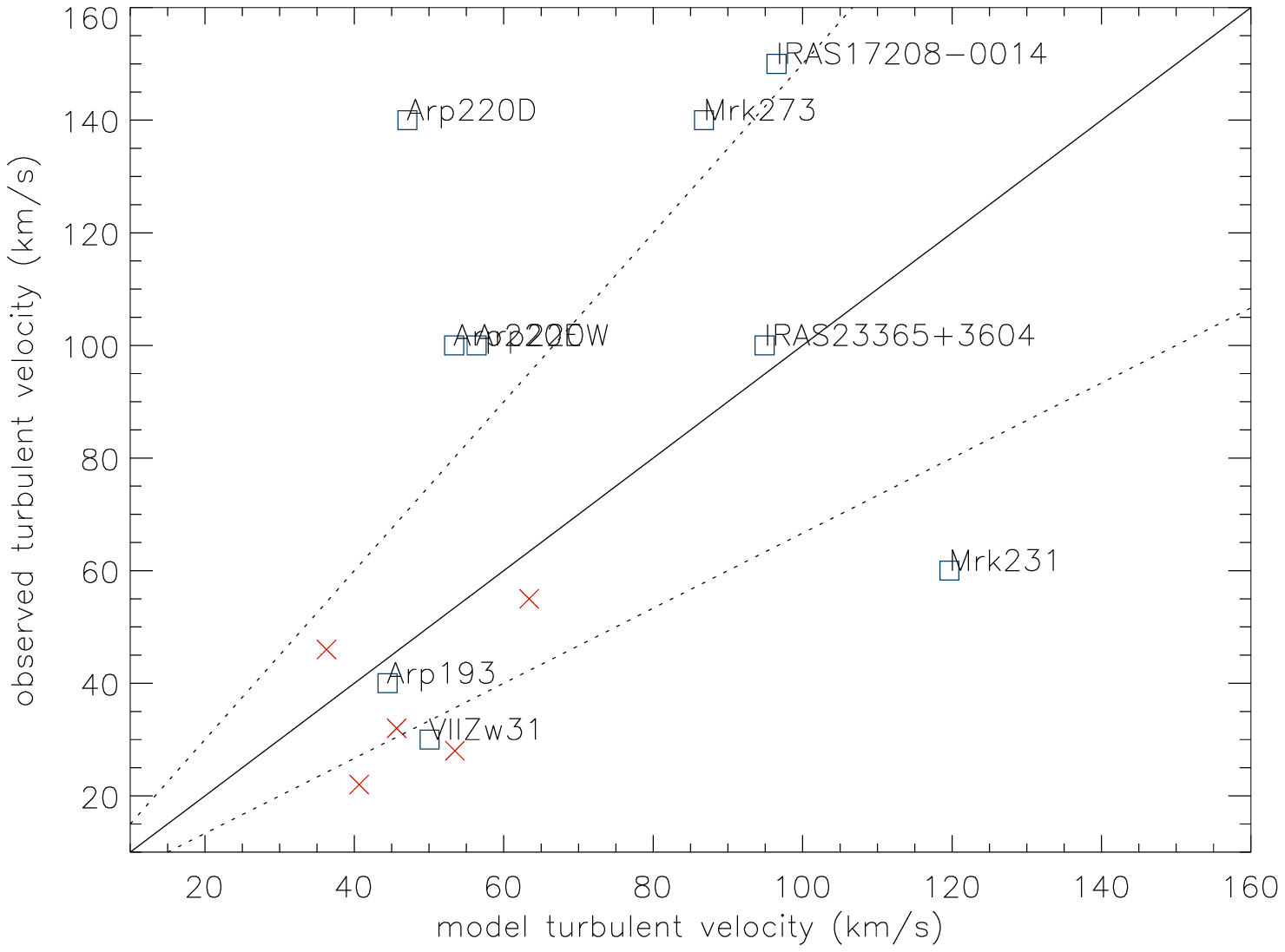}}
  \caption{Upper panel: model turbulent velocity dispersion as a function of star-formation surface density.
    The symbols are the same as in Fig.~\ref{fig:plots_HCNCO_CO}. In addition, filled triangles are the
    observed velocity dispersions from Downes \& Solomon (1998). Filled circles are observed H$\alpha$ 
    velocity dispersions from Cresci et al. (2009).
    Lower panel shows observed CO velocity dispersion (Tacconi et al. 2013) as a function of the model velocity dispersion.
    The symbols are the same as in Fig.~\ref{fig:plots_HCNCO_CO}. The solid line corresponds to equality and 
    the dotted lines to factors of $1/1.5$ and $1.5$.
  \label{fig:plots_1_vturb}}
\end{figure}

\begin{figure}
  \centering
  \resizebox{\hsize}{!}{\includegraphics{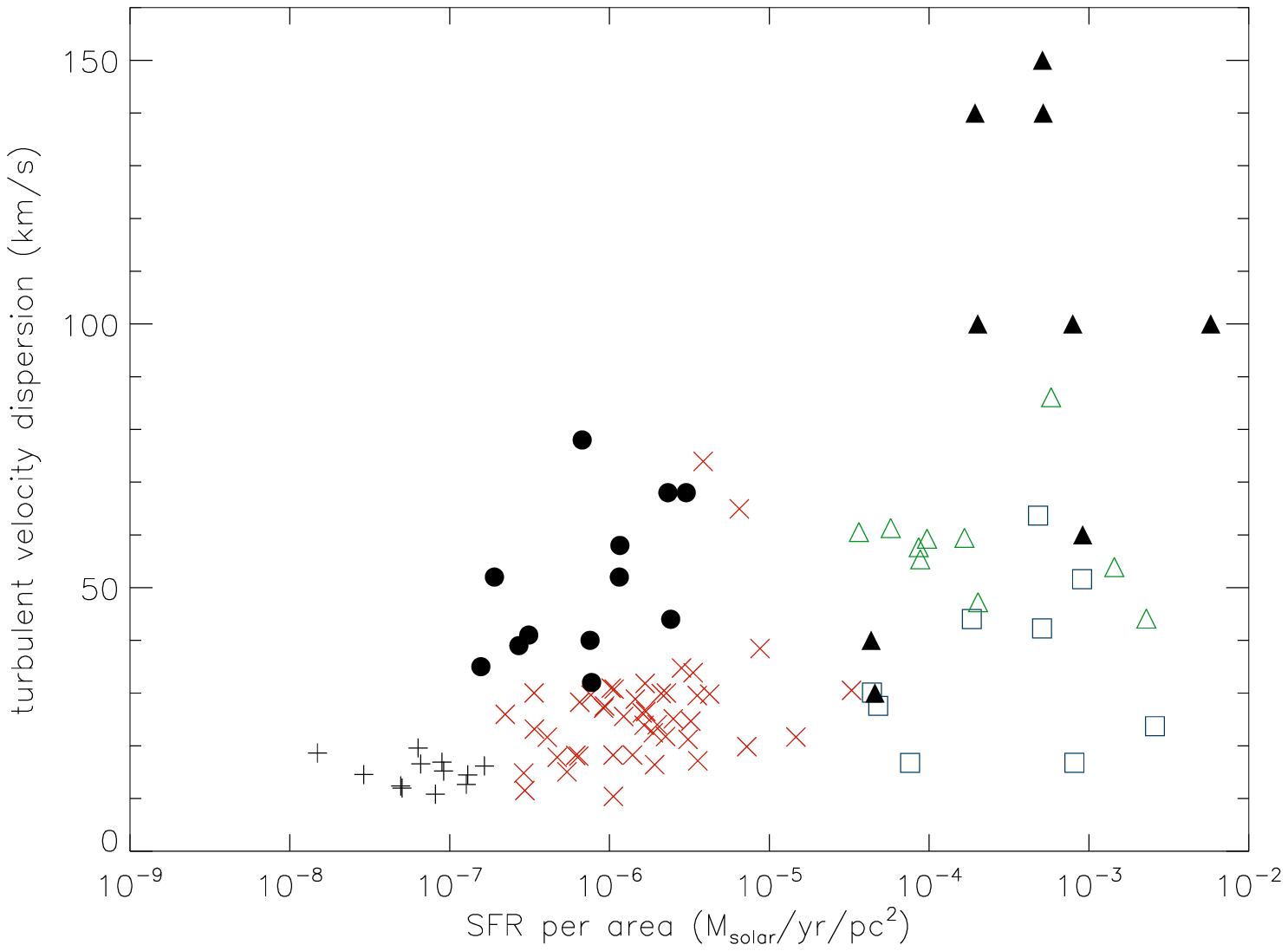}}
  \resizebox{\hsize}{!}{\includegraphics{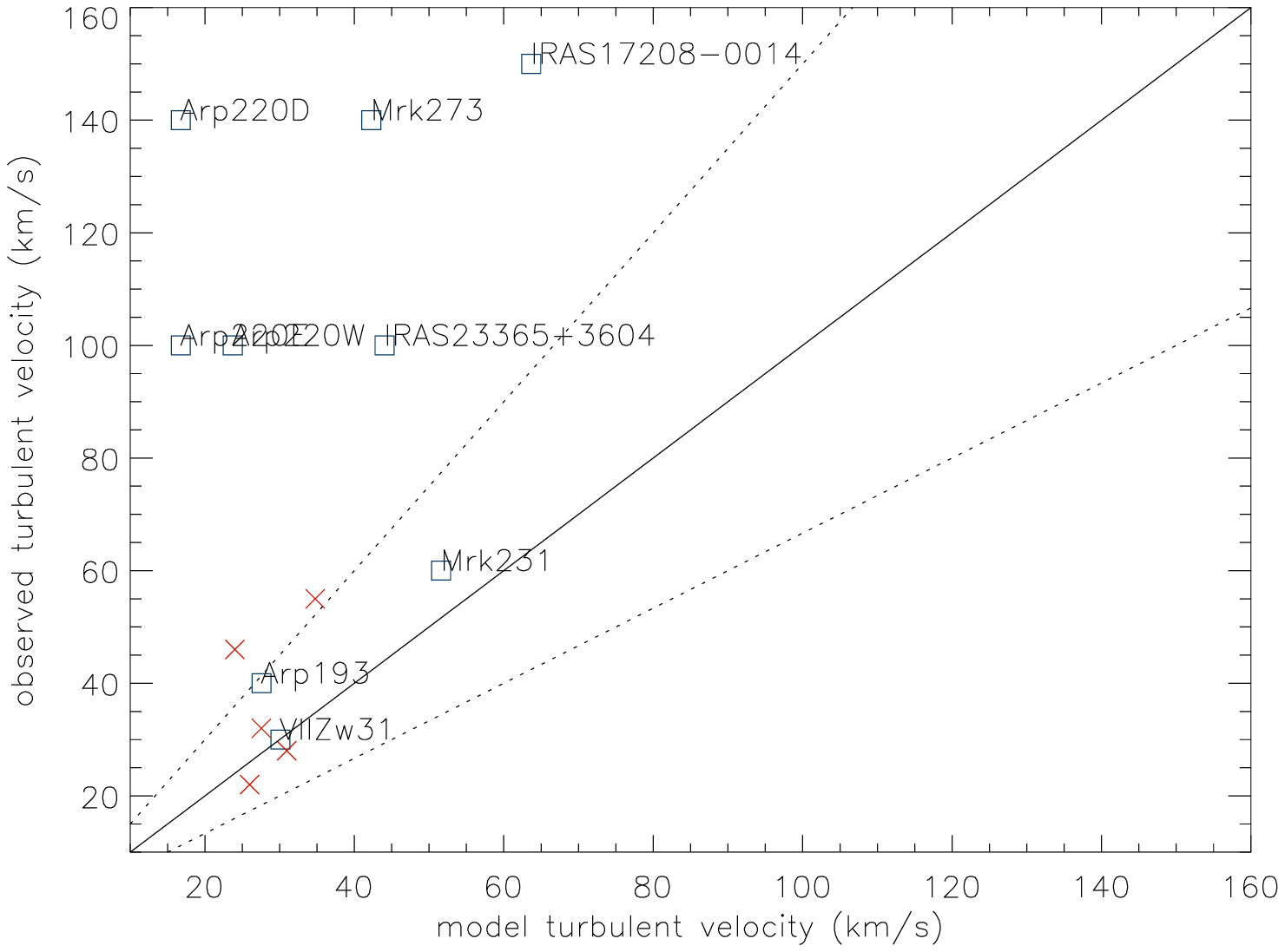}}
  \caption{As for Fig.~\ref{fig:plots_1_vturb}, but with $Q=1$ for the ULIRG, smm, and high-z star-forming galaxies.
  \label{fig:plots_1_Q_vturb}}
\end{figure}

An important finding in many high-redshift star-forming disc galaxies is the high gas velocity dispersion. Values
of $50$--$100$~km\,s$^{-1}$ (Genzel et al. 2006; Law et al. 2009; F\"orster Schreiber et al. 2009; Vergani et al. 2012; Tacconi et al. 2013)
are frequently observed. Whether or not these high-velocity dispersions are mandatory for the high-z star-forming galaxies remains to be elucidated.
To find an initial answer to this question, we compare our preferred model, $Q \sim 1.5$ (Fig.~\ref{fig:plots_1_vturb}), to the $Q \sim 1$ model 
(Fig.~\ref{fig:plots_1_Q_vturb}).
The upper panels show the comparison between the observed (H$\alpha$ and CO) and modelled gas velocity dispersions.
The galaxies within the different samples (local spiral galaxies, ULIRGs, smm, high-z star-forming galaxies) are not the same.
Clearly, the decrease of $Q$ at fixed star-formation rate from $Q \sim 1.5$ to $Q \sim 1$ has a big effect on the gas velocity 
dispersions, which decreases
by a factor of $2$ for the ULIRGs and smm-galaxies and by a factor of $1.6$ for the high-z star-forming galaxies.
A lower $Q$ decreases the velocity dispersion, but also increases the star-formation rate at fixed gas surface density.
Thus, at fixed star-formation rate, a lower $Q$ decreases both the gas surface density and the velocity dispersion.
Our preferred model ($Q \sim 1.5$) better reproduces the observed velocity dispersions.

A direct comparison of observed and modeled gas velocity dispersions for single galaxies in presented in the lower panels of
Figs.~\ref{fig:plots_1_vturb} and~\ref{fig:plots_1_Q_vturb}.
Within the preferred $Q \sim 1.5$ model the gas velocity dispersions of two ULIRGs and three high-z star-forming galaxies 
exceed the observed values by approximately a factor of two. On the other hand, the gas velocity dispersions of four ULIRGs is approximately two times lower
than the observed values. Given that the measured velocity dispersion of ULIRGs and high-z star-forming galaxies that are barely spatially resolved 
can be easily dominated by non-circular gas motions, one expects the model gas velocity dispersions to be systematically smaller than the
observed velocity dispersions. This is only the case for the $Q \sim 1$ model. 

We conclude that the preferred $Q \sim 1.5$ model better reproduces the high gas velocity dispersions observed in ULIRGs and high-star-forming galaxies.
However, if the observed linewidths are dominated by non-circular gas motions, the $Q \sim 1$ model is consistent with available observations.
We recall that the HCN(1--0) emission of the $Q \sim 1$ model is somewhat smaller than that of the preferred model leading to a less good reproduction
of the available HCN(1--0) observations (see Sect.~\ref{sec:Q}).

\subsection{Star-formation laws}

Genzel et al. (2010) and Daddi et al. (2010) found a long-lasting star-formation mode for disk galaxies and a more rapid mode for starbursts.
The two modes can be unified to a single star-formation law, if the star-formation timescale with respect to the molecular gas ($t_{\rm SF}=M_{\rm H_2}/SFR$)
observed in CO emission is assumed to be proportional to the dynamical timescale, that is, the angular velocity of the galaxy: 
$\dot{\Sigma}_* \propto \Sigma_{\rm mol} \Omega$.
Our model directly yields the local star-formation rate, the H$_2$ gas surface density, and the angular velocity.

\begin{figure}
  \centering
  \resizebox{\hsize}{!}{\includegraphics{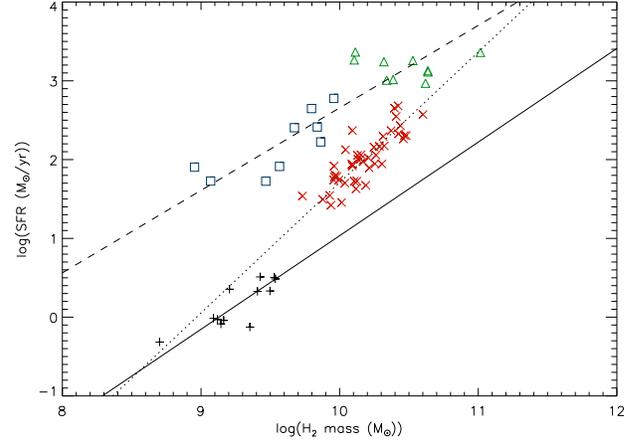}}
  \caption{Model Kennicutt-Schmidt law: star-formation rate as a function of the H$_2$ gas mass. The solid/dotted/dashed lines are
    linear bisector fits to the local spiral/ULIRG+smm/high-z star-forming galaxy samples, respectively.
  \label{fig:KSlaw}}
\end{figure}

The Kennicutt-Schmidt relation for the integrated star-formation rates and H$_2$ gas masses of our model is presented in Fig.~\ref{fig:KSlaw}.
The slopes of the relation for the local spiral galaxies and ULIRGs/smm galaxies are approximately unity ($1.0$ and $1.2$). 
We observe a factor of approximately 50 between the star-formation efficiencies $SFE_{\rm H_2}$ of the local spiral galaxies on the one hand,
and the ULIRGs/smm-galaxies on the other. The SFR--$M_{\rm H_2}$ relation of the high-z star-forming galaxies is intermediate
between those of the local spiral galaxies and ULIRGs/smm-galaxies with a slope of $1.7$. 

\begin{figure}
  \centering
  \resizebox{\hsize}{!}{\includegraphics{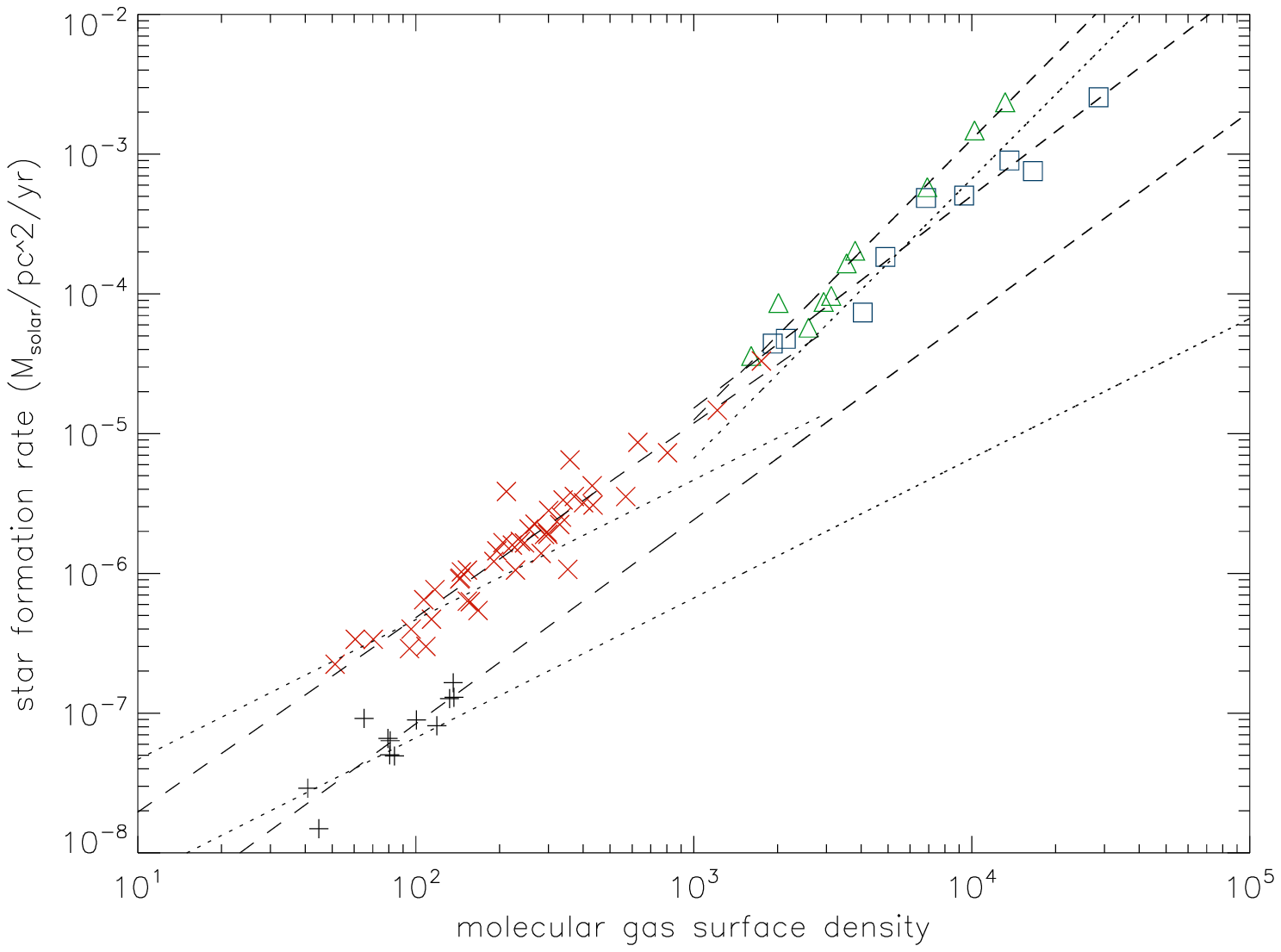}}
  \resizebox{\hsize}{!}{\includegraphics{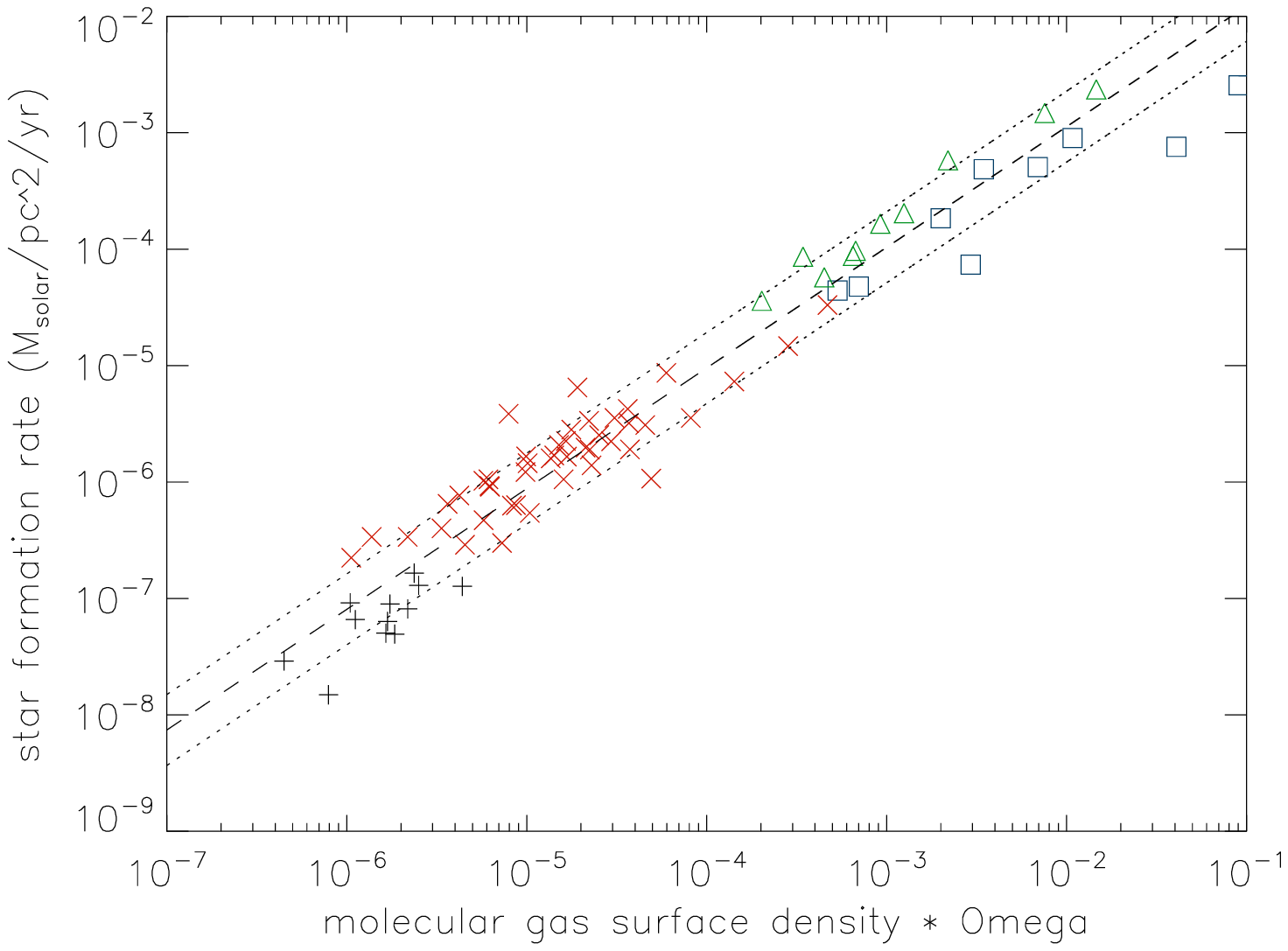}}
  \caption{Upper panel: model star-formation rate surface density as a function of the model molecular gas surface density.
    The symbols are the same as in Fig.~\ref{fig:plots_HCNCO_CO}. The dotted lines are linear and square fits to guide the eye.
    The dashed lines are robust fits to the different galaxy samples.
    Lower panel shows the model star formation per unit area as a function of the model molecular gas surface density multiplied by the angular velocity (divided by
    the dynamical timescale).
    The dashed line represents a robust bisector fit to the entire sample with its associated rms (dotted lines).
  \label{fig:plots_1_SFE}}
\end{figure}

The Kennicutt-Schmidt relation with respect to the molecular gas surface density of our model galaxies is presented in the upper panel of Fig.~\ref{fig:plots_1_SFE}. 
For the area calculation, the stellar scalelength was adopted.
The slope of the relation for the local spiral galaxies derived by the IDL routine robust\_linefit is $1.5$. 
The fitted slopes of the relations for ULIRGs and high-z star-forming galaxies
are $1.5$ and $1.4$, respectively. The slope of the relation for smm-galaxies is $2$, that of the combined ULIRG and smm-galaxy sample is $1.6$.
We observe an offset of a factor of approximately 7 between the relation of the high-z star-forming and the local spiral galaxies.

The model star formation per unit area as a function of the model molecular gas surface density multiplied by the angular velocity (or divided by
the dynamical timescale) $\Sigma_{\rm H_2} \Omega$ is presented in the lower panel of Fig.~\ref{fig:plots_1_SFE}.
For the calculation of the angular velocity $\Omega=v_{\rm rot}/R,$ we followed Daddi et al. (2010) and took 
optical radius $R=R_{25} = 4.5 \times l_*$ for the local spiral galaxies (see also Kennicutt 1998) and the half-light 
radius $R=R_{\frac{1}{2}}$ for the ULIRGs, smm-galaxies, and high-z star-forming galaxies. The slope of this relation is $1.0$ for the entire sample.
 Compared to
that of the local spiral galaxies, the molecular star-formation efficiency of
the ULIRGs, smm, and high-z star-forming galaxies is approximately twice as high. With $R=R_{\frac{1}{2}}$ for the local spiral galaxies, this ratio would increase to a factor of $5$.

We conclude that the model Kennicutt-Schmidt laws for the integrated H$_2$ masses and surface densities do not show the same slopes.
The integrated  Kennicutt-Schmidt law has a slope ofapproximately 1 for the local spirals, ULIRGs, and smm-galaxies, whereas the
slope is $1.7$ for high-z star-forming galaxies.
The model shows Kennicutt-Schmidt laws with respect to the molecular gas surface density with slopes of approximately $1.5$ for local spiral galaxies,
ULIRGs, and high-z star-forming galaxies. The slope for the smm-galaxies is approximately $2$. The model star-formation rate per unit area is, as observed, 
approximately proportional to the molecular gas surface density divided by the dynamical timescale.

\section{Conclusions \label{sec:conclusions}}

The  theory  of  clumpy  gas  disks  (Vollmer \& Beckert 2003)  provides the large-scale and small-scale properties of galactic gas disks. 
Large-scale properties considered are the gas surface
density,  density,  disk  height,  turbulent  driving  length  scale, velocity  dispersion,  gas  viscosity,  volume  filling  factor,  and
molecular  fraction.  Small-scale  properties  are  the  mass,  size, density, turbulent, free-fall, and molecular formation timescales of
the most massive self-gravitating gas clouds. These quantities depend on the stellar surface density, the angular velocity $\Omega$, the
disk radius $R$, and three additional parameters, which are the Toomre parameter $Q$ of the gas, the mass accretion rate $\dot{M}$, and the ratio
$\delta$ between  the driving length scale of turbulence and the cloud size. 
The large-scale part of the model disk is governed by vertical pressure equilibrium, a constant Toomre $Q$ parameter, conservation of the
turbulent energy flux, a relation between the gas viscosity and the gas surface density, a star-formation recipe (Sect.~\ref{sec:method}), and 
a simple closed-box model for the gas metallicity.
The small-scale part is divided into two parts according to gas density: non-self-gravitating and self-gravitating gas clouds. 
The mass fraction at a given density is determined by a density probability distribution
involving the overdensity and the Mach number (Sect.~\ref{sec:gasfrac}). Both density
regimes are governed by different observed scale relations (Sect.~\ref{sec:scaling}). The dense gas clouds are mechanically heated by turbulence. In addition,
they are heated by cosmic rays. The gas temperature is calculated through the equilibrium between gas heating and cooling via molecular
line emission (CO, H$_2$, H$_2$O; Sect.~\ref{sec:gdtemp}). The dust temperature is determined by the equilibrium between radiative heating and cooling and 
the heat transfer between gas and dust (Sect.~\ref{sec:dustemission}). The molecular line emission calculation is based on the escape probability 
formalism (Sect.~\ref{sec:lineemission}).
An important ingredient for the line emission is the area-filling factor of the gas clouds, which is a result of the small-scale part of
the analytic disk model.  The molecular abundances of individual gas clouds are determined by a detailed chemical network
involving the cloud lifetime, density, and temperature (Sect.~\ref{sec:network}). 
H$_2$ and CO dissociation in photodissociation regions
are taken into account (Sect.~\ref{sec:dissociation}). Moreover, a simple formalism for HCN infrared pumping is applied to the HCN line emission
(Sect.~\ref{sec:irpumping}). 

The stellar radiation field is constrained by the observed IR luminosity and SED. 
The normalization of the cosmic ray ionization rate is constrained by the observed HCO$^+$(1--0) emission.
The density and temperature structure of the clumpy gas disk is constrained by the observed HCN(1--0) and multi-transition CO emission.
This model is applied to samples of local spiral galaxies, ULIRGs, smm, and high-z star-forming galaxies (Sect.~\ref{sec:samples}).
 
Based on the comparison between the model results and observations available in the literature (see Table~\ref{tab:correl}) we conclude that
\begin{enumerate}
\item 
the following observed quantities are consistent with observations:
\begin{itemize}
\item
global metallicities (Fig.~\ref{fig:metallicities}), 
\item
total infrared luminosities and dust SEDs (Fig.~\ref{fig:tirlum}, Appendix~\ref{sec:seds}),
\item
dust temperatures (Fig.~\ref{fig:tdust}),
\item
H{\sc i} masses and radial profiles of the local spiral galaxies (Figs.~\ref{fig:HImasses}, \ref{fig:HIprofiles}),
\item
CO luminosities (Fig.~\ref{fig:plots_HCNCO_CO}),
\item
HCO$^+$ luminosities of the ULIRGs, smm, and high-z star-forming galaxies (Fig.~\ref{fig:plots_HCNCO_HCO}),
\item
CO SLEDs up to $J=6$ (Fig.~\ref{fig:plots_HCNCO_SLED}),
\item
CO SLEDs up to $J=12$ for the ULIRGs (Fig.~\ref{fig:ulirg_ladder}),
\end{itemize}
\item
the model HCN radial profiles are a factor $1.5$--$2$ higher than the observed profiles (Fig.~\ref{fig:THINGS_profiles}),
\item
the model HCN luminosities are a factor of $1.5$ higher/lower than the observed luminosities for the local spiral galaxies/ULIRGs (Fig.~\ref{fig:plots_HCNCO_CO}),
\item
the model HCO$^+$ luminosities of the local spiral galaxies are a factor of $\sim 3$ higher than the observed HCO$^+$ luminosities (Fig.~\ref{fig:plots_HCNCO_HCO});
the HCO$^+$ emission mainly depends on the CR ionzation rate used in the chemical network,
\item
all model conversion factors (mass-to-light and SFR-to-light) have uncertainties of a factor of two,
\item
the model CO conversion factors deduced when including CO-dark H$_2$ are 
$\alpha_{\rm CO}=4.7 \pm 1.8,\ 1.7 \pm 0.4,\ 1.4 \pm 0.7,\ 2.6 \pm 0.9$~M$_{\odot}({\rm K\,km\,s}^{-1}{\rm pc}^2)^{-1}$
for the local spirals, ULIRGs, smm, and high-z star-forming galaxies (Fig.~\ref{fig:plots_HCNCO_alphaCO});
the model CO conversion factor of the ULIRGs is a factor of two higher than the value derived by Downes \& Solomon (1998), the CO conversion factor of
the high-star-forming galaxies is a factor of two lower than assumed by Genzel et al. (2010) and Tacconi et al. (2010),
\item
the model HCN-dense gas conversion factor is $\alpha_{\rm HCN}=21 \pm 6$~M$_{\odot}({\rm K\,km\,s}^{-1}{\rm pc}^2)^{-1}$ for the local spiral galaxies and ULIRGs;  this is a factor of two higher than the value used in the literature (e.g., Gao \& Solomon 2004);
the model HCN-dense gas conversion factor is
$33 \pm 17$~M$_{\odot}({\rm K\,km\,s}^{-1}{\rm pc}^2)^{-1}$ for the smm-galaxies, and $59 \pm 21$~M$_{\odot}({\rm K\,km\,s}^{-1}{\rm pc}^2)^{-1}$ for the high-z star-forming galaxies,
\item
both, the HCN and HCO$^+$ emission trace the dense molecular gas to a factor of approximately $2$ for the local spiral galaxies, ULIRGs and smm-galaxies.
\end{enumerate}

We tested the influence of constant abundances (Sect.~\ref{sec:constabund}), Toomre $Q$ ($Q=1$ instead of $Q=1.5$; Sect.~\ref{sec:Q}), and of 
the scale parameter $\delta$ ($\delta=15$ instead of $\delta=5$; Sect.~\ref{sec:delta}). The changes in molecular emission are minor ($< 0.2$~dex).
The $Q=1$ and $\delta=15$ model overestimate the CO SLEDs of the ULIRGs. The $Q=1$ model yields a lower HCN(1--0) emission than the
$Q=1.5$ model. Since the $Q=1.5$ model already underestimates the HCN(1--0) luminosity with respect to observations, the $Q=1.5$ model
is our preferred model. 

Whereas the CO emission is robust against the variation of model parameters and chemistry, the HCN and HCO$^+$ emission is most sensitive to the chemistry of the interstellar medium.

Within the model framework $\sim 60$\,\% of the CO(1--0) emission of local spiral galaxies and high-z star-forming galaxies is emitted in
non-self-gravitating clouds. This fraction increases to $\sim 80$\,\% for ULIRGs.
Whereas $\sim 80$\,\% of the HCN(1--0) and HCO$^+$ emission originates in non-self-gravitating clouds, this fraction decreases to 
$\sim 30$\,\% for local spirals and high-z star-forming galaxies (Sect.~\ref{sec:nonself}).

The resulting CO, HCN, and HCO$^+$ line emission does not change significantly if cloud-substructure is not taken into account (Sect.~\ref{sec:substruct}).
Ignoring the cosmic ray heating (Sect.~\ref{sec:CRheat}) leads to low CO(1--0) emission from high-z star-forming galaxies and low
HCN(1--0) emission from ULIRGs.

The effect of HCN infrared pumping is small but measurable ($20$--$30$\,\%; Sect.~\ref{sec:IRpumping}).

The gas velocity dispersion varies significantly with the Toomre $Q$ parameter. The $Q=1.5$ model yields high-velocity dispersions
($v_{\rm disp} \gg 10$~km\,s$^{-1}$) consistent with available observations of high-z star-forming galaxies and ULIRGs (Fig.~\ref{fig:plots_1_vturb}). However, we note that
these high-velocity dispersions may not be mandatory for starburst galaxies (Fig.~\ref{fig:plots_1_Q_vturb}).

The model yields molecular star-formation laws ($\dot{\Sigma}_*$--$\Sigma_{\rm H_2}$) with slopes of $1.5/2$ for the local spiral galaxies,
ULIRGs, and high-z star-forming galaxies/smm galaxies. The model star-formation rate per unit area is, as observed, 
proportional to the molecular gas surface density divided by the dynamical timescale (lower panel of Fig.~\ref{fig:plots_1_SFE}).
There is a pronounced offset between the $\dot{\Sigma}_*$--$\Sigma_{\rm H_2} \Omega$ relations of the local spirals on the one hand,
and ULIRGs, smm, and high-z star-forming galaxies on the other.

We conclude that our relatively simple analytic model (Sect.~\ref{sec:model}), together with the recipes for the molecular line emission 
(Fig.~\ref{fig:bild1}), captures the essential physics of galactic clumpy gas disks.


\begin{acknowledgements}
We would like to thank the anonymous referee for his/her suggestion to include the IR radiative transfer into our model and for
improving the manuscript significantly. PG acknowledges funding by the European Research Council (Starting Grant 3DICE, grant agreement 336474, PI: V. Wakelam).
This research has made use of the SIMBAD database and the VizieR catalogue access tool, operated at CDS, Strasbourg, France.
This research has made use of the NASA/IPAC Extragalactic Database (NED) which is operated by the Jet Propulsion Laboratory, California Institute of Technology, 
under contract with the National Aeronautics and Space Administration. 
\end{acknowledgements}


\begin{appendix}

\section{Mean dust temperatures of molecular clouds \label{sec:atten}}

\begin{figure}
  \centering
  \resizebox{\hsize}{!}{\includegraphics{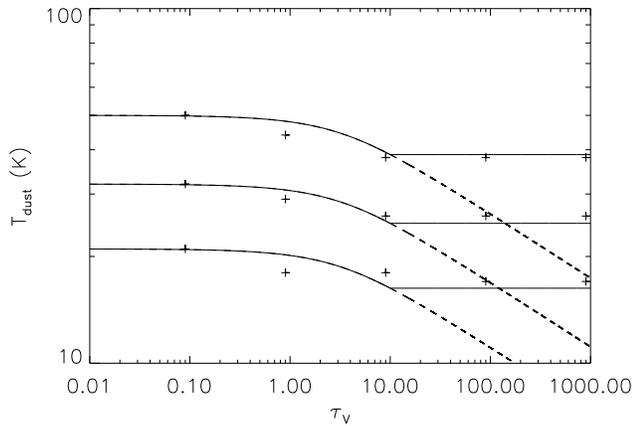}}
  \caption{Dust temperature as a function of optical depth of a molecular clouds which are illuminated by an interstellar radiation field (ISRF) 
    of $(10, 100, 1000)$ times the Galactic ISRF at the solar radius (plus signs). The solid line for $\tau_{\rm V} \le 10$ and dashed line
    for $\tau_{\rm V} > 10$ correspond to $T_{\rm dust} \propto \big(I/I(0)\big)^{1/5.5}$. The solid line for $\tau_{\rm V} > 10$ corresponds to
    $I/I(0)=0.246$.
  \label{fig:tdust_tauv}}
\end{figure}

As described in Sect.~\ref{sec:gdtemp}, we assume that the main radiative heating agent is the UV emission component of the
interstellar radiation field. For this emission component, we use an attenuation factor based on the mean extinction of a sphere 
of constant density (Eq.~\ref{eq:attenuationfactor}). For high optical depths, Eq.~\ref{eq:attenuationfactor} becomes 
$I/I(0) \sim 3 \tau_{\rm V}^{-1}$, that is, the attenuation decreases to very low values. However, when the molecular clouds become optically 
thick in the near-infrared (at $\tau_{\rm V} \sim 10$), radiative transfer effects become important: a significant infrared radiation 
field builds up, which heats the dust until deep into the molecular clouds.  
To take this effect into account, we calculated 1D radiation transfer models for spherical clouds of constant density that
are illuminated by an interstellar radiation field. For this purpose, we used the code TRANSPHERE (Dullemond et al. 2002). 
The clouds have a size of $20$~pc and densities of $n=(10^1, 10^2, 10^3, 10^4, 10^5)$~cm$^{-3}$. The clouds are illuminated
by an interstellar radiation field of $(10, 100, 1000)$ times the Galactic ISRF at the solar radius.
For each cloud model, the resulting infrared SED was fitted by a modified Planck function to
give the effective dust temperature of the cloud (Fig.~\ref{fig:tdust_tauv}). 
According to Eq.~\ref{eq:dustheating} and \ref{eq:dustcooling}, the relation between the dust temperature and $I/I(0)$ 
is $T_{\rm dust} \propto \big(I/I(0)\big)^{1/5.5}$ (solid line for $\tau_{\rm V} \le 10$ and dashed line
for $\tau_{\rm V} > 10$ in Fig.~\ref{fig:tdust_tauv}). It turned out that the effective dust temperatures obtained from
the radiative transfer models for $\tau_{\rm V} > 10$ are well-fitted by a constant attenuation factor $I/I(0)=0.246$.

\section{Galaxy samples \label{sec:asamples}}

\begin{table*}
\begin{center}
\caption{Local spiral galaxies.\label{tab:gleroy}}
\begin{tabular}{lccccccccccc}
\hline
Galaxy & $v_{\rm rot}$ & $l_{\rm flat}$ & $l_*$ & $M_*$ & $\dot{M_*}$ & L$_{\rm TIR}$ & L$'_{\rm CO(2-1)}$ & $Q^{\rm (b)}$ & $\delta^{\rm (b)}$ & $\dot{M}^{\rm (a)}$ & $M_{\rm gas}^{\rm (a)}$ \\
 & (km\,s$^{-1}$) & (kpc) & (kpc) & ($10^{10}$~M$_{\odot}$) & (M$_{\odot}$yr$^{-1}$) & ($10^{10}$~L$_{\odot}$) & ($^{(c)}$) & & & (M$_{\odot}$yr$^{-1}$) & ($10^{9}$~M$_{\odot}$) \\
\hline
   NGC628 & 217 &  0.8 &  2.2 & 1.26 & 0.81 &  0.8 &  2.3 &  3.0 &  5.0 &  0.2 &  5.4 \\
   NGC3198 & 150 &  2.7 &  3.2 & 1.26 & 0.93 &  1.0 &  1.1 &  2.0 &  5.0 &  0.3 &  8.3 \\
   NGC3184 & 210 &  2.7 &  2.4 & 2.00 & 0.90 &  1.0 &  3.0 &  2.5 &  5.0 &  0.1 &  5.6 \\
   NGC4736 & 156 &  0.2 &  1.1 & 2.00 & 0.48 &  0.6 &  1.1 &  5.0 &  5.0 &  0.1 &  1.2 \\
   NGC3351 & 196 &  0.6 &  2.2 & 2.51 & 0.94 &  0.8 &  1.9 &  6.0 &  5.0 &  0.4 &  4.0 \\
   NGC6946 & 186 &  1.3 &  2.5 & 3.16 & 3.24 &  3.2 &  8.8 &  2.0 &  5.0 &  0.4 &  9.7 \\
   NGC3627 & 192 &  1.2 &  2.7 & 3.98 & 2.22 &  2.5 &  5.5 &  2.0 &  5.0 &  0.3 &  3.3 \\
   NGC5194 & 219 &  0.8 &  2.7 & 3.98 & 3.12 &  0.0 &  9.3 &  2.0 &  5.0 &  0.3 & 11.4 \\
   NGC3521 & 227 &  1.3 &  2.9 & 5.01 & 2.10 &  3.2 &  7.6 &  2.0 &  5.0 &  0.1 &  9.4 \\
   NGC2841 & 302 &  0.6 &  4.0 & 6.31 & 0.74 &  1.3 &  2.8 &  8.0 &  5.0 &  0.3 &  8.0 \\
   NGC5055 & 192 &  0.6 &  3.2 & 6.31 & 2.12 &  2.0 &  9.8 &  3.0 &  5.0 &  0.3 &  8.8 \\
   NGC7331 & 244 &  1.2 &  3.2 & 7.94 & 3.00 &  5.0 & 11.3 &  3.0 &  5.0 &  0.4 & 11.6 \\
\hline
\end{tabular}
\begin{tablenotes}
      \item $^{\rm (a)}$ calculated quantities; $^{\rm (b)}$ assumed quantities; all other columns are input quantities from Leroy et al. (2008).
        \item $^{\rm (c)}$ in $10^8$~K\,km\,s$^{-1}$pc$^{2}$
    \end{tablenotes}
\end{center}
\end{table*}

\begin{table*}
\begin{center}
\caption{Ultraluminous infrared galaxies.\label{tab:gulirg}}
\begin{tabular}{lccccccccccc}
\hline
Galaxy Name & $v_{\rm rot}$ & $l_{\rm flat}$ & $l_*$ & $M_*$ & $\dot{M_*}$  & L$_{\rm TIR}$ & L$'_{\rm CO(1-0)}$ & $Q^{\rm (b)}$ & $\delta^{\rm (b)}$ & $\dot{M}^{\rm (a)}$ & $M_{\rm gas}^{\rm (a)}$ \\
 & (km\,s$^{-1}$) & (kpc) & (kpc) & ($10^{10}$~M$_{\odot}$) & (M$_{\odot}$yr$^{-1}$) & ($10^{11}$~L$_{\odot}$) & ($^{(c)}$) & & & (M$_{\odot}$yr$^{-1}$) & ($10^{9}$~M$_{\odot}$) \\
\hline
IRAS17208-0014 & 260 & 0.02 &  0.5 &  0.8 &  435 & 25.6 &  5.8 &    1.2 &  5.0 &  313.3 & 14.8 \\
        Mrk231 & 345 & 0.02 &  0.4 &  1.3 &  595 & 35.0 &  4.2 &    1.5 &  5.0 &  499.8 & 16.5 \\
       Arp220D & 330 & 0.02 &  0.4 &  1.2 &   52 &  3.1 &  3.7 &    2.5 &  5.0 &   15.8 &  3.8 \\
        Mrk273 & 280 & 0.02 &  0.4 &  0.9 &  253 & 14.9 &  4.6 &    1.5 &  5.0 &  182.4 &  8.4 \\
IRAS23365+3604 & 260 & 0.02 &  0.6 &  1.0 &  258 & 15.2 &  5.4 &    1.5 &  5.0 &  242.9 & 14.4 \\
       VIIZw31 & 290 & 0.02 &  1.1 &  2.2 &  164 &  9.7 &  5.5 &    1.5 &  5.0 &   28.0 & 13.7 \\
        Arp193 & 230 & 0.02 &  0.7 &  0.9 &   81 &  4.8 &  2.1 &    1.5 &  5.0 &   18.8 &  6.5 \\
       Arp220W & 300 & 0.01 &  0.1 &  1.2 &   79 &  4.7 &  0.6 &    2.0 &  5.0 &   34.8 &  1.1 \\
       Arp220E & 350 & 0.01 &  0.1 &  1.9 &   52 &  3.1 &  0.8 &    2.8 &  5.0 &   22.1 &  1.3 \\
\hline
\end{tabular}
\begin{tablenotes}
\item $^{\rm (a)}$ calculated quantities; $^{\rm (b)}$ assumed quantities; all other columns are input quantities from Downes \& Solomon (1998).
  \item $^{\rm (c)}$ in $10^9$~K\,km\,s$^{-1}$pc$^{2}$
    \item $^{\rm d}$ Arp220D, Arp220W, and Arp220E refer to the Disk, Western, and Eastern components, respectively.
    \end{tablenotes}
\end{center}
\end{table*}

\begin{table*}
\begin{center}
\caption{Submillimeter galaxies.\label{tab:gbzk}}
\begin{tabular}{lccccccccccc}
\hline
Galaxy Name & $v_{\rm rot}$ & $l_{\rm flat}$ & $l_*$ & $M_*^{\rm (c)}$ & $\dot{M_*}$ & L$_{\rm TIR}$ & L$'_{\rm CO}$ & $Q^{\rm (b)}$ & $\delta^{\rm (b)}$ & $\dot{M}^{\rm (a)}$ & $M_{\rm gas}^{\rm (a)}$ \\
 & (km\,s$^{-1}$) & (kpc) & (kpc) & ($10^{10}$~M$_{\odot}$) & (M$_{\odot}$yr$^{-1}$) & ($10^{11}$~L$_{\odot}$) & ($^{(d)}$) & & & (M$_{\odot}$yr$^{-1}$) & ($10^{9}$~M$_{\odot}$) \\
\hline
SMM J02399-013 & 590 & 0.10 &  3.5 & 10.0 & 2294 &   68.8 &   49.0 &    1.5 &  5.0 & 1927.0 & 318.5 \\
SMM J09431+470 & 295 & 0.10 &  0.9 & 10.0 & 1746 &   52.4 &   25.0 &    1.5 &  5.0 & 1117.4 &  39.1 \\
SMM J105141+57 & 457 & 0.10 &  2.1 & 10.0 & 1296 &   38.9 &   50.0 &    1.5 &  5.0 &  423.4 &  98.2 \\
SMM J123549+62 & 442 & 0.10 &  0.6 & 24.0 & 1794 &   53.8 &   40.0 &    1.5 &  5.0 &   71.8 &  17.1 \\
SMM J123634+62 & 343 & 0.10 &  2.8 & 10.0 &  930 &   27.9 &   34.0 &    1.5 &  5.0 &  737.8 & 117.0 \\
SMM J123707+62 & 317 & 0.10 &  1.9 & 24.0 & 1016 &   30.5 &   19.0 &    1.5 &  5.0 &  135.5 &  45.6 \\
SMM J131201+42 & 430 & 0.10 &  2.1 & 10.0 & 1340 &   40.2 &   30.0 &    1.5 &  5.0 &  589.6 &  99.5 \\
SMM J131232+42 & 346 & 0.10 &  1.4 & 10.0 & 1016 &   30.5 &   28.0 &    1.5 &  5.0 &  257.4 &  41.5 \\
SMM J163650+40 & 523 & 0.10 &  1.6 & 46.0 & 1772 &   53.2 &   69.0 &    1.5 &  5.0 &   59.1 &  50.8 \\
SMM J163658+41 & 590 & 0.10 &  0.5 & 52.0 & 2248 &   67.4 &   56.0 &    1.5 &  5.0 &   28.5 &  16.0 \\
\hline
\end{tabular}
\begin{tablenotes}
      \item $^{\rm (a)}$ calculated quantities; $^{\rm (b)}$ assumed quantities; all other columns are input quantities from Genzel et al. (2010).
        \item $^{\rm (c)}$ we assumed $M_*=10^{11}$~M$_{\odot}$ for galaxies whose mass is not given in Genzel et al. (2010).
          \item $^{\rm (d)}$ in $10^9$~K\,km\,s$^{-1}$pc$^{2}$
    \end{tablenotes}
\end{center}
\end{table*}

\begin{table*}
\begin{center}
\caption{High-z star-forming disk galaxies.\label{tab:gphibbs}}
\begin{tabular}{lccccccccccc}
\hline
Galaxy Name & $v_{\rm rot}^{\rm (c)}$ & $l_{\rm flat}$ & $l_*$ & $M_*$ & $\dot{M_*}$ & L$_{\rm TIR}$ & L$'_{\rm CO(3-2)}$ & $Q^{\rm (b)}$ & $\delta^{\rm (b)}$ & $\dot{M}^{\rm (a)}$ & $M_{\rm gas}^{\rm (a)}$ \\
 & (km\,s$^{-1}$) & (kpc) & (kpc) & ($10^{10}$~M$_{\odot}$) & (M$_{\odot}$yr$^{-1}$) & ($10^{11}$~L$_{\odot}$) & ($^{(c)}$) & & & (M$_{\odot}$yr$^{-1}$) & ($10^{9}$~M$_{\odot}$) \\
\hline
   EGS12004280 & 230 & 0.10 &  4.7 &  4.1 &  100 &  10.0 &   7.9 &    1.5 &   5.0 &   30.5 &   47.5 \\
   EGS12004754 & 215 & 0.10 &  6.5 &  9.3 &   53 &   5.3 &   4.7 &    1.5 &   5.0 &    5.8 &   39.2 \\
   EGS12007881 & 232 & 0.10 &  5.7 &  5.2 &   94 &   9.4 &   8.5 &    1.5 &   5.0 &   23.5 &   53.8 \\
   EGS12015684 & 233 & 0.10 &  4.0 &  4.6 &  113 &  11.3 &   5.8 &    1.5 &   5.0 &   30.5 &   41.3 \\
   EGS12023832 & 215 & 0.10 &  4.7 &  5.9 &  115 &  11.5 &   3.8 &    1.5 &   5.0 &   36.8 &   47.4 \\
   EGS12024462 & 253 & 0.10 &  8.6 &  6.0 &   78 &   7.8 &   5.8 &    1.5 &   5.0 &   12.9 &   73.5 \\
   EGS12024866 & 221 & 0.10 &  4.6 &  2.5 &   31 &   3.1 &   3.7 &    1.5 &   5.0 &    4.0 &   24.5 \\
   EGS13003805 & 387 & 0.10 &  5.7 & 17.0 &  200 &  20.0 &  24.0 &    1.5 &   5.0 &    9.7 &   70.9 \\
   EGS13004661 & 171 & 0.10 &  5.0 &  3.0 &   60 &   6.0 &   3.7 &    1.5 &   5.0 &   36.3 &   39.4 \\
   EGS13004684 & 295 & 0.10 &  5.0 & 11.0 &   42 &   4.2 &   9.5 &    1.5 &   5.0 &    1.4 &   27.7 \\
   EGS13011148 & 260 & 0.10 &  5.2 & 11.0 &   52 &   5.2 &   5.2 &    1.5 &   5.0 &    2.9 &   31.3 \\
   EGS13011155 & 296 & 0.10 &  7.8 & 12.0 &  201 &  20.1 &  13.0 &    1.5 &   5.0 &   35.2 &  106.5 \\
   EGS13011166 & 363 & 0.10 &  6.5 & 12.0 &  373 &  37.3 &  29.0 &    1.5 &   5.0 &   69.0 &  133.0 \\
   EGS13017614 & 346 & 0.10 &  4.5 & 13.0 &   88 &   8.8 &  13.0 &    1.5 &   5.0 &    2.9 &   36.0 \\
   EGS13017707 & 324 & 0.10 &  3.6 &  7.4 &  351 &  35.1 &  11.0 &    1.5 &   5.0 &   93.0 &   72.2 \\
   EGS13017843 & 227 & 0.10 &  4.2 &  4.0 &   35 &   3.5 &   6.3 &    1.5 &   5.0 &    3.3 &   22.2 \\
   EGS13017973 & 155 & 0.10 &  7.2 &  4.4 &   55 &   5.5 &   3.7 &    1.5 &   5.0 &   36.6 &   52.0 \\
   EGS13018632 & 319 & 0.10 &  1.9 &  5.2 &   82 &   8.2 &   3.9 &    1.5 &   5.0 &    3.9 &   15.1 \\
   EGS13019114 & 327 & 0.10 &  7.2 &  6.6 &   47 &   4.7 &   6.1 &    1.5 &   5.0 &    1.9 &   45.9 \\
   EGS13019128 & 194 & 0.10 &  5.2 &  4.4 &   87 &   8.7 &   4.9 &    1.5 &   5.0 &   39.6 &   48.1 \\
   EGS13026117 & 436 & 0.10 &  3.2 & 13.0 &  113 &  11.3 &  16.0 &    1.5 &   5.0 &    2.3 &   29.9 \\
   EGS13033624 & 301 & 0.10 &  5.3 &  8.9 &  148 &  14.8 &  10.0 &    1.5 &   5.0 &   17.0 &   59.6 \\
   EGS13033731 & 350 & 0.10 &  5.5 &  2.8 &   28 &   2.8 &   4.6 &    1.5 &   5.0 &    0.8 &   29.2 \\
   EGS13034339 & 299 & 0.10 &  3.0 &  6.6 &   86 &   8.6 &   6.0 &    1.5 &   5.0 &    5.4 &   24.4 \\
   EGS13034541 & 330 & 0.10 &  8.0 &  9.3 &  183 &  18.3 &   9.1 &    1.5 &   5.0 &   24.2 &  107.8 \\
   EGS13034542 & 195 & 0.10 &  4.0 &  5.2 &   61 &   6.1 &   1.8 &    1.5 &   5.0 &   11.9 &   26.7 \\
   EGS13035123 & 219 & 0.10 & 11.2 & 15.0 &   87 &   8.7 &  10.0 &    1.5 &   5.0 &   14.8 &   89.3 \\
   EGS13042293 & 167 & 0.10 &  5.2 &  3.9 &   55 &   5.5 &   3.0 &    1.5 &   5.0 &   24.2 &   36.0 \\
      zC406690 & 224 & 0.10 &  6.3 &  4.0 &  480 &  48.0 &   9.4 &    1.1 &   5.0 &  304.8 &  158.1 \\
    Q1623BX599 & 376 & 0.10 &  1.7 &  5.7 &  131 &  13.1 &   5.7 &    1.5 &   5.0 &    5.6 &   17.5 \\
    Q1700BX691 & 260 & 0.10 &  3.9 &  7.6 &   50 &   5.0 &   4.7 &    1.5 &   5.0 &    2.9 &   23.3 \\
    Q2343BX610 & 402 & 0.10 &  4.6 & 10.0 &  212 &  21.2 &  26.0 &    1.5 &   5.0 &   12.7 &   63.2 \\
    Q2343BX442 & 309 & 0.10 &  4.3 & 12.0 &  145 &  14.5 &   7.0 &    1.5 &   5.0 &   10.5 &   44.1 \\
     Q2343MD59 & 371 & 0.10 &  2.8 &  7.6 &   26 &   2.6 &  62.0 &    1.5 &   5.0 &    0.3 &   13.2 \\
  Q2346BX482se & 285 & 0.10 &  2.4 &  0.6 &   34 &   3.4 &   9.2 &    1.5 &   5.0 &    3.9 &   15.8 \\
       BzK4171 & 261 & 0.10 &  4.5 &  4.0 &  101 &  10.1 &  19.0 &    1.5 &   5.0 &   20.2 &   45.4 \\
     BzK210000 & 292 & 0.10 &  4.7 &  7.8 &  231 &  23.1 &  20.0 &    1.5 &   5.0 &   54.3 &   72.2 \\
      BzK16000 & 258 & 0.10 &  4.0 &  4.3 &   82 &   8.2 &  14.0 &    1.5 &   5.0 &   11.9 &   34.4 \\
      BzK17999 & 238 & 0.10 &  4.7 &  3.9 &  450 &  45.0 &  15.0 &    1.2 &   5.0 &  351.0 &  122.8 \\
      BzK12591 & 361 & 0.10 &  4.5 & 11.0 &  267 &  26.7 &  29.0 &    1.5 &   5.0 &   26.7 &   69.4 \\
      BzK25536 & 254 & 0.10 &  3.0 &  3.3 &   62 &   6.2 &  10.0 &    1.5 &   5.0 &    7.4 &   22.2 \\
    J2135-0102 & 381 & 0.10 &  1.5 &  1.7 &  230 &  23.0 &  12.0 &    1.5 &   5.0 &   46.0 &   27.8 \\
\hline
\end{tabular}
\begin{tablenotes}
      \item $^{\rm (a)}$ calculated quantities; $^{\rm (b)}$ assumed quantities; all other columns are input quantities from Tacconi et al. (2013).
        \item $^{\rm (c)}$ if $v_{\rm rot} < \sqrt{(M_{\rm gas}+M_*)\,G/(2\,l_*)}$ the assumed rotation velocity is $v_{\rm rot}=\sqrt{(M_{\rm gas}+M_*)\,G/(2\,l_*)}$.
          \item $^{\rm (d)}$ in $10^9$~K\,km\,s$^{-1}$pc$^{2}$
    \end{tablenotes}
\end{center}
\end{table*}

\section{Infrared SEDs \label{sec:seds}}

\begin{figure*}
  \centering
  \resizebox{\hsize}{!}{\includegraphics{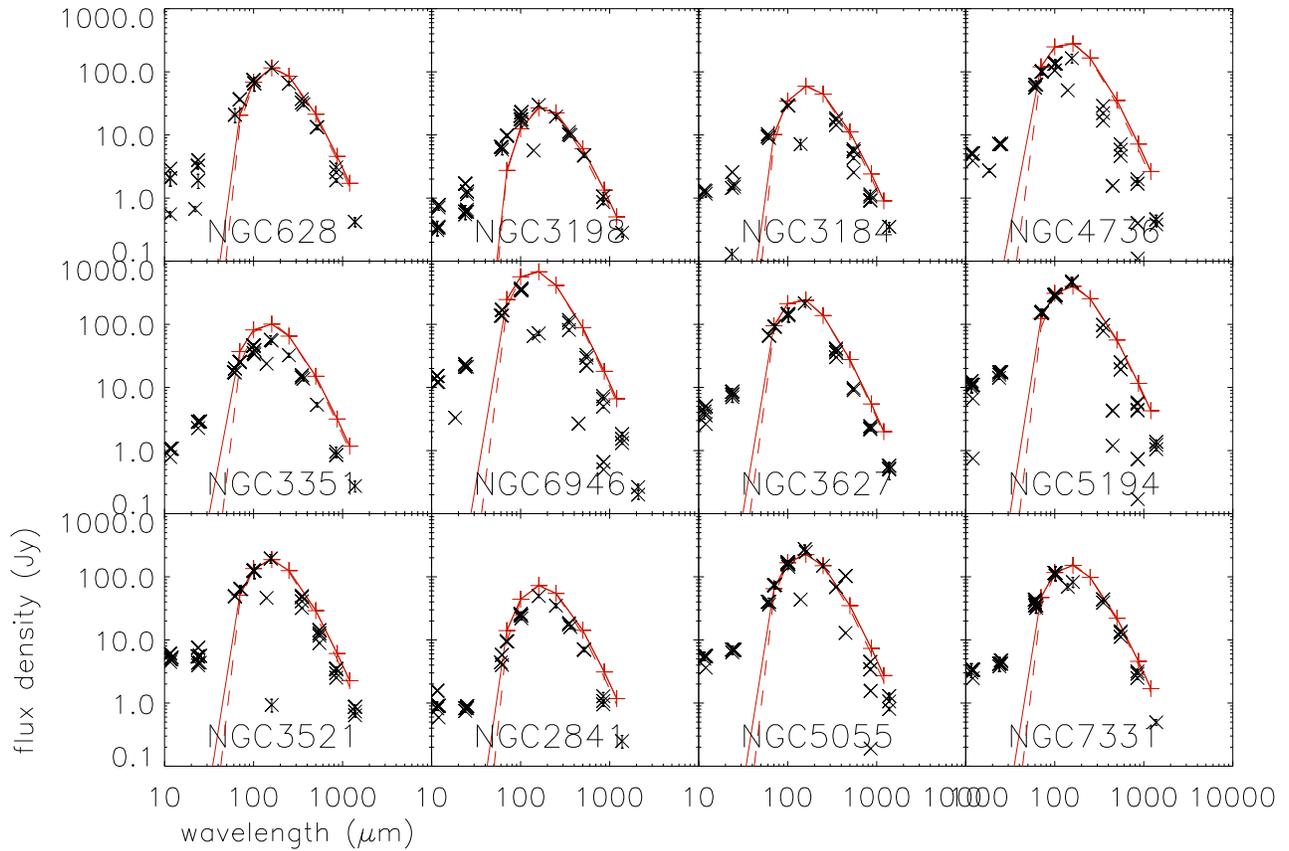}}
  \caption{Infrared SEDs of the sample of local spiral galaxies. Red plus symbols and solid line: model SED.
    Red dashed line is modified Planck fit for temperature determination. 
    Black crosses represent VizieR photometry. The errors bars are shown if present in the VizieR tables, but often barely visible.
  \label{fig:IRspectra_spirals}}
\end{figure*}

\begin{figure*}
  \centering
  \resizebox{\hsize}{!}{\includegraphics{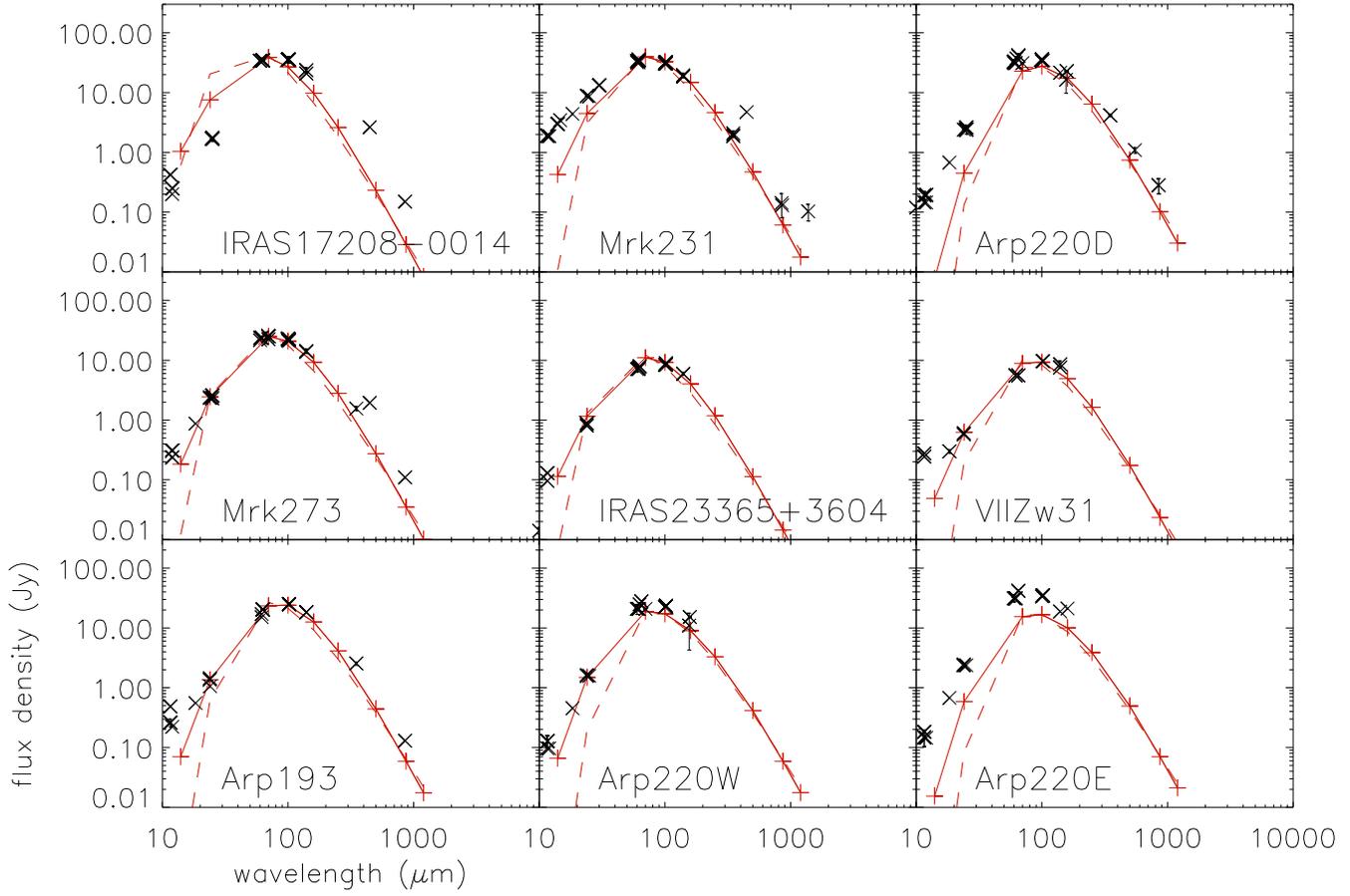}}
  \caption{Infrared SEDs of the sample of ultraluminous infrared galaxies. Red plus symbols and solid line represent model SED.
    Red dashed line is the modified Planck fit for temperature determination. 
    Black crosses represent VizieR photometry. The errors bars are shown if present in the VizieR tables, but often barely visible.
    The SED of Arp~220 was multiplied by $0.2$, $0.3$, and $0.2$ for Arp~220D, Arp~220W, and Arp~220E, respectively (see Sect.~\ref{sec:sedtir}).
  \label{fig:IRspectra_ulirgs}}
\end{figure*}

\begin{figure*}
  \centering
  \resizebox{\hsize}{!}{\includegraphics{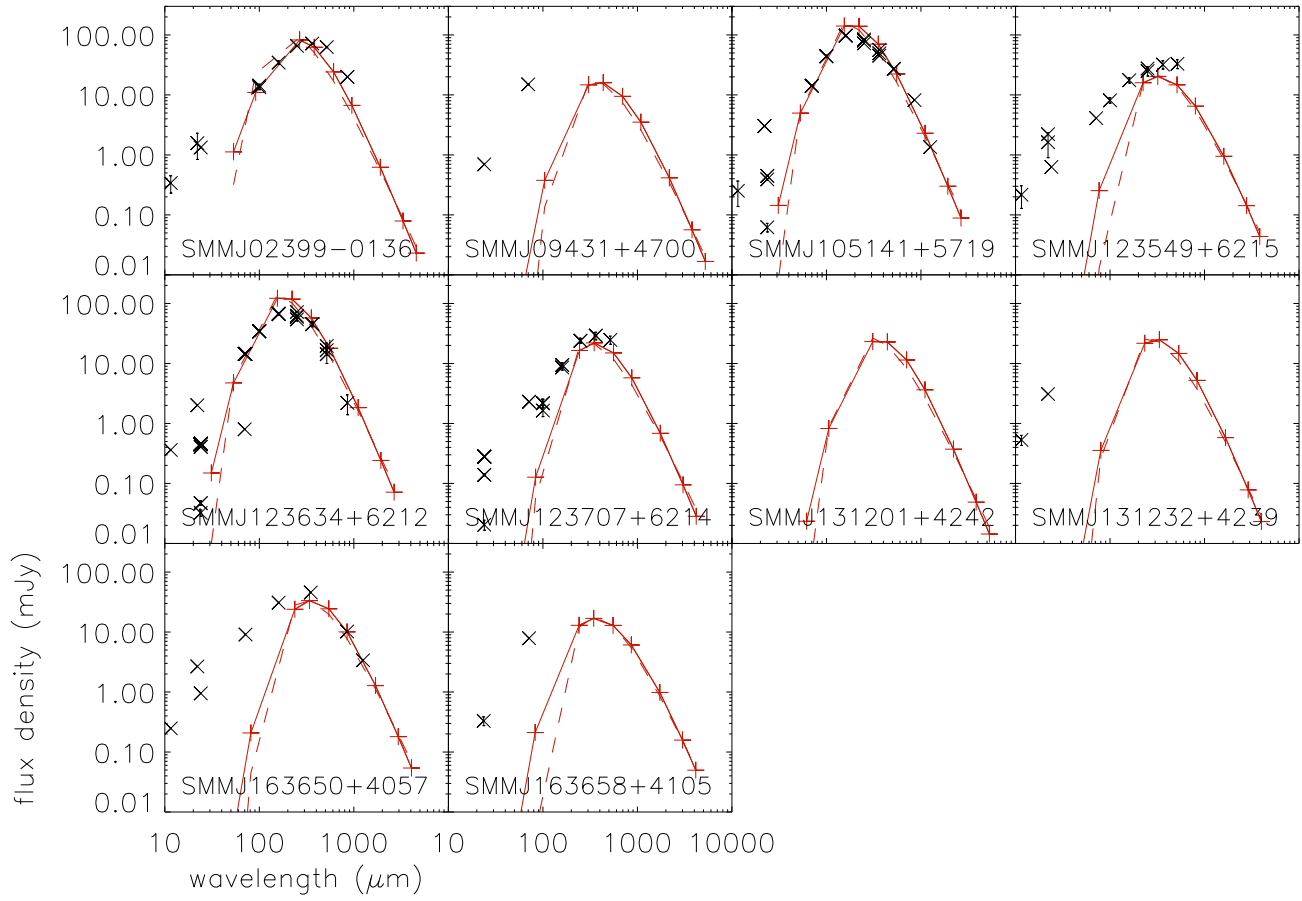}}
  \caption{Infrared SEDs of the sample of smm-galaxies. Red plus symbols and solid line are model SED.
    Red dashed line represents modified Planck fit for temperature determination. 
    Black crosses are VizieR photometry. The errors bars are shown if present in the VizieR tables, but often barely visible.
  \label{fig:IRspectra_smm}}
\end{figure*}

\begin{figure*}
  \centering
  \resizebox{\hsize}{!}{\includegraphics{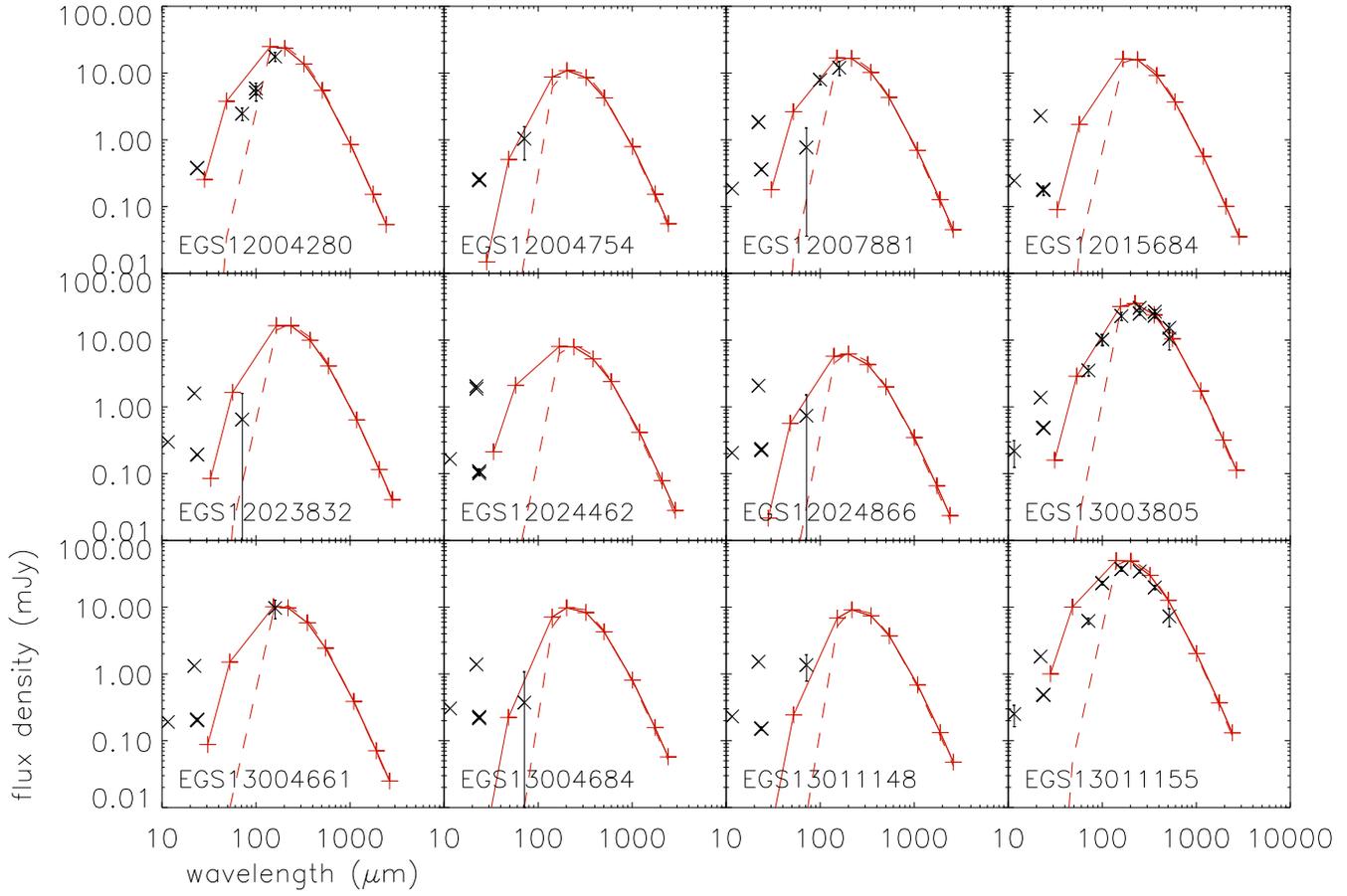}}
  \caption{Infrared SEDs of the sample of high-z star-forming galaxies. Red pluses and solid line represent model SED.
    Red dashed line is modified Planck fit for temperature determination. 
    Black crosses are VizieR photometry. The errors bars are shown if present in the VizieR tables, but often barely visible.
  \label{fig:IRspectra_phibss1}}
\end{figure*}
\setcounter{figure}{3}
\begin{figure*}
  \centering
  \resizebox{\hsize}{!}{\includegraphics{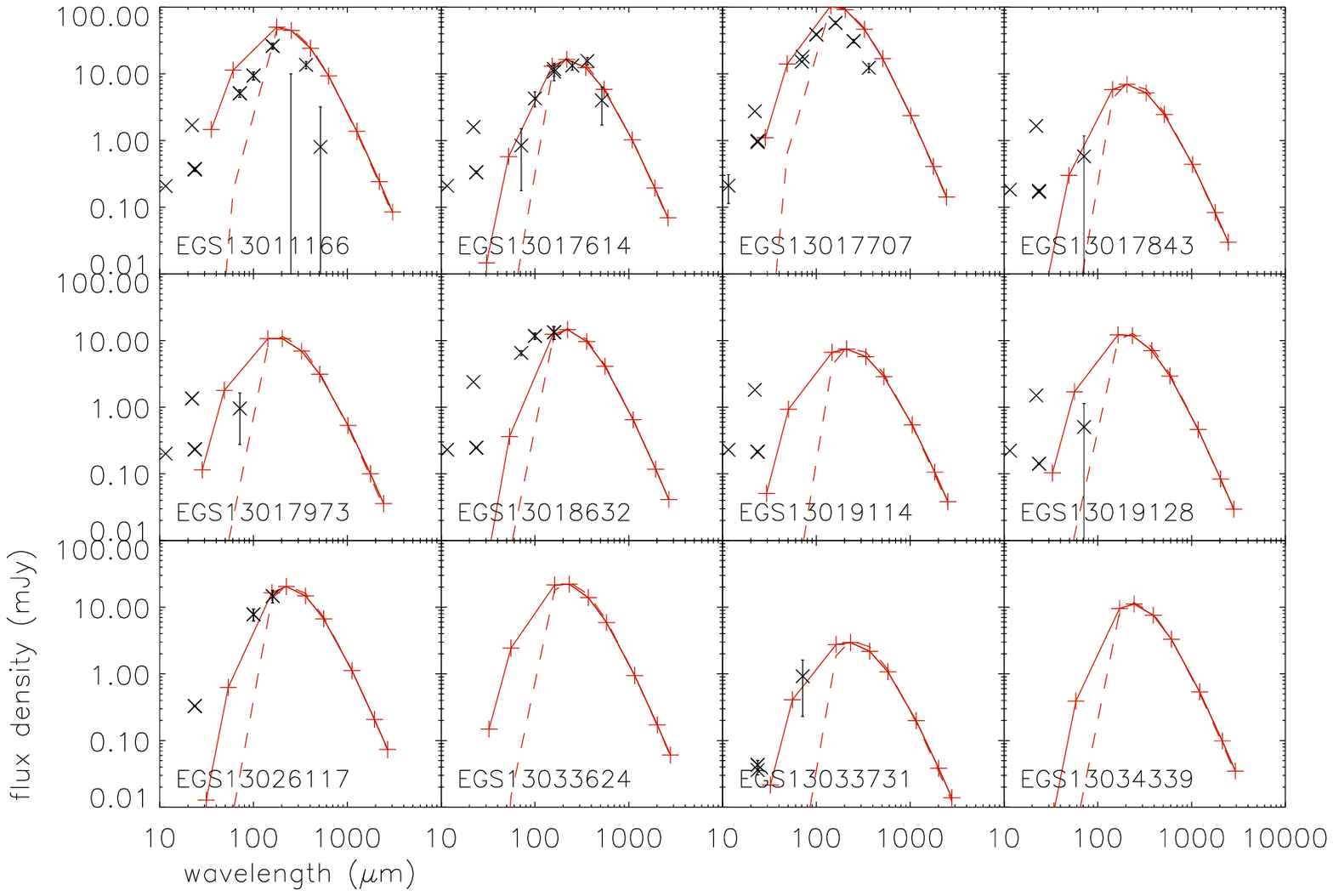}}
  \caption{Continued.
  \label{fig:IRspectra_phibss2}}
\end{figure*}
\setcounter{figure}{3}
\begin{figure*}
  \centering
  \resizebox{\hsize}{!}{\includegraphics{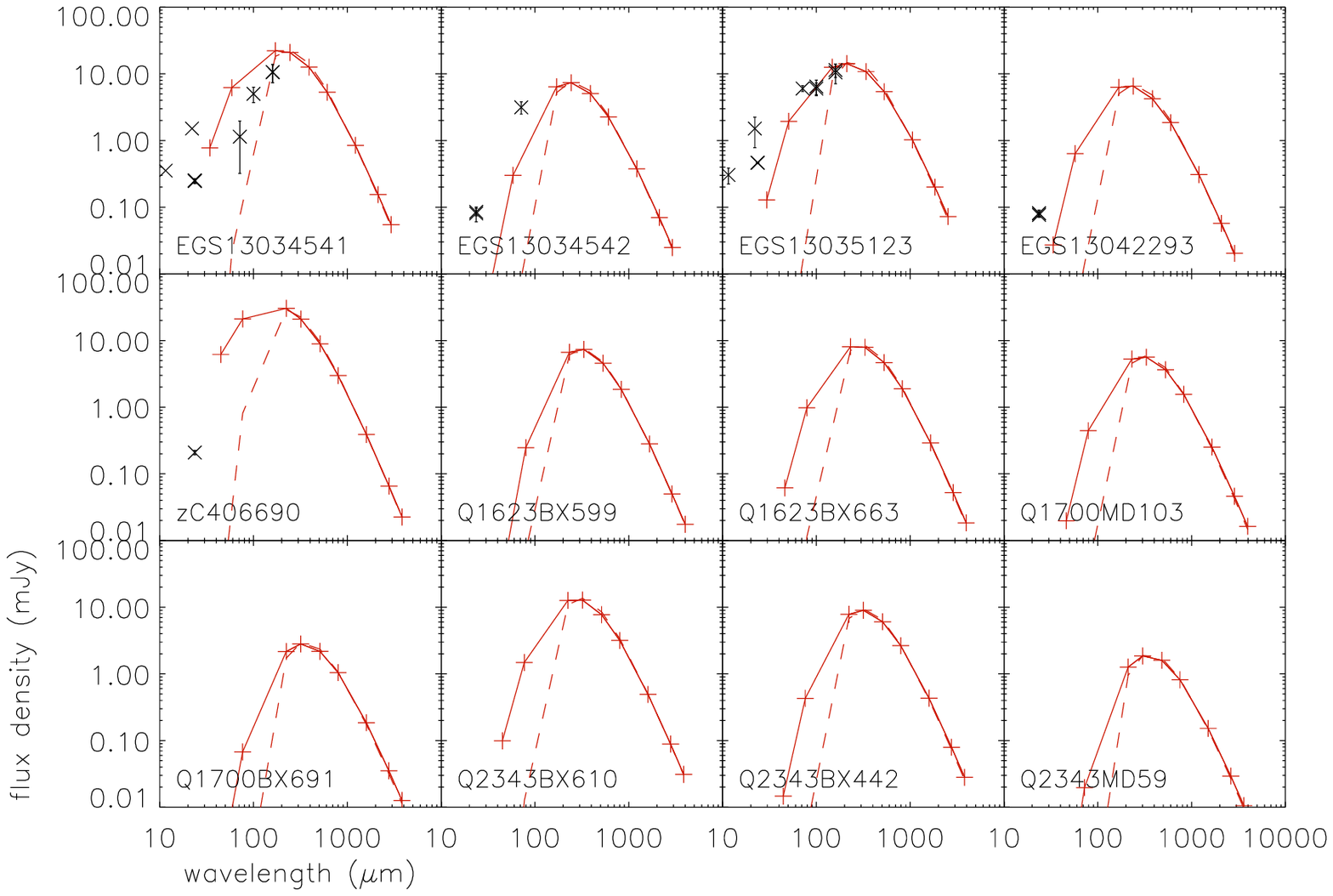}}
  \caption{Continued.
  \label{fig:IRspectra_phibss3}}
\end{figure*}
\setcounter{figure}{3}
\begin{figure*}
  \centering
  \resizebox{\hsize}{!}{\includegraphics{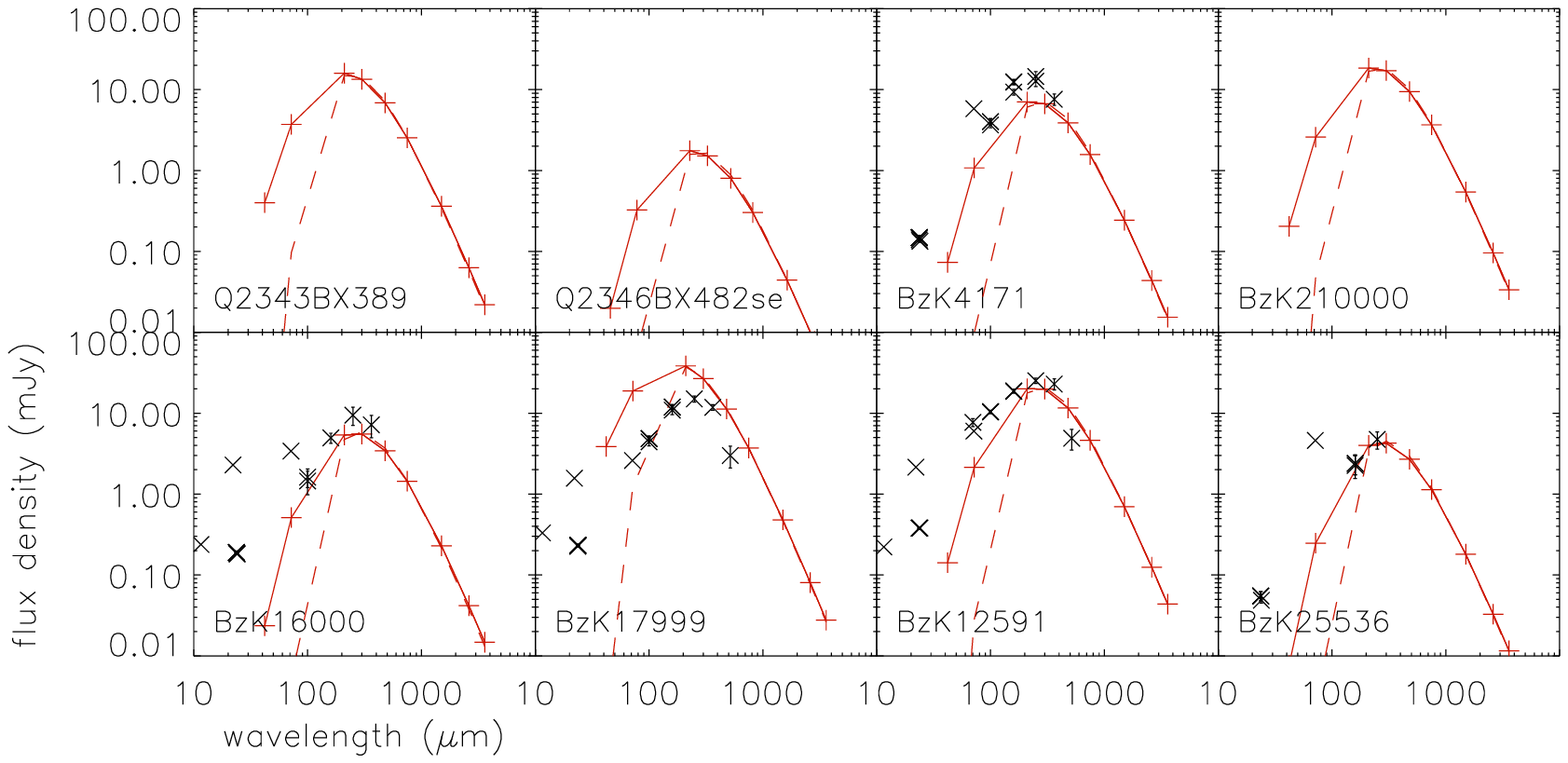}}
  \caption{Continued.
  \label{fig:IRspectra_phibss4}}
\end{figure*}

\end{appendix}


\begin{thebibliography}{}
 
\bibitem[Aalto et al.(2015)]{2015A&A...584A..42A} Aalto, S., Mart{\'{\i}}n, S., Costagliola, F., et al.\ 2015, A\&A, 584, A42 

\bibitem[Asplund et al.(2005)]{2005ASPC..336...25A} Asplund, M., Grevesse, N., \& Sauval, A.~J.\ 2005, Cosmic Abundances as Records of Stellar Evolution and Nucleosynthesis, 336, 25 

\bibitem[\protect\citeauthoryear{Ballesteros-Paredes \& Hartmann}{2007}]{Ballesteros} Ballesteros-Paredes, J. \& Hartmann, L. 2007, RMxAA, 43, 123

\bibitem[Bigiel et al.(2008)]{2008AJ....136.2846B} Bigiel, F., Leroy, A., Walter, F., et al.\ 2008, AJ, 136, 2846 

\bibitem[Bigiel et al.(2016)]{2016ApJ...822L..26B} Bigiel, F., Leroy, A.~K., Jim{\'e}nez-Donaire, M.~J., et al.\ 2016, ApJL, 822, L26 

\bibitem[Boissier et al.(2003)]{2003MNRAS.346.1215B} Boissier, S., Prantzos, N., Boselli, A., \& Gavazzi, G.\ 2003, MNRAS, 346, 1215 

\bibitem[Bolatto et al.(2013)]{2013ARA&A..51..207B} Bolatto, A.~D., Wolfire, M., \& Leroy, A.~K.\ 2013, ARA\&A, 51, 207 

\bibitem[Boulares \& Cox(1990)]{1990ApJ...365..544B} Boulares, A., \& Cox, D.~P.\ 1990, ApJ, 365, 544 

\bibitem[Bournaud et al.(2015)]{2015A&A...575A..56B} Bournaud, F., Daddi, E., Wei{\ss}, A., et al.\ 2015, A\&A, 575, A56 

\bibitem[Brinchmann et al.(2004)]{2004MNRAS.351.1151B} Brinchmann, J., Charlot, S., White, S.~D.~M., et al.\ 2004, MNRAS, 351, 1151 

\bibitem[Brouillet et al.(2005)]{2005A&A...429..153B} Brouillet, N., Muller, S., Herpin, F., Braine, J., \& Jacq, T.\ 2005, A\&A, 429, 153

\bibitem[Carilli \& Walter(2013)]{2013ARA&A..51..105C} Carilli, C.~L., \& Walter, F.\ 2013, ARA\&A, 51, 105 

\bibitem[Carroll \& Goldsmith(1981)]{1981ApJ...245..891C} Carroll, T.~J., \& Goldsmith, P.~F.\ 1981, ApJ, 245, 891 

\bibitem[Chen et al.(2015)]{2015ApJ...810..140C} Chen, H., Gao, Y., Braine, J., \& Gu, Q.\ 2015, ApJ, 810, 140 

\bibitem[Combes et al.(2013)]{2013A&A...550A..41C} Combes, F., Garc{\'{\i}}a-Burillo, S., Braine, J., et al.\ 2013, A\&A, 550, A41 

\bibitem[Conselice et al.(2009)]{2009MNRAS.394.1956C} Conselice, C.~J., Yang, C., \& Bluck, A.~F.~L.\ 2009, MNRAS, 394, 1956 

\bibitem[Cresci et al.(2009)]{2009ApJ...697..115C} Cresci, G., Hicks, E.~K.~S., Genzel, R., et al.\ 2009, ApJ, 697, 115 

\bibitem[da Cunha et al.(2013)]{2013ApJ...766...13D} da Cunha, E., Groves, B., Walter, F., et al.\ 2013, ApJ, 766, 13 

\bibitem[Daddi et al.(2007)]{2007ApJ...670..156D} Daddi, E., Dickinson, M., Morrison, G., et al.\ 2007, ApJ, 670, 156 

\bibitem[Daddi et al.(2010)]{2010ApJ...713..686D} Daddi, E., Bournaud, F., Walter, F., et al.\ 2010, ApJ, 713, 686 

\bibitem[Daddi et al.(2015)]{2015A&A...577A..46D} Daddi, E., Dannerbauer, H., Liu, D., et al.\ 2015, A\&A, 577, A46 

\bibitem[\protect\citeauthoryear{Dalcanton}{2007}]{Dalcanton} Dalcanton, J.~J. 2007, ApJ, 658, 941

\bibitem[Dale et al.(2012)]{2012ApJ...745...95D} Dale, D.~A., Aniano, G., Engelbracht, C.~W., et al.\ 2012, ApJ, 745, 95 

\bibitem[Dickey \& Lockman(1990)]{1990ARA&A..28..215D} Dickey, J.~M., \& Lockman, F.~J.\ 1990, ARA\&A, 28, 215 

\bibitem[Downes \& Solomon(1998)]{1998ApJ...507..615D} Downes, D., \& Solomon, P.~M.\ 1998, ApJ, 507, 615 

\bibitem[\protect\citeauthoryear{Draine \& Bertoldi}{1996}]{DraineBertoldi1996} Draine, B.T. \&  Bertoldi, F. 1996, ApJ, 468, 269

\bibitem[Draine \& Li(2007)]{2007ApJ...657..810D} Draine, B.~T., \& Li, A.\ 2007, ApJ, 657, 810 

\bibitem[Draine(2011)]{2011piim.book.....D} Draine, B.~T.\ 2011, Physics of the Interstellar and Intergalactic Medium by Bruce T.~Draine.~Princeton University Press, 2011.~ISBN: 978-0-691-12214-4 

\bibitem[Dullemond et al.(2002)]{2002A&A...389..464D} Dullemond, C.~P., van Zadelhoff, G.~J., \& Natta, A.\ 2002, A\&A, 389, 464 

\bibitem[Dumouchel et al.(2010)]{2010MNRAS.406.2488D} Dumouchel, F., Faure, A., \& Lique, F.\ 2010, MNRAS, 406, 2488 

\bibitem[Elbaz et al.(2007)]{2007A&A...468...33E} Elbaz, D., Daddi, E., Le Borgne, D., et al.\ 2007, A\&A, 468, 33 

\bibitem[\protect\citeauthoryear{Elmegreen}{1989}]{Elmegreen89} Elmegreen, B.G. 1989, 338, 178

\bibitem[\protect\citeauthoryear{Elmegreen \& Falgarone}{1996}]{Elmegreen} Elmegreen, B.G., Falgarone E. 1996, ApJ, 471, 816 

\bibitem[Erb et al.(2006)]{2006ApJ...647..128E} Erb, D.~K., Steidel, C.~C., Shapley, A.~E., et al.\ 2006, ApJ, 647, 128

\bibitem[\protect\citeauthoryear{Evans}{2008}]{Evans2008} Evans, N.~J., II 2008, Pathways Through an Eclectic Universe, 390, 52

\bibitem[Feruglio et al.(2015)]{2015A&A...583A..99F} Feruglio, C., Fiore, F., Carniani, S., et al.\ 2015, A\&A, 583, A99 

\bibitem[F{\"o}rster Schreiber et al.(2009)]{2009ApJ...706.1364F} F{\"o}rster Schreiber, N.~M., Genzel, R., Bouch{\'e}, N., et al.\ 2009, ApJ, 706, 1364 

\bibitem[Forbes et al.(2014)]{2014MNRAS.438.1552F} Forbes, J.~C., Krumholz, M.~R., Burkert, A., \& Dekel, A.\ 2014, MNRAS, 438, 1552 

\bibitem[\protect\citeauthoryear{Frisch}{1995}]{Frisch} Frisch U. 1995, Turbulence -- The Legacy of A.N. Kolmogorov, Cambridge University press

\bibitem[Gao \& Solomon(2004)]{2004ApJ...606..271G} Gao, Y., \& Solomon, P.~M.\ 2004, ApJ, 606, 271 

\bibitem[Genzel et al.(2006)]{2006Natur.442..786G} Genzel, R., Tacconi, L.~J., Eisenhauer, F., et al.\ 2006, Nature, 442, 786 

\bibitem[Genzel et al.(2010)]{2010MNRAS.407.2091G} Genzel, R., Tacconi, L.~J., Graci\'a-Carpio, J., et al.\ 2010, MNRAS, 407, 2091 

\bibitem[Genzel et al.(2015)]{2015ApJ...800...20G} Genzel, R., Tacconi, L.~J., Lutz, D., et al.\ 2015, ApJ, 800, 20 

\bibitem[Glover \& Clark(2012)]{2012MNRAS.426..377G} Glover, S.~C.~O., \& Clark, P.~C.\ 2012, MNRAS, 426, 377 

\bibitem[Glover \& Clark(2012)]{2012MNRAS.421....9G} Glover, S.~C.~O., \& Clark, P.~C.\ 2012, MNRAS, 421, 9 

\bibitem[Goldreich \& Kwan(1974)]{1974ApJ...189..441G} Goldreich, P., \& Kwan, J.\ 1974, ApJ, 189, 441 

\bibitem[Goldsmith(2001)]{2001ApJ...557..736G} Goldsmith, P.~F.\ 2001, ApJ, 557, 736 

\bibitem[Graci{\'a}-Carpio et al.(2006)]{2006ApJ...640L.135G} Graci{\'a}-Carpio, J., Garc{\'{\i}}a-Burillo, S., Planesas, P., \& Colina, L.\ 2006, ApJL, 640, L135 

\bibitem[Graci{\'a}-Carpio et al.(2008)]{2008A&A...479..703G} Graci{\'a}-Carpio, J., Garc{\'{\i}}a-Burillo, S., Planesas, P., Fuente, A., \& Usero, A.\ 2008, A\&A, 479, 703

\bibitem[Green \& Thaddeus(1976)]{1976ApJ...205..766G} Green, S., \& Thaddeus, P.\ 1976, ApJ, 205, 766 

\bibitem[Grogin et al.(2011)]{2011ApJS..197...35G} Grogin, N.~A., Kocevski, D.~D., Faber, S.~M., et al.\ 2011, ApJS, 197, 35 

\bibitem[Habing(1968)]{1968BAN....19..421H} Habing, H.~J.\ 1968, BAN, 19, 421 

\bibitem[Hailey-Dunsheath et al.(2012)]{2012ApJ...755...57H} Hailey-Dunsheath, S., Sturm, E., Fischer, J., et al.\ 2012, ApJ, 755, 57 

\bibitem[Hasegawa \& Herbst(1993)]{1993MNRAS.261...83H} Hasegawa, T.~I., \& Herbst, E.\ 1993, MNRAS, 261, 83 

\bibitem[Hersant et al.(2009)]{2009A&A...493L..49H} Hersant, F., Wakelam, V., Dutrey, A., Guilloteau, S., \& Herbst, E.\ 2009, A\&A, 493, L49 

\bibitem[Heyer \& Dame(2015)]{2015ARA&A..53..583H} Heyer, M., \& Dame, T.~M.\ 2015, ARA\&A, 53, 583 

\bibitem[\protect\citeauthoryear{Hollenbach \& Tielens}{1997}]{Hollenbach} Hollenbach D.J., Tielens A.G.G.M. 1997, ARA\&A, 35, 179

\bibitem[Hopkins \& Beacom(2006)]{2006ApJ...651..142H} Hopkins, A.~M., \& Beacom, J.~F.\ 2006, ApJ, 651, 142 

\bibitem[Imanishi et al.(2016)]{2016ApJ...825...44I} Imanishi, M., Nakanishi, K., \& Izumi, T.\ 2016, ApJ, 825, 44

\bibitem[\protect\citeauthoryear{Joung \& Mac Low}{2006}]{JoungMacLow} Joung, M.K.R. \& Mac Low, M.-M. 2006, ApJ, 653, 1266

\bibitem[Juneau et al.(2009)]{2009ApJ...707.1217J} Juneau, S., Narayanan, D.~T., Moustakas, J., et al.\ 2009, ApJ, 707, 1217-1232 

\bibitem[Jura(1975)]{1975ApJ...197..575J} Jura, M.\ 1975, ApJ, 197, 575 

\bibitem[Kamenetzky et al.(2016)]{2016ApJ...829...93K} Kamenetzky, J., Rangwala, N., Glenn, J., Maloney, P.~R., \& Conley, A.\ 2016, ApJ, 829, 93 

\bibitem[Kazandjian et al.(2015)]{2015A&A...574A.127K} Kazandjian, M.~V., Meijerink, R., Pelupessy, I., Israel, F.~P., \& Spaans, M.\ 2015, A\&A, 574, A127 

\bibitem[Karim et al.(2011)]{2011ApJ...730...61K} Karim, A., Schinnerer, E., Mart{\'{\i}}nez-Sansigre, A., et al.\ 2011, ApJ, 730, 61 

\bibitem[Kartaltepe et al.(2012)]{2012ApJ...757...23K} Kartaltepe, J.~S., Dickinson, M., Alexander, D.~M., et al.\ 2012, ApJ, 757, 23 

\bibitem[Kennicutt(1998)]{1998ARA&A..36..189K} Kennicutt, R.~C., Jr.\ 1998, ARA\&A, 36, 189 

\bibitem[Kennicutt \& Evans(2012)]{2012ARA&A..50..531K} Kennicutt, R.~C., \& Evans, N.~J.\ 2012, ARA\&A, 50, 531 

\bibitem[Kilerci Eser et al.(2014)]{2014ApJ...797...54K} Kilerci Eser, E., Goto, T., \& Doi, Y.\ 2014, ApJ, 797, 54 

\bibitem[Klaas et al.(2001)]{2001A&A...379..823K} Klaas, U., Haas, M., M{\"u}ller, S.~A.~H., et al.\ 2001, A\&A, 379, 823 

\bibitem[Knudsen et al.(2007)]{2007ApJ...666..156K} Knudsen, K.~K., Walter, F., Weiss, A., et al.\ 2007, ApJ, 666, 156 

\bibitem[\protect\citeauthoryear{K\"{o}ppen \& Edmunds}{1999}]{Koeppen} K\"{o}ppen, J. \& Edmunds, M.~G. 1999, MNRAS, 306, 317

\bibitem[\protect\citeauthoryear{Kregel et al.}{2002}]{Kregel} Kregel, M., van der Kruit, P.~C., de Grijs, R. 2002, MNRAS, 334, 646

\bibitem[Krumholz \& Thompson(2007)]{2007ApJ...669..289K} Krumholz, M.~R., \& Thompson, T.~A.\ 2007, ApJ, 669, 289 

\bibitem[Krumholz et al.(2008)]{2008ApJ...689..865K} Krumholz, M.~R., McKee, C.~F., \& Tumlinson, J.\ 2008, ApJ, 689, 865 

\bibitem[\protect\citeauthoryear{Krumholz et al.}{2009}]{Krumholz09} Krumholz, M.~R., McKee, C.~F., \& Tumlinson, J.\ 2009, ApJ, 693, 216

\bibitem[Krumholz(2012)]{2012ApJ...759....9K} Krumholz, M.~R.\ 2012, ApJ, 759, 9 

\bibitem[Krumholz et al.(2011)]{2011ApJ...731...25K} Krumholz, M.~R., Leroy, A.~K., \& McKee, C.~F.\ 2011, ApJ, 731, 25 

\bibitem[Kulkarni \& Heiles(1987)]{1987ASSL..134...87K} Kulkarni, S.~R., \& Heiles, C.\ 1987, Interstellar Processes, 134, 87 

\bibitem[Law et al.(2009)]{2009ApJ...697.2057L} Law, D.~R., Steidel, C.~C., Erb, D.~K., et al.\ 2009, ApJ, 697, 2057 

\bibitem[\protect\citeauthoryear{Leroy et al.}{2008}] {Leroy} Leroy, A.K., Walter, F., Brinks, E., et al. 2008, AJ, 136, 2782

\bibitem[Leroy et al.(2009)]{2009AJ....137.4670L} Leroy, A.~K., Walter, F., Bigiel, F., et al.\ 2009, AJ, 137, 4670 

\bibitem[\protect\citeauthoryear{Lin \& Pringle}{1987}]{LinPringle} Lin, D.N.C. \& Pringle J.E. 1987, ApJ, 320, L87

\bibitem[Lombardi et al.(2010)]{2010A&A...519L...7L} Lombardi, M., Alves, J., \& Lada, C.~J.\ 2010, A\&A, 519, L7 

\bibitem[Mac Low(1999)]{1999ApJ...524..169M} Mac Low, M.-M.\ 1999, ApJ, 524, 169

\bibitem[\protect\citeauthoryear{Mac Low \& Klessen}{2004}]{MacLow}  Mac Low M.-M. \& Klessen R.S. 2004, RvMP, 76, 125

\bibitem[Madau et al.(1998)]{1998ApJ...498..106M} Madau, P., Pozzetti, L., \& Dickinson, M.\ 1998, ApJ, 498, 106 

\bibitem[Magdis et al.(2012)]{2012ApJ...760....6M} Magdis, G.~E., Daddi, E., B{\'e}thermin, M., et al.\ 2012, ApJ, 760, 6 

\bibitem[Magnelli et al.(2014)]{2014A&A...561A..86M} Magnelli, B., Lutz, D., Saintonge, A., et al.\ 2014, A\&A, 561, A86 

\bibitem[Maloney(1993)]{1993ApJ...414...41M} Maloney, P.\ 1993, ApJ, 414, 41 

\bibitem[Mathis et al.(1983)]{1983A&A...128..212M} Mathis, J.~S., Mezger, P.~G., \& Panagia, N.\ 1983, A\&A, 128, 212 

\bibitem[McGaugh(1991)]{1991ApJ...380..140M} McGaugh, S.~S.\ 1991, ApJ, 380, 140 

\bibitem[McLeod et al.(1993)]{1993ApJ...412..111M} McLeod, K.~K., Rieke, G.~H., Rieke, M.~J., \& Kelly, D.~M.\ 1993, ApJ, 412, 111 

\bibitem[Mac Low et al.(1998)]{1998PhRvL..80.2754M} Mac Low, M.-M., Klessen, R.~S., Burkert, A., \& Smith, M.~D.\ 1998, Physical Review Letters, 80, 2754 

\bibitem[Moustakas \& Kennicutt(2006)]{2006ApJ...651..155M} Moustakas, J., \& Kennicutt, R.~C., Jr.\ 2006, ApJ, 651, 155 

\bibitem[Nagao et al.(2012)]{2012A&A...542L..34N} Nagao, T., Maiolino, R., De Breuck, C., et al.\ 2012, A\&A, 542, L34 

\bibitem[Narayanan \& Krumholz(2014)]{2014MNRAS.442.1411N} Narayanan, D., \& Krumholz, M.~R.\ 2014, MNRAS, 442, 1411 

\bibitem[Nelson \& Langer(1997)]{1997ApJ...482..796N} Nelson, R.~P., \& Langer, W.~D.\ 1997, ApJ, 482, 796 

\bibitem[Neufeld \& Kaufman(1993)]{1993ApJ...418..263N} Neufeld, D.~A., \& Kaufman, M.~J.\ 1993, ApJ, 418, 263 

\bibitem[Neufeld et al.(1995)]{1995ApJS..100..132N} Neufeld, D.~A., Lepp, S., \& Melnick, G.~J.\ 1995, ApJS, 100, 132

\bibitem[Nguyen et al.(1992)]{1992ApJ...399..521N} Nguyen, Q.-R., Jackson, J.~M., Henkel, C., Truong, B., \& Mauersberger, R.\ 1992, ApJ, 399, 521 

\bibitem[Noeske et al.(2007)]{2007ApJ...660L..47N} Noeske, K.~G., Faber, S.~M., Weiner, B.~J., et al.\ 2007a, ApJL, 660, L47 

\bibitem[Noeske et al.(2007)]{2007ApJ...660L..43N} Noeske, K.~G., Weiner, B.~J., Faber, S.~M., et al.\ 2007b, ApJL, 660, L43 

\bibitem[Oey et al.(2007)]{2007ApJ...661..801O} Oey, M.~S., Meurer, G.~R., Yelda, S., et al.\ 2007, ApJ, 661, 801 

\bibitem[Ostriker et al.(1999)]{1999ApJ...513..259O} Ostriker, E.~C., Gammie, C.~F., \& Stone, J.~M.\ 1999, ApJ, 513, 259 

\bibitem[Padoan et al.(1997)]{1997MNRAS.288..145P} Padoan, P., Nordlund, A., \& Jones, B.~J.~T.\ 1997, MNRAS, 288, 145 

\bibitem[Padovani \& Galli(2013)]{2013ASSP...34...61P} Padovani, M., \& Galli, D.\ 2013, Cosmic Rays in Star-Forming Environments, 34, 61

\bibitem[Pannella et al.(2015)]{2015ApJ...807..141P} Pannella, M., Elbaz, D., Daddi, E., et al.\ 2015, ApJ, 807, 141 

\bibitem[Papadopoulos(2010)]{2010ApJ...720..226P} Papadopoulos, P.~P.\ 2010, ApJ, 720, 226 

\bibitem[Papadopoulos et al.(2012)]{2012MNRAS.426.2601P} Papadopoulos, P.~P., van der Werf, P.~P., Xilouris, E.~M., et al.\ 2012, MNRAS, 426, 2601 

\bibitem[Pilbratt et al.(2010)]{2010A&A...518L...1P} Pilbratt, G.~L., Riedinger, J.~R., Passvogel, T., et al.\ 2010, A\&A, 518, L1 

\bibitem[Planck Collaboration et al.(2014)]{2014A&A...571A..11P} Planck Collaboration, Abergel, A., Ade, P.~A.~R., et al.\ 2014, A\&A, 571, A11 

\bibitem[Polk et al.(1988)]{1988ApJ...332..432P} Polk, K.~S., Knapp, G.~R., Stark, A.~A., \& Wilson, R.~W.\ 1988, ApJ, 332, 432 

\bibitem[Pope et al.(2013)]{2013ApJ...772...92P} Pope, A., Wagg, J., Frayer, D., et al.\ 2013, ApJ, 772, 92 

\bibitem[\protect\citeauthoryear{Pringle}{1981}]{Pringle} Pringle J.E. 1981, ARA\&A, 19, 137

\bibitem[Queyrel et al.(2012)]{2012A&A...539A..93Q} Queyrel, J., Contini, T., Kissler-Patig, M., et al.\ 2012, A\&A, 539, A93

\bibitem[Quiroga(1983)]{1983Ap&SS..93...37Q} Quiroga, R.~J.\ 1983, Ap\&SS, 93, 37 

\bibitem[R{\'e}my-Ruyer et al.(2014)]{2014A&A...563A..31R} R{\'e}my-Ruyer, A., Madden, S.~C., Galliano, F., et al.\ 2014, A\&A, 563, A31


\bibitem[Rodighiero et al.(2011)]{2011ApJ...739L..40R} Rodighiero, G., Daddi, E., Baronchelli, I., et al.\ 2011, ApJL, 739, L40 

\bibitem[Ruaud et al.(2015)]{2015MNRAS.447.4004R} Ruaud, M., Loison, J.~C., Hickson, K.~M., et al.\ 2015, MNRAS, 447, 4004 

\bibitem[Rupke et al.(2008)]{2008ApJ...674..172R} Rupke, D.~S.~N., Veilleux, S., \& Baker, A.~J.\ 2008, ApJ, 674, 172

\bibitem[Sakamoto et al.(2010)]{2010ApJ...725L.228S} Sakamoto, K., Aalto, S., Evans, A.~S., Wiedner, M.~C., \& Wilner, D.~J.\ 2010, ApJL, 725, L228 

\bibitem[Sakamoto et al.(2014)]{2014ApJ...797...90S} Sakamoto, K., Aalto, S., Combes, F., Evans, A., \& Peck, A.\ 2014, ApJ, 797, 90 

\bibitem[Salim et al.(2007)]{2007ApJS..173..267S} Salim, S., Rich, R.~M., Charlot, S., et al.\ 2007, ApJS, 173, 267 

\bibitem[Sargent et al.(2012)]{2012ApJ...747L..31S} Sargent, M.~T., B{\'e}thermin, M., Daddi, E., \& Elbaz, D.\ 2012, ApJL, 747, L31 

\bibitem[Schneider et al.(2015)]{2015A&A...578A..29S} Schneider, N., Csengeri, T., Klessen, R.~S., et al.\ 2015, A\&A, 578, A29 

\bibitem[Sch{\"o}ier et \ al.(2005)]{2005A&A...432..369S} Sch{\"o}ier, F.~L., van der Tak, F.~F.~S., van Dishoeck, E.~F., \& Black, J.~H.\ 2005, A\&A, 432, 369

\bibitem[Scoville \& Solomon(1974)]{1974ApJ...187L..67S} Scoville, N.~Z., \& Solomon, P.~M.\ 1974, ApJ, 187, L67 

\bibitem[Scoville(2013)]{2013seg..book..491S} Scoville, N.~Z.\ 2013, Secular Evolution of Galaxies, ed: Falc{\'o}n-Barroso, J. and Knapen, J.~H., p.491 

\bibitem[Scoville et al.(2015)]{2015ApJ...800...70S} Scoville, N., Sheth, K., Walter, F., et al.\ 2015, ApJ, 800, 70 

\bibitem[Semenov et al.(2010)]{2010A&A...522A..42S} Semenov, D., Hersant, F., Wakelam, V., et al.\ 2010, A\&A, 522, A42 

\bibitem[Shapley et al.(2004)]{2004ApJ...612..108S} Shapley, A.~E., Erb, D.~K., Pettini, M., Steidel, C.~C., \& Adelberger, K.~L.\ 2004, ApJ, 612, 108 

\bibitem[Shirley(2015)]{2015PASP..127..299S} Shirley, Y.~L.\ 2015, PASP, 127, 299 

\bibitem[Sofia \& Meyer(2001)]{2001ApJ...558L.147S} Sofia, U.~J., \& Meyer, D.~M.\ 2001a, ApJL, 558, L147 

\bibitem[Sofia \& Meyer(2001)]{2001ApJ...554L.221S} Sofia, U.~J., \& Meyer, D.~M.\ 2001b, ApJL, 554, L221 

\bibitem[\protect\citeauthoryear{Solomon et al.}{1987}]{Solomon} Solomon, P.M., Rivolo, A.R., Barrett, J., Yahil, A. 1987, ApJ, 319, 730

\bibitem[Speagle et al.(2014)]{2014ApJS..214...15S} Speagle, J.~S., Steinhardt, C.~L., Capak, P.~L., \& Silverman, J.~D.\ 2014, ApJS, 214, 15 

\bibitem[Stone et al.(1998)]{1998ApJ...508L..99S} Stone, J.~M., Ostriker, E.~C., \& Gammie, C.~F.\ 1998, ApJL, 508, L99 

\bibitem[Suchkov et al.(1993)]{1993ApJ...413..542S} Suchkov, A., Allen, R.~J., \& Heckman, T.~M.\ 1993, ApJ, 413, 542 

\bibitem[Symeonidis et al.(2013)]{2013MNRAS.431.2317S} Symeonidis, M., Vaccari, M., Berta, S., et al.\ 2013, MNRAS, 431, 2317 

\bibitem[Tacconi et al.(2013)]{2013ApJ...768...74T} Tacconi, L.~J., Neri, R., Genzel, R., et al.\ 2013, ApJ, 768, 74 

\bibitem[Tecza et al.(2004)]{2004ApJ...605L.109T} Tecza, M., Baker, A.~J., Davies, R.~I., et al.\ 2004, ApJ, 605, L109 

\bibitem[Thilker et al.(2002)]{2002AJ....124.3118T} Thilker, D.~A., Walterbos, R.~A.~M., Braun, R., \& Hoopes, C.~G.\ 2002, AJ, 124, 3118 

\bibitem[\protect\citeauthoryear{Thornton et al.}{1998}]{Thornton} Thornton, K., Gaudlitz, M., Janka, H.-Th., \& Steinmetz, M. 1998, ApJ, 500, 95

\bibitem[\protect\citeauthoryear{Tielens \& Hollenbach}{1985}]{TielensH} Tielens A.G.G.M. \& Hollenbach D. 1985, ApJ, 291, 722

\bibitem[Usero et al.(2015)]{2015AJ....150..115U} Usero, A., Leroy, A.~K., Walter, F., et al.\ 2015, AJ, 150, 115 

\bibitem[van der Kruit \& Searle(1982)]{1982A&A...110...61V} van der Kruit, P.~C., \& Searle, L.\ 1982, A\&A, 110, 61 

\bibitem[van der Tak et al.(2007)]{2007A&A...468..627V} van der Tak, F.~F.~S., Black, J.~H., Sch{\"o}ier, F.~L., Jansen, D.~J., \& van Dishoeck, E.~F.\ 2007, A\&A, 468, 627

\bibitem[Veilleux et al.(2005)]{2005ARA&A..43..769V} Veilleux, S., Cecil, G., \& Bland-Hawthorn, J.\ 2005, ARA\&A, 43, 769 

\bibitem[Vergani et al.(2012)]{2012A&A...546A.118V} Vergani, D., Epinat, B., Contini, T., et al.\ 2012, A\&A, 546, A118 

\bibitem[\protect\citeauthoryear{Vollmer \& Beckert}{2002}]{VollmerBeckert} Vollmer B., Beckert T. 2002, A\&A, 382, 872

\bibitem[\protect\citeauthoryear{Vollmer \& Beckert}{2003}]{Vollmer} Vollmer, B. \& Beckert, T. 2003, A\&A, 404, 21, (VB03)

\bibitem[Vollmer \& Leroy(2011)]{2011AJ....141...24V} Vollmer, B., \& Leroy, A.~K.\ 2011, AJ, 141, 24 

\bibitem[Wakelam et al.(2015)]{2015ApJS..217...20W} Wakelam, V., Loison, J.-C., Herbst, E., et al.\ 2015, ApJS, 217, 20 

\bibitem[Walter et al.(2008)]{2008AJ....136.2563W} Walter, F., Brinks, E., de Blok, W.~J.~G., et al.\ 2008, AJ, 136, 2563

\bibitem[Wei{\ss} et al.(2007)]{2007A&A...467..955W} Wei{\ss}, A., Downes, D., Neri, R., et al.\ 2007, A\&A, 467, 955

\bibitem[Whitaker et al.(2012)]{2012ApJ...754L..29W} Whitaker, K.~E., van Dokkum, P.~G., Brammer, G., \& Franx, M.\ 2012, ApJL, 754, L29

\bibitem[Wolfire et al.(1993)]{1993ApJ...402..195W} Wolfire, M.~G., Hollenbach, D., \& Tielens, A.~G.~G.~M.\ 1993, ApJ, 402, 195 

\bibitem[Wolfire et al.(2003)]{2003ApJ...587..278W} Wolfire, M.~G., McKee, C.~F., Hollenbach, D., \& Tielens, A.~G.~G.~M.\ 2003, ApJ, 587, 278 
 
\bibitem[Wolfire et al.(2010)]{2010ApJ...716.1191W} Wolfire, M.~G., Hollenbach, D., \& McKee, C.~F.\ 2010, ApJ, 716, 1191 

\bibitem[Wong \& Blitz(2002)]{2002ApJ...569..157W} Wong, T., \& Blitz, L.\ 2002, ApJ, 569, 157 

\bibitem[Wuyts et al.(2011)]{2011ApJ...742...96W} Wuyts, S., F{\"o}rster Schreiber, N.~M., van der Wel, A., et al.\ 2011, ApJ, 742, 96 

\bibitem[Yang et al.(2010)]{2010ApJ...718.1062Y} Yang, B., Stancil, P.~C., Balakrishnan, N., \& Forrey, R.~C.\ 2010, ApJ, 718, 1062


\end{thebibliography}
\end{document}